\documentclass[9pt]{book}
\usepackage{titlesec}
\titleformat{\chapter}[display]
{\normalfont\bfseries\filcenter}
{\LARGE\thechapter}
{1ex}
{\titlerule[2pt]
\vspace{2ex}%
\LARGE}
[\vspace{1ex}%
{\titlerule[2pt]}]
\usepackage{sidenotes}
\usepackage{amsmath}
\usepackage{mwe}
\usepackage{nicefrac}
\usepackage{dsfont}
\usepackage{fontenc}

\usepackage{esint}
\usepackage[paperwidth=155mm,paperheight=235mm]{geometry}
\usepackage{float}
\usepackage{epstopdf}
\usepackage{fancyhdr}
\usepackage{amsfonts}
\usepackage{setspace}
\usepackage{booktabs}
\usepackage{pdfpages}
\usepackage{ragged2e}

\usepackage{xcolor}
\definecolor{gray}{gray}{0.5}
\usepackage[format=plain,
            labelfont={it},
            textfont=it]{caption}
\newcommand{\chapterend}{
 \par
    \vspace{\stretch{.5}}
    \begin{center}
    \includegraphics[scale=.1]{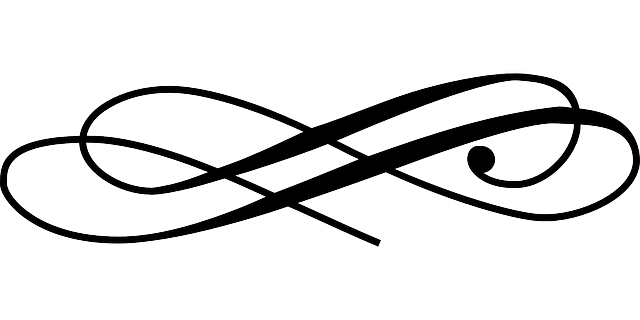}
    \end{center}
    \vspace{\stretch{0.1}}
}

\newcommand*\cleartoleftpage{%
\clearpage
\ifodd\value{page}\hbox{}\newpage\fi}

\renewcommand{\sffamily}{\rmdefault}

\titleformat{\chapter}[display]
    {\normalfont\huge\bfseries}
    {\chaptertitlename\ \thechapter}
    {10pt}
    {\Huge}
    [\vspace{0.5ex}\titlerule]

\titlespacing*{\chapter}{0pt}{-50pt}{20pt}

\makeatletter
\renewcommand*{\cleardoublepage}{\clearpage\if@twoside \ifodd\c@page\else
\hbox{}%
\thispagestyle{empty}%
\newpage%
\if@twocolumn\hbox{}\newpage\fi\fi\fi}
\makeatother

\titleformat{\chapter}[display]
  {\LARGE\bfseries}{\huge\thechapter.}{20pt}{\LARGE}

\makeatletter
\let\ps@plain\ps@empty
\makeatother

\begin{document}
\includepdf{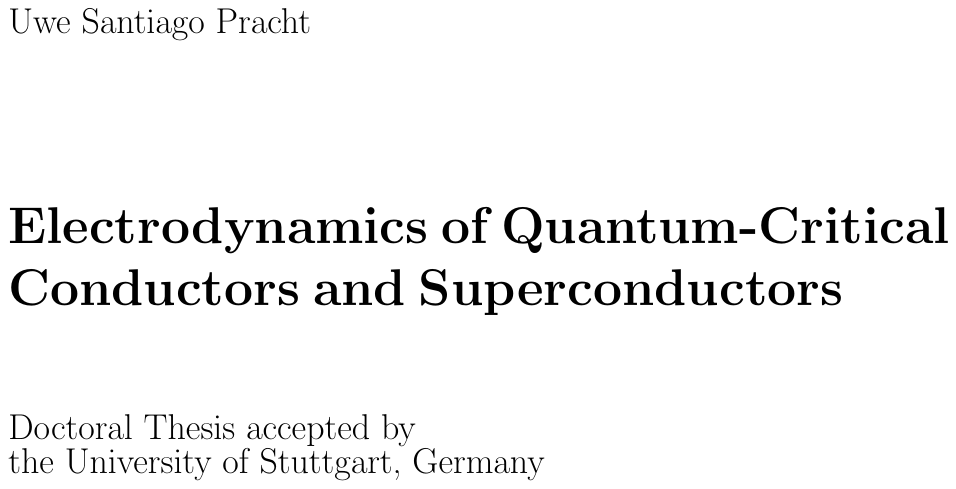}
\begin{spacing}{1.0}

\cleardoublepage
\thispagestyle{empty}
\begin{flushright}
\footnotesize{
Dedicated to \\
\textbf{Markus M\"{u}ller, Rainer Veith, and Karl Weber}\\
who initiated a passion for nature and math\\
right when it was needed the most.\\[8pt]
\emph{}}
\end{flushright}

\cleardoublepage

\section*{Abstract}
\addcontentsline{toc}{chapter}{Abstract}
\thispagestyle{empty}
\begin{flushright}
\footnotesize{
The stars are indifferent\\
to astronomy\\[8pt]
\emph{Nada Surf}}
\end{flushright}

Any truly remarkable physical theory bears on simple concepts. Although this conjecture naturally withdraws from strict mathematical approval, it has become a working hypothesis, a mantra: If an answer isn't simple enough, one hasn't got to the core of the question yet. This claim may strike surprising and the associated ambition even sound ridiculous given the shape of modern physics; crowded with disparate models, intricate categories, toy models, rules and even more exceptions on the one side, and overwhelmingly complex mathematical models maybe not even linked to the real world on the other side. Yet at the same time, in retrospect, most complicated problems tend to unwind into miraculously simple concepts. Gravity is a property of curved space time, wave and particle are just two manifestations of the same entity, the time-inverted negative-energy solutions of Dirac's equation are antiparticles, or - the conceptual framework of the presented work - spontaneous breaking of symmetry giving rise to superconductivity and collective excitations. 

In the most fundamental setting going back to Landau, we understand superconductivity in terms of a complex function $\Psi(x,t)$, the \emph{order parameter}. If $\Psi$ is zero, the system is in the disordered normal state, if $\Psi$ acquires finite values, the system turns superconducting and evolves a long range order we call phase lock. We determine $\Psi$ by constructing a Lagrangian $\mathcal{L}[\Psi]$ suited to meet our requirements for, e.g., the conservation of charge, and finding its minimum. From here, it only took only a few of the brightest minds of the 20$^\mathrm{th}$ century, to understand the implications: Symmetry breaking causes the superconducting energy gap (Nambu) and leads to angular excitation modes within the degenerate ground-state manifold (Goldstone). Those modes may disappear giving way to massive gauge bosons of electroweak interaction (Glashow, Salam, Weinberg). The same mechanism applies to condensed matter, where it renders photons massive, that is the Meissner effect of superconductivity (Anderson). In the symmetry group of the Standard Model, left-overs from gauge transformations are massive bosons we nowadays call Higgs bosons (Englert, Brout, Higgs). In certain superconductors, Higgs- and Goldstone-like excitations of $\Psi$ become well-defined and visible (Varma, Auerbach, Benfatto, and many more). Working out and  scrutinizing the implications of the above break-troughs, theoretical and experimental physicists are kept busy to this very day with an end not in sight yet.  

This PhD project is dedicated to the experimental study of materials which, neglecting subtleties for a moment, share a similar phase diagram despite being chemically and structurally different: disordered NbN, granular Al, and the Heavy-Fermion metal CeCoIn$_5$. In these compounds, superconductivity ($\Psi>0$) can continuously be controlled, suppressed, and eventually replaced by a new ground state (with $\Psi=0$) by turning up a non-thermal parameter such as disorder or magnetic fields. This transition between ground states may take place even at absolute zero temperature, where the quantum nature of the electronic system is the only source of critical fluctuations thus coining the term \emph{quantum phase transition}. Residing at zero temperature, the quantum phase transition naturally escapes from direct observation. The emergent quantum-critical fluctuations, however, may affect the metallic-, insulating-, or superconducting states at elevated temperatures quite drastically leading to new states of matter beyond our understanding of canonical solid-state- or condensed-matter systems such as Cooper-pair insulators or hidden Fermi liquids. Understanding the relation between these enigmatic states of matter and quantum criticality and how it gives rise to exotic phenomena and new states of matter is one of the prime intellectual and experimental challenges of solid state physics at date. The experimental approach employed within this work is conceptually very simple: We shine coherent THz radiation on a thin film of the material to be studied, measure the amplitude and phase shift of the transmitted light, and calculate the dynamical conductivity as function of the photons' energy. Repeating this experiment at different frequencies and temperatures we get a handle on how the electronic state, its single-particle- and collective excitations change when the systems are tuned towards quantum criticality. Although the systems studied differ when it comes to the details of superconductivity, quantum criticality, and emerging phenomena, the insights we obtain contribute pieces to a puzzle which, once completed, may lead to a unified - and maybe even stunningly simple - picture of quantum-critical superconductors. \\

The first material we studied are thin films of superconducting NbN with variable degrees of homogeneously distributed disorder. Bearing on celebrated works of Anderson \emph{et al.}, we know that moderate disorder does not significantly affect superconductivity, while for extreme disorder, the electrons tend to localize forming an insulator. In between the antagonizing superconductor and insulator extrema, there must be a region where the electrons cannot decide whether to pair up and superconduct or to localize. In spatial dimensions $D<3$, this region is shrunk to a quantum critical point (QCP), where $T_c=0$ and the system undergoes a superconductor-insulator transition (SIT). Although the microscopic mechanism leading to the eventual destruction of superconductivity remains the central open problem, there is no doubt that with the gradual cease of $\Psi$, fluctuations of its phase- and amplitude degrees-of-freedom are of growing importance to understand the peculiar superconducting state in approach of the QCP. By expanding previous tunneling spectroscopy measurements by Chand \emph{et al.} towards optical spectroscopy, we systematically compare the superconducting energy gap $2\Delta$ as it appears in the tunneling density-of-states (DOS) with the spectral gap $\Omega$ in the dissipative conductivity $\sigma_1(\nu)$. Using a effective pair-breaking ansatz to fit the tunneling spectra, we obtain the Green's functions for a given degree of disorder used to calculate a prediction $\sigma_1(\nu)$ based on the value of $2\Delta$ from tunneling. Not unsurprisingly, for only little disorder the inferred prediction agrees well with the optical measurement. However, as disorder increases we identify a growing mismatch between the $2\Delta$-based prediction and the measured $\sigma_1(\nu)$: The spectral gap $\Omega$ is progressively suppressed below $2\Delta$ and a growing amount of spectral weight piles up on top of the anticipated $\sigma_1(\nu)$ curve. We demonstrate that this gap-edge absorption channel can quantitatively not be attributed to disorder-induced broadening of the DOS, and instead suggest an alternative explanation. Given its particular properties, we argue that NbN in approach of quantum criticality can be understood within a relativistic bosonic $O$(2) 2D-field theory, for which Auerbach \emph{et al.} envisioned an optically active amplitude-excitation of $\Psi$ called Higgs mode whose energy, the Higgs mass $m_\mathrm{H}$, vanishes critically at the QCP. By isolating the sub-gap absorption from the measured $\sigma_1$ spectra, we identify the twofold nature of the spectral gap $\Omega$. Far from criticality, $\Omega$ measures the superconducting gap $2\Delta$, whereas in approach of the SIT it is identified with $m_\mathrm{H}<2\Delta$. Furthermore, we find qualitative agreement in the dispersion of the isolated sub-gap mode with Quantum Monte Carlo simulations of the Higgs-mode conductivity by Trivedi \emph{et al.} substantiating the interpretation that the anomalous low-energy electrodynamics on disordered NbN are due to a amplitude-excitation of $\Psi$. 
For the sake of completeness, however, we also address counter arguments calling the applicability of the $O$(2) model into question and leaving space for future speculations.\\

The second material under study is granular Al, i.e. thin films composed of nano-scaled Al grains coupled across thin insulating barriers into macroscopic arrays. While bulk Al is a conventional superconductor with $T_{c0}=1.19$\,K, the nature of granular Al is strikingly different: as function of decoupling (measured by the normal-state resistivity), $T_c$ is first \emph{enhanced} up to around $3T_{co}$ at a few 100\,$\mu\Omega$cm before it is suppressed at higher resistivities until superconductivity eventually ceases shaping a superconducting dome in the phase diagram. If this transition is a direct superconductor-insulating quantum phase transition with a QCP as in homogeneously disordered thin films or (given the 3D character of the films) crosses through an inter-metallic phase, is an open question. In any case, the superconducting properties of granular Al strikingly resemble those of prototypical quantum critical superconductors such as high-$T_c$ cuprates and pnictides, low-dimensional organic and heavy-fermion metals. Though being known for half a century, and despite several theoretical proposals, the underlying mechanism enhancing $T_c$ in granular Al has not been identified yet. One prominent mechanism discussed, relies on the nano-scaled size of the grains: Similar to what happens in atoms or nuclei in form of sharp energy levels, the electronic density of states tends to discretize as the electrons are spatially confined. Indeed, this can boost the pairing interaction in isolated grains - but also in macroscopic arrays thereof? To identify the origin of the superconducting dome in granular Al, we use optical spectroscopy to trace the evolution of the characteristic energy scales, that is the pairing amplitude $\Delta$ and the superfluid stiffness $J$ measuring the strength of phase lock, as a function of grain coupling. We show that, starting from well-coupled grains, $\Delta$ grows with progressive grain decoupling, causing the increasing of $T_c$. As the grain-coupling is suppressed further, $\Delta$ saturates while $T_c$ decreases, concomitantly with a sharp decline of $J$. Measurements of the  temperature evolution of $\sigma_1(\nu)$ reveals a spectral gap persisting in to the normal state on the phase-driven side of the dome while absent on the other. Using a generalized Mattis-Bardeen theory we show that this pseudogap smoothly evolves into the superconducting gap as temperature falls below $T_c$. A second peculiar finding is a sub-gap absorption similar to the one observed in disordered NbN situated at sub-gap energies. Given the nature of granular Al, a interpretation in terms of the Higgs mode is unlikely. Instead, bearing in mind the extremely low superfluid stiffness, we suggest to interpret the excess absorption as optically-active Goldstone modes of $\Psi$. We \emph{quantitatively} substantiate this idea with realistic calculations of the temperature- and frequency dependence of Goldstone modes within the disordered $XY$ model. Our studies not only identify the mechanism shaping the superconducting dome in granular Al, open a route to deterministically enhance bulk-superconductivity by nano engineering, and reveal a pseudogap and Goldstone modes but also establish granular Al as particular simple model to look at enigmatic phenomena in unconventional superconductors from a new perspective.\\

The third materials subject to investigation, is the heavy-fermion superconductor CeCoIn$_5$. Other than for NbN and granular Al, where the peculiar superconducting properties arise from a inherently conventional pairing mechanism, superconductivity in CeCoIn$_5$ is unconventional and intimately related to quantum criticality. The magnetic interaction and eventual hybridization between conduction electrons and localized Ce$^{3+}$ moments is the heart of why electrons below $\sim 40$\,K turn into heavy quasiparticles (QP), i.e. keeping the electrons' quantum numbers, but acquiring effective masses exceeding the band mass drastically. While the experimental status quo leaves little doubt, that the strongly interacting quantum liquid formed by the heavy QP is \emph{not} a canonical Fermi liquid, two natural questions remains open: what kind of quantum liquid is it then, and how does it relate to quantum criticality? Both are of great importance given that, at low temperatures, those heavy QP are paired up via an mechanism that is essentially unknown and condense into an unconventional $d$-wave superconducting state with even more enigmatic FFLO sub-phase. No matter from which side one wants to tackle the question of superconductivity and quantum criticality in CeCoIn$_5$, a thorough understanding of the non-FL heavy-fermion state is a prerequisite. In this work, we measured the dynamical conductivity $\sigma_1+i\sigma_2$ by dc-transport and phase-sensitive THz transmission spectroscopy in the HF regime at equal energy scales $k_BT\sim h\nu$ on a CeCoIn$_5$ thin film of unprecedented quality significantly advancing previous THz studies. Within the generalized Drude formalism we calculate the effective mass enhancement which approximately equals the inverse renormalization, $m^\ast/m_b\sim Z^{-1}$, and disentangle the transport and optical relaxation rates $\Gamma(\nu,T)\propto \rho_1(\nu,T)$ from the QP one. As expected for a non-FL, we find a strong and non-trivial $T$- and $\nu$-dependence of $Z$. At the same time, however, the extracted $\Gamma^\ast$ displays approximate $T^2$- and $\nu^2$ behavior - a typical characteristic of a FL. As we demonstrate, this putative paradox unwinds in the context of \emph{hidden} FL, a special kind of non-FL: Behind the anomalous behavior of resistivity, spin susceptibility, specific heat etc. at odds with canonical FL theory, one has well-defined \emph{resilient} QP obeying $\Gamma^\ast\propto \nu^2,T^2$ whose existence, however, is hidden behind an energy-dependent renormalization. To solidify our interpretation we revisit the previously reported anomalous $T$-dependencies of the electronic specific heat, the magnetic susceptibility and spin-lattice relaxation time to our measurements of $Z(T)$, and prove them consistent with the hidden FL scenario. Triggered by the strong temperature and frequency dependence of $m^\ast/m_b$ at approximately equal energy scales, we furthermore search for scaling behavior of this quantity as typical signature of quantum criticality. Although a theory for quantum criticality from resilient QP has not been worked out yet, we propose and experimentally verify an approximate $\nu/T$ scaling ansatz inspired from scaling analysis of the quantum-critical HF metals CeCu$_{6-x}$Au$_x$. Our results open a surprising new perspective on the nature of the heavy fermion state in CeCoIn$_5$, encourage a careful reconsideration of other non-FL systems, and hopefully stimulate theorists to work out the existing models of quantum criticality in HF systems towards $m^\ast/m_b$ scaling and resilient QP.

This thesis is structured as follows. Chapter \ref{Field theory of superconductivity} discusses a model of superconductivity based on fieldtheoretical considerations to familiarize the reader with the concept of $\Psi$, collective modes thereof, and the basic superconducting phenomenology from the perspective of $\Psi$. In addition, more on a technical account, the reader is introduced to the Usadel equation and the relevant Green's functions which, later on, are used to model the density of states and dynamical conductivity of disordered superconductors. Although the theoretical background is dispensable for the experimental chapters, the author strongly recommends not to skip it as a proper understanding of the field theoretical concepts (and language) helps to cast the presented results into a broader context. Each of the experimental chapters 2, 3, and 4 starts with a short summary of the results, an introduction covering basic theoretical aspects of the physical background to acquaint the reader with the particular system, a review of the essential experimental studies, and an embedding of the presented research into the scientific status quo. Subsequently, details and the relevant aspects (or issues) concerning the measurements, sample preparation and -handling are addressed briefly. (Detailed information concerning the measurement techniques and -procedures, the characterization of the low-temperature performance as well as the data treatment and  analyses is provided in the Appendix.) The main part of the experimental chapter- discussion and interpretation - is prepended with introduction to particular theoretical framework beyond Ch.\,1 in which the experimental data is subsequently interpreted. In more detail, the Higgs mode in the context of the bosonic $O$(2) mode in Sec.\,\ref{Sec:NbNHiggs}, pseudogap and Goldstone modes within the disordered $XY$ model in Sec.\,\ref{The pseudogap for phase-driven superconductivity} and \ref{Goldstone modes}, and the theory of hidden Fermi liquids in Sec.\,\ref{FLNFL}. The discussion is closed with an outlook on future experimental studies and theoretical challenges. This structure may appear somewhat cumbersome, yet it eases the understanding as it allows to follow the sequence of arguments directly without the frequent back and forth of the typical theory-materials-experiment-analysis-discussion structure.\\

Uwe S. Pracht, Winter 2016

\cleardoublepage
\section*{Acknowledgments} 
\addcontentsline{toc}{chapter}{Acknowledgments}
\thispagestyle{empty}
\begin{flushright}
\footnotesize{
I can't believe I got so far\\
With a head so empty\\[8pt]
\emph{The Thermals}}
\end{flushright}
At the time I take down these final notes, the year 2016 is about to end - just as a journey initiated in early 2013 with the decision for a PhD in experimental physics, condensed matter and superconductivity. Aside from all dead ends, nonsensical data, referees who \emph{just don't understand the paper}, and endless days and nights in the lab, these four year of PhD research were, more than anything else, the following: a great opportunity to learn, to gather experience, to meet people at distant places, and to have to the unique possibility to contribute a small piece to this huge puzzle, to unravel some of natures fascinating mysteries. Opportunities, I certainly don't take for granted and which I am truly thankful for. At the same time, I am grateful for all the support I received over the years. At no stage, this was a one-man-project and what follows is no obligation but true pleasure.   \\

First of all, I would like to thank Prof. Martin Dressel. I very much enjoyed my time at your institute and I am grateful for the trust you put in me. Not only regarding the THz lab, but also when it came to presenting our work and representing our group at conferences all across Europe. You were always very open concerning the way I structured my research, and the emphases and collaborations I chose. Without doubt, this trust and independence was a vital motivation for my work. \\

In sincerly thank Prof. Tilman Pfau and Prof. Guy Deutscher who kindly accepted to co-referee this work.\\

I would like to thank Dr. Marc Scheffler, for being my supervisor for over five years. During this time, your guidance, shared knowledge, and support in so many regards were indispensable for my work. I learned a lot from all those discussions in our offices, scrawling all over the white boards, sometimes even getting somewhere and sometimes even unraveling something of importance (e.g. finding the reason for the superconducting dome in granular Al on January, 30$^\mathrm{th}$, 2015). \\

Casting my experimental results into insightful phys- ics would not have been possible without the steady exchange with theorists. In particular, I would like to thank Dr. Lara Benfatto. I very much liked our work on granular Al and the discussions in Grenoble, Geneva, Stuttgart, and Leiden. You have the truly wonderful gift to unwind theory such that experimentalists can work with it! Also, I thank Prof. Assa Auerbach for our illuminating exchange on the Higgs mode and for proof-reading the corresponding parts of this thesis, and Dr. Yakov Fominov for all the detailed explanations, supplementary math on the Green's functions and Usadel equation, and critically reading Sec.\,1.4 and 1.5 of this work.\\

Special thanks go out to Dr. Nimrod Bachar, who dragged me into the granular Al field. It's been a pleasure working with you in the lab, chasing ghosts, wrapping our minds around theory, writing papers, and letting thoughts run freely over a beer or two. I hope you stay open minded and keep your happy-go-lucky attitude!\\  

I also thank Dr. Boris Gorshunov and Dr. Artem Pronin for the steady hands-on support regarding the technical issues of the THz machine. Although, sometimes, hands off the table would've been the better choice - \emph{what should happen? It's just high voltage} - but never mind :)\\

Over the years, I received support from numerous colleagues from all over the world, always open-minded, uncomplicated, and helpful. Thanks to Prof. Nandini Trivedi, Prof. Aviad Frydman, Dr. Jernej Mravlje, Prof. Peter W\"{o}lfle, Dr. Eli Farber, Prof. Guy Deutscher, Dr. Sangita Bose, Prof. Pratap Raychaudhuri, Dr. Eduard Driessen, Prof. Antonio Garcia-Garcia, Dr. Elena Zhukova, Prof. Peter Armitage, Prof. Dirk Manske, Prof. Christoph Strunk, Dr. Kostya Illin, Dr. Ina Schneider, Prof. Michael Siegel, Dr. Tatjana Baturina, Prof. Yuji Matsuda, Prof. Takasada Shibauchi, Dr. Florence Levy-Bertrand, Tomohiro Ishii, Dr. Marta Autore, Dr. Marie-Aude Measson, and Dr. Daniel Sherman.\\  

I thank Gabi Untereiner for 4-point contacting \emph{challenging} samples, cutting substrates, the steady supply of all sorts of lab consumables, and (also important) numerous early-morning chats.  \\

Then, I would like to thank my faithful PhD companions Dr. David Neubauer, Dr. Eric Heintze, Dr. Conrad Clauss, Markus Thiemann, and Dr. Ralph H\"{u}bner. \begin{marginfigure}
\includegraphics[width=\marginparwidth]{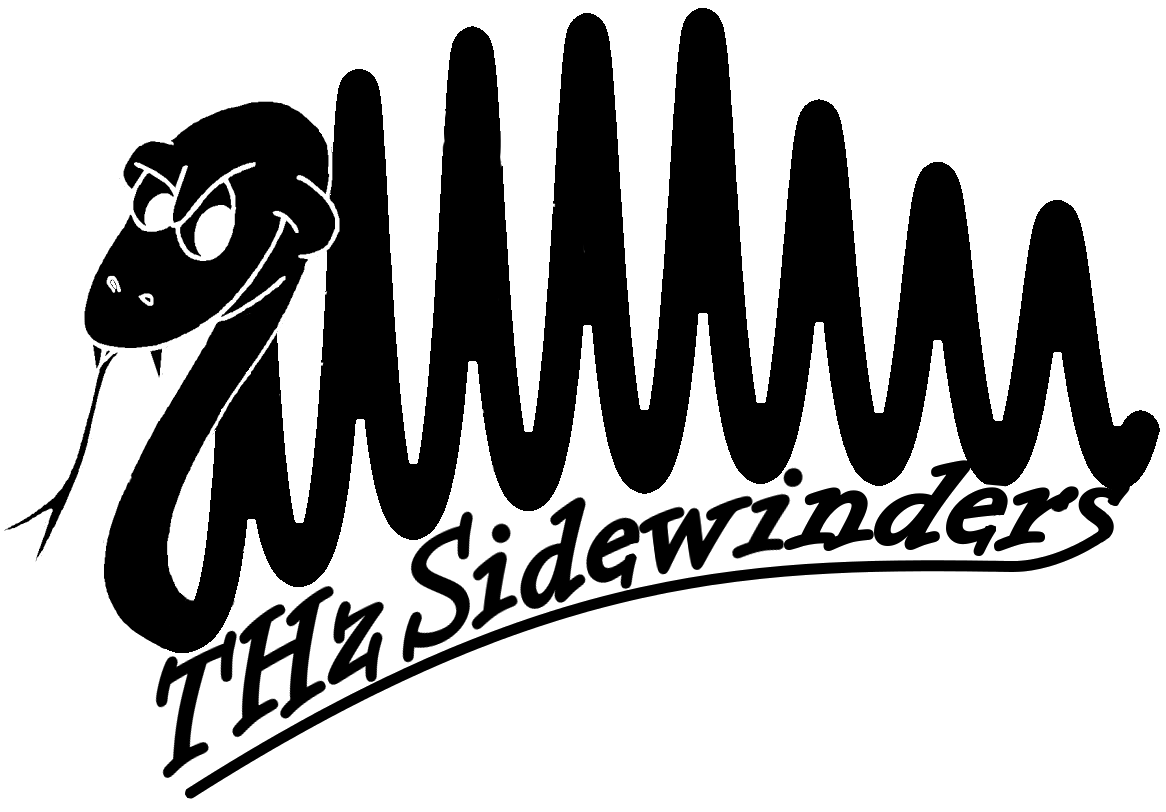}
\end{marginfigure} Needless to say, without your company my time here would have been much less pleasurable. I guess it will be the time spent on the balcony, near the coffee machine, in the basement, and the \emph{Palast} that, in retrospect, will constitute this 'great time when we did our PhD'. \\

Similarly, I should thank my first-semester colleagues Dr. Dennis Dast, Dr. Christoph Tresp, Dr. Matthias Schmitt, Dr. Daniel Haag and Dr. Roland Bek for all those fantastic years between lectures, the CIP pool, and the \emph{Palmenhof}. I also thank my non-physics friends for teaming up against the - apparently omnipresent - challenges of a PhD: Dr. Steffen Heinz, Dr. Marcel G\"{u}nter, Dr. Ayse Julius, and Dr. Wolfgang Merkle.\\

I thank my students MSc. Lena Daschke, MSc. Julian Simmendinger, MSc. Subash Panthi, MSc. Sanu Mishra, MEd. Karsten Paulokat, BSc. Julian Uhl, and BSc. Jonathan Gusko for contributing singular aspects to the big picture and helping me to keep the lab running!\\

I also thank the Studienstiftung des deutschen Volkes for a three-year scholarship providing not only the financial prerequisite for my PhD studies, but also insights to state-of-the-art social, educational, and political research and valuable contacts to industry. \\
 
Finally and most important, I thank you, Ines, for everything you did for me and for us. For not being interested in broken symmetry, Cooper pairs, and all this stuff, for being a point of rest, for always being there, listening, caring, laughing, loving.\\   
\begingroup
\thispagestyle{empty}
\tableofcontents
\endgroup


\cleardoublepage
\chapter{Field-theoretical basics \\on superconductivity}\label{Field theory of superconductivity}
\thispagestyle{empty}
\begin{flushright}
\footnotesize{
Vor hundert Jahren ging es los,\marginnote{\footnotesize \textcolor{gray}{It has started one hundered years ago,\\ like a revolution, \\ destination and direction unknown, \\ everything old was burnt to the ground.}}\\
Es war wie eine Revolution.\\
Ziel und Richtung unbekannt,\\
Alles Alte wurde niedergebrannt.\\[8pt]
\emph{Die Toten Hosen}}
\end{flushright}
\emph{Starting with general considerations  quantum field theories (QFT) and spontaneous symmetry breaking therein in Sec.\,\ref{Some preliminary prerequisites} and \ref{SSB}, we will motivate a field-theoretical description of electrically neutral and charged superfluids in Sec.\,\ref{Sec:FieldTheoforSC} akin to the early ideas of Ginzburg and Landau. Using the language of QFT we will set up model obeying global $U(1)$ gauge symmetry,  whose breaking leads to the physics of superfluids. Turning to charged superfluids, we will promote the global $U(1)$ symmetry to a local one and examine the consequences of a symmetry-breaking superconducting ground state and associated collective excitations of the order parameter field. Subsequently, we will make step-by-step simplifications testing our QFT aiming for a low-energy model reproducing some of the most crucial properties of superconductors: how gradients of the phase field lead to currents, how the topology of the order parameter explains dissipation free dc-transport, and how the Higgs mechanism at play in a superconductors turns photons massive leading to the Meissner-Ochsenfeld effect. In Sec.\,\ref{Sec:Green}, we will introduce a Green's-function formalism to calculate experimentally accessible properties of superconductors. This will be done on basis of the Usadel equation allowing for pair-breaking beyond the canonical Bardeen-Cooper-Schrieffer theory. We will discuss how to solve the equation and how to obtain the quasiparticle density-of-states and the dynamical conductivity in Sec.\,\ref{Density of states and dynamical conductivity}.}
\clearpage{}
\section{A preliminary outline}\label{Some preliminary prerequisites}

Before we consider symmetry breaking in gauge theories and mainly the construction and study of Lagrangians in greater detail, we will sketch how these steps are part of the entire route that eventually leads to a particular physical theory for superconductivity. We will do this for \emph{bosonic} superconductors. The justification for this lies in the observation that granular and disordered superconductors can often be described in terms of effectively bosonic theories, that is neglecting the composite character of Cooper pairs. The entire procedure follows a six-point plan:     
\begin{enumerate}
\item{Decide, which energies, interactions, conservation laws the eventual theory should properly account for. }
\item{Based on that, write down\sidenote{\footnotesize{This is very similar to assembling the Hamiltonian in quantum mechanics.}} the Lagrangian $\mathcal{L}$ in terms of bosonic\sidenote{\footnotesize{The Cooper pairs are understood as excitations of this field.}} fields $\psi(x)$, where $x=(t,\mathbf{x})$ denotes a point in space-time.}
\item{Quantize the fields by mode expansion in creation and annihilation operators
\begin{equation}
\hat{\psi}(x)=\frac{1}{(2\pi)^\frac{3}{2}} \int \frac{\mathrm{d}\mathbf{p}}{\sqrt{E_\mathbf{p}}}\big(\hat{a}_\mathbf{p}\mathrm{e}^{ipx}+\hat{a}^\dagger_\mathbf{p}\mathrm{e}^{-ipx}\big)
\end{equation}
and postulate commutator relations.}
\item{Infer the normalized generating functional $\mathcal{Z}_0[J]=Z[J]/Z[0]$ by calculating the path integral (here omitting the hat-symbol\sidenote{\footnotesize{As bosonic fields commute, we can simplify the calculations by treating the operator values fields $\hat{\psi}$ as classical Heisenberg fields. This is not possible for anti-commuting fermionic fields where one has to resort to Grassmannian fields.}})
\begin{equation}
Z[J]=\int \mathcal{D}\psi\mathrm{e}^{\frac{\mathrm{i}}{2}\int \mathrm{d}^4x\,\mathcal{L}[\psi(x)]+\mathrm{i}\int\mathrm{d}^4x\,J\psi(x)}
\end{equation}
where $\mathcal{D}\psi$ integrates over all possible field configurations and $J(x)$ is a auxiliary \emph{source} function.}
\item{Infer the Greens functions $\mathcal{G}$ by differentiating $\mathcal{Z}[J]$ with respect to $J$.}
\item{Use $\mathcal{G}$ to calculate the desired physical properties\sidenote{\footnotesize{e.g. the density of states or the dynamical conductivity.}}.}
\end{enumerate}
We will focus on the first two and and last step, while some calculations are simplified and motivated with the intermediate ones at the back of our minds. The field-theoretical approach to superconductivity we will sketchily review in this first Chapter is based on the excellent introduction to quantum field theory by Lancaster and Blundell \cite{Lancaster2014}.

\section{Spontaneous $U$(1) symmetry breaking}\label{SSB}

Symmetry is the property of a physical object to not change itself under a symmetry transformation. The objects we will encounter and study regarding their symmetry retaining or breaking properties are ground-state or vacuum expectation values (VEV) of quantum fields\sidenote{\footnotesize{i.e. something we will later on identify as the superconducting order parameter.}} quantum fields and Lagrangians. The study of symmetry behavior turns out to be a versatile approach to describe numerous physical systems using one and the same concept despite their seeming difference.
In greater detail, we will consider continuous symmetries, that is the symmetry transformation is characterized by a variable with continuous spectrum of real values that allows a smooth transformation between initial and final states\sidenote{\footnotesize{In contrast to e.g. the discrete mirror, parity or time-reversal symmetries.}}. According to the Noether theorem, any continuous symmetry of the Lagrangian implies the existence of a conserved physical property\sidenote{\footnotesize{For instance, the isotropy of time and space leads to the conservation of energy and momentum, respectively.}}. Here, we will focus on the transformations that can be represented by the elements of the unitary group $U(1)$. These transformations can be thought of rotating an object, e.g. a (quantum) field $\psi(x)$, in the complex plane by an angle $\alpha$ such that 
\begin{equation}
\psi\rightarrow \mathrm{e}^{i\alpha}\psi.\label{eq:global}
\end{equation}      
If the object under consideration remains unchanged it is said to preserve (or, otherwise, break) \emph{global} $U(1)$ symmetry as we apply the same change at all space-times. Physically speaking\sidenote{\footnotesize{An example is the theory of electromagnetism which is $U(1)$ invariant and conserves the electric charge $q$. More exotic charges, e.g. color, require global symmetries beyond $U(1)$.}}, global $U(1)$ invariance implies the conservation of charge. Symmetry considerations and the demand for certain conservation laws are important guides to define a (quantum) field theory (QFT) for a given system, that is, casting the essential interactions and energies into the Lagrangian

The global symmetry is promoted to a \emph{local} one, if the $U(1)$ transformation may be different for each space-time point, 
\begin{equation}
\psi\rightarrow \mathrm{e}^{i\alpha(x)}\psi,\label{eq:local}
\end{equation}    
leaving the transformed object invariant. In general, the demand for local $U(1)$ invariance imposes stronger constrictions to a QFT than its global version, and we will see below, that it requires \emph{gauge fields} to guarantee the invariance\sidenote{\footnotesize{Accordingly, theories that are locally $U(N)$ invariant are called gauge theories.}}. In a sense, we can think of the gauge field as a device that repairs the damage caused by the $U(1)$ operation in order to preserve symmetry at each space-time point $x$ accordingly. A more physical  role of the gauge field is to mediate the interaction between the particle excitations of the quantum fields carrying the charge related to the global symmetry\sidenote{\footnotesize{Note that \emph{local} always implies \emph{global}, but not vice versa.}}. This offers an enlightening new perspective: \emph{if} we want the particles to interact by exchange of gauge bosons, \emph{then} we have to set up the Lagrangian such that it has a local symmetry. Before we come to the consequences of promoting global to local symmetries, we will examine what happens, when global $U(1)$ symmetry is spontaneously\sidenote{\footnotesize{Symmetry breaking is spontaneous, if broken and unbroken phases are connected smoothly. This has to be contrasted with explicitly broken symmetries such as the handedness of chiral molecules.}} broken. \\
We will start with the Lagrangian the problem of superfluidity in a gas of neutral bosons. Owing to their vanishing charge, we can neglect electromagnetic interactions and the coupling to external electromagnetic fields. The Lagrangian is of the form kinetic energy minus potential energy and reads\sidenote{\footnotesize{In what follows, we use the standard notation
\begin{equation}
\partial_\mu=(\partial_t,\nabla)\nonumber
\end{equation}
and 
\begin{equation}
\partial^\mu=(\partial_t,-\nabla)^\mathrm{T}.\nonumber
\end{equation}
Consequently, 
\begin{equation}\partial_\mu\partial^\mu=\partial^2=\partial^2_t-\nabla^2.\nonumber
\end{equation} }}
\begin{equation}
\mathcal{L}=\partial^\mu\psi^\dagger\partial_\mu\psi-V(\psi,\psi^\dagger).
\end{equation}
We expressed $\mathcal{L}$ in terms of classical bosonic Heisenberg fields $\psi(x)$ rather than operator-valued fields $\hat{\psi}(x)$ of a proper quantum-field description - a simplification owing to the commuting algebra of bosonic operators. To get a handle on the potential $V$, we assume only small variations of the system around its minimum so that we can express $V(\psi,\psi^\dagger)$ in a power series 
\begin{equation}
V=\gamma^2\psi\psi^\dagger+\frac{1}{2}\lambda(\psi\psi^\dagger)^2\label{eq:potential}
\end{equation}
where $\lambda>0$. This general form fulfills the symmetry requirement, $V(\psi)=V(-\psi)$, and it provides a stable minimum for $|\psi,\psi^\dagger|\to\infty$.
Assembling kinetic and potential energies we obtain a scalar complex field theory 
\begin{equation}
\mathcal{L}=\partial^\mu\psi^\dagger\partial_\mu\psi-\gamma^2\psi\psi^\dagger-\frac{1}{2}\lambda(\psi\psi^\dagger)^2\label{eq:fullSFforL}
\end{equation}
\begin{marginfigure}
    \includegraphics[width=\marginparwidth]{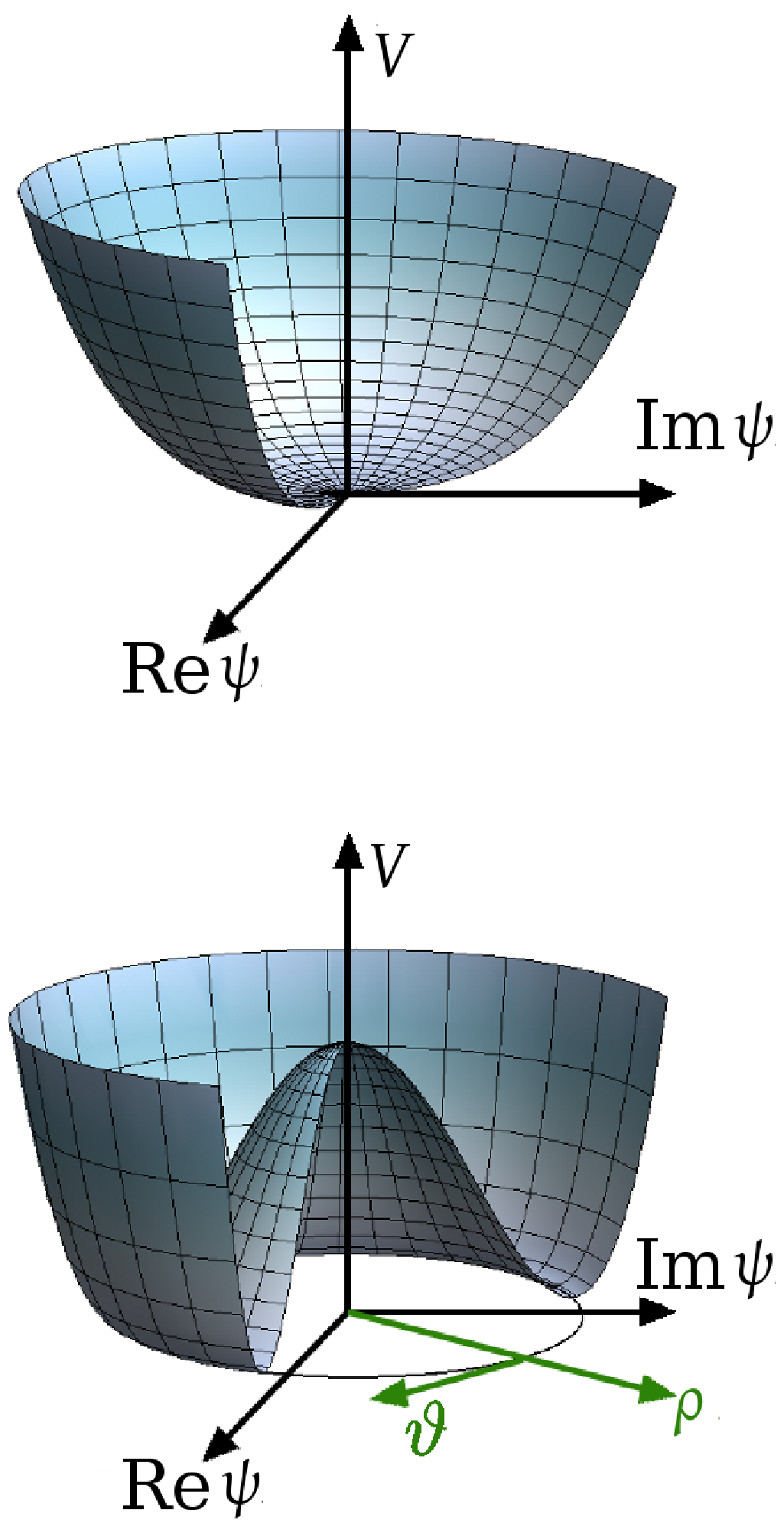}
    \caption{Potential energy with symmetric (top) and broken-symmetry (bottom) ground states. $U(1)$ operations can be visualized as rotations around the $V$  axis.\label{fig:potential}}
\end{marginfigure}
with a self-interaction term governed by the coupling constant $\lambda$.  For a positive mass term $\mu^2>0$ the potential has a unique minimum at $\psi=0$. For $\gamma^2<0$ the potential acquires a Mexican-hat shape, see Fig.~\ref{fig:potential}, and an infinite number of degenerate minima on a circle with radius $r$. We can easily check the global $U(1)$ invariance by transforming the fields according to Eq.~(\ref{eq:global}). The phase factors are not affected by the derivatives and cancel when multiplied with their complex-conjugates. Note that this holds true irrespective of the sign of the mass term. The symmetry of $\mathcal{L}$ can be captured by the polar representation $\psi=\sqrt{\rho(x)}\mathrm{e}^{\mathrm{i}\vartheta(x)}$ with a real longitudinal scalar field $\rho$ and a real angular scalar field $\vartheta$. The ground state of the QFT is found upon minimizing the potential energy. For $\gamma^2>0$ the system resides in the unique minimum at the origin, so that the vacuum expectation value (VEV) of $\rho$ takes the value $\langle\rho\rangle=0$ at every point in space-time. Just as $\mathcal{L}$ itself, the VEV is $U(1)$ invariant. For $\mu^2<0$, the ground state has to spontaneously pick a particular minimum on the circle of minima, e.g. the one with $\langle\vartheta\rangle=\vartheta_0$. Once this has happened, the symmetry is broken: a global $U(1)$ transformation shifts the angular field from one VEV $\vartheta_0$ to another one  $\vartheta_0+\alpha$. In the next section, we will study the consequences of this spontaneous symmetry breaking (SSB) in a superfluid.

\section{Field theory approach to neutral and charged superfluids}\label{Sec:FieldTheoforSC}

In condensed matter physics, we are usually in the non-relativistic limit where the mass energy $mc^2$ is by far the biggest energy scale. In the time evolution, we can factor out this fast oscillation by re-defining\sidenote{\footnotesize{We use natural units in which $h=c=1$}} the field $\psi=\frac{1}{\sqrt{2m}}\Psi\mathrm{e}^{\mathrm{i}mt}$. Inserting this form in Eq.~(\ref{eq:fullSFforL}) and performing the derivatives gives
\begin{eqnarray}
\mathcal{L}&=&\frac{1}{2m}\partial_t \Psi^\dagger\partial_t \Psi+\mathrm{i}\Psi^\dagger\partial_t \Psi
-\frac{1}{2m}\nabla\Psi^\dagger\nabla\Psi\nonumber\\
&&-(\gamma^2-m^2)\Psi^\dagger\Psi-\frac{\lambda}{2}\left(\Psi^\dagger\Psi\right)^2. \label{eq:LGoldstone1}
\end{eqnarray}
Expressing $\Psi$ as above in terms of polar coordinates, Eq.~(\ref{eq:LGoldstone1}) is rewritten to
\begin{eqnarray}
\mathcal{L}&=&\frac{1}{2m}\left(\frac{(\partial_t \rho)^2}{4\rho}+(\partial_t \vartheta)^2\rho\right)+\frac{i}{2}\partial_t \rho-\rho\partial_t \vartheta\nonumber\\&&-\frac{1}{2m}\left(\frac{(\nabla \hat{\rho})^2}{4\rho}+\rho(\nabla \vartheta)^2\right)-\mu\rho-\frac{\lambda}{2}\rho^2\label{eq:LGoldstone2}
\end{eqnarray}
where we have defined $\mu=\gamma^2-m^2$. This result can be simplified. First we note that we explicitly separated out the rapid oscillations and, thus, we can consider the remaining time dependence as weak. Thus, with the huge $1/m$ prefactor, the first term in Eq.~(\ref{eq:LGoldstone1}) can be neglected\sidenote{\footnotesize{Even the typically large energy scales in condensed matter such as the Fermi energy are still $\sim 10^5$ times smaller than the electron rest mass.}}.  Furthermore, also the bare time derivative of $\rho$ can be omitted, as it does not affect the particle dynamics\sidenote{\footnotesize{This can be understood when looking at the action 
\begin{equation}
S=\int \mathrm{d}\mathbf{x}\int\mathrm{d}t\mathcal{L}\nonumber
\end{equation}
where $\partial_t \rho$ turns out to be irrelevant for the physics as the fields vanish for $|x|\to\infty$.}}.
The Lagrangian we obtained is unstable towards symmetry breaking: For $\mu^2<0$ we have a Mexican-hat potential with infinite number of minima along a circle with radius $\langle\rho\rangle=\mu^2/\lambda\equiv n$. For simplicity, we chose the minimum with $\langle\vartheta\rangle=0$. In order to study the excitations of the fields from their mimimal values we make a coordinate transform to the minimum defining  $\sqrt{\rho}\mathrm{e}^{\mathrm{i}\vartheta}\to(\sqrt{n}+h)\mathrm{e}^{\mathrm{i}\vartheta}$  with a new longitudinal field $h$. Up to bilinear terms and neglecting constant terms and the total time derivative $\partial_t\vartheta$, we arrive at a field theory for superfluids reading
\begin{equation}
\mathcal{L}=-\frac{1}{2m}(\nabla h)^2-2\lambda n h^2-(2\sqrt{n}\partial_t\vartheta)h-\frac{n}{2m}(\nabla \vartheta)^2+...\label{eq:LGoldstone3}
\end{equation}
The real scalar field $h$ describes spin-zero bosons with mass $\sqrt{2\lambda n}$ forming the superfluid (e.g. $^4$He atoms). From the QFT point of view, these bosons are finite-energy longitudinal modes above the ground state. This is reflected in the non-zero mass term of Eq.~(\ref{eq:LGoldstone3}).  The excitations of the scalar field $\vartheta$ are  Goldstone modes\sidenote{\footnotesize{The Goldstone mode of the superfluid is the zero-sound mode. Another (approximate) solid-state Goldstone mode are magnons, i.e. collective spin-excitations with vanishingly small energy at long wavelengths.}} in angular direction not leaving the ground state manifold. Consequently, these modes should be excited at zero energy and, indeed, there is no term in  Eq.~(\ref{eq:LGoldstone3}) proportional to $\vartheta^2$ - the modes are massless. Here we can state another important finding: The Goldstone-mode term is proportional to $n$ being the radius of the ground-state circle. If the ground state does not break $U(1)$ symmetry, then this manifold is a singular point at $n=0$ and the Goldstone mode disappears. Turning this argument around captures the essence of the Goldstone theorem: Breaking a continuous global $U(1)$ symmetry leads to the emergence of massless excitations. \\
A superconducting condensate is a superfluid of electrons that are bound to bosonic pairs via an essentially electromagnetic interaction\sidenote{\footnotesize{In fact, the interaction must not necessarily be electromagnetic: quarks carrying a color charge may (at least in theory) form color-superconducting pairs through the exchange of gluons, i.e. via a strong-force interaction.}} mediated by photons. We include photons by gauging the QFT of the superfluid making it invariant under local $U(1)$ transformations. This is done in two steps. First, we take Eq.~(\ref{eq:fullSFforL}) and add the Lagrangian
\begin{equation}
\mathcal{L}^\prime=-\frac{1}{4}(\partial_\mu A_\nu-\partial_\mu A_\nu)(\partial^\mu A^\nu-\partial^\mu A^\nu)\label{eq:PhotonL}
\end{equation}
 describing the dynamics of the photon vector field $A_\mu=(A_0,\mathbf{A})$. Second, we include the interaction between $A_\mu$ and the superconducting charge field $\psi$ via minimal coupling by replacing the derivatives by covariant derivatives\sidenote{\footnotesize{reading $\partial_t\rightarrow\partial_t+iqA_0$ and $\nabla\rightarrow\nabla+iq\mathbf{A}$ when expanded.}} 
\begin{equation}
\partial_\mu\rightarrow D_\mu=\partial_\mu+\mathrm{i}qA_\mu
\end{equation}
where the coupling strength between electrons and photons is given by the charge $q$. The full Lagrangian we obtain is
\begin{eqnarray}
\mathcal{L}&=&(\partial^\mu-\mathrm{i}qA^\mu)\psi^\dagger(\partial_\mu-\mathrm{i}qA_\mu)\psi-\gamma^2\psi\psi^\dagger-\frac{1}{2}\lambda(\psi\psi^\dagger)^2\nonumber\\
&&-\frac{1}{4}(\partial_\mu A_\nu-\partial_\mu A_\nu)(\partial^\mu A^\nu-\partial^\mu A^\nu)\nonumber\\
&=&D^\mu\psi^\dagger D_\mu\psi-\gamma^2\psi\psi^\dagger-\frac{1}{2}\lambda(\psi\psi^\dagger)^2-\frac{1}{4}\mathcal{F}_{\mu\nu}\mathcal{F}^{\mu\nu}\nonumber\\
&&\label{eq:HiggsL1}
\end{eqnarray} 
where we have introduced the electromagnetic field tensor $\mathcal{F}_{\mu\nu}=\partial_\mu A_\nu-\partial_\mu A_\nu$. This field theory is invariant under local $U(1)$ transformation, Eq.~(\ref{eq:local}), if we demand\sidenote{\footnotesize{That this is a meaningful transformation is due to the \emph{freedom of gauge}: If we put the such-transformed field $A^\prime_\mu$ into $\mathcal{L}^\prime$, the terms with $\partial_\mu\alpha$ add up to zero and leave the physics unaffected.}} the field $A_\mu$ to transform as 
\begin{equation}
A_\mu\rightarrow A^\prime_\mu=A_\mu-\frac{1}{q}\partial_\mu\alpha .\label{eq:freedom}
\end{equation} 
We can always find a scalar field $\alpha$ such that it guarantees gauge invariance and keeps the form of the resulting field equations\sidenote{\footnotesize{via the Euler-Lagrange equation}}. 
We can now resort to our result obtained for the neutral superfluid in Eq.~(\ref{eq:LGoldstone2}), where we again neglect the bilinear derivative, replace the derivations by covariant ones and add $\mathcal{L}^\prime$
\begin{eqnarray}
\mathcal{L}&=&\mathrm{i}\Psi^\dagger(\partial_t+iqA_0) \Psi-\frac{1}{2m}(\nabla-iq\mathbf{A})\Psi^\dagger(\nabla+iq\mathbf{A})\Psi\nonumber\\
&&-\mu\Psi^\dagger\Psi-\frac{\lambda}{2}\left(\Psi^\dagger\Psi\right)^2-\frac{1}{4}\mathcal{F}_{\mu\nu}\mathcal{F}^{\mu\nu}. \label{eq:LHiggsk2}
\end{eqnarray}
Here, the field $\Psi$ can again be expressed in scalar longitudinal and angular fields, $\rho$ and $\vartheta$. When expanding the derivatives, we encounter terms proportional to $A_0+\frac{1}{q}\partial_t\vartheta$ and $\mathbf{A}+\frac{1}{q}\nabla\vartheta$. We use the freedom of gauge according to Eq.~(\ref{eq:freedom}) and define a new field $C_\mu=A_\mu+\frac{1}{q}\partial_\mu\vartheta$. Expansion leads to\sidenote{\footnotesize{Note that we have chosen the mass term for the $\rho$ field to be negative (giving a positive sign thereof in Eq.~(\ref{eq:LHiggs3})). This means, that we will have a broken-symmetry ground state.}} 
\begin{eqnarray}
\mathcal{L}&=&\frac{\mathrm{i}}{2}\partial_t \rho-q\rho C_0-\frac{1}{2m}\left(\frac{(\nabla \rho)^2}{4\rho}+q^2\rho\mathbf{C}^2\right)\nonumber\\
&&+\mu\rho-\frac{\lambda}{2}\rho^2-\frac{1}{4}\mathcal{F}_{\mu\nu}\mathcal{F}^{\mu\nu}. \label{eq:LHiggs3}
\end{eqnarray}
The total time derivative can again be neglected. As in the superfluid, we modify the longitudinal field by a small perturbation from the ground state replacing $\sqrt{\rho(x)}\rightarrow \sqrt{n}+h(x)$. 
After some simplifications\sidenote{\footnotesize{i.e. omitting constant terms, neglecting interaction terms that are cubic (or higher) in the fields, and treating the perturbation $h$ as much smaller than $\sqrt{n}$ .}} we obtain the Lagrangian for superconductors reading
\begin{equation}
\mathcal{L}=-\frac{1}{2m}(\nabla h)^2 -2q\sqrt{n}C_0 h-2\mu h^2-\frac{q^2n}{2m}\mathbf{C}^2-\frac{1}{4}\mathcal{F}_{\mu\nu}\mathcal{F}^{\mu\nu} \label{eq:LHiggs4}.
\end{equation}
If we compare this result to Eq.~(\ref{eq:LGoldstone3}), we see that the Goldstone mode of the superfluid has disappeared. At the same time, the field $C_\mu$ has acquired a mass $\sqrt{q^2n/2m}$. The initially\sidenote{\footnotesize{Note that before symmetry breaking there was no bilinear term in $A_\mu$.}} massless gauge field $A_\mu$ is said to have absorbed the Goldstone mode to become massive when we gauged the theory going from Eq.~(\ref{eq:LHiggsk2}) to Eq.~(\ref{eq:LHiggs3}). This is the heart of the \emph{Higgs mechanism}: the freedom of gauge is used to let the angular component $\vartheta$ of $\Psi$ disappear giving mass to the photon field $A_\mu$. It is to be noted that the same mechanism works - up to mathematical subtleties -  identically for non-abelian gauge theories such as the electroweak unification of Abdus and Salam: The role of $\psi$ is played by the four-component Higgs field and we replace the simple $U(1)$ symmetry by the more complex $SU(2)\times U(1)$. Here, partial\sidenote{\footnotesize{that is, \emph{most} of the symmetry is broken, while a small subgroup of symmetry remains intact.}} symmetry breaking of the Higgs-field's VEV and the vanishing angular modes give rise to massive $W^\pm$ and $Z^0$ gauge bosons for the weak force while the photon (belonging to the unbroken subgroup) for electromagnetism remains massless. The transversal degrees of freedom (related to the unbroken subgroup) of the Higgs field are massive excitations, the celebrated Higgs bosons\sidenote{\footnotesize{While the electroweak unification explains how gauge bosons may become massive, the other particles of the Standard Model acquire their masses by coupling their fields to the Higgs field.}}. Though our treatment of the charged $U$(1) superfluid in 3+1D includes a stunning realization of the Higgs \emph{mechanism} in solid state physics, it does not herald a new massive excitation \emph{mode}. Under special circumstances, however, this may change: In the context of, e.g., superconductivity in quantum-critical NbN, see Chapter~\ref{Sec:NbN}, we will reexamine some of the above considerations and see that there is a massive Higgs mode in the charged $O$(2) superfluid in 2+1D in the relativistic limit. \\
In the next part, we will derive some of the fundamental and well-known properties of superconductors from our field theoretical approach.

\section{A low-energy Lagrangian for superconductivity}\label{Sec:LowEforSC}
To pull some superconducting phenomenology from our so-far Lagrangian we need to simplify Eq.~(\ref{eq:LHiggs4}). This can be done using a mathematical trick known as integrating out the energetic field $h$. We do this following the road map we set out at the beginning of this chapter and insert $\mathcal{L}$ back into the action\sidenote{\footnotesize{which is simply the space-time integral $S=\int\mathrm{d}^4x\mathcal{L}$.}} $S$ and the generating functional $Z$
\begin{eqnarray}
Z&=&\iiint \mathcal{D}\mathbf{C}\,\mathcal{D}C_0\,\mathcal{D}h\,\mathrm{e}^{\frac{\mathrm{i}}{2}S[\mathbf{C},C_0,h]}\nonumber\\
&=&\iiint \mathcal{D}\mathbf{C}\,\mathcal{D}C_0\,\mathcal{D}h\,\mathrm{exp}\Big[\frac{\mathrm{i}}{2}\int \mathrm{d}^4x(-\frac{1}{2m}(\nabla h)^2\nonumber\\
&&-2q\sqrt{n}C_0 h-2\mu h^2-\frac{q^2n}{2m}\mathbf{C}^2-\frac{1}{4}\mathcal{F}_{\mu\nu}\mathcal{F}^{\mu\nu}\Big]\nonumber\\
&&\label{eq:ZHiggs1}
\end{eqnarray}
where the path (functional) integral measure $\mathcal{D}[...]$ integrates the functional $S$ over all configurations of the respective fields it depends on - just as the ordinary integral measure $\mathrm{d}x$ integrates a function $f(x)$ over all values $x$. While the mathematical handling of path integrals is often complicated, a tedious calculation shows that once a Lagrangian can be cast into a Gaussian integral, the path integral can be solved exactly\sidenote{\footnotesize{In short, this calculation relies on the solvable Gaussian integral 
\begin{equation*}
\int\mathrm{d}x\mathrm{e}^{-\frac{a}{2}x^2+bx}
\end{equation*}
which evaluates to $\sqrt{2\pi/a}\,\mathrm{e}^{b^2/2a}$.  For any bi-quadratic Lagrangian 
\begin{eqnarray*}
\mathcal{L}&=&\mathbf{\Phi}^\mathrm{T} \mathbf{K}(x)\mathbf{\Phi}\\
&&+2\mathbf{b}^\mathrm{T}(x)\mathbf{\Phi},
\end{eqnarray*}
where $\mathbf{\Phi}$ and $\mathbf{b}$ are $N$-component vectors and $\mathbf{K}$ is a symmetric $N\times N$ matrix, the path integral 
\begin{equation*}
\int \mathcal{D}\mathbf{\Phi}\,\mathrm{e}^{\frac{\mathrm{i}}{2}\int \mathrm{d}^4x\,\mathcal{L}[\mathbf{\Phi}]}
\end{equation*}
takes values
\begin{eqnarray*}
\mathrm{e}^{-\frac{\mathrm{i}}{2}\int \mathrm{d}^4x\,\mathbf{b}^\mathrm{T}\mathbf{K}^{-1}\mathbf{b}}\\
\times\frac{(2\pi\mathrm{i})^N}{\sqrt{\mathrm{det}\mathbf{K}}}
\end{eqnarray*}
.}}. Thus, we consider the part of the action just containing $h$ and find after partial integration
\begin{eqnarray}
S[h]&=&\int\mathrm{d}^4x\Big[-\frac{1}{2m}(\nabla h)^2 -2q\sqrt{n}C_0 h-2\mu h^2\Big]\nonumber\\
&=&-\frac{1}{2m}\Big[h\nabla h\Big]^\infty_{-\infty}\nonumber\\
&&+\int\mathrm{d}^4x\Big[\frac{1}{2m}h\nabla^2 h -2q\sqrt{n}C_0 h-2\mu h^2\Big]\nonumber\\
&=&\int\mathrm{d}^4x\Big[h\left(\frac{1}{2m}\nabla^2 -2\mu\right) h -2q\sqrt{n}C_0 h\Big]\nonumber\\
&&\label{eq:Shaction}
\end{eqnarray}
where we demanded the field $h$ to vanish for $|x|\to\infty$. Inserting this action back in Eq.~(\ref{eq:ZHiggs1}) and splitting $S=S[h]+S[\mathbf{C},C_0]$ gives
\begin{eqnarray}
Z&=&\iint \mathcal{D}\mathbf{C}\,\mathcal{D}C_0\int\mathcal{D}h\nonumber\\
&&\mathrm{exp}\left(\frac{\mathrm{i}}{2}\int\mathrm{d}^4x\Big\{h\big[\frac{1}{2m}\nabla^2 -2\mu\big] h -2q\sqrt{n}C_0 h\Big\}\right)\nonumber\\
&&\times\mathrm{exp}\left(-\frac{\mathrm{i}}{2}\int\mathrm{d}^4x\Big\{\frac{q^2n}{2m}\mathbf{C}^2+\frac{1}{4}\mathcal{F}_{\mu\nu}\mathcal{F}^{\mu\nu}\Big\}\right).\nonumber\\
&&
\end{eqnarray}
Assigning $\mathbf{K}\to\frac{1}{2m}\nabla^2-2\mu$ and $\mathbf{b}\to -q\sqrt{n}C_0$ according to above scheme, we obtain after integration
\begin{eqnarray}
Z&=&\frac{2\pi\mathrm{i}}{\sqrt{\mathrm{det}\mathbf{A}}}\iint \mathcal{D}\mathbf{C}\,\mathcal{D}C_0\nonumber\\
&&\mathrm{exp}\left(-\frac{\mathrm{i}}{2}\int\mathrm{d}^4x\,q\sqrt{n}C_0\frac{1}{\frac{1}{2m}\nabla^2-2\mu}q\sqrt{n}C_0\right)\nonumber\\
&&\times\mathrm{exp}\left(-\frac{\mathrm{i}}{2}\int\mathrm{d}^4x\Big\{\frac{q^2n}{2m}\mathbf{C}^2+\frac{1}{4}\mathcal{F}_{\mu\nu}\mathcal{F}^{\mu\nu}\Big\}\right)\nonumber\\
&&
\end{eqnarray} 
with the Lagrangian 
\begin{equation}
\mathcal{L}= \frac{q^2n}{2\mu}C_0^2-\frac{q^2n}{2m}\mathbf{C}^2-\frac{1}{4}\mathcal{F}_{\mu\nu}\mathcal{F}^{\mu\nu}\label{eq:LHiggs2}
\end{equation}
where we have neglected $\frac{1}{2m}\nabla^2$ against $2\mu$. In the same fashion we can also integrate-out the $C_0$ field so that we eventually end up with the low-energy Lagrangian only containing the vector potential $\mathbf{C}$ and higher-order derivatives thereof
\begin{equation}\label{eq:LowBosonL}
\mathcal{L}=-\frac{q^2n}{2m}\mathbf{C}^2+\mathrm{derivatives}.
\end{equation}
With this simple Lagrangian we can now turn to some real-life implications. In more detail, we are interested how the fundamental properties of superconductors can be deduced from the field theory we have constructed.
\subsection*{Currents}
In general, the current is obtained by performing a functional derivative of the Lagrangian with respect to the gauge field, i.e. we calculate
\begin{eqnarray}
\mathbf{j}=\frac{\delta\mathcal{L}}{\delta\mathbf{C}}.
\end{eqnarray}
To account for the freedom of gauge, we unwind the gauge we performed in Eq.~(\ref{eq:LHiggs3}) setting $\mathbf{C}\to \mathbf{C}-\frac{1}{q}\nabla\vartheta$ and obtain
\begin{equation}
\mathcal{L}=-\frac{q^2n}{2m}\mathbf{C}^2-\frac{n}{2m}(\nabla \vartheta)^2+\frac{qn}{m}\mathbf{C}\nabla\vartheta.
\end{equation}
Performing the functional derivative\sidenote{\footnotesize{Alternatively, one can also derive with respect to $\nabla\vartheta$ to obtain the particle current, which after multiplication with the charge $q$ is the electrical current.}} yields
\begin{equation}
\mathbf{j}=\frac{qn}{m}\left(\nabla\vartheta - q\mathbf{C}\right)\label{Eq:London}
\end{equation}
which is the London equation. The current in a superconductor can arise from coupling to a gauge field (i.e. upon light irradiation) or as result of a phase gradient of the superconducting order parameter. A close examination of the second mechanism explains the dissipation-free nature of transport as we will see in the next paragraph.

\subsection*{Dissipation-free transport}
We can understand the vanishing transport resistance of a superconductor as a consequence of the finite vacuum expectation value of the order parameter $\Psi$ and its topology \cite{Lancaster2014}. Remember, that in the superconducting state, we can write the order parameter as $\Psi=\sqrt{n}\mathrm{e}^{\mathrm{i}\vartheta}=\mathrm{const.}$ throughout the system. We can think of the condensate to life inside a box of dimension $L\times L\times L$ where we impose periodic boundary conditions. In, e.g., $x$-direction, this implies that $\Psi(x=0)=\Psi(x=L)$. A current in $x$-direction is induced if we supply the charge carriers in this direction with momentum $p$, that is, by applying a Lorentz-boost transformation  $\Psi\to\mathrm{e}^{\mathrm{i}px}\Psi$. At the boundaries, we therefore obtain $\Psi(x=0)=\mathrm{e}^{\mathrm{i}pL}\Psi(L)$ and thus $\sqrt{n}\mathrm{e}^{\mathrm{i}\vartheta}=\sqrt{n}\mathrm{e}^{\mathrm{i}(\vartheta+pL)}$. In other words, as we walk along the condensate in $x$-direction, the current imposes a twist $t(x)=px$ to the ground state phase $\vartheta$ that must fulfill $pL=2\pi k$ with $k=1,2,3,...$ to ensure periodic boundary conditions. A phase gradient, as shown above, always leads to a current. 

Demanding the carrier flow to slow down on the way from $x=0$ to $x=L$, i.e. to be resistive, is equal to demanding the phase gradient to be zero. This, however, is problematic as the current itself necessitates a phase twist of least $2\pi$ between $x=0$ and $L$ as shown above. The only way to achieve $\nabla\vartheta=0$ is by a rupture of the condensate which is energetically extremely costly and thus unlikely\sidenote{\footnotesize{The probability can be estimated by the width of the zero-frequency $\delta$-function of the dynamical conductivity $\sigma_1(\omega)$ we will deal with later on.}} to happen: the current just keeps on flowing. The creation of vortices, being topological local defects in the phase field, may remove the phase twist and vortex movement can dissipate current.

\subsection*{Meissner-Ochsenfeld effect}
We have seen that after symmetry is broken, the vector gauge field $A^\nu$ has acquired a mass term. The relevant part of the full Lagrangian Eq.~(\ref{eq:LHiggs4}) reads\sidenote{\footnotesize{remember that $\mathcal{F}_{\mu\nu}=\partial_\mu A_\nu-\partial_\mu A_\nu$}}
\begin{equation}
\mathcal{L}_A=-\frac{1}{4}\left(\partial_\mu A_\nu-\partial_\nu A_\mu\right)\left(\partial^\mu A^\nu-\partial^\nu A^\mu\right)-\frac{q^2 n}{2m}A_\nu A^\nu.
\end{equation}
We can find the dynamics of this field by applying the Euler-Lagrange equation
\begin{equation}
\partial_\mu\frac{\partial\mathcal{L}}{\partial (\partial_\mu A_\nu)}-\frac{\partial \mathcal{L}}{\partial A_\nu}=0
\end{equation}  
resulting in the differential equation of motion or \emph{Proca} equation
\begin{equation}
-\frac{1}{4}\partial_\mu\left(\partial^\mu A^\nu-\partial^\nu A^\mu\right)+\frac{q^2 n}{2m} A^\nu=0
\end{equation}
which can be cast\sidenote{\footnotesize{By usage of the Lorenz gauge $\partial_\mu A^\mu=0$}} into 
\begin{equation}
\left(\partial_\mu\partial^\mu-\frac{2q^2n}{m}\right)A^\nu=0.
\end{equation}
In the static limit, $\partial^0 A^0=0$ we can concentrate on the space-like components and rewrite this result using the gradient operator
\begin{equation}
\nabla^2\mathbf{A}=\frac{2q^2 n}{m}\mathbf{A}\label{VectorHiggs}
\end{equation}
which is solved by an exponentially decaying vector potential measured from the boundary of the superconductor and the very essence of the Meissner-Ochsenfeld effect\sidenote{\footnotesize{To make a closer contact to the usual notion of the Meissner-Ochsenfeld effect, one can rewrite the left side of Eq.\,(\ref{VectorHiggs}) using the identity $\nabla^2 \mathbf{A}=\nabla(\nabla \mathbf{A})-\nabla\times\nabla\times\mathbf{A}$, the Lorenz gauge, and Eq.\,(\ref{Eq:London}) leading to $\nabla^2\mathbf{A}\propto \mathbf{B}$.}}. In analogy with the massive $W^\pm$ and $Z^0$ bosons of the short-range weak interaction, it is often said the photons (i.e. the quanta of the field $A^\nu$) can penetrate the superconductor only on very short distances as they rapidly decay by virtue of their mass they acquire in the interior of the superconductor \cite{Lancaster2014}.  
\clearpage

\section{Green functions and the Usadel equation}\label{Sec:Green}
Once the Green functions of a given system are known, basically any physical quantity may be derived from them. For any practical application, the exact form of the Green functions must be determined by solving equations which contain the systems characteristics and account for appropriate boundary conditions. The most fundamental equation for the case of superconductivity was first derived\sidenote{\footnotesize{Other than Bardeen, Cooper, and Schrieffer, who started with a second-quantization approach to superconductivity, Gor'kov worked with quantum field theoretical Green function techniques - a formalism very popular among Russian theorists at this time - and solved the problem almost at the same time. Though credit usually is given to BCS,  major advances in our understanding of superconductivity are based on the work of Gor'kov \cite{Cooper2011}.}} by Gor'kov \cite{Gorkov58,Gorkov59}. The Gor'kov equation, however, is too complicated to be used for practical applications. Eilenberger simplified the Gor'kov equation for the case $\Delta\ll E_F$ (or equivalently $\xi\gg\lambda_F$), where the highly oscillatory dynamics $\sim\mathrm{exp}(iE_Ft/\hbar)$ can be removed from the equations and taking (quasi-classically) averaged Greens functions \cite{eilenberger68}. The resulting Eilenberger equations were simplified even more by Usadel \cite{Usadel70} when applied to systems in the dirty limit where electron transport is of rather diffusive than ballistic nature. Here, pairing takes place between electrons traveling on diffusive rather than ballistic trajectories so that, loosely speaking, the constitutes of an Cooper pair are expected to scatter multiple times before losing phase coherence \cite{Fominov2016}. The particles momentum therefore may be averaged over all directions rendering the problem essentially isotropic. The obtained Usadel equation for the superconducting matrix Greens function $\check{g}$ reads\sidenote{\footnotesize{What follows is mainly taken from Ref.~\cite{Fominov2011}. For a better understanding, however, some of the calculations of  Ref.~\cite{Fominov2011} are carried out in greater detail here.}} \cite{Fominov2011,Rammer86,Larkin86}        
\begin{equation}\label{eq:Usadel}
iE[\hat{\tau}_3\otimes\hat{\sigma}_0, \check{g}]-\Delta[\hat{\tau}_1\otimes\hat{\sigma}_0, \check{g}]+i[\widehat{\Sigma},\check{g}]=0
\end{equation}  
where $[\cdot,\cdot]$ is a commutator, $\tau_i$ and $\sigma_i$ are the Pauli matrices of the $2\times 2$ Nambu- and spin-spaces\sidenote{\footnotesize{Note that  $\hat{\tau}_i$ and $\hat{\sigma}_i$ are defined on different Hilbert spaces, the tensorial product $\hat{\tau}_i\otimes\hat{\sigma}_j$ thus is a 4x4 matrix on the product space $N\otimes S$. }} $N$ and $S$, and $\Delta$ is the pairing energy. The term $[\widehat{\Sigma},\hat{g}]$ is a self-energy term to renormalize the spectrum due to, e.g., paramagnetic-impurity or spin-orbit scattering, where time-reversal symmetry is not preserved.
Still, this formalism can be applied to strong potential scattering off non-magnetic impurities, where the Anderson theorem breaks down, although potential scattering is fully time-reversal symmetric \cite{Fominov2016}. The equivalence of strong potential scattering and the (paramagnetic) pair-breaking  parameter is established in a two step procedure: First, potential disorder is linked to fluctuations of the pairing field $\Delta(\mathbf{r})$ \cite{Feigelman2012} and second these fluctuations are related to an effective pair-breaking parameter \cite{Larkin1971}.    \\
The Usadel equation is formally solved by the matrix Green function
\begin{equation}
\check{g}=\begin{pmatrix}
\hat{g}^R & \hat{g}^K \\ 0 & \hat{g}^A
\end{pmatrix}
\end{equation}
that comprises different types of Green functions, i.e. the retarded (R), advanced (A) and Keldysh (K) Greens functions, which can be transformed into each other 
\begin{eqnarray}
\hat{g}^R&=&\hat{\tau}_3\otimes\hat{\sigma}_0G+\hat{\tau}_1\otimes\hat{\sigma}_0F\label{retarded}\\
\hat{g}^A&=&-\hat{\tau}_3\otimes\hat{\sigma}_0\hat{g}^{R\dagger}\hat{\tau}_3\otimes\hat{\sigma}_0\label{advanced}\\
\hat{g}^K&=&\left(\hat{g}^R-\hat{g}^A\right)\times\tanh(E/2k_\mathrm{B}T)\label{keldysh}
\end{eqnarray} 
and eventually expressed in terms of the normal (quasiparticle) and the anomalous (pair) Green functions, $G$ and $F$. These functions are one-particle propagators of the type $\langle\psi_2\psi_1^\dagger\rangle$ or $\langle\psi_2^\dagger\psi_1\rangle$ for $G$ and $\langle\psi_2\psi_1\rangle$ or $\langle\psi_2^\dagger\psi_1^\dagger\rangle$ for $F$. 
We see, that the normal propagator describes creation (annihilation) and subsequent annihilation (creation) of a particle (hole) and gives non-zero contributions in any case, whereas the anomalous propagator describes the creation (annihilation) of two particles or holes which can only be non-zero in the case of pairing, i.e. the superconducting state. Despite the complicated structure of the full matrix Green function, it is effectively just two kinds of Green functions, $G$ and $F$, that carry the physical information. \\
The retarded Green function obeys the normalization constraint $\hat{g}^R\hat{g}^R=\hat{\tau}_0\otimes\hat{\sigma}_0=1$. Expressing $\hat{g}^R$ in terms of $G$ and $F$ yields
\begin{eqnarray}
\hat{\tau}_0\otimes\hat{\sigma}_0&=&\hat{g}^R\hat{g}^R\nonumber\\
&=&\hat{\tau}_0^2\otimes\hat{\sigma}_0^2G^2+\hat{\tau}_0^2\otimes\hat{\sigma}_0^2F^2\nonumber\\
&&+(\hat{\tau}_3\hat{\tau}_1)\otimes\hat{\sigma}_0^2FG+(\hat{\tau}_1\hat{\tau}_3)\otimes\hat{\sigma}_0^2GF\nonumber\\
&=&\hat{\tau}_0\otimes\hat{\sigma}_0G^2+\hat{\tau}_0\otimes\hat{\sigma}_0F^2\nonumber\\
&&+i\hat{\tau}_2\otimes \hat{\sigma}_0FG-i\hat{\tau}_2\otimes\hat{\sigma}_0GF\nonumber\\
&=&\hat{\tau}_0\otimes\hat{\sigma}_0(G^2+F^2)
\end{eqnarray}
from which the common parametrization $G=\cos(\theta)$ and $F=\sin(\theta)$ follows. The angle $\theta(E,\Delta)$ is a complex function termed proximity angle. The whole problem of finding a set of Green functions solving the Usadel equation is transferred to the determination of the function $\theta(E,\Delta)$.  
Once $\theta(E,\Delta)$ is found, all superconducting properties (such as the density of quasi-particle or pair states, the coherence functions, the current density, or the dynamical conductivity) can be calculated following standard procedures. 
The Usadel equation (\ref{eq:Usadel}) may now be simplified drastically. With the Green functions $G$ and $F$ defined above, the quasiparticle commutator becomes
\begin{eqnarray}
iE[\hat{\tau}_3\otimes\hat{\sigma}_0,\hat{g}]&=&iE[\hat{\tau}_3\otimes\hat{\sigma}_0,\hat{\tau}_3\otimes\hat{\sigma}_0G]\nonumber\\
&&+iE[\hat{\tau}_3\otimes\hat{\sigma}_0,\hat{\tau}_1\otimes\hat{\sigma}_0F]\nonumber\\
&=&iE\big((\hat{\tau}_3\otimes\hat{\sigma}_0)(\hat{\tau}_1\otimes\hat{\sigma}_0F)\nonumber\\
&&-(\hat{\tau}_1\otimes\hat{\sigma}_0F)(\hat{\tau}_3\otimes\hat{\sigma}_0)\big)\nonumber\\
&=&iE\left(\hat{\tau}_3\hat{\tau}_1\otimes\hat{\sigma}_0F-\hat{\tau}_1\hat{\tau}_3\otimes\hat{\sigma}_0F\right)\nonumber\\
&=&-2E(\hat{\tau}_2\otimes\hat{\sigma}_0F)\nonumber\\
&=&-2E(\hat{\tau}_2\otimes\hat{\sigma}_0\sin\theta) \label{CommEnergy}
\end{eqnarray}
and the pair commutator gives in similar fashion
\begin{eqnarray}
\Delta[\hat{\tau}_1\otimes\hat{\sigma}_0, \hat{g}]&=&-2i\Delta(\hat{\tau}_2\otimes\hat{\sigma}_0G)\nonumber\\
&=&-2i\Delta(\hat{\tau}_2\otimes\hat{\sigma}_0\cos\theta) \label{CommDelta}.
\end{eqnarray}
The self-energy upon strong potential or spin-flip scattering (with scattering time $\tau_s$ and $\xi$ a dimensionless parameter describing the 'strength' of a single magnetic impurity) is
\begin{equation}
\widehat{\Sigma}_s=\frac{i}{2\tau_s}\frac{1}{\hat{\tau}_0\otimes\hat{\sigma}_0+\xi^2\hat{\tau}_3\otimes\hat{\sigma}_0\check{g}\hat{\tau}_3\otimes\hat{\sigma}_0\check{g}}\hat{\tau}_3\otimes\hat{\sigma}_0\check{g}\hat{\tau}_3\otimes\hat{\sigma}_0\label{self}.
\end{equation}
The entire calculation of the the self-energy term in Eq. (\ref{eq:Usadel}) is lengthy, but straight forward\sidenote{\footnotesize{The central manipulations rely on the series expansion
\begin{equation*}
\frac{1}{1-\hat{M}}=1-\hat{M}+\hat{M}^2-...
\end{equation*}
where $\hat{M}$ is a matrix in $N\otimes S$ space.}}. Here, we will just state the result
\begin{equation}
i[\widehat{\Sigma},\check{g}]=-\frac{i}{\tau_s}\hat{\tau}_2\otimes \hat{\sigma}_0\frac{\sin2\theta}{1+\xi^4+2\xi^2\cos2\theta} \label{CommSelf}.
\end{equation}
We are now equipped to rephrase the Usadel equation (\ref{eq:Usadel}) in terms of the complex proximity angle $\theta(E)$ using the commutators. Note, that all commutators have the identical $N\otimes S$ structure, so that each component of the actual $4\times4$ matrix equation (\ref{eq:Usadel}) reads
\begin{equation}\label{Usadel}
iE\sin\theta+\Delta\cos\theta-\frac{1}{2\tau_s}\frac{\sin2\theta }{1+\xi^4+2\xi^2\cos 2\theta}=0.
\end{equation}\\
This version of the Usadel equation is solvable for $\theta$, however, any solving procedure suffers from the periodicity of trigonometric functions and hence is numerically cumbersome. A more straight representation of Eq.~(\ref{Usadel}) is found using renormalized energy and order parameter,
\begin{eqnarray}
\tilde{E}&=&E+\frac{i}{2\tau_s}\frac{\cos\theta }{1+\xi^4+2\xi^2\cos 2\theta}\\
\tilde{\Delta}&=&\Delta-\frac{1}{2\tau_s}\frac{\sin\theta }{1+\xi^4+2\xi^2\cos 2\theta}
\end{eqnarray}
\begin{figure}[t!]
\begin{centering}
\includegraphics[width=\textwidth]{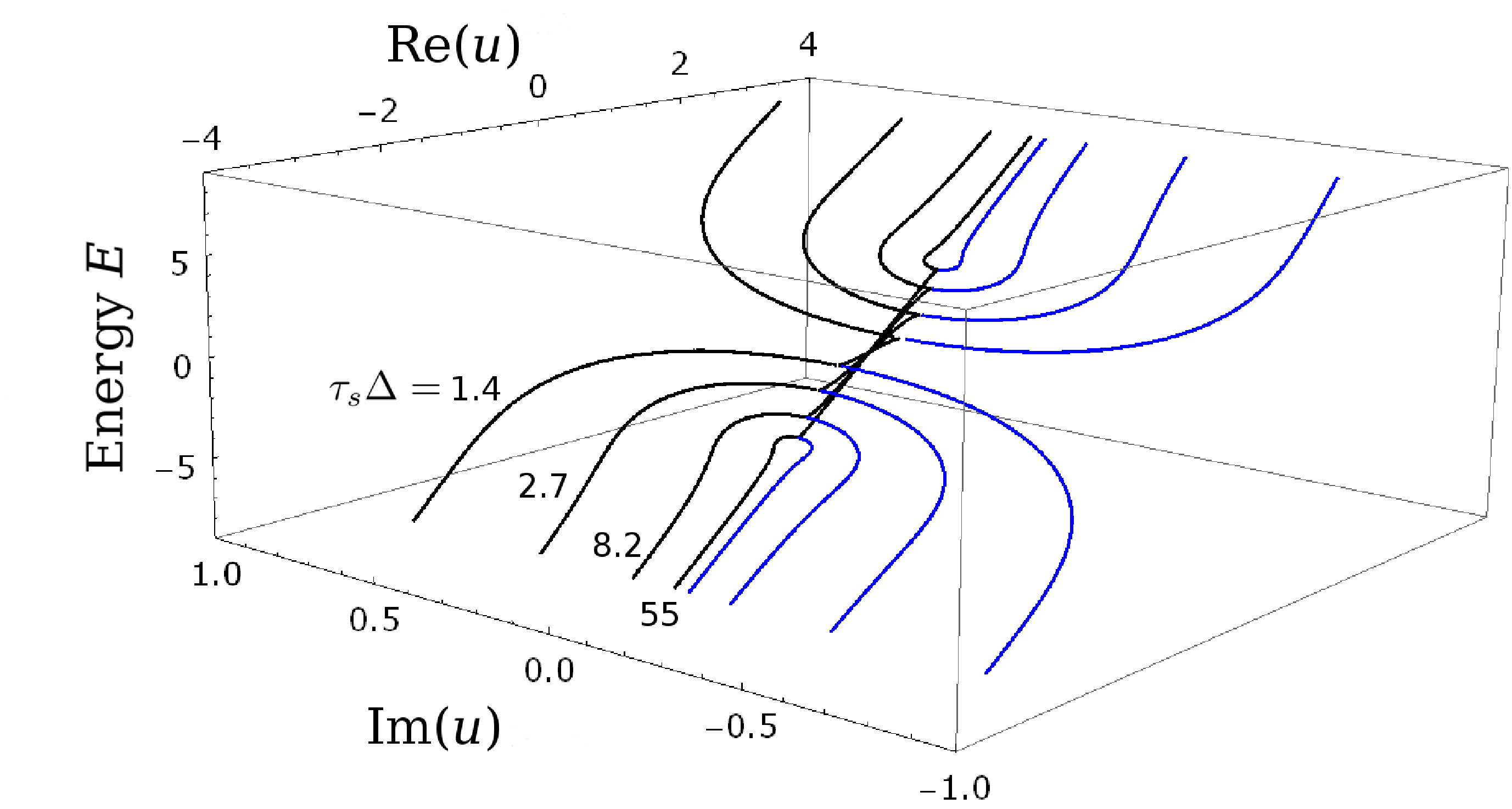}
\caption{\label{fig:UsadelSolutions}\textbf{Complex solutions of Usadel equation for various pair-breaking values $\tau_s$.} Purely real and complex solutions correspond to a vanishing and finite quasiparticle density-of-states (DOS), respectively. With increasing pair-breaking scattering $\tau_s$, the energy window with purely real solutions shrinks leading to a reduction of the band edge $\Omega_g$ despite a constant pairing amplitude $\Delta$. Note that only the branch with $\mathrm{Im} (u)>0$ yields a physically meaningful dynamical conductivity.}
\end{centering}
\end{figure} 
\noindent so that Eq.~(\ref{Usadel}) becomes
\begin{equation}
i\tilde{E}\sin\theta+\tilde{\Delta}\cos\theta=0.
\end{equation}
The normal and anomalous Greens functions $G=\cos\theta$ and $F=\sin\theta$ can now be parametrized in terms of a renormalized paramater  $u=\tilde{E}/\tilde{\Delta}$,  
\begin{eqnarray}
G&=&\frac{u}{\sqrt{u^2-1}}\label{GreenG}\\
F&=&\frac{i}{\sqrt{u^2-1}}\label{GreenF}
\end{eqnarray}
which is readily obtained by solving
\begin{equation}\label{Usadelu}
u\left(1+\frac{1}{(1+\xi^2)^2\tau_s\Delta}\frac{\sqrt{1-u^2}}{u^2-\left|\frac{1-\xi^2}{1+\xi^2}\right|^2}\right)-\frac{E}{\Delta}=0.
\end{equation}
Finding a solution $u$ of Eq.~(\ref{Usadelu}) is simpler than directly solving Eq.~(\ref{Usadel}), but still some care must be taken in order to filter the correct \emph{physical} solution from the \emph{mathematical} solutions. The solutions to Eq.~(\ref{Usadelu}) as function of $E$ are plotted in Fig.~\ref{fig:UsadelSolutions} for various pair breaking parameters  $\tau_s\Delta=1.4-55$, while $\Delta=3.7$ and $\xi=10^{-4}$ are kept constant. Common to all sets of solutions for a given $\tau_s$ is that there are energy regimes, where the solutions are either purely real or complex. While all real solutions are unique, this does not hold for complex ones:  Fig.~\ref{fig:UsadelSolutions} displays how the solution set splits into two branches (colored black and blue) once a $\tau_s$-specific energy is reached. Though being general solutions, only the ones with $\mathrm{Im}(u)>0$ lead to physically meaningful Green functions in the given context as discussed in the next paragraph.
\section{Density of states and dynamical conductivity}\label{Density of states and dynamical conductivity}
\begin{figure}[b!]
\begin{centering}
\includegraphics[width=\textwidth]{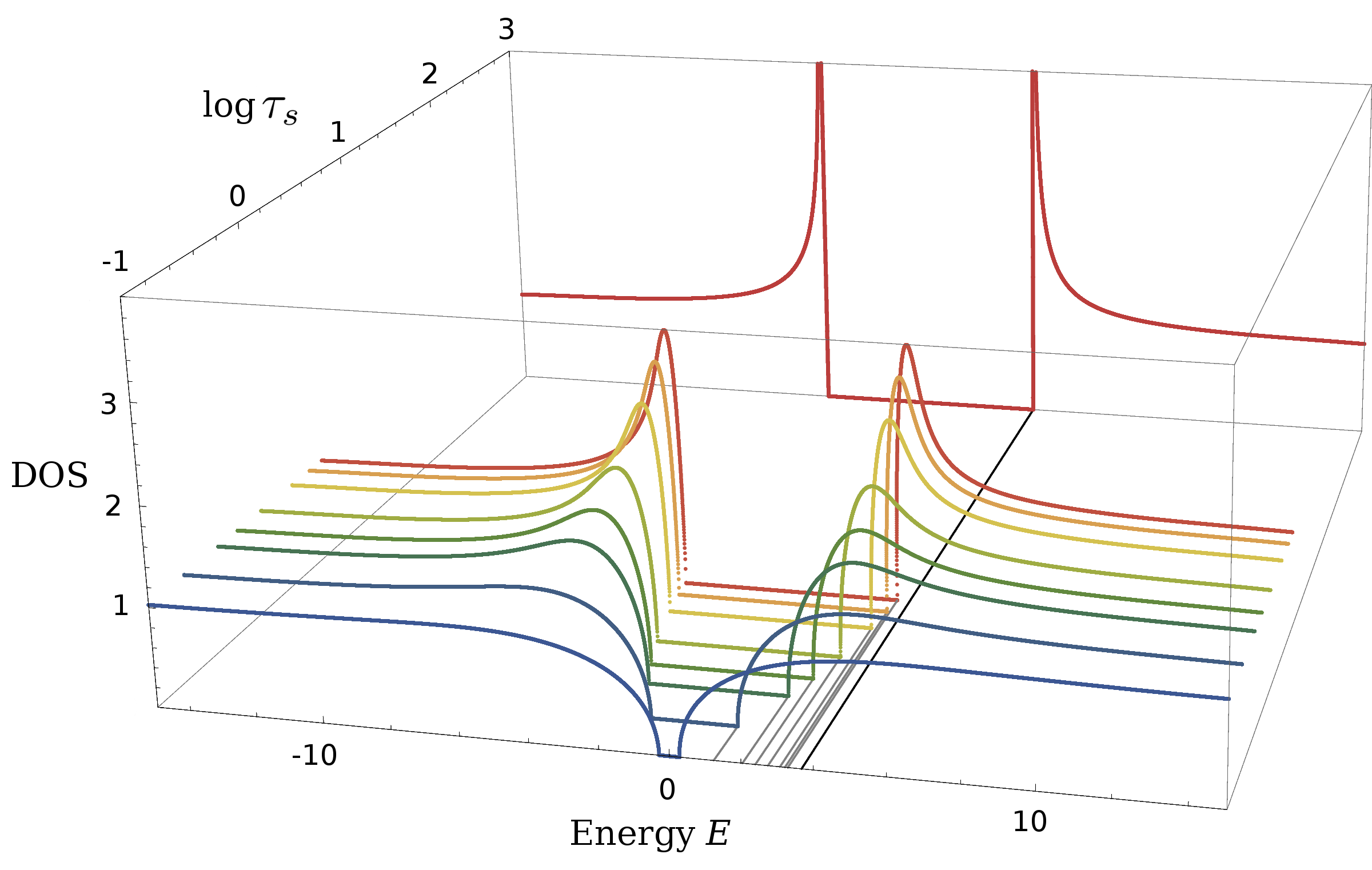}
\caption{\label{fig:Fominov_DOS}\textbf{Quasiparticle density of states for various pair-breaking values $\tau_s$}. Although the order parameter $\Delta$ is constant for each curve, the band edge $\Omega_g$ decreases considerably below the BCS value $\Omega_g=2\Delta$ with increasing pair-breaking strength. At the same time, the coherence peaks are smeared out. For the strongest pair-breaking ($\tau_s\Delta=1.4$) the coherence peaks are barely present and the spectral gap is with $0.04\times 2\Delta$ almost closed.}
\end{centering}
\end{figure}
On general grounds, the quasiparticle density of states $\mathcal{D}$ is defined as 
\begin{eqnarray}\label{eq:DOS}
\mathcal{D}&\equiv&\frac{1}{8}\mathrm{tr}\left((\hat{\tau}_3\otimes\hat{\sigma}_0)[\hat{g}^R-\hat{g}^A]\right)
\end{eqnarray}
We can use the above expressions (\ref{advanced}) and (\ref{retarded}) to formulate $\mathcal{D}$ in terms of $G$ and $F$ in order to express Eq.\,(\ref{eq:DOS}) in terms of the proximity angle
\begin{eqnarray}
8\mathcal{D}&=&\mathrm{tr}\big\{(\hat{\tau}_3\otimes\hat{\sigma}_0)[\hat{\tau}_3\otimes\hat{\sigma}_0G+\hat{\tau}_1\otimes\hat{\sigma}_0F\nonumber\\
&&+\hat{\tau}_3\otimes\hat{\sigma}_0G^\dagger-\hat{\tau}_1\otimes\hat{\sigma}_0F^\dagger]\big\}\nonumber\\
&=&\mathrm{tr}\left\{(\hat{\tau}_3\otimes\hat{\sigma}_0)[\hat{\tau}_3\otimes\hat{\sigma}_02\mathrm{Re}G+\hat{\tau}_1\otimes\hat{\sigma}_02i\mathrm{Im}F]\right\}\nonumber\\
&=&\mathrm{tr}\left\{\hat{\tau}_3^2\otimes\hat{\sigma}_0^22\mathrm{Re}G+\hat{\tau}_3\hat{\tau}_1\otimes\hat{\sigma}_02i\mathrm{Im}F\right\}\nonumber\\
&=&\mathrm{tr}\left\{\hat{\tau}_0\otimes\hat{\sigma}_02\mathrm{Re}G+(-i)\hat{\tau}_2\otimes\hat{\sigma}_02i\mathrm{Im}F\right\}\nonumber\\
&=&8\mathrm{Re}G\nonumber\\
\mathcal{D}&=&\mathrm{Re}\left(\cos\theta\right)\label{eq:DOS}
\end{eqnarray}
In the second last step we have taken the partial trace over the spin space ($4\mathrm{Re}G$) and multiplied with the partial space over the Nambu space (2). The second term vanished due to the traceless Pauli matrix $\hat{\tau}_2$. Apparently, the quasiparticle DOS is simply the real part of the normal Green function. Similarly, on finds the density of pairs, $\mathcal{P}$, as the imaginary part of the anomalous Green function upon replacing $\hat{\tau}_3$ by $\hat{\tau}_1$ in the calculation leading to Eq.~(\ref{eq:DOS}). Both play a crucial role in the dynamical conductivity. The quasiparticle DOS for various pair-breaking parameters $\tau_s$ is shown in  Fig.~\ref{fig:Fominov_DOS}. In the BCS limit ($\tau_s\Delta=10^3$) the well-known sharp band edge at $\Omega_g=\Delta$ and pronounced coherence peaks are recovered. For moderate pair-breaking, $\tau_s\Delta=54$, the peaks are rounded off and $\Omega_g$ shifts slightly below $\Delta$, and consequently, reduces the spectral gap below the BCS value $2\Delta$. Further increase of pair breaking progressively smears out the peaks and relocates states below $\Delta$. At extreme pair-breaking, $\tau_s\Delta=1.4$, the spectral gap is almost completely closed and the coherence peaks are smeared entirely into the quasiparticle continuum. In this limit, the dynamical conductivity is strongly modified compared to the BCS case, as we will see below.  \\   

The response kernel $Q(\omega)$, that links the supercurrent to the electromagnetic vector potential reads \cite{Fominov2011}
\begin{equation}
Q(\omega)=-\frac{i\sigma_0}{8}\int\limits_{-\infty}^\infty\mathrm{d}E\,\mathrm{tr}\left[(\hat{\tau}_3\otimes\hat{\sigma}_1)\check{g}(E)(\hat{\tau}_3\otimes\hat{\sigma}_0)\check{g}(E-\hbar\omega)\right]
\end{equation}
from which the conductivity follows directly as $\hat{\sigma}(\omega)=iQ/\omega$. After expressing the matrix Green function $\check{g}$ in terms of the normal and anomalous ones and taking the traces over Nambu- and spin spaces we arrive at
\begin{eqnarray}
\hat{\sigma}(\omega)&=&\frac{\pi e^2 n_s}{m}\delta(\omega)\nonumber\\
&&-\frac{\sigma_0}{2\omega}\int\limits_{0}^\infty\mathrm{d}E\Big\{\tanh\frac{ E-\frac{\hbar\omega}{2}}{2k_\mathrm{B}T}\big[G^{+}G^{-}_1-iF^{+}F^{-}_2\big]\nonumber\\
&&-\tanh\frac{E+\frac{\hbar\omega}{2}}{2k_\mathrm{B}T}\big[(G^{-})^*G^{+}_1+i(F^{-})^*F^{+}_2\big]\Big\}\nonumber\\
&&\label{FominovCond}
\end{eqnarray}
where we understand subscripts 1 and 2 to denote real and imaginary parts and used the short-hand notation $\{G,F\}^\pm=\{G,F\}(E\pm\hbar\omega/2)$. Note also that we included the delta-response of the superfluid manually.

\chapterend

\chapter{Experimental studies on disordered NbN thin films}\label{Sec:NbN}
\thispagestyle{empty}
\begin{flushright}
\footnotesize{
Now, what do you own the world?\\
How do you own disorder, disorder?\\
Somewhere between the sacred silence and sleep\\
Disorder, disorder, disorder\\[8pt]
\emph{System of a Down}}
\end{flushright}
\emph{Content of this chapter are optical studies on NbN thin films on approach of the superconductor-insulator quantum phase transition (SIT) providing experimental evidence for the existence of the Higgs mode. We will start our discussion in Sec.\,\ref{SIT} with some theoretical considerations regarding superconductivity and disorder followed by a review of hallmark tunneling-spectroscopy studies on homogeneously disordered SIT materials in Sec.\,\ref{SITExp}. Subsequently in Sec.\,\ref{Sec:ExpNbN}, we will discuss measurements of the dynamical conductivity of a series of NbN films covering the range from clean to strongly disordered and compare the results with measurements of the (differential) tunneling conductance. We will see that the mutual analysis within a model inspired by the paramagnetic pair-breaking theory of Abrikosov and Gor'kov reveals a discrepancy between both spectroscopies that cannot be explained by pair-breaking effects only but rather calls for for radically new ideas. Given the specific nature of short coherence-length quasi-2D NbN near quantum criticality we will in Sec.\,\ref{Sec:NbNHiggs} present an explanation involving a excitation of the superconducting order parameter namely the Higgs mode - an condensed-matter realization of what is known as Higgs boson in particle physics yet sharing the same origin: spontaneous symmetry breaking. We will discuss experimental results justifying this interpretation, yet also open questions concerning the visibility of the Higgs mode in disordered systems, Sec.\,\ref{NbNremarks}.    }
\clearpage

\section{The superconductor-insulator \\quantum phase transition}\label{SIT}
A prototypical quantum phase transition in condensed matter is the transition between superconducting and insulating ground states of an electronic system at zero temperature in two spatial dimensions (2D). This superconductor - insulator transition (SIT) can be tuned by various parameters, most importantly lattice impurities acting as source for weak localization effects or potential traps strongly localizing charge carriers. The special appeal of the SIT and associated phenomena lies in its intermediate position framed by two of the most fundamental paradigms of condensed matter physics, namely the Anderson theorem for superconductivity and the Anderson localization each of which breaks down at the SIT giving rise to intriguing physics vivified by the interplay of localization and condensation. In what follows, we will briefly examine both paradigms and qualitatively introduce the two fundamental scenarios possibly leading to the cease of superconductivity at the quantum critical point\sidenote{\footnotesize{For an comprehensive review of this topic the interested reader is referred to the reviews of Gantmakher and Dolgopolov \cite{gant10} and Lin, Nelson and Goldman \cite{goldman15}. Also, the textbook by Dobrosavljevic, Trivedi, and Valles \cite{Dobrosavljevic2012} provides a valuable collection of articles introducing theoretical approaches and experimental hallmarks.    }}.  

\subsection{The insulating side: scaling theory of conductivity}
Maybe the most stunning aspect of the SIT in 2D lies in the direct transition between two ground states, which could not be any more different. On the one side, a coherent many-body
ground state composed of delocalized Cooper pairs (a superconductor) and, on the other side, a ground state with incoherent and localized quasiparticles (an insulator) presumably \emph{without} an intermediate ground state of incoherent but delocalized quasiparticle states (a metal). To understand the absence of this intermediate metallic state one can employ the \emph{single parameter scaling} argument \cite{Abrahams1979} suggested by Abrahams, Anderson, Licciardello, and Ramakrishnan\sidenote{\footnotesize{also known as the \emph{gang of four}}} which is an elegant application of the renormalization-group ideas: Start with a cube of spatial dimension $d=1,2,3$ and side lengths $L$, then find expressions for the conductance, define a scaling function thereof and study the general implications by sending $L\to\infty$. 
\begin{figure}[b!]
\begin{centering}
\includegraphics[width=\textwidth]{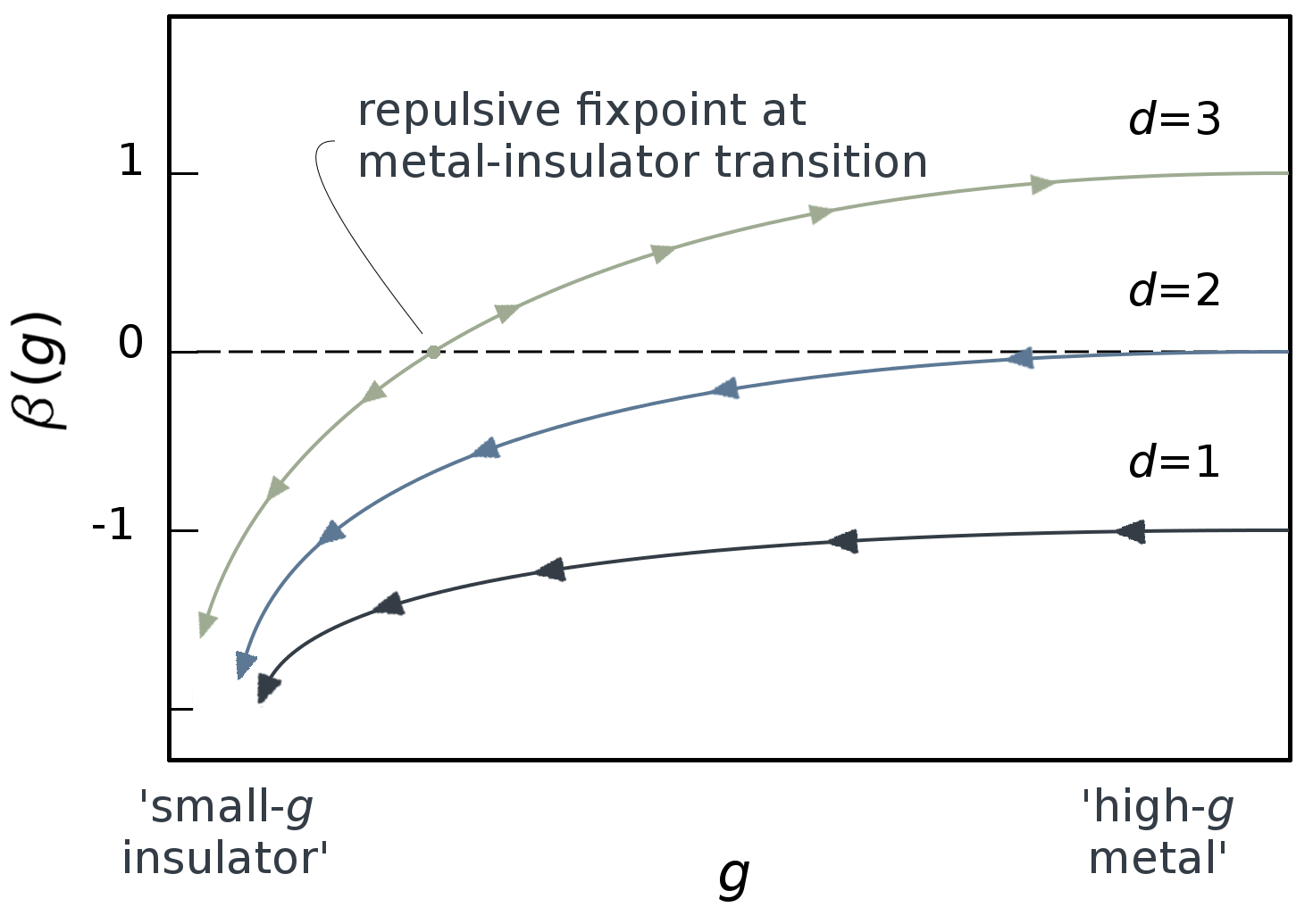}
\caption{\label{fig:single-parameter-scaling}single-parameter-scaling function $\beta$ of a d-dimensional cube $L^d$ versus conductance interpolating between logarithmic decay (insulating limit) and saturation (metallic limit) in 1,2, and 3 spatial dimensions. Arrows indicate the renormalization flow in the limit $L\to\infty$. For $d=1$ and 2 any amount of disorder will always favor the insulating ground state over the metallic one, whereas for $d=3$ the system has a metal-to-insulator transition at an repulsive fix point (green dot).}
\end{centering}
\end{figure}
For a metal in the high-conductivity ($\sigma$) limit the conductance $g$ is simply 
\begin{equation}
g_m=\sigma L^{d-2}\label{eq:condMet}
\end{equation} 
which is easily verified for the dimensions above. For insulators in the low-conductivity localization limit, one can expect a conductance which is exponentially damped on a characteristic length scale $\xi$ 
\begin{equation}
g_i=g_ce^{-\nicefrac{L}{\xi}}.\label{condIns}
\end{equation}
While at intermediate regimes the particular form of $g$ surely depends on details of electron transport and incipient  localization, we will see that the asymptotic limits Eqs.~(\ref{eq:condMet}) and (\ref{condIns}) are sufficient to understand the essential physics. For a renormalization treatment removing the explicit dependence on  $L$, Abrahams \emph{et al.} suggested the scaling function
\begin{equation}\label{eq:Abrahamscaling}
\beta(g)\equiv\frac{\mathrm{d}~\mathrm{ln}g}{\mathrm{d}\mathrm{ln}L}.
\end{equation}   
Inserting the explicit expressions Eq.~(\ref{eq:condMet},\ref{condIns}) gives the limits
\begin{eqnarray}
\lim_{g\to\infty}\beta(g)&=&d-2,\\
\lim_{g\to 0}\beta(g)&=&\mathrm{ln}\left(\frac{g(L)}{g_c}\right).
\end{eqnarray} 
Figure~\ref{fig:single-parameter-scaling} displays schematically $\beta(g)$ for $d=$1,2, and 3 dimensions. In the insulating limit, all curves collapse irrespective of dimensionality, while in the metallic limit, the asymptotes are different and only the 3d case features a zero-crossing at a critical conductance $g_c$. Now we consider the \emph{renormalization flow} upon coarse graining, indicated by the arrows in Fig.~\ref{fig:single-parameter-scaling}. For small disorder, we will obtain a conductance $g_0>g_c$,  and sending $L\to\infty$ for $d=3$ will increase the conductance as $g=\sigma L$ pushing the system towards the metallic limit, whereas for $g_0<g_c$ at high disorder the conductance will rather decrease dragging the system towards to insulating regime. The system is said to \emph{flow} away from the repulsive fix point at $g_c$ either towards metallic or insulating ground states depending on the specific disorder. This is the so-called  metal-to-insulator quantum phase transition (MIT) at the mobility edge $g_c$ purely driven by disorder. In $d=1$ and 2 the situation is distinctly different. Here, increasing $L$ will leave the conductance unaffected $g=\sigma$ ($d=2$) or reduce it as $g=\sigma/L$ ($d=1$). The drag towards the insulator cannot be compensated by an attractive metallic limit so that even the smallest amount of disorder will unavoidably cause a flow towards the insulating ground state. In other words, in dimensions $d<3$ \emph{no} metallic ground state is possible at $T=0$. Note that the above reasoning does not rely on specific details of $g(L)$. It can be shown, that quantum corrections to transport will affect the specific functional forms, but there is \emph{never} a mobility edge in less than 3 spatial dimensions \cite{Abrahams1979}. 

\subsection{The superconducting side: Anderson theorem}\label{Sec:Anderson}
Although superconductivity as ground state of electronic systems is limited to comparably low energy scales $T_c\sim 10$\,K, its ubiquitous appearance as favorable ground state of numerous materials ranging from single crystals to amorphous films results from its insensitivity against disorder. While the above scaling arguments rules out a 2d metal, this insensitivity can qualitatively be seen as reason for 2d superconductors - a celebrated result first obtained by Anderson \cite{anderson59} and commonly referred to as Anderson theorem. A comprehensive discussion of the Anderson theorem is lengthy and beyond the scope of this work. Instead, we will only sketch the succession of arguments leading to the conclusion\sidenote{\footnotesize{What follows is based on the more comprehensive discussion in Ref.~\cite{Lee2009}. }}.\\

\noindent The electron-phonon interaction term $H$ of the BCS Hamiltonian $H_0$ for clean superconductors reads\sidenote{\footnotesize{with the usual notion of fermionic creation and annihilation operators $\hat{c}^\dagger,\hat{c}$ and a constant interaction potential $V$}}
\begin{equation}
H=V\sum_\mathbf{k}\hat{c}_{\mathbf{k}\uparrow}^{\dagger}\hat{c}_{\mathbf{-k}\downarrow}^{\dagger}
\hat{c}_{-\mathbf{k}'\downarrow}\hat{c}_{\mathbf{k}'\uparrow}
\end{equation}
pairing states with opposite momenta and spin, i.e. $(\mathbf{k},\uparrow)$ and $(-\mathbf{k},\downarrow)$. The BCS mean-field approximation diagonalizes this Hamiltonian and leads to the well-known energy spectrum $E_\mathbf{k}=\sqrt{(\epsilon_\mathbf{k}-\mu)^2+|\Delta|^2}$ with $\epsilon_\mathbf{k}$ and $\mu$ the band dispersion and chemical potential and the BCS self-consistency equation
\begin{equation}
\Delta=VN(0)\int d\xi \frac{\Delta}{2\sqrt{(\xi-\mu)^2+|\Delta|^2}}\label{eq:selfconsBCS}.
\end{equation}
In disordered systems, the lattice translational symmetry is lifted and scattering off impurities violates conservation of momentum such that $\mathbf{k}$ is no longer a good quantum number and Cooper pairs do no longer result from the above pairing scheme. Anderson showed, that the BCS condition for pairing can be generalized beyond $|\mathbf{k},\uparrow\rangle$ and $|-\mathbf{k},\downarrow\rangle$ to exact eigenstates $|\alpha\rangle$ of the disordered Hamiltonian $H'$ and their time-reversed counterparts $\hat{T}|\alpha\rangle$ yet restoring the energy spectrum and self-consistency equation above. 
 Starting point is the general electron-phonon interaction operator 
\begin{equation}
H=V\sum_{\mathbf{k},\mathbf{k}',\mathbf{q}}\sum_{\sigma,\sigma'}\hat{c}_{\mathbf{k}'+\mathbf{q},\sigma'}^{\dagger}\hat{c}_{\mathbf{k}-\mathbf{q},\sigma}^{\dagger}
\hat{c}_{\mathbf{k},\sigma}\hat{c}_{\mathbf{k}',\sigma'}\label{Hcumb}.
\end{equation}
 Transformation\sidenote{\footnotesize{via the Fourier integral \begin{equation}\hat{c}_{\mathbf{k},\sigma}=\int d\mathbf{r} \hat{\psi}(\mathbf{r})e^{i\mathbf{kr}}\nonumber\end{equation}}} of the $\mathbf{k}$-space operators to real-space operators $\hat{\psi}(\mathbf{r})$, subsequent integration, and spin summation yields
\begin{equation}\label{eq:Hgeneral}
H=2V\int d \mathbf{r} \hat{\psi}_\uparrow^\dagger(\mathbf{r})\hat{\psi}_\downarrow^\dagger(\mathbf{r})\hat{\psi}_\downarrow(\mathbf{r})\hat{\psi}_\uparrow(\mathbf{r}).
\end{equation}     
In general, the orbitals $\hat{\psi}(\mathbf{r})$ will be complicated functions of $\mathbf{r}$ for a randomly disordered lattice. For the sake of the argument, however, the actual form is not required given the \emph{existence} of a set of eigenstates $|\alpha\rangle$ (and corresponding orbitals $\hat{\phi}_\alpha(\mathbf{r})$) diagonalizing the Hamiltonian for the disordered problem. Transformation into this new basis\sidenote{\footnotesize{via the spectral representation \begin{equation}\hat{\psi}_\sigma(\mathbf{r})=\sum_{\alpha}\hat{\phi}_\alpha(\mathbf{r})\hat{c}_{\alpha,\sigma}\nonumber\end{equation}}}, the Hamiltonian Eq.~(\ref{eq:Hgeneral}) can be cast into the BCS mean-field form with the only difference that the operators create and annihilate (time-reversed) eigenstates $|\alpha\rangle$ instead of momentum eigenstates $|\mathbf{k}\rangle$. After diagonalization and Bogoliubov transformation\sidenote{\footnotesize{that is, introducing new fermionic operators creating and annihilating quasiparticles (instead of electrons) which allow to transform the complicated 4-operator term in Eq.\,(\ref{Hcumb}) into a harmonic oscillator.}} one finds a similar spectrum $E_\alpha=\sqrt{(\epsilon_\alpha-\mu)^2+|\Delta|^2}$ which still contains information about the specific choice of eigenstates in terms of the eigenvalues $\epsilon_\alpha$. To obtain the self-consistency equation, however, any reference to the basis is lost when the summation over eigenvalues $\epsilon_\alpha$ is replaced by an continuous integral so that the BCS result Eq.~(\ref{eq:selfconsBCS}) is exactly restored. Consequently, disorder has no impact on the superconducting properties. In other words, if impurity scattering were to destroy a Cooper pair it would need to lift the requirement of time-reversal symmetry. Scattering off an potential, however, preserves the spin orientation and hence no pair-braking takes place, \emph{as long as} the scatterer does not carry a magnetic moment: in this case, time-reversal symmetry is broken and superconductivity may be strongly suppressed \cite{Abrikosov58,Abrikosov59,Abrikosov60}.

\subsection{The quantum-critical regime}
The Anderson theorem and the Anderson localization sketched above determine the physics far from the SIT at moderate and extreme amounts of disorder, respectively. As both asymptotic regimes are continuously connected, there must be a regime of disorder, where both paradigms break down and where, pictorially speaking, electrons cannot decide whether to pair up and condense or to get localized. \\
How can we escape Anderson's theorem? The important assumption Anderson made was that the orbitals $\hat{\phi}_\alpha(\mathbf{r})$ of the disordered system are extended in space. Only then the mean-field approximation in accordance with BCS works. Clearly, spatial extension and localization tendencies antagonize compromising the prerequisites of Anderson's theorem. An important measure to put a number on disorder is the Ioffe-Regel parameter $k_F\ell$, i.e. the product of the Fermi wave vector $k_F$ and the electron mean free path $\ell$. For marginal disorder, $k_F\ell\gg 1$ and the Anderson theorem holds, while for extreme disorder\sidenote{\footnotesize{In this case, electrons scatter from \emph{each} lattice site, i.e. the mean free path is the lattice constant.}} $k_F\ell\approx 1$ and the system turns insulating signaling the ultimate break down of Anderson's theorem.  While this explains the conceptual demand for an SIT, it does not provide an answer to the question how superconductivity actually ceases. Over the past years, two models have been put forward by theory, the so-called fermionic and the bosonic scenarios we will now briefly discuss.
\subsubsection*{The amplitude-driven or fermionic SIT}
\begin{figure}
\begin{centering}
\includegraphics[scale=0.08]{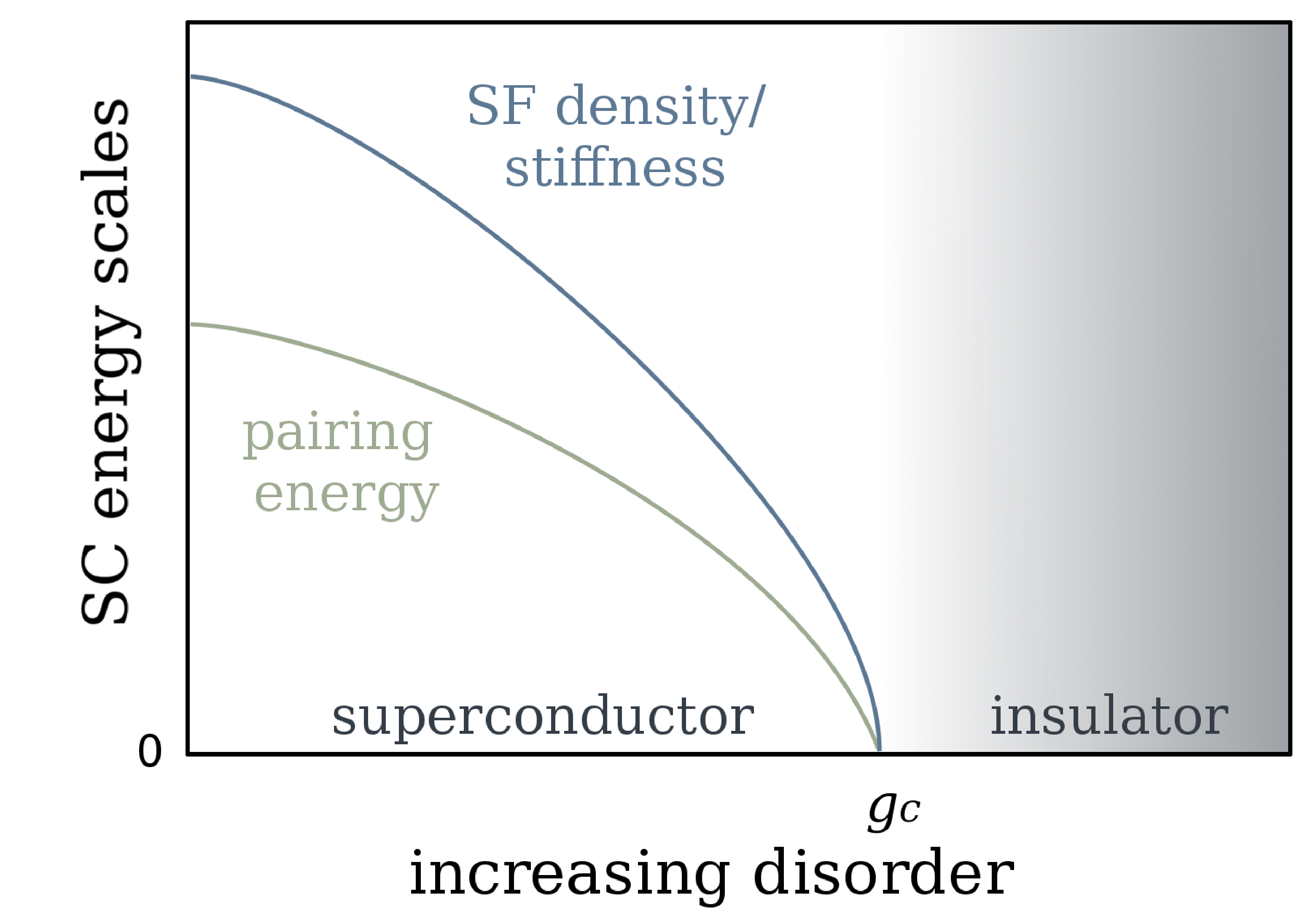}
\includegraphics[scale=0.08]{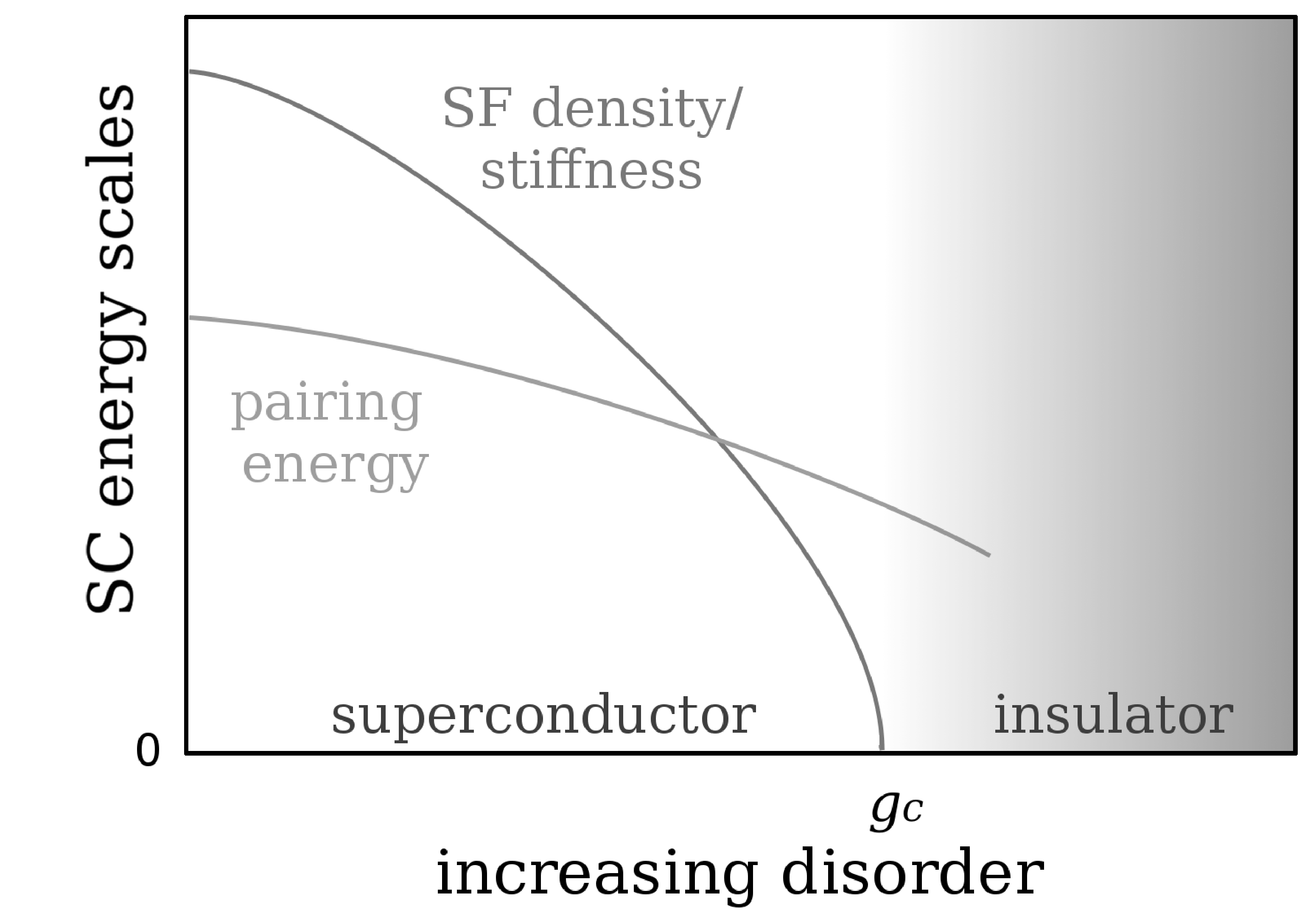}
\caption{\label{fig:fermionic_scenarios} \textbf{Energy scales at the SIT} within the fermionic (left) and the bosonic (right) scenario for the SIT. In the first case, pairing amplitude and superfluid density (or stiffness) go to zero at the QCP for critical disorder $g_c$ and the insulator contains localized fermions. In the bosonic scenario, the SIT is marked by a loss of superfluid coherence while pairing remains robust into the Cooper-pair (or Bose-) insulator.}
\end{centering}
\end{figure}

In a seminal work \cite{Fin94} Finkelstein has shown that disorder tends to renormalize Coulomb interaction between electrons such that the screening becomes less efficient: the attractive Cooper interaction pairing up electrons is challenged by the repulsive Coulomb interacting lowering the energy gain upon condensation. At a critical disorder, Coulomb repulsion is strong enough to overcompensate the pairing interaction and the bound state is no longer favorable. Using a diagrammatic renormalization approach, Finkelstein demonstrated that the amplitude of the order parameter vanishes uniformly at the SIT yet. At the same time, also the superfluid density and -stiffness go to zero where the latter always remains the greater energy scale, see Fig.~\ref{fig:fermionic_scenarios}. The insulating side contains fermionic quasiparticles which in 2D immediately localize forming a hard insulator whereas in 3D tend the insulating behavior may be weaker according to the scaling of conductivity discussed above. The fermionic scenario has been successfully applied to explain the suppression of $T_c$ in homogeneously disordered ultra-thin films of, e.g., amorphous MoGe \cite{Graybeal84}  or poly-crystalline TiN \cite{sacepe08} and MoC \cite{Samuely2016,Zemlicka2015}. At the same time - and not unsurprisingly - it fails to properly describe experimental results in the very vicinity of the SIT as it neglects quantum fluctuations relevant at low energies.  
\subsubsection*{The phase-driven or bosonic SIT} 
In the bosonic scenario global coherence necessary to form a phase-locked macroscopic superfluid condensate is lost at the SIT, see Fig.~\ref{fig:fermionic_scenarios}. In more detail, Fisher \emph{et al.} have shown in a pioneering work \cite{Fisher90} that long-range phase fluctuations among a charged (2$e$) bosonic superfluid are enhanced in approach of the SIT reducing the superfluid stiffness and finally destroying superconductivity. As a natural consequence, the bosonic model allows pairs to survive into the insulating side, where they, however, have lost a global phase coherence and become localized. This model features an intriguing charge-vortex duality which predicts a universal resistance $R_c=h/4e^2$ right at the SIT. Indeed, early dc-transport studies \cite{Haviland89} of ultra-thin $a-$Bi films quench condensed on Ge reveal a SIT at the predicted universal resistance. Later on the prediction was found at odds with other systems, e.g. ultra-thin MoGe films: Although magneto-transport measurements confirmed the localization trend of Cooper pairs at the field-induced SIT, the importance of phase fluctuations and long-range Coulomb interaction \cite{Yazdani95}, the obtained critical resistance seems sample-dependent and is scattered widely around $h/4e^2$. The discrepancy is commonly attributed to the missing fermionic degrees of freedom which naturally arises treating the Cooper pairs as hard-core charge-2$e$  bosons without internal fermionic structure.\\
The limited applicability of both models calls for a unified \emph{amplitude-phase} theory equally capturing Coulomb interactions and phase fluctuations each of which are the central aspects of the respective models. At the time of writing this thesis, such a theory is yet to be developed.\\

In the next section we will discuss some of the hallmark experiments elucidating the SIT to conceptually outline what a successful theoretical treatment would have to account for.

\section{A brief review of tunneling \\studies on the SIT}\label{SITExp}
We will restrict our discussion to  SIT systems relevant for the scope of this thesis, namely TiN and NbN. These materials can be produced in various ways as thin film with a homogeneous amorphous or polycrystalline structure and random distribution of atomic-scale disorder. The distance to the SIT can be tuned either by composition (i.e. varying $N$ concentration) or thickness. The arguably most striking result was obtained by measurements of the local tunneling conductance. Far from criticality, the energy gap $\Delta$ is uniform in space as expected for ordinary s-wave superconductors. This changes drastically in approach of the SIT where - despite structural homogeneity - measurements reveal a strongly inhomogeneous superconducting state. For TiN, Sacepe \emph{et al.} resolved a spatially fluctuating energy gap $\Delta$ \cite{sacepe08} while for NbN inhomogeneities are identified by variations of the coherence-peak height and zero-bias anomalies \cite{kamlapure13}. In both cases, the emergent superconducting domains exist on length scales comparable to a few times the superconducting coherence length of $\sim10$ nm, i.e. much larger than the atomic-scale disorder potential traps. The emergent electronic inhomogeneity tends to localize the Cooper pairs into weakly coupled superconducting puddles. The global superconducting phase, which, in the clean case, acquires its lock to a constant value by virtue of perfect delocalization, is subject to increasingly strong fluctuations, as decomposition into weakly coupled puddles increases. The long-range phase coherence is eventually destroyed when the charging energy\sidenote{\footnotesize{which can be thought of the energy to be paid for hopping from puddle to puddle.}} of the puddles by far exceeds the Josephson energy\sidenote{\footnotesize{which is a measure for the energy pay off due to hopping.}}. This peculiar inhomogeneity of the superconducting state is in perfect agreement with calculations within the Bogoliubov de-Gennes model \cite{Ghosal2001,Ghosal1998} and favors the bosonic scenario of the SIT. 
This view is strengthened by tracing the energy towards the SIT. Comparison of $T_c$ and tunneling gap $\Delta$ revealed a less strong decay of the latter towards criticality leaving an anomalously large gap for films where $T_c$ was almost zero \cite{sacepe08}. The evolution of both quantities suggests $\Delta$ to survive across the SIT forming a gapped Bose insulator. Similar measurements using planar junctions and disordered 2D InO films \cite{sherman12} compared nominally insulating and superconducting film in the very vicinity of the SIT characterized by dc-transport down to mK temperatures. Surprisingly, both the insulating and superconducting tunneling spectra display a clear gap with almost the same amplitude. The coherence peaks, however, only appear for the latter which was viewed as evidence for robust pairing without global phase coherence in agreement with the bosonic models. In a subsequent publication \cite{sherman14} the insulating gap was conceptually explained by partial screening of electronic interactions due to the nearby metallic tunneling electrode pushing the sample back on the superconducting side to explain an alleged mismatch with optical THz measurements. This argument, however, is questioned by a screening length much smaller than the tunneling barrier. Furthermore, if the electrode effectively pushes an insulating sample superconducting, also the coherence peaks should reappear what is, however, not the case.
Another consequence of strong phase fluctuations manifests at the transition to the normal state. Tunneling studies \cite{sacepe10,kamlapure13} revealed a pseudogap in the density of states surviving up to several times $T_c$. The gap, however, is not flanked by coherence peaks reminiscent of the superconducting gap. Their absence is understood as result of the vanishing superconducting order, while pair correlations remain present. In this sense, upon cooling preformed Cooper pairs exist at temperatures much higher than $T_c$ but form a globally coherent condensate only below $T_c$. While soon after this discovery, the analogy to the pseudogap in unconventional high-temperature superconductors such as the cuprates has been stressed, the comparison is delicate as the driving forces - disorder and correlation - are not quite the same. For instance, a recent  study \cite{Benhabib2015} assembles experimental evidence that the pseudogap in various cuprates is not restricted to the regime of the phase diagram where phase fluctuations delimit superconductivity (e.g. Bi-2201 or LSCO) but also appears (in case of e.g. Bi-2201) above amplitude-driven transitions\sidenote{\footnotesize{This paradigmatic view of the $T_c-$dome in unconventional superconductors emerging from an amplitude-phase crossover suggested by Emery and Kivelson \cite{emery94} is, however, challenged by the overwhelmingly complex superconductivity in these compounds.}}.   
Although tunneling studies arguably yielded the most valuable insights how energy scales behave towards criticality, the above mentioned discrepancy with optical measurements in the THz frequency range is troubling. The situation is even more alarming as also systematic studies \cite{coumou13,driessen12} bringing together resonant  microwave and tunneling measurements on  disordered TiN and NbTiN point towards the insufficiency of theories purely relying on the effects of pair-breaking. In what follows, we present comprehensive measurements of the dynamical conductivity at THz frequencies and compare them to tunneling measurements and discuss a reasonable solution to the above problem on basis of a new kind of excitation invisible to tunneling spectroscopy.

\section{Measurements of the \\dynamical conductivity}\label{Sec:ExpNbN}
We now turn to the optical spectroscopy measurements performed on superconducting NbN thin films in appraoch of the SIT. We performed comprehensive measurements of the complex transmission amplitude in the frequency range $\nu=2 - 40$\,cm$^{-1}$ (i.e. energies $E=h\nu=0.25-5$\,meV), which is well suited to study the dynamics of (strong-coupling) superconducting NbN films\sidenote{\footnotesize{Information on sample growth and -characteristics are found in Sec. \ref{sampleNbN}}} with critical temperatures ranging between $T_c=15-3$\,K and an estimated spectral gap of $4.2k_BT_c=0.9-5.4$\,meV. Measurements were performed in the normal state slightly above $T_c$ and the regime of superconducting fluctuations and in the superconducting states well below $T_c/2$, where the mean-field BCS energy gap is at almost 100\% of its zero-temperature value. The spectra of the transmission amplitude and phase were fitted to Fresnel equations via real and imaginary parts of the complex dielectric function $\hat{\epsilon}$ without any particular microscopic model for the charge dynamics\sidenote{\footnotesize{See Sec.\,\ref{sec:singlepeak} for more details of this so-called single-peak analysis.}}. The complex dynamical conductivity follows directly via $\hat{\sigma}=2\pi i\epsilon_0 \nu(\hat{\epsilon}-1)$.
Prior to the optical studies, similar NbN films with a thickness of 50nm have been fabricated as planar Ag/NbN tunneling junctions. Comprehensive measurements of the (differential) tunneling conductance have been performed by M. Chand and the tunneling spectra displayed below are originally published in Ref. \cite{Chand2012PhD}.   

Before we discuss the experimental results, it is instructive to clarify, how superconducting energy scales manifest in both the optical properties and the tunneling conductance $G$. At zero temperature, the charge current $I$ between the electrodes through the tunneling barrier sets in as soon as the Fermi level of the Ag electrode is biased by a voltage $|V|\geq\Delta/e$. As a consequence of the diverging density of states right at $\pm\Delta$, the slope of the conductance $G=I/V$ is initially infinite before it saturates to the tunneling barriers inverse Ohmic resistance. The density of states which is proportional to the differential conductance $\mathrm{d}I/\mathrm{d}V$ displays a gap of the width $2\Delta$ and thus provides a clean access to the superconducting pairing energy. While here, the amplitude of $\Delta$ is measured by shifting chemical potentials, in optical spectroscopy it is probed by the quasiparticle excitations across the gap: At zero temperature, the finite-frequency response of a superconductor sets in at $\nu=2\Delta/h$ where quasiparticles are lifted cross the DOS gap which manifests as an onset in the real part of the dynamical conductivity $\hat{\sigma}(\nu)=\sigma_1(\nu)+\mathrm{i}\sigma_2(\nu)$. While in the canonical BCS picture the onset of electromagnetic absorption at the \emph{spectral} gap $\Omega$ equals the superconducting energy gap $2\Delta$ picture, additional excitations\sidenote{\footnotesize{provided that the excitation process is a scalar susceptibility and couples to $\hat{\sigma}$ linearly.}} may open absorption channels at frequencies $\nu<2\Delta/h$ shifting $\Omega$ below $2\Delta$. Consequently, the combination of both transport and optical spectroscopic techniques is a promising route to identify excitations of the superfluid beyond ordinary quasiparticle breaking.\\
\begin{marginfigure}
\begin{centering}
\includegraphics[width=\marginparwidth]{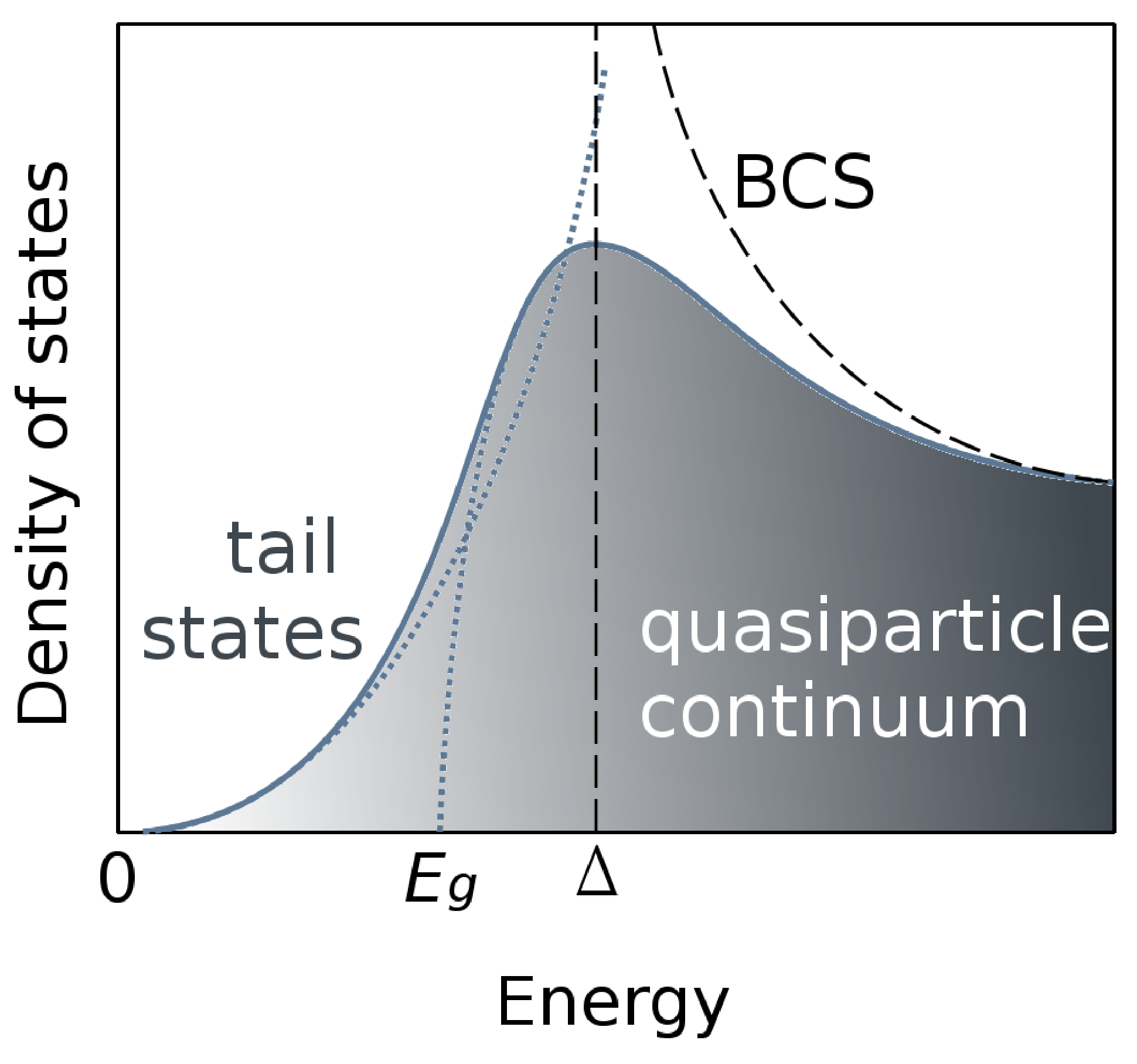}
\caption{Sketch of the disorder-smeared density of states including sub-gap tail states. The dashed line is the BCS solution. Note that the tail states are actually much less significant than displayed here.}
\end{centering}
\end{marginfigure}

\noindent The comparison is done according to the scheme described below.
\begin{enumerate}
\item{Match an optical with a tunneling spectrum for NbN with (approximately\sidenote{\footnotesize{Any NbN film produced as electrode within a tunneling junction cannot be used for an optical measurement. The relation between the structural properties and the $T_c$, i.e. the reproducibility, is sufficiently high for the given growth procedure, such that the comparison of two NbN films linked by $T_c$ is justified \cite{PratapPers}  }}) the identical $T_c$}
\item{Fit the $dI/dV$ spectrum:
\begin{enumerate}
\item{Solve the Usadel equation (\ref{Usadelu}) for a given pair breaking parameter $\tau$ (with $\xi=0$) and coupling\sidenote{\footnotesize{The temperature dependence of $\Delta$ is assumed to be of BCS type}} $c=2\Delta/k_BT_c$ for energies $|E|\geq E_g$ where
\begin{equation}
\frac{E_g}{\Delta(T)}=\left[1-\left(\frac{1}{\tau\Delta(T)}\right)^{2/3}\right]^{3/2}
\end{equation}
is the renormalized hard band gap \cite{Fominov2011} and $T$ is the temperature of the tunneling measurement.}
\item{Calculate the real part of the energy dependent normal Green function $G(E)$, i.e. the density of states $\mathcal{D}(E)$ via Eq.~(\ref{GreenG}).}
\item{Include - if required - additional sub-gap states $N_\mathrm{tail}(E)$
\begin{equation}
\mathcal{D}_\mathrm{tail}(E)=\mathrm{exp}\left[\left(\frac{E_g-E}{\Gamma}\right)^{5/4}\right]\label{TailStates}
\end{equation}
 for $|E|< E_g$ where $\Gamma$ measures the width of the tail accounting for local $\Delta$ inhomogeneities due to random impurity configurations \cite{Feigelman2012}.}
\item{Connect $\mathcal{D}(|E|\geq E_g)$ and $\mathcal{D}_\mathrm{tail}(E<E_g)$ smoothly by adjusting the relative weight to generate a continuous function $\tilde{\mathcal{D}}(E)$, $\forall E$}
\item{Calculate the functional form of $dI/dV$ at finite temperatures by convoluting $\tilde{\mathcal{D}}(E)$ with the Fermi function $f(E,T)$
\begin{equation}
\frac{dI}{dV}\propto \frac{d}{dV}\int\limits_{-\infty}^{\infty}dE\tilde{\mathcal{D}}(E)\big(f(E+eV)-f(E)\big)
\end{equation}
where the integration range is chosen such, that the result does not change any more within standard numerical precision \cite{Tinkham2004}}
\item{Adjust $\tau,c,$ and $\Gamma$ and iterate (a-e) to find the optimal fit.}
\end{enumerate}
\item{Use the obtained $\tau$ and $c$ to solve the Usadel equation for $\Delta(T^\prime)$ with $T^\prime$  the temperature of the optical measurement.}
\item{Calculate the complex normal and anomalous Green's functions $G(E)$ and $F(E)$ via Eqs.~(\ref{GreenG},\ref{GreenF})}
\item{Calculate the complex conductivity $\hat{\sigma}(\nu)$ via Eq.~(\ref{FominovCond})}
}
\end{enumerate}
Within the canonical BCS picture, the parameters required to fit the tunneling measurement should also yield a proper description of the optical measurement\sidenote{\footnotesize{At the writing of this thesis, we are not aware of a theory for the electromagnetic response including the tail states $N_\mathrm{tail}$. Compared to $\mathrm{Re}G$, however, the contribution of $N_\mathrm{tail}$ to the overall den
sity of states is very small here so that negligence thereof will not affect $\hat{\sigma}(\nu)$ substantially.}}.\\
\begin{figure}[b!]
\begin{center}
\includegraphics[width=\textwidth]{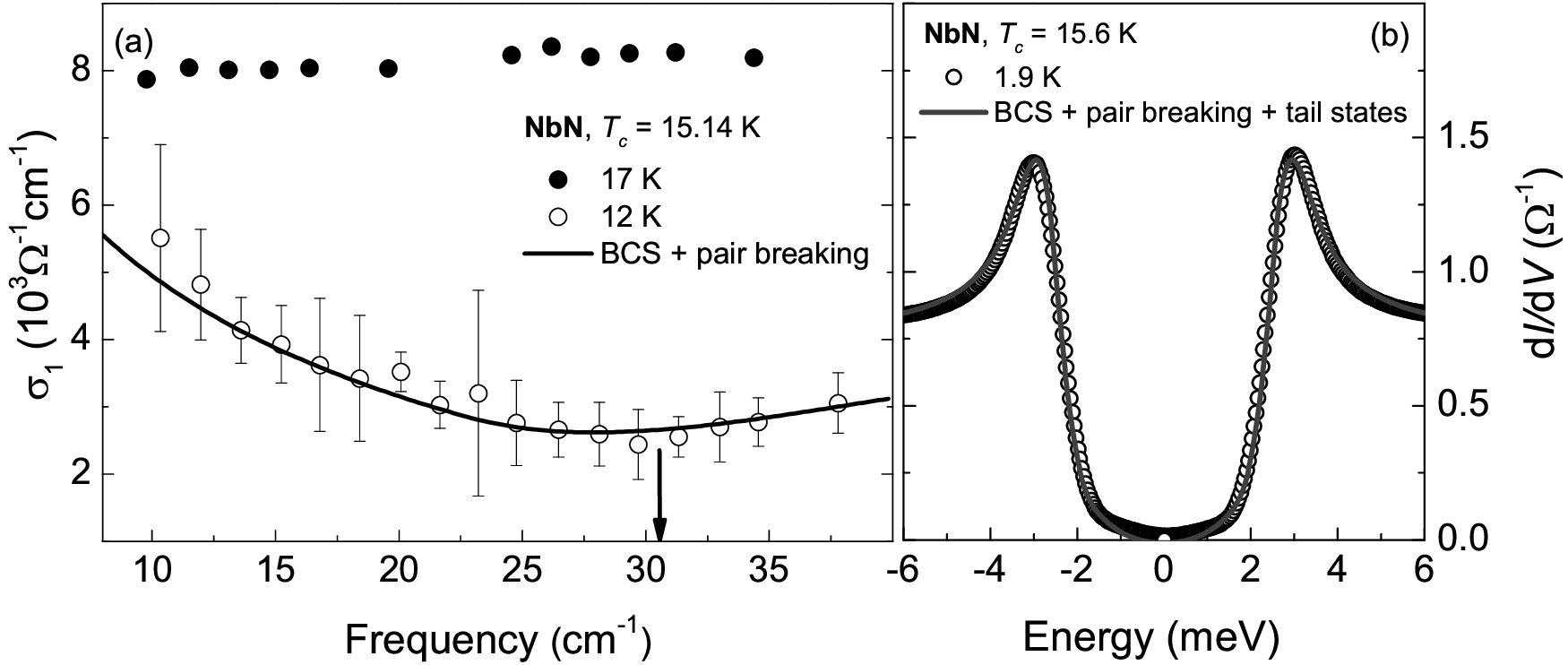}
\caption{\label{Fig_NbN15600_sigma1+DOS}\textbf{Optical and tunneling spectroscopy on clean NbN.} (a) real part $\sigma_1(\nu)$ of the dynamical conductivity in the normal (17\,K) and superconducting states (12\,K). The arrow indicates the spectral gap estimated from the minimum of $\sigma_1(\nu)$. (b) Differential tunneling conductance $dI/dV$ as function energy at 1.9\,K. Solid lines are based on Green's functions calculated for the same pair breaking $\tau$ and  ratio $\Delta/k_BT_c$. To fit the in-gap part of $dI/dV$, exponentially decaying tail states have been added, which are not taken into account for $\sigma_1$.}
\end{center}
\end{figure}
\begin{figure}[b!]
\includegraphics[width=\textwidth]{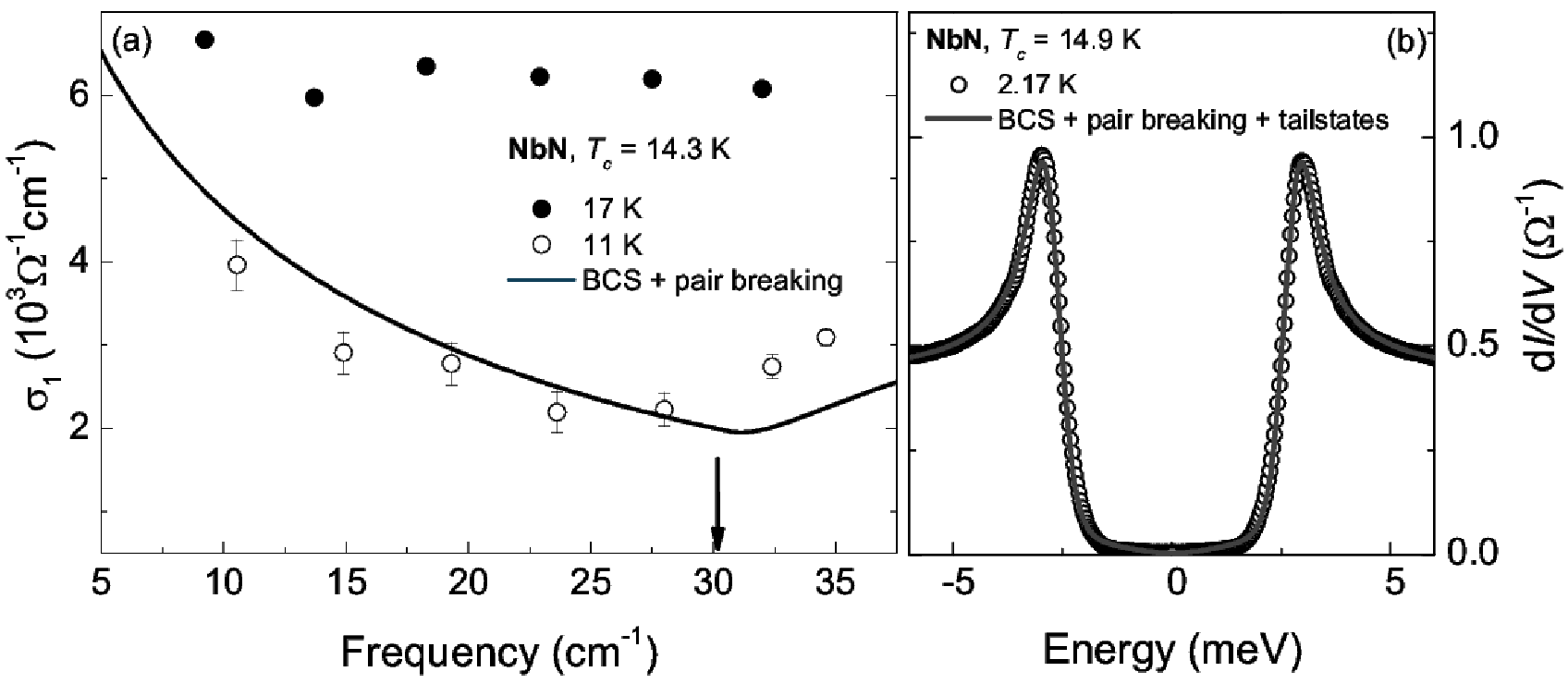}
\caption{\label{Fig_NbN14300_sigma1+DOS}\textbf{Optical and tunneling spectroscopy on clean NbN.} (a) real part $\sigma_1(\nu)$ of the dynamical conductivity in the normal (17\,K) and superconducting state (11\,K). The arrow indicates the spectral gap estimated from the minimum of $\sigma_1(\nu)$. (b) Differential tunneling conductance $dI/dV$ as function energy at 2.17\,K. Solid lines are based on Green functions calculated for the same pair breaking $\tau$ and ratio $\Delta/k_BT_c$.}
\end{figure}
Figure \ref{Fig_NbN15600_sigma1+DOS} compares measurements of the real part of the dynamical conductivity $\sigma_1(\nu)+\mathrm{i}\sigma_2(\nu)$ of a sample with $T_c=15.14$\,K with a measurement of the differential tunneling conductance $dI/dV$ of a sample with $T_c=15.6$\,K.  While the tunneling measurement is performed at 1.9\,K well below $T_c$, the displayed $\sigma_1(\nu)$ spectrum is taken at 12\,K much closer to $T_c$. The reason is that the fully opened gap is located at frequencies above the experimentally accessible range, while the elevated temperature allows a direct read-out of $\Omega$ and hence comparison with $2\Delta$ from the tunneling measurement. Though being in the clean limit, the coherence peaks in the $dI/dV$ spectrum are smeared out substantially beyond thermal broadening. The optimal fit is obtained for $\tau\Delta_0=25.3$ and a coupling of $c=2\Delta_0/k_BT_c=4.15$ and a minor $N_\mathrm{tail}$ contribution. The corresponding prediction for $\sigma_1(\nu)$ invoking the same $\tau$ and $c$ is in very good agreement with the actual experimental result. The minimum in $\sigma_1(\nu)$ lies at $\sim31$\,cm$^{-1}$ (denoted with the blue arrow) which gives a coupling $2\Delta_0/k_BT_c=4.1$ with the BCS temperature-dependence of $\Delta$ \cite{Chand2012PhD}. A similar result is found for another pair of clean-limit samples with $T_c=14.3$\,K and 14.9\,K for the optical and tunneling measurement, respectively, shown in Fig.~\ref{Fig_NbN14300_sigma1+DOS}. The minimum in the predicted $\sigma_1(\nu)$ curve is in good agreement with the onset of optical absorption beyond quasi particle dynamics. At sub-gap frequencies, the experimental dispersion matches the predicted one, while above the spectral gap, the rise is stronger than expected. 
\begin{figure}[h!]
\begin{center}
\includegraphics[width=\textwidth]{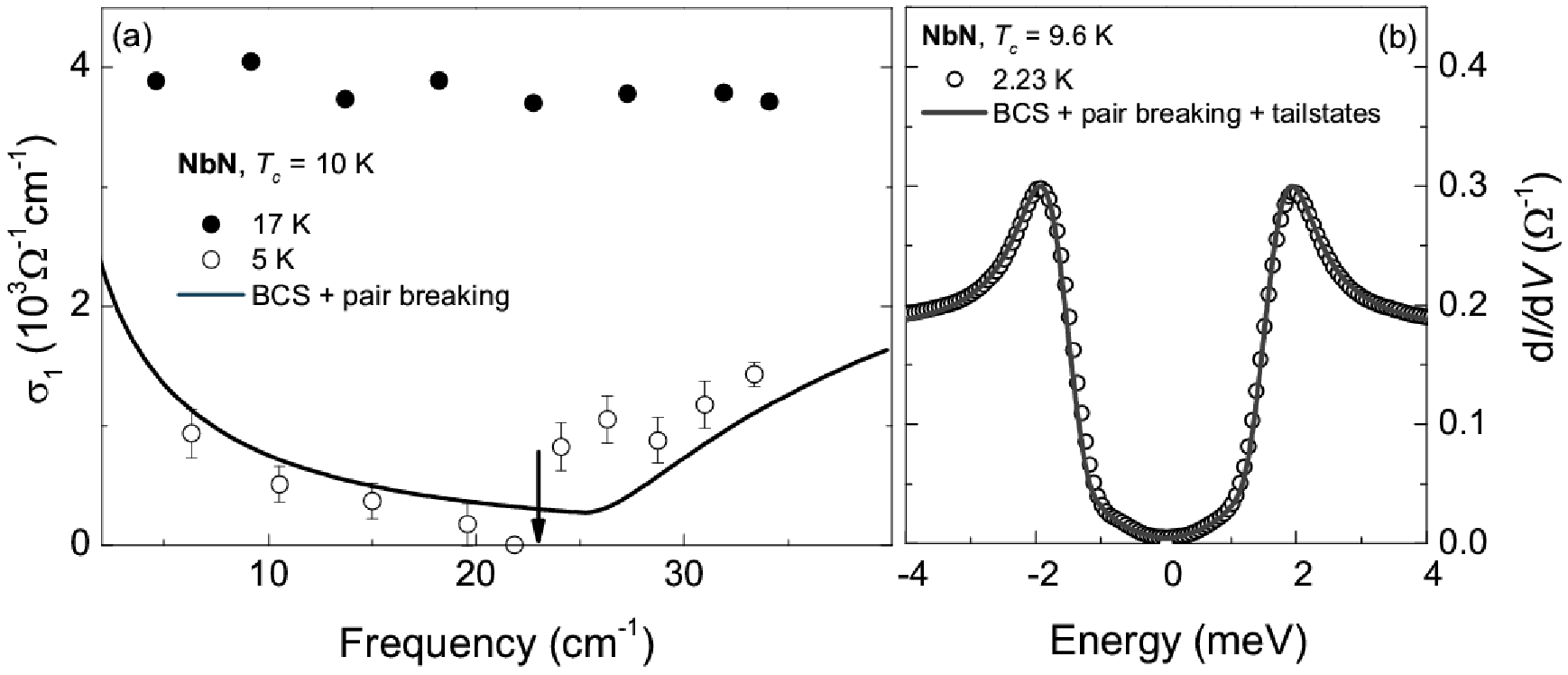}
\includegraphics[width=\textwidth]{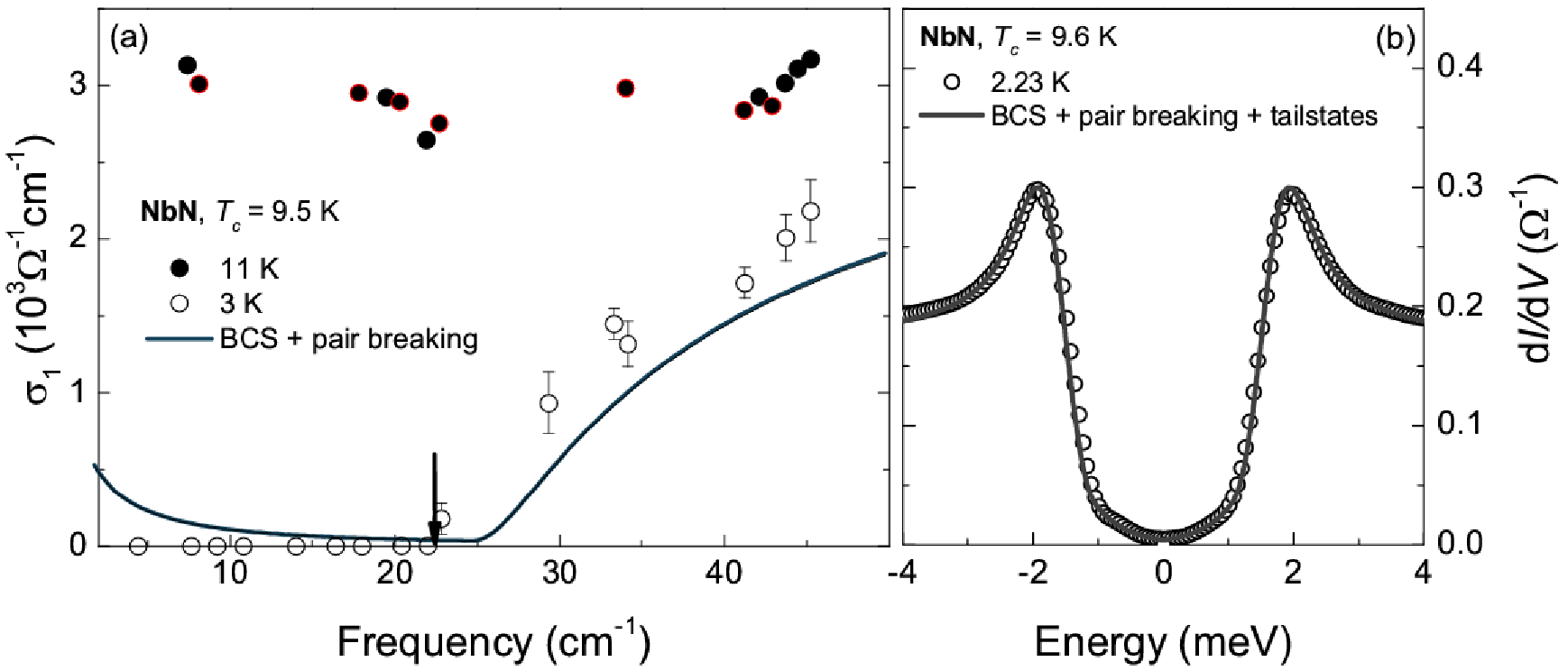}
\includegraphics[width=\textwidth]{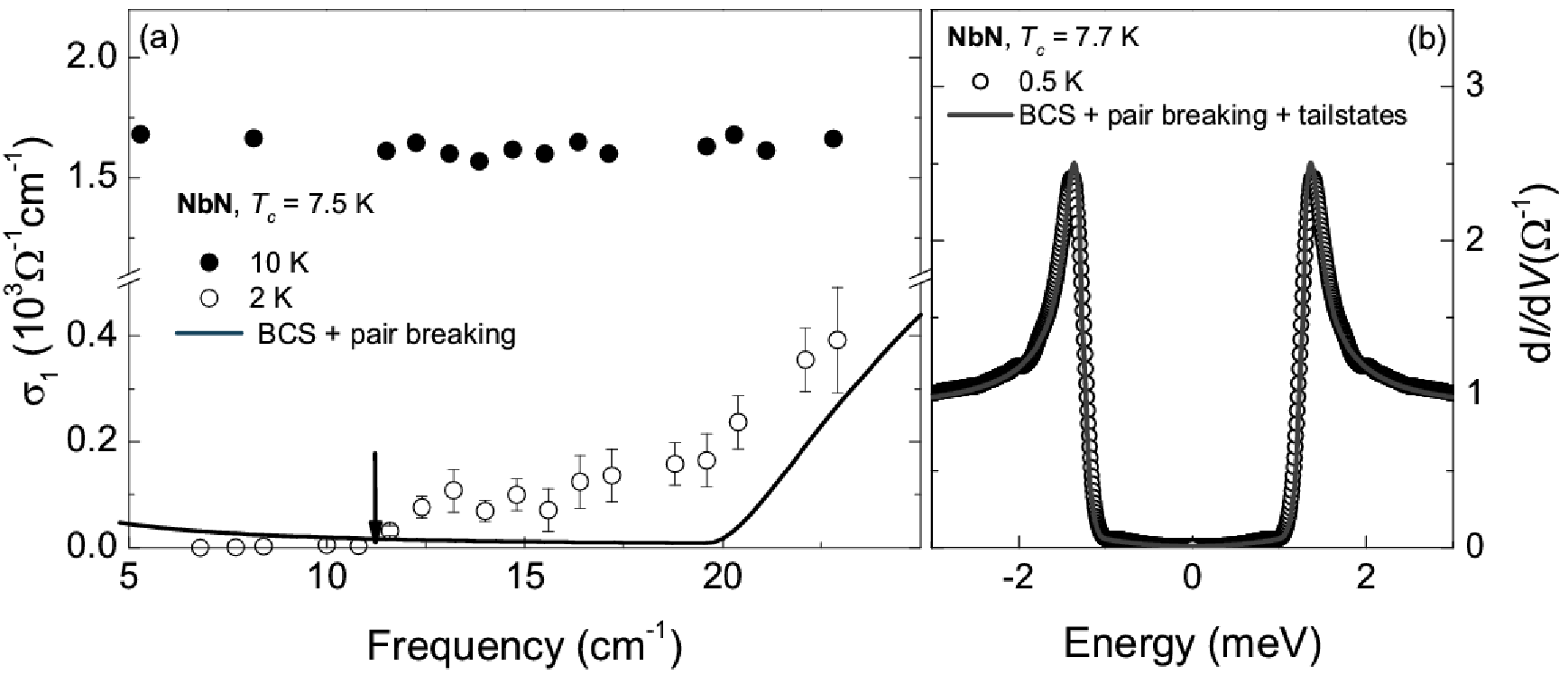}
\caption{\label{Fig_NbN10000-9500-7500_sigma1+DOS}\textbf{Optical and tunneling spectroscopy on moderately disordered NbN.} Panels (a) show real part $\sigma_1(\nu)$ of the dynamical conductivity in the normal and superconducting state. The arrows indicate the spectral gap estimated from the kink of $\sigma_1(\nu)$. Panels (b) display the differential tunneling conductance $dI/dV$ as function energy. Solid lines are based on Green functions calculated for the same pair breaking $\tau$ and ratio $\Delta/k_BT_c$.}
\end{center}
\end{figure}
Comparing samples with $T_c=10$\,K and 9.5\,K for the optical and tunneling measurement, respectively, reveals an absorption threshold $\Omega$ that is shifted below the predicted one exceeding the range of experimental uncertainty, see the top panel of Fig.~\ref{Fig_NbN10000-9500-7500_sigma1+DOS}. While again for sub-gap frequencies the experimental dispersion agrees with the predicted form, an increasing excess absorption evolves around $\Omega$ and at higher frequencies. \\
\begin{figure}[t!]
\begin{center}
\includegraphics[width=\textwidth]{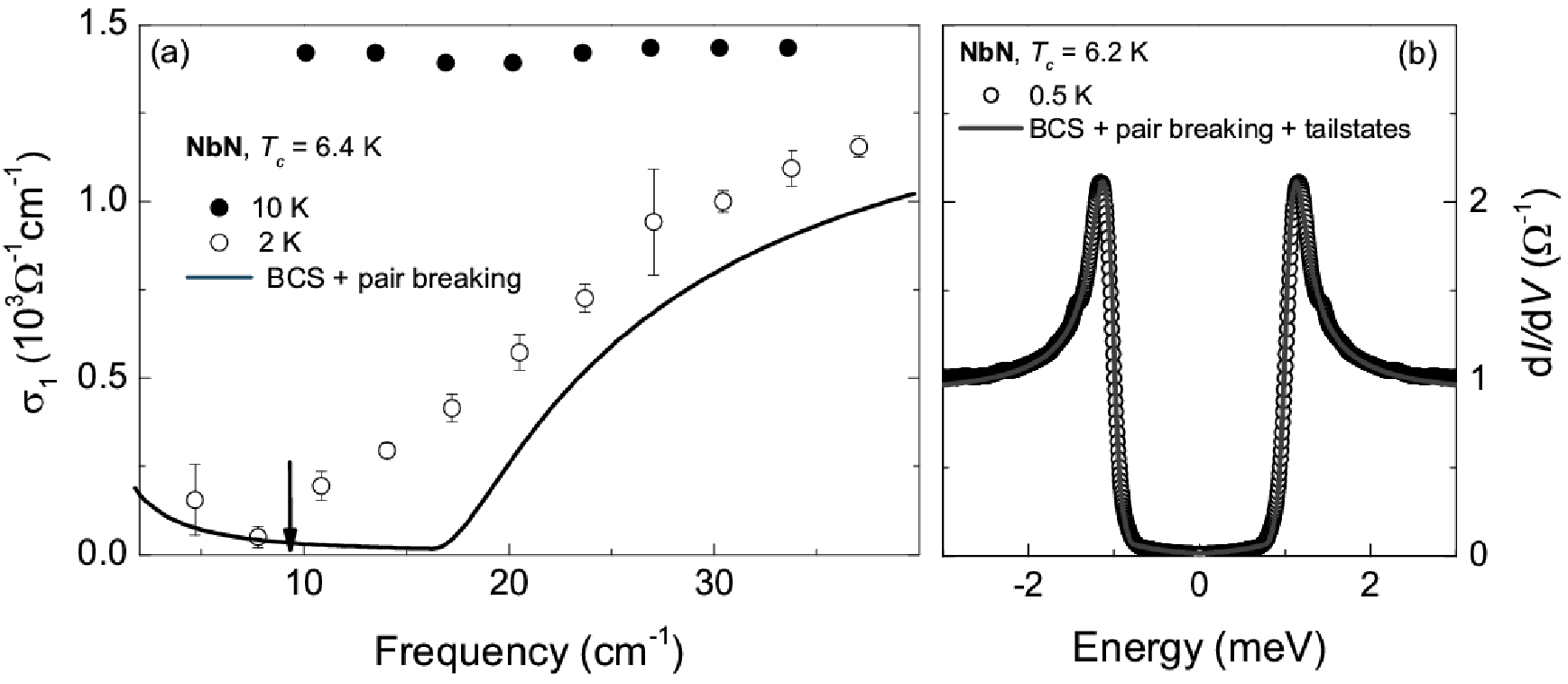}
\vspace{0.3cm}
\includegraphics[width=\textwidth]{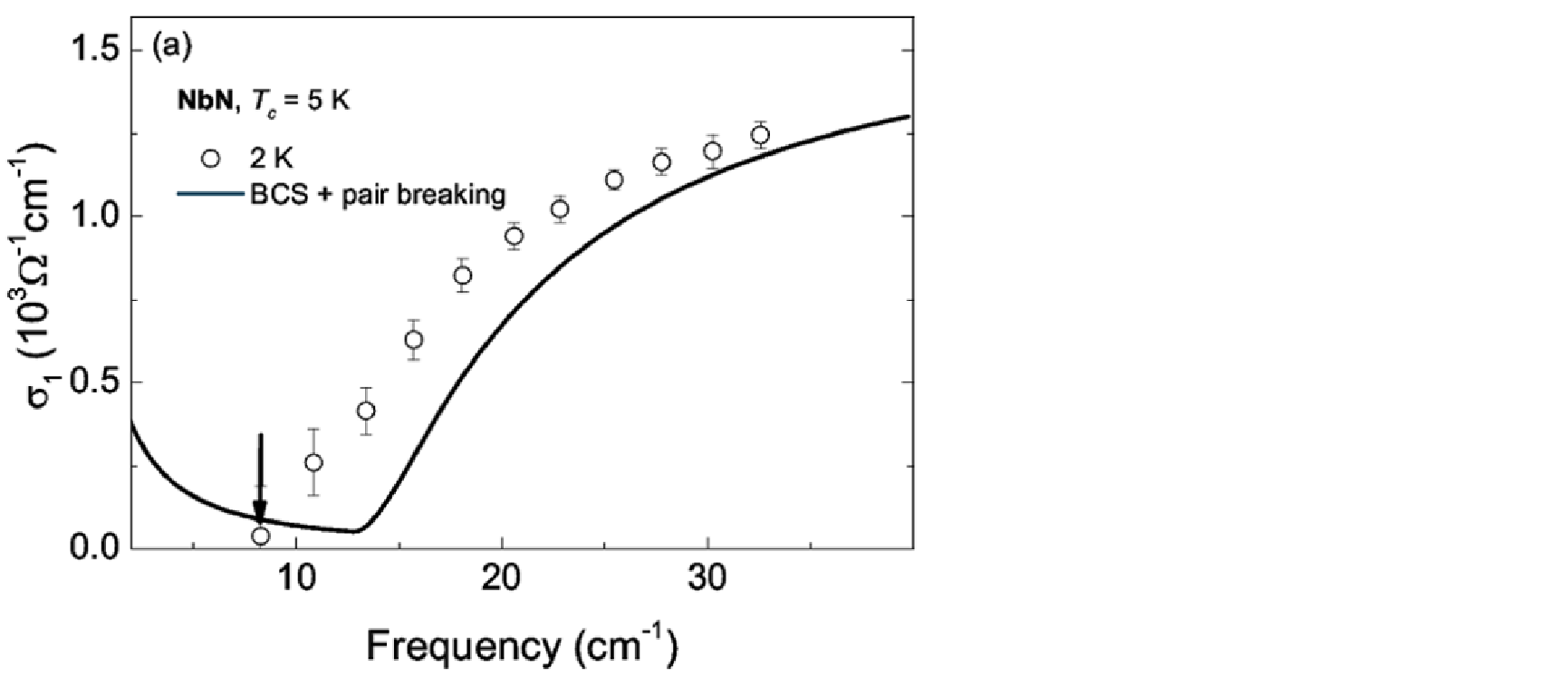}
\vspace{0.3cm}
\includegraphics[width=\textwidth]{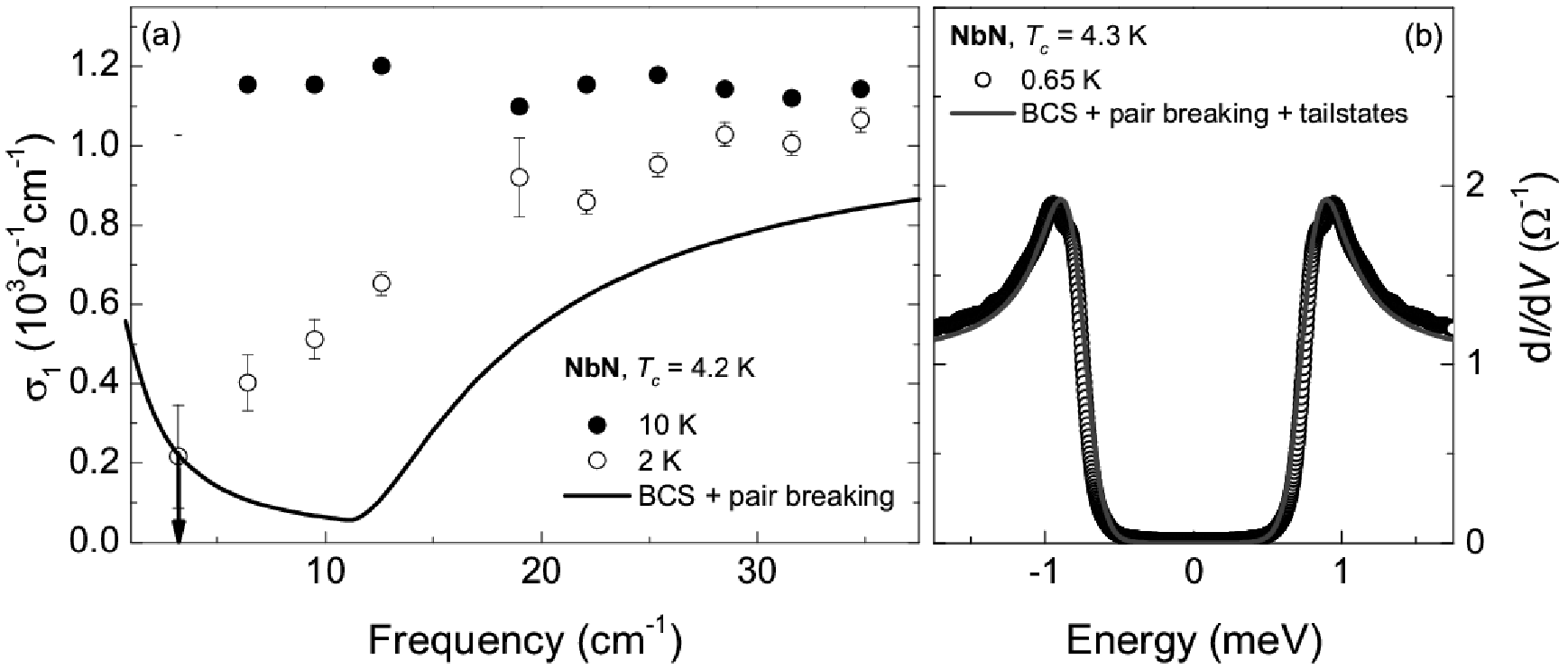}
\vspace{0.1cm}
\caption{\label{Fig_NbN6400-5000-4200_sigma1+DOS}\textbf{Optical and tunneling spectroscopy on disordered NbN.} Panels (a) show real part $\sigma_1(\nu)$ of the dynamical conductivity in the normal and superconducting state. The arrows indicate the spectral gap estimated from the kink of $\sigma_1(\nu)$. Panels (b) display the differential tunneling conductance $dI/dV$ as function energy. Solid lines are based on Green functions calculated for the same pair breaking $\tau$ and ratio $\Delta/k_BT_c$.}
\end{center}
\end{figure}
Both the suppression of $\Omega$ with respect to $2\Delta$ predicted from the tunneling measurement and the accumulation of additional spectral weight beyond the quasiparticle absorption, become increasingly pronounced as $T_c$ is reduced in approach of the SIT, as displayed in the mid and bottom panels of Fig.~\ref{Fig_NbN10000-9500-7500_sigma1+DOS} and Fig.~\ref{Fig_NbN6400-5000-4200_sigma1+DOS}. Down to $T_c=7.5$\,K the excess conductivity is concentrated around frequencies corresponding to $2\Delta$ and becomes vanishingly small towards the low-frequency limit. As for the $T_c=6.4$\,K, sample, the additional conductivity spans over the entire frequency range.  The discrepancy between the anticipated and the actual $\sigma_1(\nu)$ becomes increasingly worse as both functional form and absolute values are concerned when going to the lowest-$T_c$ sample, where $\sigma_1(\nu)$ could meaningfully be calculated.  \\
\begin{figure}[b!]
\begin{center}
\includegraphics[width=\textwidth]{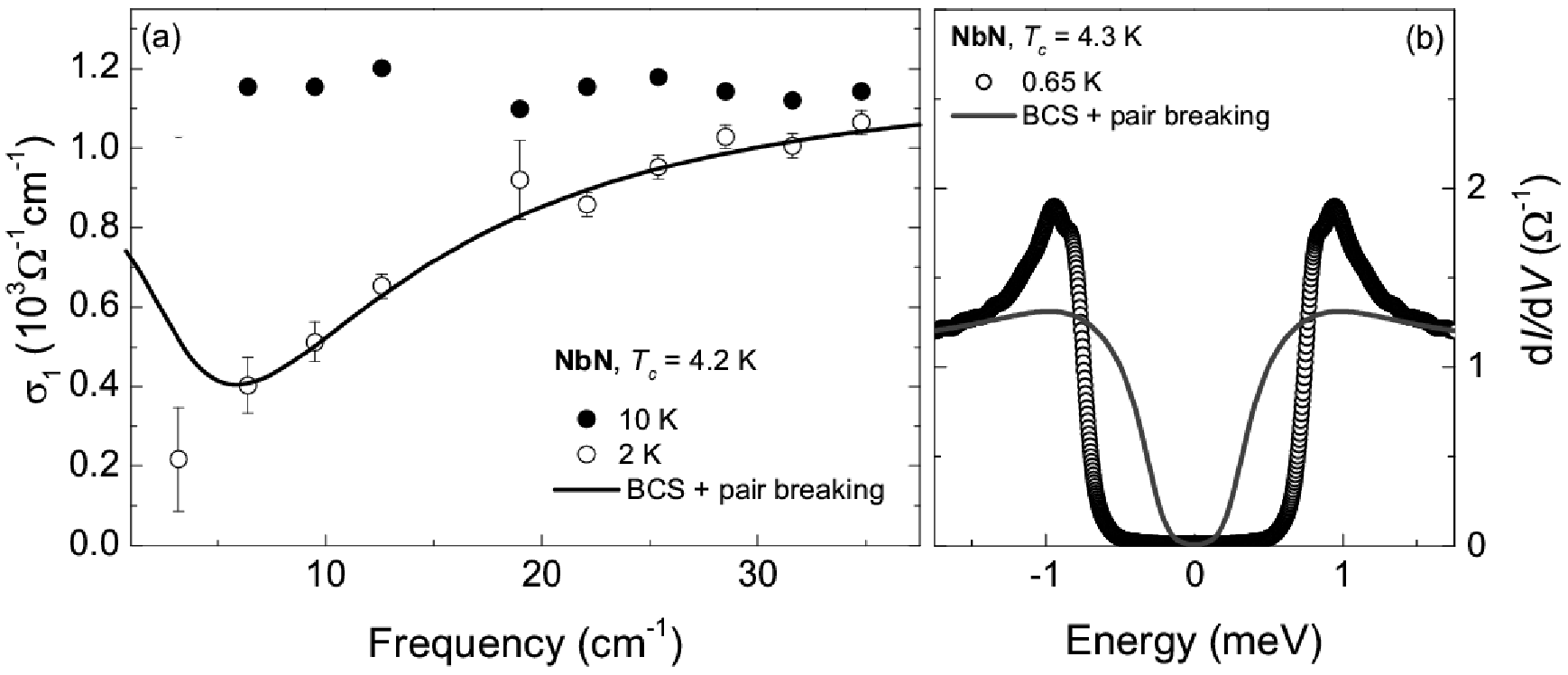}
\caption{\label{Fig_NbN4200_sigma1+DOS_INVERTED}\textbf{Inverted analysis routine} applied to a strongly disordered sample with $T_c=$4.2 (a) and 4.3\,K (b). The pair breaking parameter is chosen such that it  yields a fit of $\sigma_1(\nu)$ at 2\,K. This strong pair breaking fails to generate a satisfying description of $dI/dV$ regarding both the coherence peaks and the width of the gap.}
\end{center}
\end{figure}

As mentioned above, the above fits of the $dI/dV$ spectra to quantify the pair breaking strength incorporated exponential sub-gap tail states, which are not included in the Green's functions $G$ and $F$ from which $\sigma_1(\nu)$ was derived. Piecing together a continuous function $\tilde{\mathcal{D}}(E)$ from both the hard-gapped continuum contribution and the tail states to fit $dI/dV$ is associated with a certain degree of freedom as finite temperatures also smear out a hard gap similar as tail states do. Consequently, the lack of tail states in the calculation of $\sigma_1(\nu)$ should carefully be examined in consideration of the presented discrepancy. This can be done by inversion of the above analysis: $\sigma_1(\nu)$ is fitted to find a suited pair breaking strength, which thereupon is used to calculate a prediction for $dI/dV$. In Fig.\,\ref{Fig_NbN4200_sigma1+DOS_INVERTED}(a) we exemplary re-plot the experimental results on the samples with $T_c=4.2$ and 4.3\,K together with a fit of $\sigma_1(\nu)$. The required pair-breaking is 10 times stronger than in the previous fit of $dI/dV$ displayed in Fig.~\ref{Fig_NbN6400-5000-4200_sigma1+DOS}(c). Except for the lowest frequencies, this choice of pair breaking reproduces the experimental $\sigma_1(\nu)$ very well. At the same time, however, it completely fails to yield a satisfying description of $dI/dV$ displayed in panel (b) regarding both the height and shape of the coherence peaks and the width of the gap. It is clear, that this discrepancy cannot be accounted for by the missing tail states which could only narrow the gap further, and consequently, the effect of tail states on $\sigma_1(\nu)$ can be ruled out as origin of the excess conductivity.\\
\begin{figure}
\begin{center}
\includegraphics[width=\textwidth]{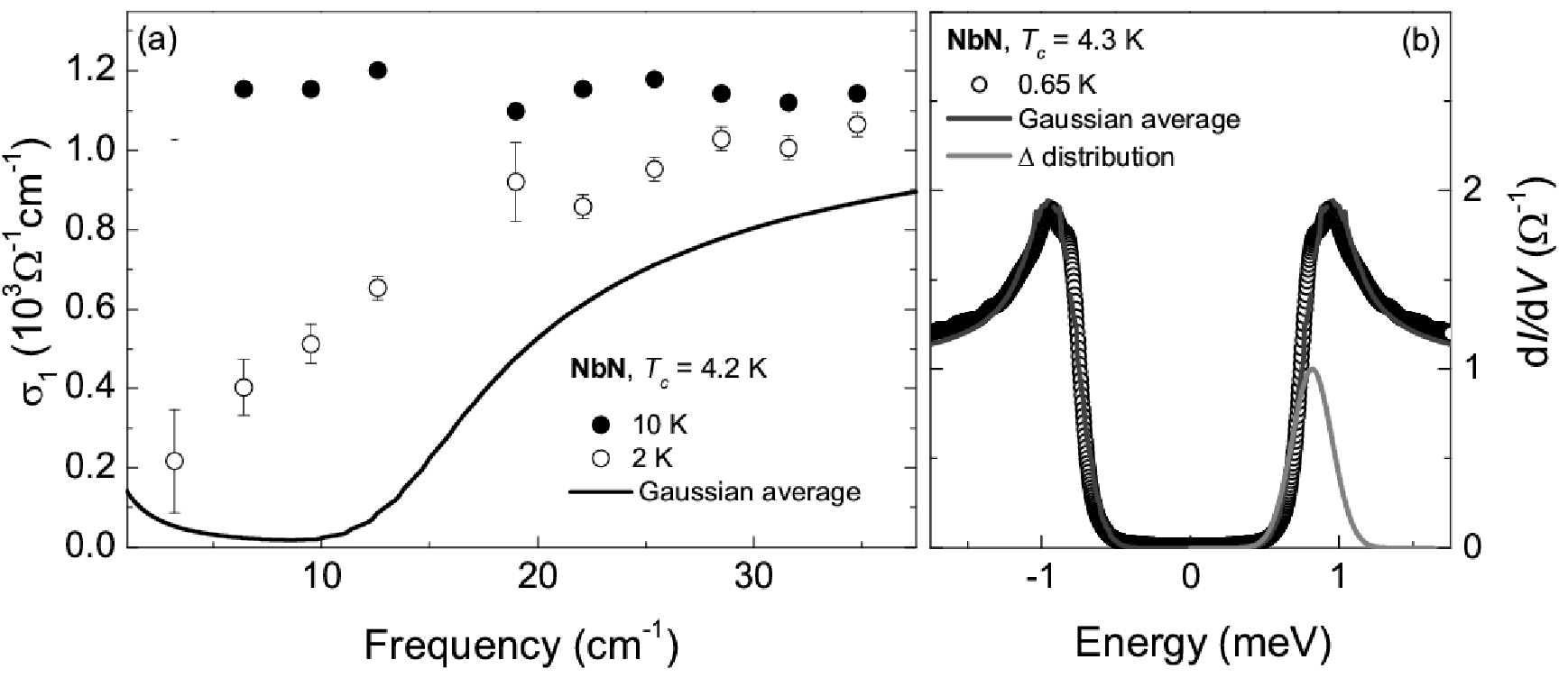}
\caption{\label{Fig_NbN4200_sigma1+DOS_AVERAGED}\textbf{Curves from Gaussian distributed local gaps $\Delta(\mathrm{r})$} applied to a strongly disordered sample with $T_c=4.2$ (a) and 4.3\,K (b). The fit of the $dI/dV$ measurement is the average of 81 $\mathcal{D}(E)$ curves calculated from a Gaussian distribution of $\Delta(\mathrm{r})$ values. The corresponding average of $\sigma_1(\nu)$ curves  cannot reproduce the experimental result.}
\end{center}
\end{figure}
What is the influence of an emergent electronic inhomogeneity? The above $dI/dV$ spectra are measured with planar tunneling junctions. Local tunneling measurements, however, revealed an emergent electronic inhomogeneity on a length scale of the coherence length much greater than the lateral dimension of the planar junctions. In that sense, the smeared $dI/dV$ spectra could also be modeled as an average of locally varying $\Delta(\mathrm{r})$ and thus $\mathcal{D}(E,\mathbf{r})$ curves. In fact, assuming a Gaussian distribution of $\Delta$ values gives a fairly good approximation of the $dI/dV$ spectrum, see Fig.~\ref{Fig_NbN4200_sigma1+DOS_AVERAGED}, although after closer examination it falls short compared to the pair-breaking model. Employing the identical $\Delta(\mathrm{r})$ distribution for an averaged $\sigma_1(\nu)$ curve yields a curve completely at odds with the actual measurement. Attributing the mismatch between $dI/dV$ and $\sigma_1(\nu)$ measurements can thus not be accounted for by a simple averaged response due to electronic inhomogeneity.\\

The failure of both pair-breaking and an averaged response due to a distribution of $\Delta(\mathrm{r})$ values as possible underlying scenarios accounting for the apparent mismatch between optical and tunneling spectroscopy calls for an alternative effect at play. In the next section, we will see that three specific properties of NbN - the short coherence length of a few nanometer, the quasi-2D character and the vicinity to quantum criticality - make this material a promising candidate for a excitation best known from the Standard Model of particle physics serving as possible explanation for puzzling observation above.

\section{The Higgs mode and the Higgs mass}\label{Sec:NbNHiggs}
The striking resemblance of both the theory of the Higgs boson in high-energy physics and superconductivity gives rise to a natural question: is there a Higgs-boson like excitation in superconductors? Clearly, on the one hand an affirmative answer is challenged by half a century of uncountable spectroscopic measurements \emph{not} revealing a superconducting Higgs mode. On the other hand, the successful and elegant treatment of superconducting phenomena by virtue of field theories and symmetry breaking is a strong pleading for a Higgs mode. Hence, a more constructive approach is not to ask \emph{if} but \emph{where} there is a Higgs mode, and \emph{how} it can be detected experimentally. 
In recent years, works of Auerbach \emph{et al.} gave a particular clear account to these questions as we will review below.

A instructive way to understand the elusive nature of the Higgs mode is to reconsider under which circumstances the particle analogue appears. The well-defined Higgs boson appearance as a sharp resonance is predicted within a \emph{relativistic} and \emph{bosonic} broken-symmetry QFT. Although Cooper pairs as constituents of superconducting condensate can be viewed as composite bosons, the fermionic nature of the underlying quasiparticles usually plays an important role. Qualitatively, this can be seen as consequence of the pairing taking place in $\mathbf{k}$- rather than real space. The coherence length $\xi$, as a measure for the quasiparticle's real-space distance, is for conventional superconductors usually much larger than the inter-atomic distance so that bosonic treatments are not applicable. While Bose-Einstein condensed (BEC) neutral superfluids contain tightly-bound pairs and bosonic theories apply, they do not require local $U$(1) invariance and gauge fields and thus the Higgs mechanism does not occur.
\begin{marginfigure}
\begin{centering}
\includegraphics[width=\marginparwidth]{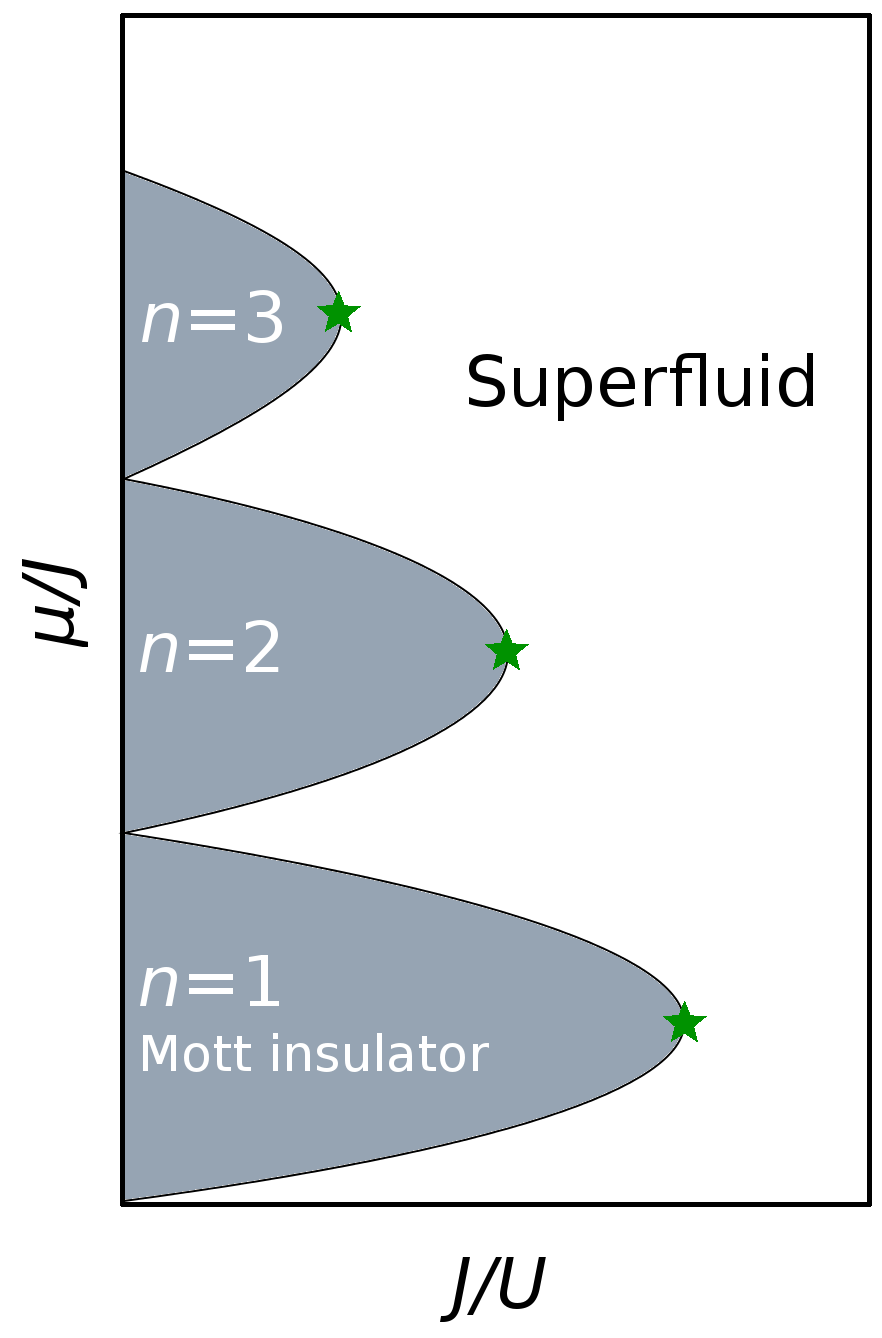}
\caption{\label{Fig:BHM}Phase diagram of lattice bosons within the Bose-Hubbard model. The tips of the lobes are points with emergent relativistic dynamics. Adopted from Ref. \cite{Gazit2015}}.
\end{centering}
\end{marginfigure}

The search for a suited system with a Higgs mode therefore resorts to effectively bosonic charged superfluids, i.e. short coherence-length superconductors. Thin films of NbN are such candidates: For films of around 50\,nm thickness, Hall- and magneto-resistance measurements \cite{Chockalingam2008} revealed a coherence length of $\xi\sim 5$\,nm, which is just around one order of magnitude larger than the lattice constant of NbN and much smaller than the penetration depth $\sim 200$\,nm rendering disordered NbN thin-films a prototypical short-coherence superconductor and ideal testbed for effectively bosonic theories. 

BEC-like pairs can be treated as lattice bosons within the Bose-Hubbard model (BHM)
\begin{equation}
\mathcal{H}=-J\sum\limits_i (b_i^\dagger b_j+b_j^\dagger b_i)-\mu \sum\limits_i n_i-U \sum\limits_i n_i(n_1-1)\label{BHM}
\end{equation}
where $b,b^\dagger$, and $n=b^\dagger b$ are bosonic annihilation, creation, and number operators, $J$ and $U$ are hopping and onside-repulsion energies, and $\mu$ is the chemical potential. The phase diagram of this model features insulating and superfluid phases depending on the filling (bosons per lattice site) and the kinetic energy, see Fig. \ref{Fig:BHM}.
By means of a Hubbard-Stratonovich transformation the BHM Hamiltonian can be cast into a field theory with an imaginary-time action reading \cite{Sengupta2005}
\begin{equation}
\mathcal{S}_{BHM}=\int\mathrm{d}\tau \mathrm{d}^d x \mathcal{L}_B
\end{equation} 
with the Lagrangian
\begin{equation}
\mathcal{L}_{BHM}=K_0\psi^\dagger \partial_\tau \psi + K_1|\partial_\tau\psi|^2+K_2|\nabla \psi|^2+r|\psi|^2+u|\psi|^4
\end{equation}
which is identical with the one we discussed in Sec.~\ref{Sec:FieldTheoforSC}. Clearly, (imaginary) time and space dependencies are not treated on equal footings. It is the linear-time derivative in the first term that spoils relativistic dynamics\sidenote{\footnotesize{The situation is similar to non-relativistic (Schr\"{o}dinger) single-particle quantum mechanics \begin{center}$(2im\partial_t-\partial_\mathrm{r}^2)\psi=0$\end{center} compared to the relativistic (Dirac) version \begin{center}$(i\gamma^\mu\partial_\mu -m)\psi=0$\end{center}} }. In a seminal work \cite{Fisher1989}, however, Fisher {et al.} have demonstrated that $K_0$ can be related to $r$ via $K_0=-\frac{\partial r}{\partial \mu}$, and further, that $K_0$ vanishes at the tips of the Mott lobes, see Fig.~\ref{Fig:BHM}. Consequently, the Lagrangian becomes symmetric in space and time coordinates and henceforth describes relativistic bosons. The $T=0$ superfluid-Mott insulator quantum phase transition (QPT) within the BHM has a prototypical condensed-matter realization in form of the superconductor-insulator quantum phase transition (SIT) observed in granular and disordered superconductors such NbN.\\

In recent years, the dynamics of the BHM in the relativistic limit have been studied intensively by Auerbach \emph{et al.} in terms of $O(N)$ field theories\sidenote{\footnotesize{
Note that the description in terms of a the $U$(1) complex scalar model is here equivalent to the $O$(2) real vector model \cite{Auerbach2015}}} with an action \cite{Sachdev1999}
\begin{equation}
\mathcal{S}[\phi]=\int \mathrm{d}\tau\mathrm{d}^2x\left\{\frac{1}{2}(\partial_\mu \phi )^2+\frac{\mu}{2}|\phi|^2+g|\phi|^4\right\} 
\end{equation}
in 2+1 space-time where $\phi$ is an $N-$component real vector. This model has turned out to be a powerful approach to study dynamics and collective modes in quantum critical condensed matter systems, in particular the superfluid-Mott insulator transition \cite{Fisher1989} for $N=2$ and Neel-singlet transition of dimerized Heisenberg antiferromagnets \cite{Chakravarty1989} for $N=3$. We will now review some of the results \cite{Podolsky2011,Gazit2013,Gazit2013B,Gazit2015} obtained by Auerbach \emph{et al.} relevant for this work.     
\begin{itemize}
\item{Relativistic dynamics are essential for the appearance of a long-lived Higgs mode. In the (Gross-Pitaevskii) limit of non-relativistic dynamics, the collective excitations of the order parameter are coupled massless amplitude-phase phonons, whereas for the relativistic case the $O(2)$ model predicts one massless phase mode \emph{and} one massive amplitude (Higgs) mode.}
\item{The aforementioned visibility in the dynamical conductivity does not apply to conventional weak-coupling superconductors, where the BCS Hamiltonian commutes with the (Cooper-pair) current operator leaving the spectral gap essentially open. In the strong coupling limit,  the coupling between charges and photons happens by gauging the theory as discussed in Sec.~\ref{Sec:FieldTheoforSC}. Via minimal coupling the gauge field $\mathbf{A}$ is introduced and one obtains 
\begin{equation}
\mathcal{L}^{O(2)}[\phi,\mathbf{A}]=\frac{1}{2}(\partial_\mu \phi )^2-\frac{\mu}{2}|\phi|^2+g|\phi|^4+\mathcal{L}_\mathrm{em}[\phi,\mathbf{A}]
\end{equation}
where\sidenote{\footnotesize{Here, we parametrize the order parameter field $\phi$ as \begin{center}$\phi=\begin{pmatrix}\langle\phi\rangle_0+\sigma \\ \pi\end{pmatrix}$\end{center} where $\sigma$ are the longitudinal fluctuations around the ground state $\langle\phi\rangle_0$ and $\pi$ is the phase field. Note, that a real two-component vector is equivalent to a complex scalar order parameter description.
}}
\begin{equation}
\mathcal{L}_\mathrm{em}=e\mathbf{A}(\nabla\pi)\left\{|\langle\phi\rangle_0|^2+2\sigma\langle\phi\rangle_0+...\right\}
\end{equation}
contains the relevant electron-light interactions. Currents are obtained by functional derivatives of $\mathcal{L}_\mathrm{em}$ with respect to $\mathbf{A}$. To lowest order, this yields two diagrams: a phase mode $e\nabla \pi$ and a coupled phase-amplitude mode $2e\nabla \pi\sigma$. }
\item{The visibility of the Higgs mode is closely related to the symmetry of the probe. The longitudinal susceptibility of the $O(N)$ field theory in two spatial dimensions diverges at low energies so that the spectral signature of the Higgs mode is washed out into a broad shoulder. However, any scalar susceptibility probing the square of the order parameter rather than its direction, is finite at all frequencies and displays a peak right at the Higgs mass $m_H$ (an energy scale) which manifests as the onset of electromagnetic absorption measured by the dissipative conductivity. }
\item{To understand the physical implications of the above phase-amplitude diagram, we recall the result of Sec.~\ref{Sec:LowEforSC}, i.e. gradients of the superconducting phase, $\nabla \pi$, generate currents and hence cause density fluctuations of the condensate. In 3+1 dimensions, these are massive plasmons with a threshold energy given by plasma frequency $\omega_p$. In turn, while being a sharp resonance in 3+1 dimensions, the Higgs mass $m_H$ is shifted to energies $\hbar \omega_p+m_H$ where the Higgs modes are completely overshadowed by electronic interband excitations and experimental identification is a hopeless task. To the contrary, the plasmons are \emph{not} gapped in 2+1 dimensions due to the nature of Coulomb interaction and can be excited at arbitrarily low energies. The excitation threshold for the conductivity is here determined\sidenote{\footnotesize{Incorporation of higher-order diagrams shows, that the threshold at $\hbar\omega=m_H$ is actually a soft gap closing as $\omega^5$ in leading order suppressed further by a tiny numerical prefactor.}} by $m_H$. With $\xi\sim 5$ and $d\sim 20$\,nm, the NbN films under study are, strictly speaking, not in the actual 2+1 limit, but  with $\xi$ and $d$ being of the same order, the dimensional crossover is close such that traces of 2+1 physics are likely to be observed.}
\end{itemize}
In conclusion, the short-coherence superconductor NbN allows an effective bosonic treatment within the BHM model (and related field theories) and together with the emergent relativistic dynamics in approach of a quantum phase transition is a natural candidate to search for the Higgs mode in a solid-state system. Furthermore, being a scalar susceptibility, the optical conductivity is expected to display an onset of finite $\sigma_1(\nu)$ (for $T=0$) or a minimum (for $T>0$) at $m_H$ and therefore is a suited probe for the Higgs mode. \\

\begin{figure}[t!]
\begin{center}
\includegraphics[width=\textwidth]{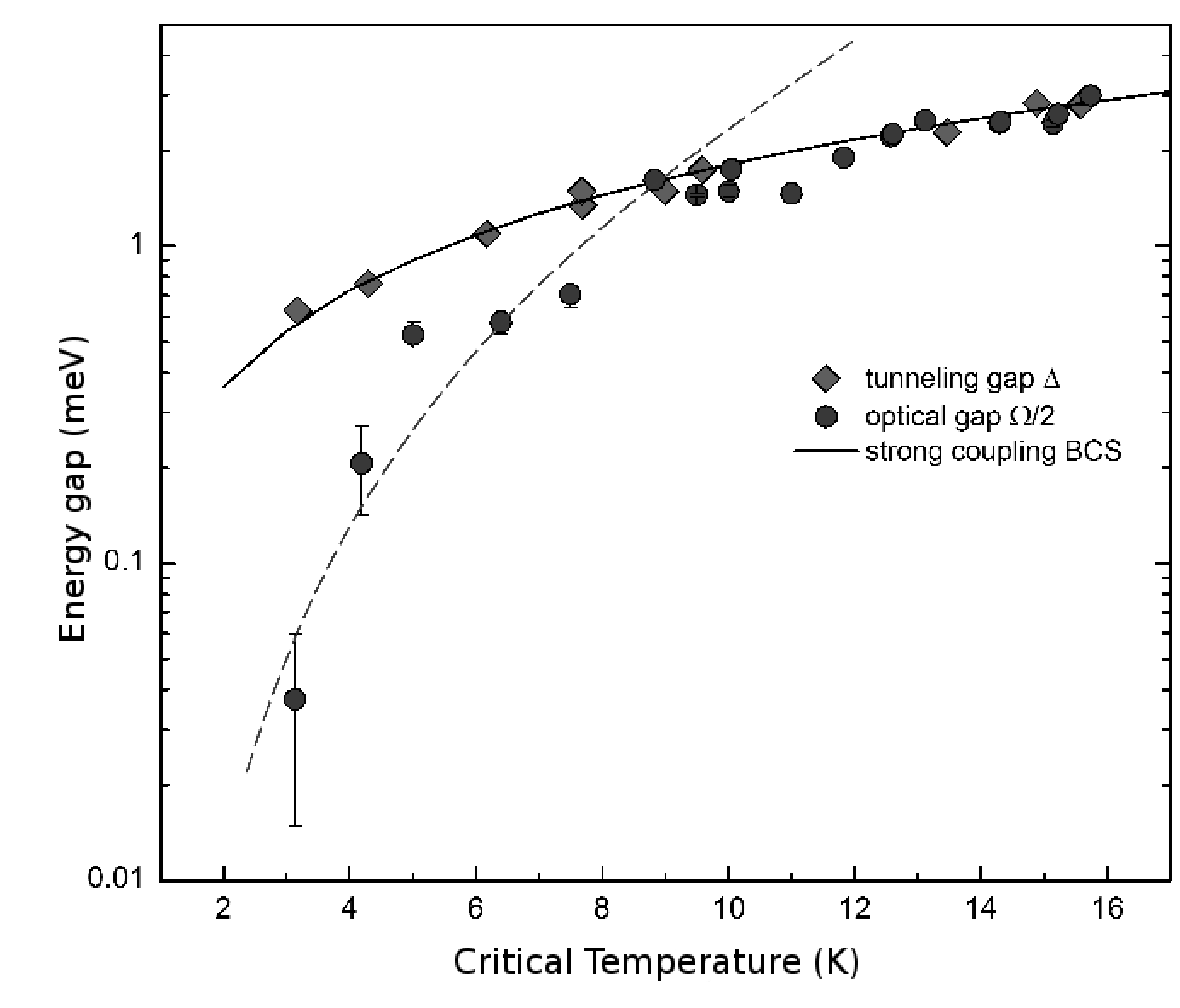}
\caption{\label{Fig:Densities}\textbf{Superconducting energy scales towards criticality} Comparison between energy gap $\Delta$ obtained from tunneling measurements and (half of) optical absorption threshold $\Omega/2$ taken as the minimum in $\sigma_1(\nu)$ for NbN films approaching the SIT. In the clean limit (high $T_c$), both energy scales are approximately the same. While the decrease of $\Delta$ follows $T_c$ in agreement with the strong-coupling BCS reduction (black line), the reduction of $\Omega/2$ speeds up towards criticality. The dashed line is a guide to the eye.   }
\end{center}
\end{figure}
Coming back to the experimental results, we can assemble characteristic features of the $\sigma_1(\nu)$ spectra as function of $T_c$. We start with the energy gap $\Delta$ obtained from tunneling and the spectral gap $\Omega$ being read out from $\sigma_1(\nu)$, see Fig.~\ref{Fig:Densities}(a). The tunneling gap $\Delta$ is reduced following the strong-coupling BCS prediction indicated by the solid line with $\Delta/k_BT_c=2.1$ down to the lowest $T_c\approx 3$\,K. Starting with clean samples far from criticality, corresponding\sidenote{\footnotesize{Note that the spectral gap for pair excitations lies at $\Omega=2\Delta$.}} spectral gap $\Omega/2$ also follows the BCS strong-coupling behavior down to $T_c\approx 9$\,K. Further approach of criticality then pushes $\Omega/2$ below $\Delta$. This relative suppression becomes stronger as $T_c$ goes down. At the lowest-$T_c$ sample\sidenote{\footnotesize{Here, $\sigma_1(\nu)$ was measured by means of microwave Corbino spectroscopy. Taken from Ref.~\cite{Mondal13}.}} the deviation amounts to nearly one order of magnitude. This implies, that while tunneling spectroscopy probes the superconducting energy gap $\Delta$ for all distances from criticality, below a certain $T_c$, another energy scale appears in the optical absorption spectrum. Far far from criticality, this additional scale is above $\Delta$, so that $\Omega$ also measures $\Delta$ (or rather $2\Delta$). The reduction below the strong-coupling curve at around 9\,K signals the appearance of a new energy scale we interpret as the Higgs mass $m_H$.\\
The Higgs mode is a critical mode of the QCP meaning that $m_H$ vanishes as $T_c\to 0$ at the SIT. Within a bosonic scenario of the SIT\sidenote{\footnotesize{which seems likely to be realized in NbN, where superconductivity at small order parameters is governed by phase fluctuations \cite{chock09,chand2012}}.}, the reduction of $m_H$ below $\Delta$ is a natural consequence, as the pairing energy remains finite across the transition. The increasing reduction of $m_H$ in approach of criticality is in qualitative agreement with the theoretical prediction of the $O(2)$ model, where $m_H\propto (\delta g)^{0.67}$ with $\delta g$ being a dimensionless parameter measuring the distance to the QCP. How $\delta g$ relates to $T_c$ is not a priori clear, so that we can only qualitatively confirm the predicted decay, while affirmation of the exponent remains an open problem.  \\
So far, we have considered only the real part of the complex conductivity. In what follows, we employ the inductive response of the superfluid, $\sigma_2(\nu)$, as a check for internal consistency of the above interpretation. \\
For any system with a constant total carrier density $n_e$, the spectral weight defined as
\begin{equation}
\mathfrak{s}=\int\limits_0^\infty d\omega \sigma_1(\omega)=\frac{\pi n_ee^2}{2 m}\label{eq:SpectralWeight}
\end{equation}
is strictly conserved\sidenote{\footnotesize{Given that the bandmass $m$ also remains constant which in first approximation holds true for non-correlated metals.}}
as it follows from the fundamental Kramers-Kronig relations for causal response functions. The opening of the superconducting gap in $\sigma_1(\nu)$ goes along with a redistribution of spectral weight such that the 'missing' spectral weight at finite frequencies, i.e. the quasi particle contribution $\sigma_{1,QP}$, is compensated by the zero-frequency $\delta$-response, namely the superfluid contribution $\sigma_{1,SF}$, thus
\begin{eqnarray}
n_e&=&\frac{2m}{\pi e^2}\int\limits_0^\infty d\omega \sigma_{1,n}(\omega)\nonumber\\
&=&\frac{2m}{\pi e^2}\int\limits_0^\infty d\omega \left\{\frac{e^2\pi n_s}{m}\delta(\omega)+\sigma_0\times\sigma_{1,QP}(\omega)\right\}\nonumber\\
&=&n_s+\frac{2m\sigma_0}{\pi e^2}\int\limits_0^\infty d\omega \sigma_{1,QP}(\omega)\label{eq:SpectralWeight2}
\end{eqnarray}    
where $n_s$ is the superfluid density, $\sigma_{1,n}$ the normal-state conductivity, and $\sigma_{1,QP}$ the normalized quasiparticle conductivity scaled with the normal-state dc-conductivity $\sigma_0$. Equation (\ref{eq:SpectralWeight2}) holds true for any system where the charge dynamics follows BCS theory. It further states that if $\sigma_0$ is reduced $\sigma_0\to \alpha\sigma_0$ with $|a|<1$, then both $n_e$ and $n_s$ are reduced by the same factor $\alpha$. In turn, the appearance of superconducting excitations beyond BCS theory associated with finite-frequency spectral weight calls for a reduction of thereof in a disparate spectral range. For the Higgs mode, being a excitation of the superfluid condensate, a reduction of the superfluid spectral weight is most likely. Indeed, by quantum Monte Carlo (QMC) simulations of the disordered XY-model, Trivedi \emph{et al.} have shown a direct correspondence between the excess spectral weight arising from the Higgs mode and the reduction of the superfluid weight \cite{Swa14}. \\
The superfluid dominates $\sigma_1(\nu\approx 0) \approx \delta(\nu)$ in the zero-frequency limit and therefore, following Kramers-Kronig relations, also determines $\sigma_2$ at small, but finite frequencies $\nu < 2\Delta/h$. Considering the Kramers-Kronig transform for the superfluid contribution
\begin{equation}
\sigma_2(\omega\approx0)=-\frac{2}{\pi}\mathrm{P}\int\limits_0^\infty\frac{\mathrm{d}\omega^\prime\omega}{\omega^{\prime 2}-\omega^2}\frac{\pi n_s e^2\delta(\omega^\prime)}{2m}
\end{equation} 
one finds the relation
\begin{equation}
n_s=\frac{2\pi m}{e^2}\nu\sigma_2(\nu)\Bigr|_{\nu = 0} \label{ns}
\end{equation}
using $\omega=2\pi\nu$ and $\delta(\omega)=\frac{1}{2\pi}\delta(\nu)$. In practice, we multiply the measured $\sigma_2(\nu)$ with frequency and extrapolate to $\nu\to 0$ to estimate $n_s$. This procedure is comparably reliable as the dependence $\sigma_2(\nu)\propto 1/\nu$ cancels the $n_s\propto \nu$ dependency so that the right side of Eq. (\ref{ns}) becomes constant in the low-$\nu$ limit. In Fig.~\ref{Fig:sig2NbN} we display the product $\sim \nu\sigma_2$ for the NbN films studied in this work. The zero-frequency extrapolation is taken as the mean value of the two left-most data points and shown as red stars.

\begin{figure}[h!]
\begin{center}
\includegraphics[width=\textwidth]{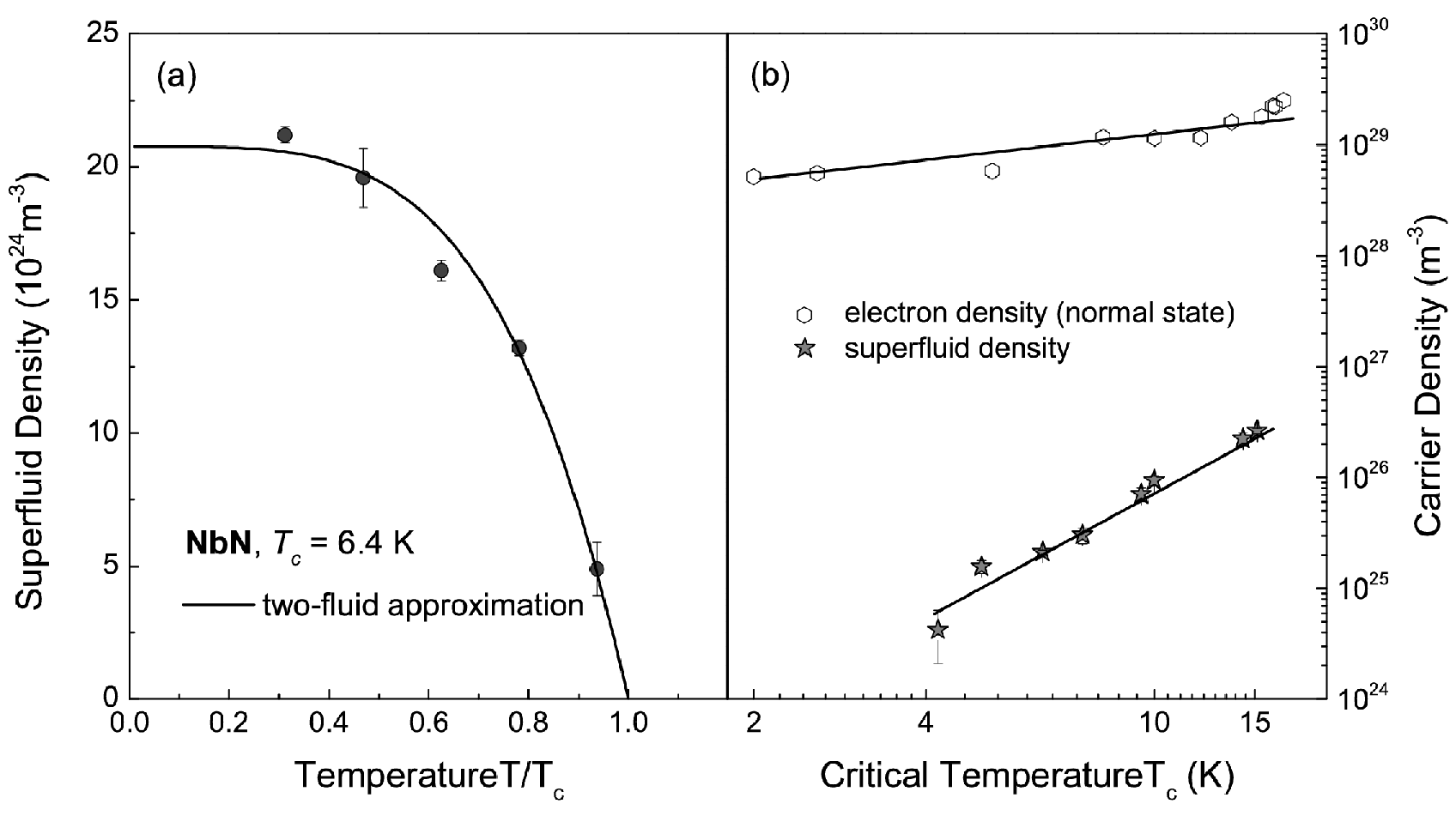}
\caption{\label{Fig:Densities}\textbf{Superfluid and total carrier densities} (a) temperature dependence of the superfluid density $n_S$ for a sample with $T_c=6.4$\,K. The solid line is a fit to the two-fluid approximation giving a zero-temperature extrapolation of $n_s(0)=2.07\times 10^{25}$\,m$^{-3}$. (b) Comparison between the total carrier density $n_e$ \cite{Chand2012PhD} from normal-state Hall measurements and the zero-temperature superfluid density obtained from $\sigma_2(\nu)$. Note the stronger decay of $n_s$ compared to $n_e$ in agreement with a redistribution of spectral weight weakening the superfluid contribution towards criticality.  Lines are guides to the eye.  }
\end{center}
\end{figure}
\begin{figure}[t!]
\begin{center}
\includegraphics[width=\textwidth]{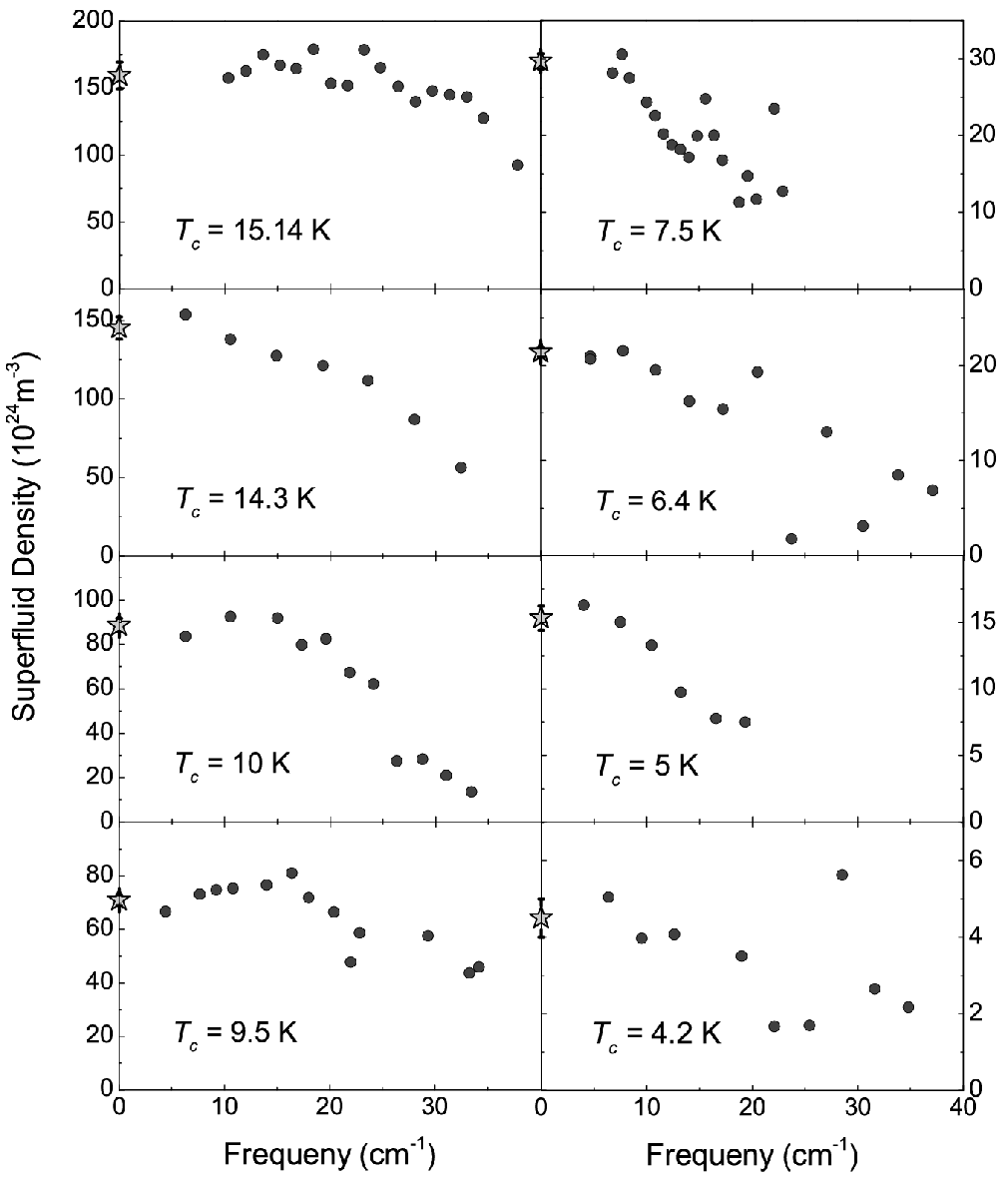}
\caption{\label{Fig:sig2NbN}\textbf{Determination of the superfluid density} The figure assembles experimental results on the inductive response of NbN samples under study multiplied with frequency, $\nu\sigma_2$ which according to Eq.~(\ref{ns}) measures the superfluid density in the limit $\nu\to 0$ (shown as red stars)  }
\end{center}
\end{figure}
To account for the different relative temperatures $T/T_c$ of the various measurements, we correct to thus-obtained value of $n_s(T)$ by assuming the two-fluid approximation \cite{Dressel2002}
\begin{equation}
\frac{n_s(T)}{n_s(0)}\approx 1-t^4\label{Eq:2fluid}
\end{equation}  
which only relies on the relative temperature $t=T/Tc$. The applicability of this approximation is shown for a representative sample with $T_c=6.4$\,K in Fig.~\ref{Fig:Densities}(a), where a fit according to Eq.~(\ref{Eq:2fluid}) yields a satisfying description of the temperature dependence\sidenote{\footnotesize{Note that the zero-temperature extrapolation of $n_s$ and the value at the lowest measurement temperatures differs by less than 5\% so that the actual form of $n_s(T)$ plays only a minor role for $n_s(0)$.}}.\\
Figure~\ref{Fig:Densities}(b) compares $n_s(0)$ with the total carrier density $n_e$ obtained by Hall measurements in the normal state \cite{Chand2012PhD}. As the resistivity of the films is increased towards criticality, both $T_c$ and $n_e$ are reduced. According to Eq.\,(\ref{eq:SpectralWeight2}) the decline of $n_e$ should cause the same reduction of $n_s$. Experimentally, however, we find $n_s$ to be reduced approximately 10 times stronger than $n_e$ by nearly two orders of magnitude. The progressive suppression of  superfluid spectral weight is in agreement with the redistribution thereof at finite frequencies due to the emergence of the Higgs mode. Note that this finding also proves the superconducting origin of the excessive dynamical conductivity in $\sigma_1(\nu)$ and rules out a redistribution of spectral weight from higher energies due to, e.g., a change in the plasma frequency.\\
\begin{figure}[b!]
\begin{center}
\includegraphics[width=\textwidth]{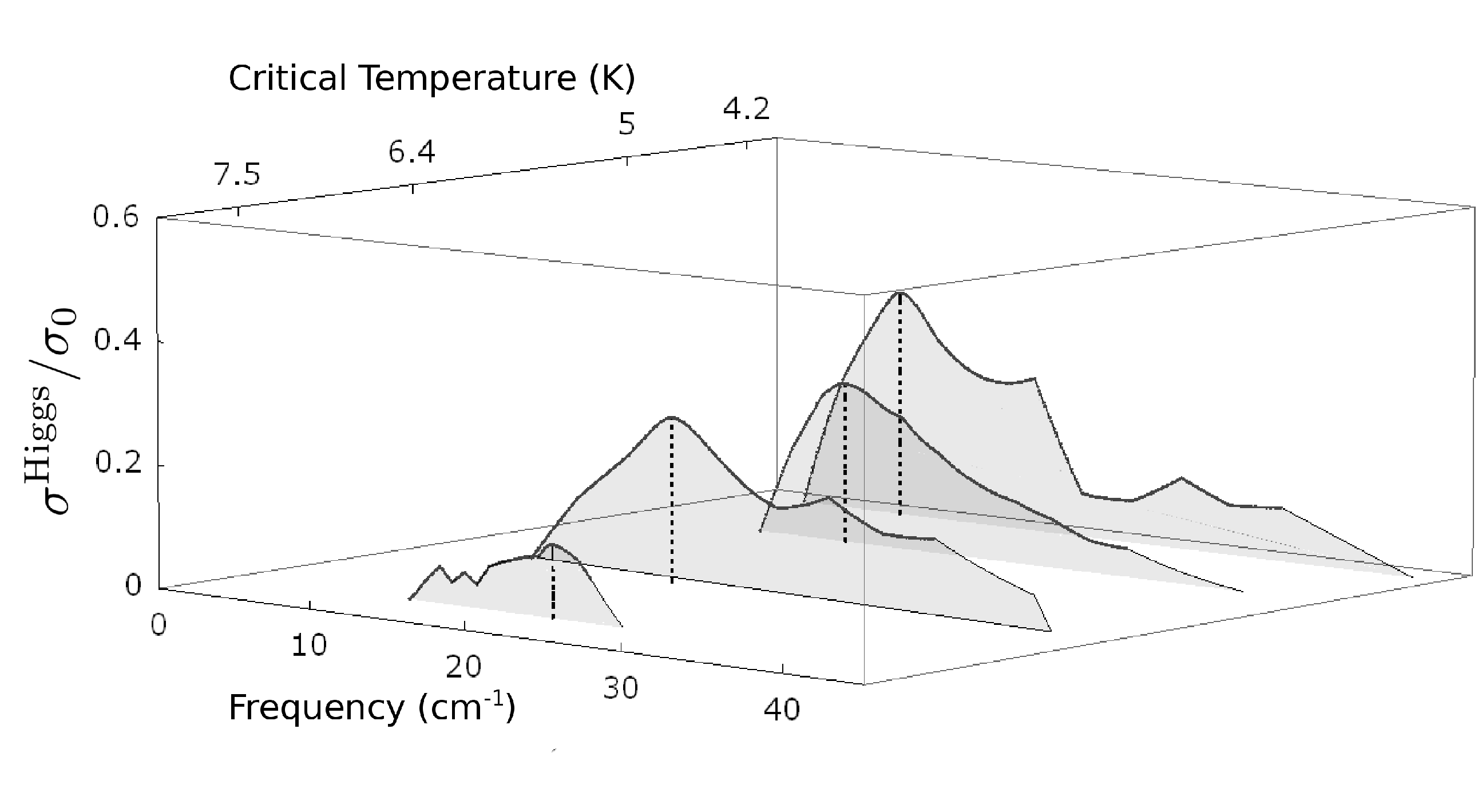}\\
\vspace{1cm}
\includegraphics[width=\textwidth]{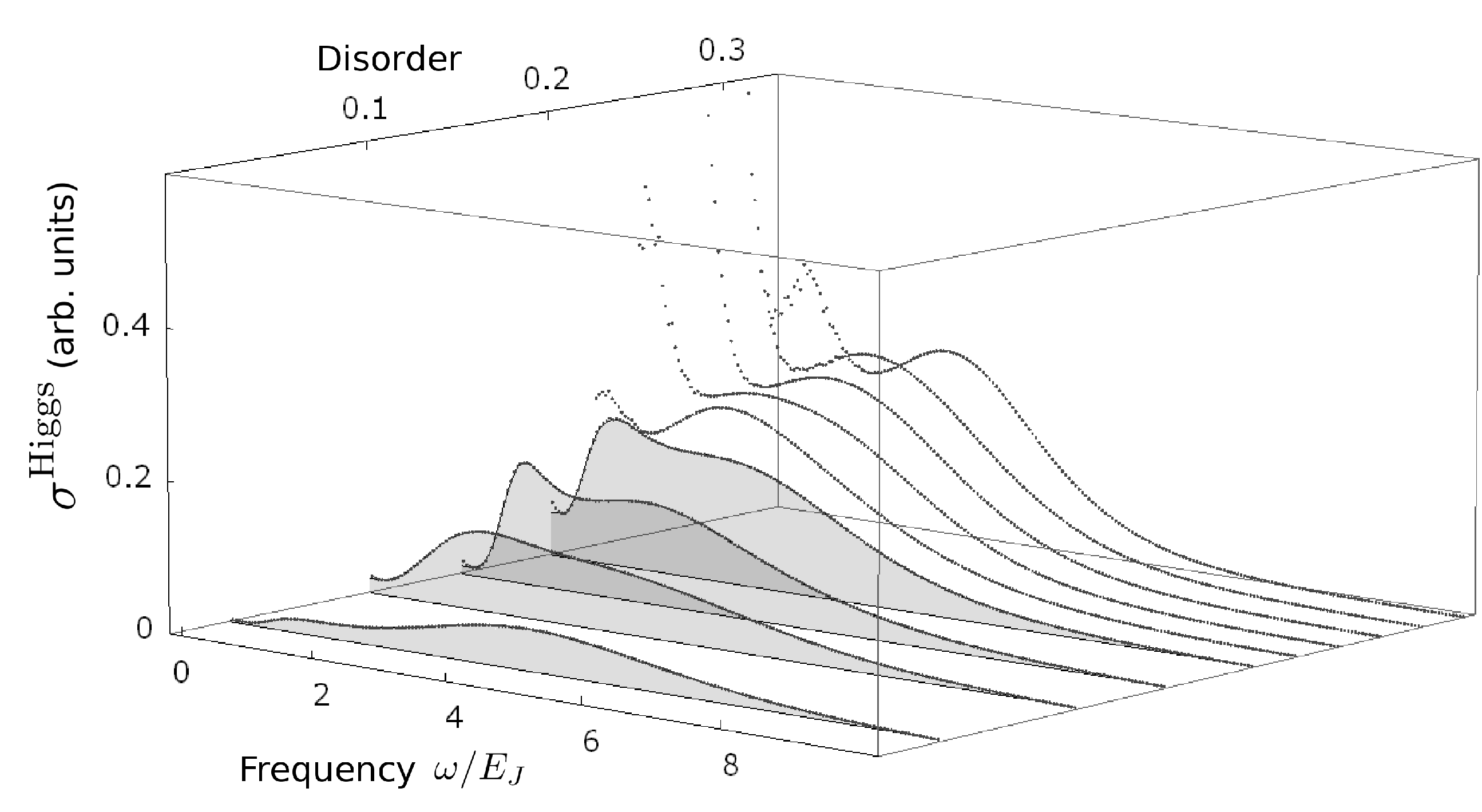}
\caption{\label{Fig:NbNHiggsExp}\textbf{Higgs conductivity in experiment and theory.} Top panel: excessive conductivity extracted by subtraction of the quasiparticle contribution from the experimental $\sigma_1(\nu)$ for various films towards criticality. As $T_c$ goes down, the mode grows in spectral weight and acquires a maximum shifting towards lower energies. Bottom panel: Higgs conductivity calculated from QMC simulations of the disordered XY Hamiltonian \cite{Swa14} at various disorder values. Shaded curves are chosen for comparison based on line shape similarities to the experiment.}
\end{center}
\end{figure}
Finally, we address the dispersion of the Higgs mode. For this we subtract the predicted $\sigma_1(\nu)$ curves based on the tunneling pair-breaking strength from the actual measurements. The upper panel of Fig.\,\ref{Fig:NbNHiggsExp} displays $\sigma^\mathrm{Higgs}(\nu)$ extracted for four samples\sidenote{\footnotesize{Although traces of excessive conductivity appear in samples with $T_c$ up to 14.3\,K, a \emph{meaningful} isolation of $\sigma^\mathrm{Higgs}$ only works for the samples shown in Fig.\,\ref{Fig:NbNHiggsExp} due to the finite data spacing.}} with $T_c=7.5-4.2$\,K.  Although the studied spectral range allows to capture the major part of the Higgs mode (displayed as thick blue lines), it´s connection to the quasi particle curve at high frequencies (thin black lines) is not fully recovered for all samples. To calculate the spectral weight of the Higgs mode for a quantitative comparison with the reduction in $n_s$, an extrapolation is required, whose particular form, however, strongly affects the resulting integral. Unfortunately, the inaccessible high-frequency tail renders a quantitative spectral weigh analysis impossible. Nevertheless, some important connections can be made to theory. The lower panel of Fig.\,\ref{Fig:NbNHiggsExp} displays the theoretical prediction of the bare Higgs mode conductivity obtained from Quantum Monte-Carlo studies of  the disordered XY Hamiltonian \cite{Swa14} at various distances from the SIT measured in terms of the disorder parameter $p$. To relate the energy scale of the simulation, the Josephson energy $E_J$, to experiment, one can estimate $E_J$ from the clean-sample $T_c$ as $E_J=2k_BT_c/\pi$ \cite{Trivedi2016} yielding$\sigma^\mathrm{Higgs}/\sigma_0$ $E_J\approx 0.8$\,meV. In experimental units, the energy scale in the lower panel of Fig.\,\ref{Fig:NbNHiggsExp} would then cover frequencies $0-64$\,cm$^{-1}$ similar to the measurement range. Comparing the measured and simulated $\sigma^\mathrm{Higgs}$ (highlighted by colored areas beneath corresponding curves) reveals a fundamental resemblance: starting far from criticality, both amplitude and width of the mode increase towards the SIT. At the same time, a pronounced peak evolves that grows in amplitude and shifts towards lower energies. A close examination, however, also reveals differences: While the clean onset of absorption at $m_H$ of the $O(2)$ model is somewhat washed out by disorder, the measured $\sigma^\mathrm{Higgs}$ displays a clear onset at finite energies. This points towards the important question of the interplay between disorder, the SIT, and the Higgs mode we will address in the next section.\\

In conclusion, by systematic comparison of tunneling and optical measurements on a set of NbN films tuned towards the SIT we unraveled an energy scale that shifts below the superconducting energy gap $\Delta$ associated with the emergence of an enhanced conductivity at energies around $2\Delta$. Neither disorder enhanced pair-breaking effects nor an inhomogeneous spatial gap distribution could explain the evolution of the additional spectral weight and shift of absorption threshold towards criticality. Instead, we suggest that the vicinity to quantum criticality and the short coherence length of the quasi-2D films allow a treatment within the fully relativistic bosonic $O(2)$ field theory. Within this framework, we interpret the additional absorption channel beyond quasi particle dynamics as the collective amplitude mode of the superconducting order parameter, namely the Higgs mode, and the new energy scale diving below $\Delta$ as the Higgs mass $m_H$. The appearance of the Higgs mode goes hand in hand with an anomalous reduction of superfluid density and is in good agreement with the line shape predictions obtained from Quantum Monte Carlo simulations of the disordered $XY$ Hamiltonian. The observation of the softening Higgs mode is a  direct proof that the SIT is a quantum critical point  in which a diverging timescale is detected \cite{sherman15}. Evidently, the vicinity to the QPT offers a unique opportunity to study the nature of the low energy collective excitations in superconductors. As a prototype of quantum criticality, the findings presented here also have implications on broader questions about the effects of interactions and disorder in condensed matter, and to related questions in interacting cold atoms and quantum statistical mechanics.

\section{Some concluding remarks on the Higgs mode in disordered systems} \label{NbNremarks}
In the previous sections we considered NbN films tuned to quantum criticality by increasing disorder. Indeed, disorder certainly increases as measured by the increasing normal state resistivity and, e.g., the Ioffe-Regel parameter $k_F\ell$ \cite{Chand2012PhD}. Considering disorder, however, as the underlying mechanism \emph{driving} the thin films from superconducting to insulating phases, may lead to troublesome inconsistencies in the above interpretation that deserve a closer examination. In what follows, we will first argue, that disorder increases, but not necessarily causes the SIT, and, by this, justify the applicability of the disorder-free $O(2)$ field theory on which the above interpretation is essentially build on.\\
\begin{marginfigure}
\includegraphics[width=\marginparwidth]{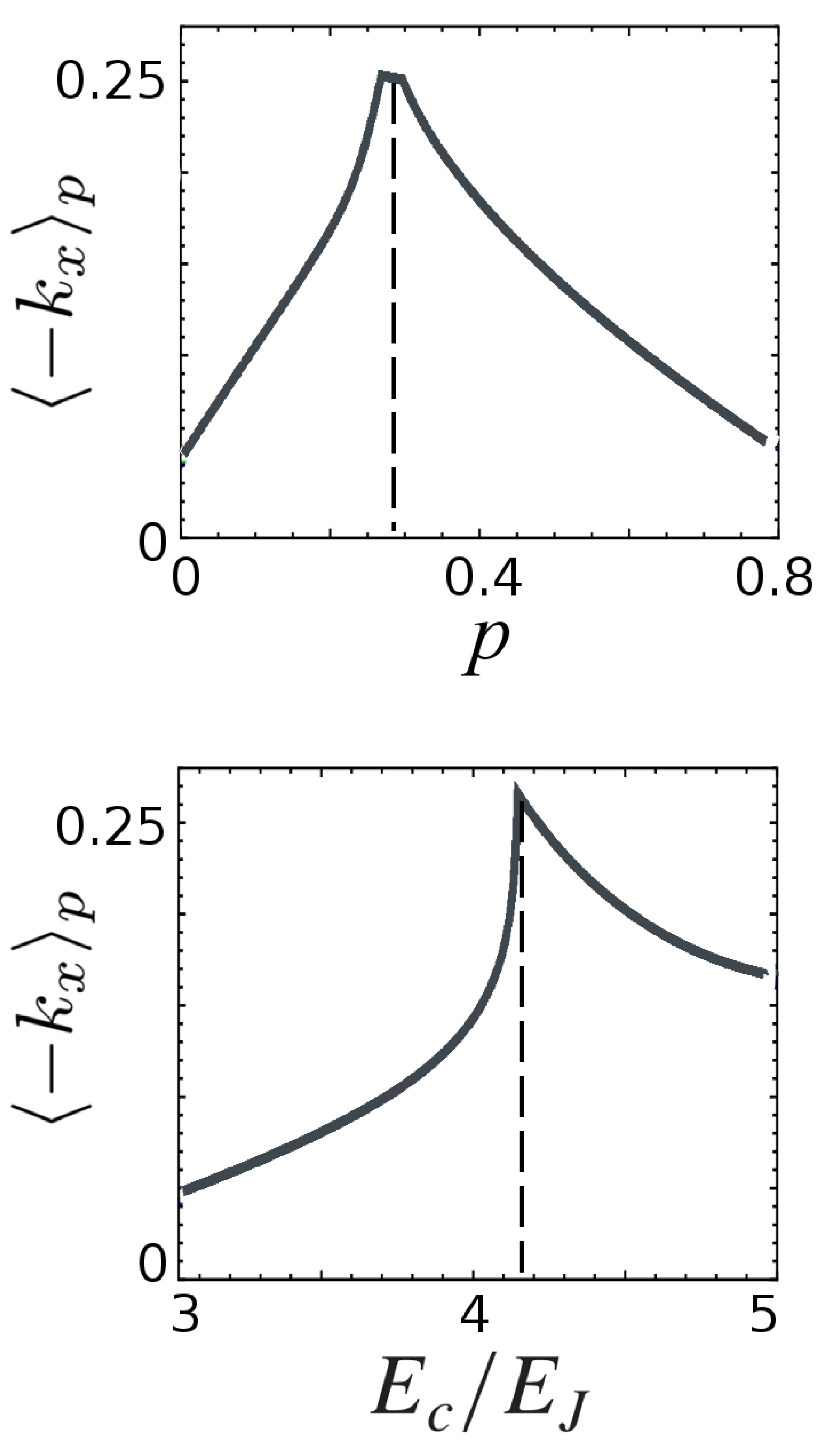}
\caption{\label{disorder_p}Average total kinetic energy obtained from simulations of the XY Hamiltonian as functions of disorder $p$ and the Coulomb-to-Josephson energy ratio $E_C/E_J$. The SIT (dashed lines) can be realized by tuning $p$ or $E_C/E_J$. Adopted from Ref.~\cite{Swa14}.}
\end{marginfigure}
The experimentally detected Higgs modes, displayed in the upper panel of Fig.\,\ref{Fig:NbNHiggsExp}, were qualitatively mapped to simulations of the \emph{disordered} $XY$ Hamiltonian. In particular, the highlighted curves in the lower panel of Fig.\,\ref{Fig:NbNHiggsExp} correspond to disorder parameters $p=$0.025, 0.1, 0.15, and 0.2 which have to be compared to the value at the SIT, $p_c=0.337$. While the assignment $T_c-p$ was done here considering similarities of experiment and theory, establishing a direct relation between both parameters is delicate. A possible route includes the average kinetic energy along the $x$-bonds in the disordered $XY$ square lattice, $\langle-k_x\rangle$, given as \cite{Swa14}
\begin{equation}
\langle-k_x\rangle_p=\frac{1}{\pi}\int\limits_{-\infty}^\infty d\omega\left\{n_s\delta(\omega)+\sigma^\mathrm{Higgs}(\omega,p)\right\}
\end{equation}
which relates to the experimentally accessible sheet resistances $R_\square$ of clean and disordered systems as \cite{Trivedi2016}
\begin{equation}
\frac{R_\square(\mathrm{disordered})}{R_\square(\mathrm{clean})}=\frac{\langle-k_x\rangle_p}{\langle-k_x\rangle_{p=0}}.\label{eq:AvKinEn}
\end{equation}

The left-sided ratios of Eq.~(\ref{eq:AvKinEn}) for the samples under study are approximately 1.45,~1,5,~1.66, and 2.43 in approach of the SIT. The corresponding $p$-value, see Fig.~\ref{disorder_p} are approximately $p=$0.029,~0.033,~0.043, and 0.092, i.e. systematically lower than the anticipated values above. This could, on the one hand, suggest that with the critical disorder $p_c=0.337$, the samples under study are still far from the SIT which appears rather unlikely considering the massive suppression of $T_c$. Alternatively, the SIT in thin films of NbN might not necessarily be driven by disorder, but instead the competition between Coulomb and Josephson energies, i.e. localization and mobility of charges. This second route envisions a SIT even for zero disorder, when the system falls apart into superconducting  islands decoupled from each other, so that Cooper pairs are localized. This scenario of an emergent electronic inhomogeneity is supported by local measurements of the order parameter amplitude $\Delta(\mathrm{r})$. In addition, the sharpness of the SIT in NbN favors a transition driven by $E_C/E_J$ instead of pure disorder, which would rather destroy the QCP and turn the direct SIT into a blurred superconductor-metal-insulator transition \cite{Auerbach2015}. Unfortunately, we cannot directly compare our excess conductivity shown in Fig.~\ref{Fig:NbNHiggsExp} to simulations for a $E_C/E_J$ driven SIT as these data have been made public only in parts. The main difference \cite{Swa14} between the predictions for $\sigma^\mathrm{Higgs}$ from the $p$ and $E_C/E_J$ scenarios, however, is a well-defined absorption threshold at the Higgs mass all the way to the SIT in case of the latter in agreement with both our experimental findings and the results of the relativistic $O(2)$ field theory. In the light of these considerations, the SIT in thin films of NbN is likely to be driven by a localization/delocalization mechanism, where Cooper pairs become trapped in emergent superconducting islands towards the QCP.       
The subordinate role of disorder is important also from another point of view: As discussed previously, the Higgs mode appears from a Lagrangian with relativistic dynamics, i.e. obeying Lorentz invariance. The latter is characterized by a dynamical critical exponent of $z=1$. For the disordered superfluid-Mott transition in bosonic systems, one finds $z=1.65$ \cite{Svistunov2014} which consequently spoils relativistic dynamics and existence of a well-defined Higgs mode. In addition, Benfatto \emph{et. al} have shown that once disorder is introduced, the amplitude and phase fluctuations of the order parameters are mixed so that one can no longer speak of separate dispersive amplitude and phase modes \cite{Cea2015,cea14}. At the same time, the Higgs mode considered here is a $\mathbf{q}=0$ mode and so it is not strongly affected by the lattice and lattice disorder. Furthermore, near the clean QCP, the Higgs mode provides the dominant contribution to the spectral weight which cannot suddenly disappear once weak disorder is introduced. At the time of writing this thesis, there is no consensus on the relative importance of amplitude and phase modes in weakly disordered systems. While calculations up to the random-phase-approximation (RPA) level \cite{Cea2015,cea14} put emphasis on the latter, the more fundamental QMC simulations \cite{Swa14} favor the first\sidenote{\footnotesize{Although only the RPA analysis allows to clearly disentangle phase and amplitude contributions, which is not possible in the QMC simulations.}}.\\
\begin{figure}[b!]
\begin{center}
\includegraphics[width=\textwidth]{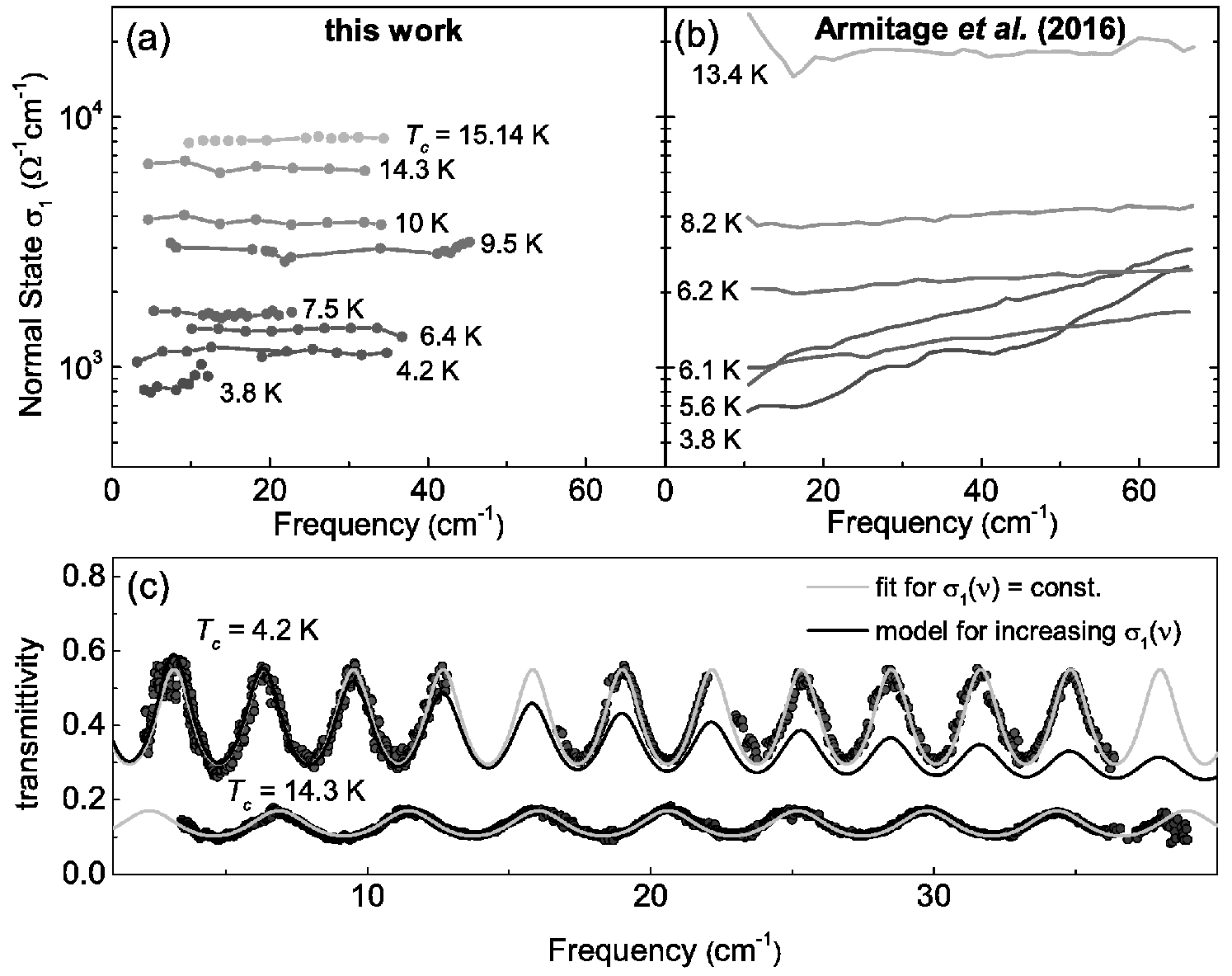}
\caption{\label{Fig:normalstatecomparison}\textbf{Normal-state optical data} for various NbN films measured by (a) frequency domain THz spectroscopy (this work) and (b) time-domain THz spectroscopy (values taken from \cite{Cheng2016}). While the experimental studies presented in this work reveal a frequency-independent $\sigma_{1,n}$, a growing rise with frequency is reported in Ref.~\cite{Cheng2016}. (c) Raw transmittivity spectra for two films (this work) far and close to the SIT together fitted by a frequency-independent Drude conductivity (yellow) and a curve assuming a dispersion similar to the $T_c=3.8$\,K sample of panel (b).}
\end{center}
\end{figure}
Finally, we note similar THz time-domain measurements as discussed in this chapter have been conducted by Armitage \emph{et al.} \cite{Cheng2016} shortly after we had published our main results. While the main experimental finding, i.e. the impossibility to explain the optical results on pure basis of pair-breaking induced from tunneling measurements, is identical with what is presented in detail in Sec.~\ref{Sec:ExpNbN}, the experimental outcomes concerning the normal state conductivity $\sigma_{1,n}(\nu)$ differ significantly: In this work,  $\sigma_{1,n}(\nu)$ reveals (within the experimental error bars) a flat Drude response which is seemingly at odds with the results of Ref.~\cite{Cheng2016}, where $\sigma_{1,n}(\nu)$ grows with frequency towards criticality, see Fig.~\ref{Fig:normalstatecomparison}(a,b). The authors of Ref.~\cite{Cheng2016} attribute this to the localization tendencies of the charge carriers and restrict their analysis to the normalized superconducting spectra to meet the requirement for a dispersion less normal state withing the standard BCS expression for $\sigma_1(\nu)$. Such a normalization is not required for the data presented and discussed in this work. To rule out a systematic error stemming from our $\hat{\sigma}(\nu)$ analysis, we can directly compare the raw transmittivity spectra, $\mathcal{T}(\nu)$, of two films far ($T_c=14.3$\,K) and close ($T_c=4.2$\,K) to the SIT, see Fig.~\ref{Fig:normalstatecomparison}(c). Irrespective of $T_c$, the $\mathcal{T}(\nu)$ spectra do not feature any frequency dependence beyond the Fabry-Perot pattern in favor of a frequency independent film conductivity, which is supported by a perfect Drude fit (yellow lines) with a relaxation rate at eV energies or higher. For comparison, another  $\mathcal{T}(\nu)$ curve is shown (black line) based on a conductivity rise similar of the $T_c=3.8$\,K sample displayed in   Fig.~\ref{Fig:normalstatecomparison}(b) clearly at odds with our measurement. Although the spectral range of Ref.~\cite{Cheng2016}  is greater by a factor of two so that a frequency dependence can more easily be traced, the different results and how they may be related to an technical origin, remain an open problem. \\

At about the same time the results of this chapter were published, another superconducting excitation also termed \emph{Higgs mode} has been reported by Matsunaga \emph{et al.} by measuring the dynamics of the energy gap $\Delta$ of the BCS superconductor Nb$_{1-x}$Ti$_x$N in the non adiabatic regime after an intense THz pulse \cite{matsunaga13}. These studies revealed an oscillatory behavior of $\Delta(t)$ quickly leveling off to the equilibrium value $\Delta_0$, which can be understood using the BCS Hamiltonian expressed in terms of pseudo-spins introduced by Anderson. This \emph{Higgs mode}, however, is qualitatively distinct from the relativistic \emph{Higgs mode} of the $O(N)$ field theory. While the mode reported in Ref.~\cite{matsunaga13} resides exactly at $2\Delta$ and is a consequence of the single-particle DOS of weak-coupling  BCS superconductors\sidenote{\footnotesize{Pictorially, the weak-coupling mode can be understood as a coherent \emph{breathing mode} of Cooper pairs oscillating around their center of mass \cite{Manske2015}.}}, the quantum-critical mode presented in this work is a strong-coupling mode with an energy solely determined by the distance to the QCP not related to $\Delta$ at all \cite{Auerbach2015}. The question, which of both modes should meaningfully be interpreted as the Higgs particle analogue, is not only of semantic nature. The Higgs particle is a bosonic excitation of longitudinal components of the Higgs field which obeys relativistic dynamics. The strong-coupling Higgs mode emerges from the essentially identical Lagrangian with no fermionic degrees of freedom and becomes well-defined only below $2\Delta$ towards criticality. Here, the composite nature of Cooper pairs and the superconducting scale $\Delta$ becomes irrelevant, which is clearly not the case for the weak-coupling mode residing exactly at $2\Delta$. Furthermore, the strong-coupling mode follows from the $O(N)$ model as an excitation of the order parameter field $\psi$, while the weak-coupling mode is an oscillation of the energy gap $\Delta$, which, strictly speaking, is not the same\sidenote{\footnotesize{In fact, using $\Delta$ as the superconducting order parameter is a reasonable approximation for BCS superconductors, which, however, may break down in presence of strong pair-breaking as shown early after BCS by Abrikosov and Gorkov, when the energy gap vanishes, while superconductivity persists \cite{Abrikosov58,Abrikosov59,Abrikosov60}.}}. In that sense, linking the strong coupling mode to the Higgs particle is not only reasonable, but also allows a fascinating view on the universe:  When crossing from insulator to superconductor within the $O(N)$ model, the ground state of the symmetric order-parameter field $\psi$ starts with an expectation value of  $\langle\psi\rangle_0=0$ and there is no Higgs mode, while in the superconducting broken-symmetry phase  $\langle\psi\rangle_0\neq 0$ and a Higgs mode emerges. In the identical sense, the universe passed through a critical point shortly after the Big Bang, when the temperature dropped below the electroweak unification scale of about $250$\,GeV. Today, the ground state of the Higgs field breaks symmetry giving rise to massive Higgs bosons, which are - so to speak - nothing else but critical modes of the early-universe phase transition.
\chapterend
\newpage
\cleardoublepage

\chapter{\label{Sec:Al}Experimental studies on granular Al thin films}
\thispagestyle{empty}
\begin{flushright}
\footnotesize{
Was auch immer wir jetzt wissen,\marginnote{\footnotesize{\textcolor{gray}{Whatever we know,\\is certainly not correct,\\whatever we  do,\\certaily doesn' t matter. }}}\\
ist mit Sicherheit nicht richtig.\\
Was auch immer wir jetzt machen,\\
ist mit Sicherheit egal\\[8pt]
\emph{Die Nerven}}
\end{flushright}
\emph{
Content of this chapter are experimental studies on the transport- and dynamical conductivity of superconducting granular Al, i.e. films composed of nanoscaled grains of aluminum coupled into a macroscopic array. We will start our discussion with a short introduction to superconductivity in confined geometries at the nanoscale governed by the interplay of phase-number uncertainty and the shell effect. Section\,\ref{Superconductivity in granular Al} will provide both an introduction to the superconductivity of granular Al with emphasis on the well-known yet enigmatic superconducting dome in the phase diagram and a broader context aiming for a unified picture of granular and unconventional superconductors by virtue of a common fundamental interplay of superconducting energy scales. In Sec.\,\ref{Measurements of the transport- and  dynamical conductivity} we will start with a discussion of the resistive transitions and -fluctuations of various samples with different resistivity, and afterwards focus on dynamical conductivity $\hat{\sigma}(\nu)$ and, in Sec.\,\ref{Enhanced Cooper pairing versus suppressed coherence}, identify the energy gap $\Delta$ and the superfluid stiffness $J$ as underlying energy scales shaping the superconducting dome in terms of an  amplitude-phase crossover with a pseudogap feature discussed as natural consequence thereof in Sec.\,\ref{The pseudogap for phase-driven superconductivity}. In the following Sec.\,\ref{Goldstone modes} we will closely examine a low-energy absorption evident in several samples, which we identify as the collectiive excitation of the phase field, the supeconducting Goldstone mode, within a nearly parameter free microscopic model.}

\clearpage

\section{Some considerations on nano-scale superconductivity }
\subsection{The number-phase uncertainty}\label{number-phase}
The hallmark of a superconducting condensate is the constant phase of the many-body wave function across the system. For nanoscaled superconductors, however, the phase-lock is not a priori prevailed due to the confinement of carriers. This is captured in the important number-phase uncertainty which is a fundamental property of coherent many-body states. In what follows, we will briefly introduce this uncertainty relation and discuss the implications for superconductivity at the nanoscale. \\
We start with the Lagrangian Eq.~(\ref{eq:LHiggs3}) for superconductivity as introduced in Sec. \ref{Sec:FieldTheoforSC}, where we maintain\sidenote{\footnotesize{i.e. we rewind the gauge transformation applied to Eq~(\ref{eq:LHiggs2})}} the phase field $\vartheta(x)$ 
\begin{eqnarray}
\mathcal{L}&=&\frac{\mathrm{i}}{2}\partial_t \rho-q\rho \left(A_0+\frac{1}{q}\partial_t \vartheta\right)\nonumber\\
&&-\frac{1}{2m}\left(\frac{(\nabla \rho)^2}{4\rho}+q^2\rho\left(\mathbf{A}-\frac{1}{q}\nabla \vartheta\right)^2\right)\nonumber\\
&&+\mu\rho-\frac{\lambda}{2}\rho^2-\frac{1}{4}\mathcal{F}_{\mu\nu}\mathcal{F}^{\mu\nu} 
\end{eqnarray}   
Note that this expression reflects the Ginzburg-Landau functional for classical fields $\rho(x)$ and $\vartheta(x)$. In order to obtain an expression relating the uncertainties associated with the density field $\rho$ and the phase $\vartheta$, we must replace the classical fields by operator-valued fields $\hat{\rho}(x)$ and $\hat{\vartheta}(x)$. This is formally done by imposing commutator relations \cite{Lancaster2014} reading\sidenote{\footnotesize{with the time-like component of the momentum density\begin{equation}\hat{\Pi}^\mu_{\hat{\phi}}\equiv\frac{\partial_\mu \mathcal{L}}{\partial(\partial_\mu\hat{\phi})}\nonumber\end{equation} where $\hat{\phi}=\hat{\rho},\hat{\vartheta}$}}
\begin{eqnarray}
\left[\hat{\rho}(\mathbf{x},t),\hat{\Pi}_\rho^0(\mathbf{y},t)\right]&=&i\delta^{(3)}(\mathbf{x}-\mathbf{y})\label{eq:Comm1}\\
\left[\hat{\vartheta}(\mathbf{x},t),\hat{\Pi}_\vartheta^0(\mathbf{y},t)\right]&=&i\delta^{(3)}(\mathbf{x}-\mathbf{y})\label{eq:Comm2}
\end{eqnarray} 
Using the above definitions and the superfluid Lagrangian, the momentum densities are easily calculated as $\hat{\Pi}_\rho^0(\mathbf{y},t)=\frac{i}{2}$ and $\hat{\Pi}_\vartheta^0(\mathbf{y},t)=-\hat{\rho}(\mathbf{y},t)$ and Eq.~{\ref{eq:Comm2}} reads
\begin{equation}
-\left[\hat{\vartheta}(\mathbf{x},t),\hat{\rho}(\mathbf{y},t)\right]=i\delta^{(3)}(\mathbf{x}-\mathbf{y})
\end{equation}
We integrate this expression over positions $\mathbf{y}$ so that the density field is converted into the number operator $\hat{N}$ giving the commutator
\begin{equation}
\left[\hat{\vartheta}(\mathbf{x}),\hat{N}\right]=-i\label{uncertaintycommutator}
\end{equation}
for all times $t$. Equation (\ref{uncertaintycommutator}) states that the phase and the particle number of the superconducting condensate are conjugate variables, i.e. there is a uncertainty relation\sidenote{\footnotesize{which, however, is formally not a \emph{quantum} uncertainty relation as $\hat{\vartheta}$ is not Hermitian and has no continuous single-valued eigenvalue spectrum \cite{Lancaster2014}. In a more rigorous way, treating $\widehat{\sin(\vartheta)}$ as operator-valued gives a proper yet less insightful quantum-uncertainty relation \cite{Carruthers1968}.}} between both such that \cite{Lancaster2014}
\begin{equation}
\Delta N \Delta \vartheta \geq \frac{1}{2}\label{numberphase}
\end{equation}
This is an important result for many-body condensates linking the number of condensed particles and the global phase of the coherent wave function. In case of an ordinary superconductor, the low-temperature BCS ground state is macroscopically occupied so that $\Delta N$ is large \cite{Lancaster2014}. Consequently,  the uncertainty in the phase, $\Delta\vartheta$ is vanishingly small. This has to be contrasted with the extreme conditions of individual nanoscale grains, where the number of electrons in, e.g., a 2\,nm grain of Al is of the order of a few hundred, which is many orders of magnitude less than in bulk Al. Assembled to a macroscopic yet electrically decoupled array, the particle number is subject to negligible fluctuations and the ensemble average is $\Delta N\approx 0$. In turn, the uncertainty with the phase, $\Delta \vartheta$ must be large. Obviously, the requirement for a phase lock on macroscopic scale cannot be met so that a true superconducting condensate ceases to exist in such a confined system. Surprisingly, this does not necessarily imply the concomitant absence of a finite pairing gap $\Delta$ as we will discuss below.    
\subsection{The shell effect}
\begin{marginfigure}
\begin{centering}
\includegraphics[width=\marginparwidth]{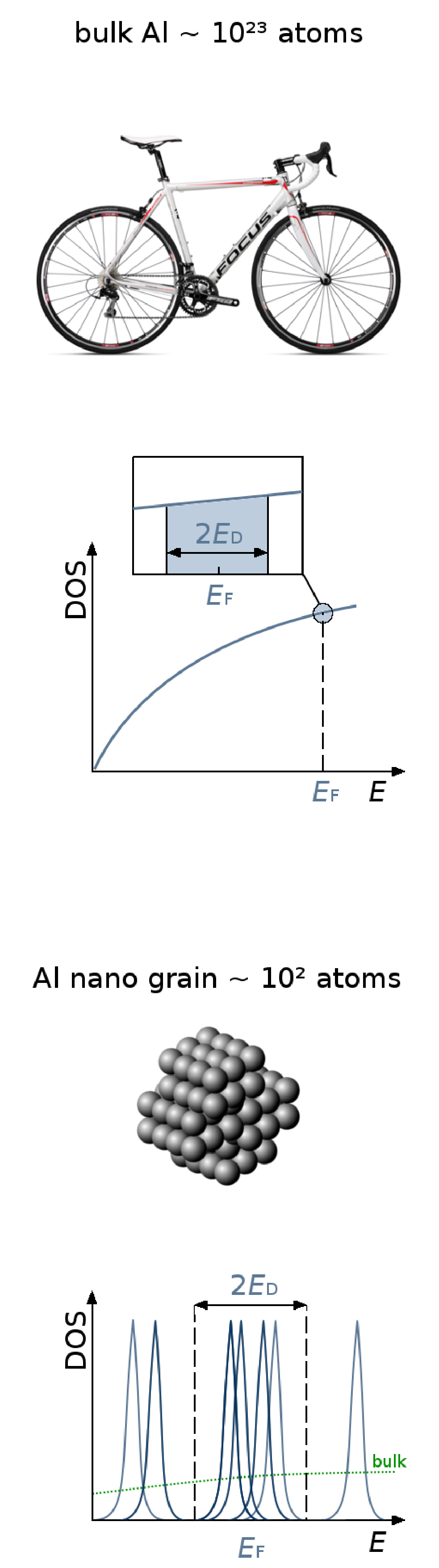}
\caption{\label{sketch_atom_metal_grain}(Schematic view) For a bulk piece of Al (e.g. major parts of the author's road bike), the DOS is that of a quasi-free electron gas and approximately constant in the pairing window $E_F\pm E_D$. For an Al nanograin composed of a few $100$ atoms, the DOS is a series of (possibly degenerated) sharp shell levels and might be substantially enhanced within $E_F\pm E_D$ compared to the bulk (green, dotted), enhancing the pairing efficiency.}
\end{centering}
\end{marginfigure}

 A single Al atom contains 13 electrons which are strongly localized around the nucleus. This spatial confinement allows only certain discrete energy eigenstates, which are referred to as orbitals, or less precise, shells. If assembled in lattices of macroscopic size, the spatially more extended orbitals may overlap so that the occupying electrons move freely from atom to atom. The dispersion relation is approximately that of a free electron gas, $E=(\hbar |\mathbf{k}|)^2/2m$, and the density of states (DOS) $\mathcal{D}$ is a smooth function $\mathcal{D}(E)\propto \sqrt{E}$ and essentially flat in the vicinity of the Fermi energy $E_F$, see Fig.\,(\ref{sketch_atom_metal_grain}). In individual nanoscaled grains, the electrons delocalization is naturally limited by the size of the grain.
 Consequently, the smooth and continuous DOS of the bulk stemming from complete delocalization, is replaced by a set of discrete (and maybe highly degenerate) energy levels, or, in analogy with nuclear or atomic physics, \emph{shells}, defined by the size (and symmetry) of the grain. Stronger confinement favors sharper discretization and higher symmetry (e.g. for spherical grains) enhances level degeneracy.  
As already done in the original work of BCS, the explicit form of $\mathcal{D}(E)$ appearing in the self-consistency equation for the pairing energy given by \cite{mayoh14}
\begin{equation}
1=\frac{\lambda}{2}\int\limits_{-E_D}^{E_D}\frac{1}{\sqrt{E^2+\Delta^2}}\frac{\mathcal{D}(E)}{\mathcal{D}_\mathrm{bulk}(E_F)}\tanh{\left(\frac{\sqrt{E^2+\Delta^2}}{2k_BT}\right)}dE\label{SelfCons}
\end{equation}
(where $\lambda$ is the BCS coupling constant and $E_D$ is the Debye energy cut-off) is simplified such, that $\mathcal{D}(E)$ is taken constant within the pairing window $\pm E_D$ and removed from the integral. While this approximation is valid in most metallic superconductors, it is deemed to fail for systems, where, e.g. as result of the shell effect, $\mathcal{D}(E)$ is strongly influenced by the size and shape of nanograins. Indeed, by combining the BCS self-consistency equation with a semi classical expression for $\mathcal{D}(E)$ and the interaction matrix elements, Garcia-Garcia \emph{et al.} have shown, that small changes in the electron density or shape of an individual nanograin lead to substantial changes  of $\Delta$ as result of the shell effect \cite{Garcia2008}. Similar results for finite-size pairing gap fluctuations where achieved by Olofsson \emph{et al.} \cite{Olofsson2008} in the broader context of ultra-cold Fermi gases. 

First works \cite{Blatt1963,Parmenter68b,Parmenter68} envisioning enhanced superconductivity in confined geometries date back to the 1960's. Yet only half a century later \cite{bose10}, the shell effect and the size-dependent impact on $\Delta$ in individual nanograins was confirmed experimentally by Bose \emph{et al.} : The tip of a scanning tunneling microscope (STM) was placed above quench-condensed (semi-spherical) isolated grains of Sn of different height $h$ and the tunneling conductance was measured well below the bulk-$T_c$ giving the amplitude of $\Delta$. While small modifications of $h$ imposed to large grains leave $\Delta$ approximately unchanged, small grains experience tremendous fluctuations of $\Delta$ upon small variations of $h$. The increase in $\Delta$-fluctuations with decreasing $h$ is explained by ever-sharper electronic shells shifting in (enhancement) and out (suppression) the pairing window around $E_F$ in perfect agreement with theory. Surprisingly, the identical experiment performed on Pb nano grains revealed no fluctuations but only a gradual suppression of $\Delta$. The absence is explained by a stronger level broadening due to the short coherence length and quasiparticle lifetime, so that no substantial enhancement of $\mathcal{D}(E_F)$ is realized, yet confirming the interpretation in terms of the shell effect. 
\begin{marginfigure}
\begin{centering}
\includegraphics[width=\marginparwidth]{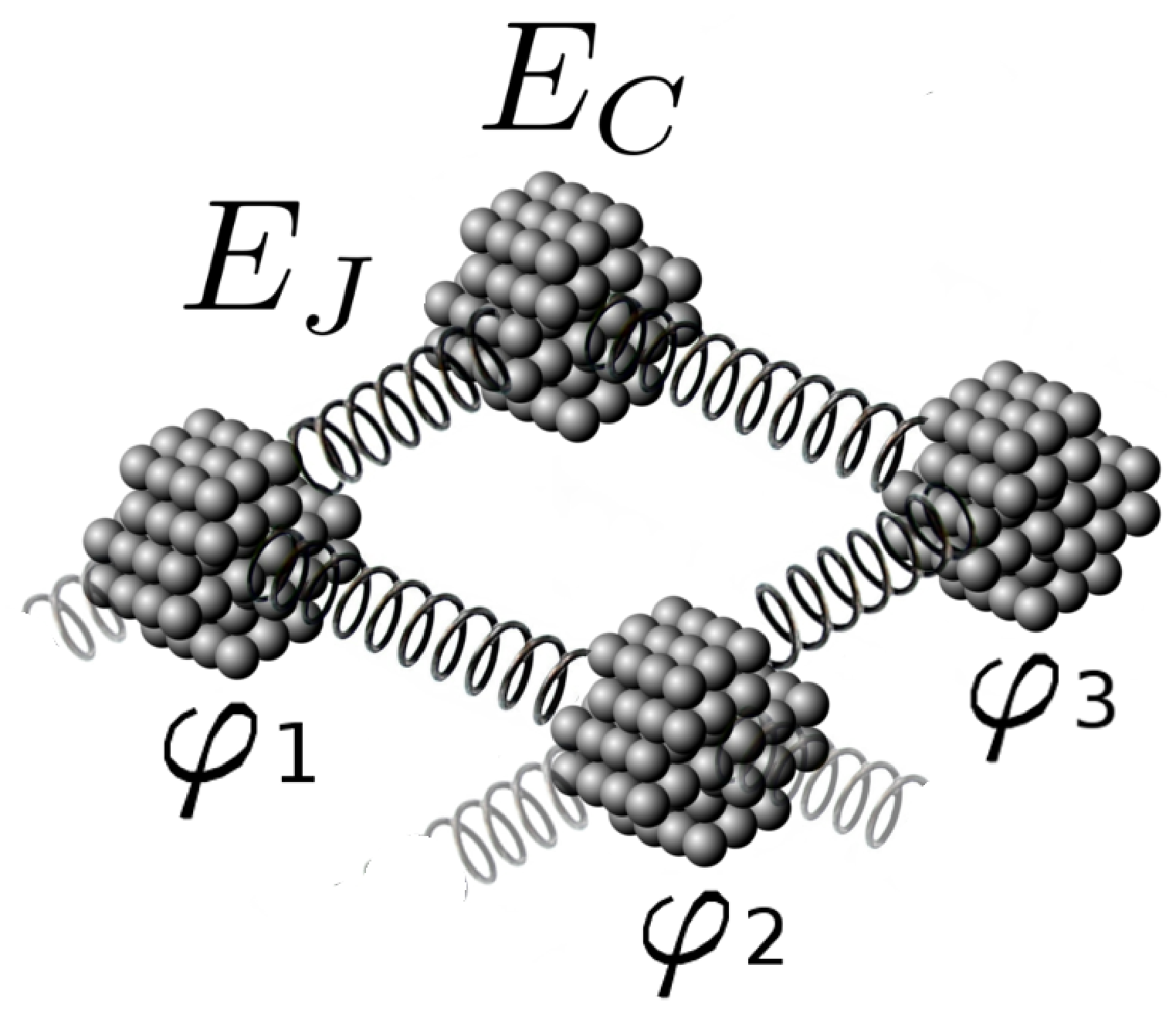}
\caption{\label{sketch_array} Upon fine tuning the Josephson coupling $E_J$ and Coulomb blockade $E_c$ of n nanograin array, a coherent superfluid condensate may exist with similar local phases $\varphi_i$, while the shell effect still enhance $\Delta$ in each grain, so that $T_c$ is enhanced compared to the bulk (i.e. the strongly-coupled limit)}
\end{centering}
\end{marginfigure}
By virtue of the number-phase uncertainty Eq.\,(\ref{numberphase}) those Sn or Pb nanoislands cannot sustain a macroscopically coherent condensate as the number of charge carriers is fixed and the phase of different grains is completely disordered. Still, it envisions a possible route to controlled enhancement of macroscopic superconductivity by engineering arrays of nanograins sufficiently small to enhance the pairing efficiency yet electronically coupled to let Cooper pairs tunnel between grains. Obviously, two antagonizing effects have to be brought together: strong grain coupling (i.e. the Josephson energy exceeds the Coulomb blockade, $E_J\gg E_C$) favor a rigid phase, which is measured in terms of the superfluid stiffness $J$, but weakens the shell effect, whereas poor coupling ($E_J\ll E_C$) enhances $\Delta$ but suppresses phase coherence. The situation is sketched in Fig.\,\ref{sketch_array} where the limit $E_J\gg E_C$ implies $\varphi_1\approx ...\approx \varphi_4$  while for $E_J\ll E_C$ the local phase fields can be greatly different. It is not a priori clear if there is a regime of coupling, where $T_c$ can actually be increased over the bulk value, or if the inter-grain coupling overcompensates the shell effect. Based on their work on isolated nanograins \cite{Garcia2008}, Mayoh and Garcia-Garcia considered nanograins assembled in Josephson-coupled 3D arrays of different geometry. By tuning down the grain coupling, they could show, theoretically,  that indeed $T_c$ is first enhanced, passes a maximum at optimal coupling, and eventually dwindles to zero. \\

In the next section we will introduce the most well-known system of this kind, \emph{granular} Al thin films, and see, if the above scenario sheds some light on a problem that has withdrawn from solution for more than half a century.

\section{Superconductivity in granular aluminum}\label{Superconductivity in granular Al}
\subsection{The phenomenology}
\begin{figure}[b!]
\begin{centering}
\includegraphics[width=\textwidth]{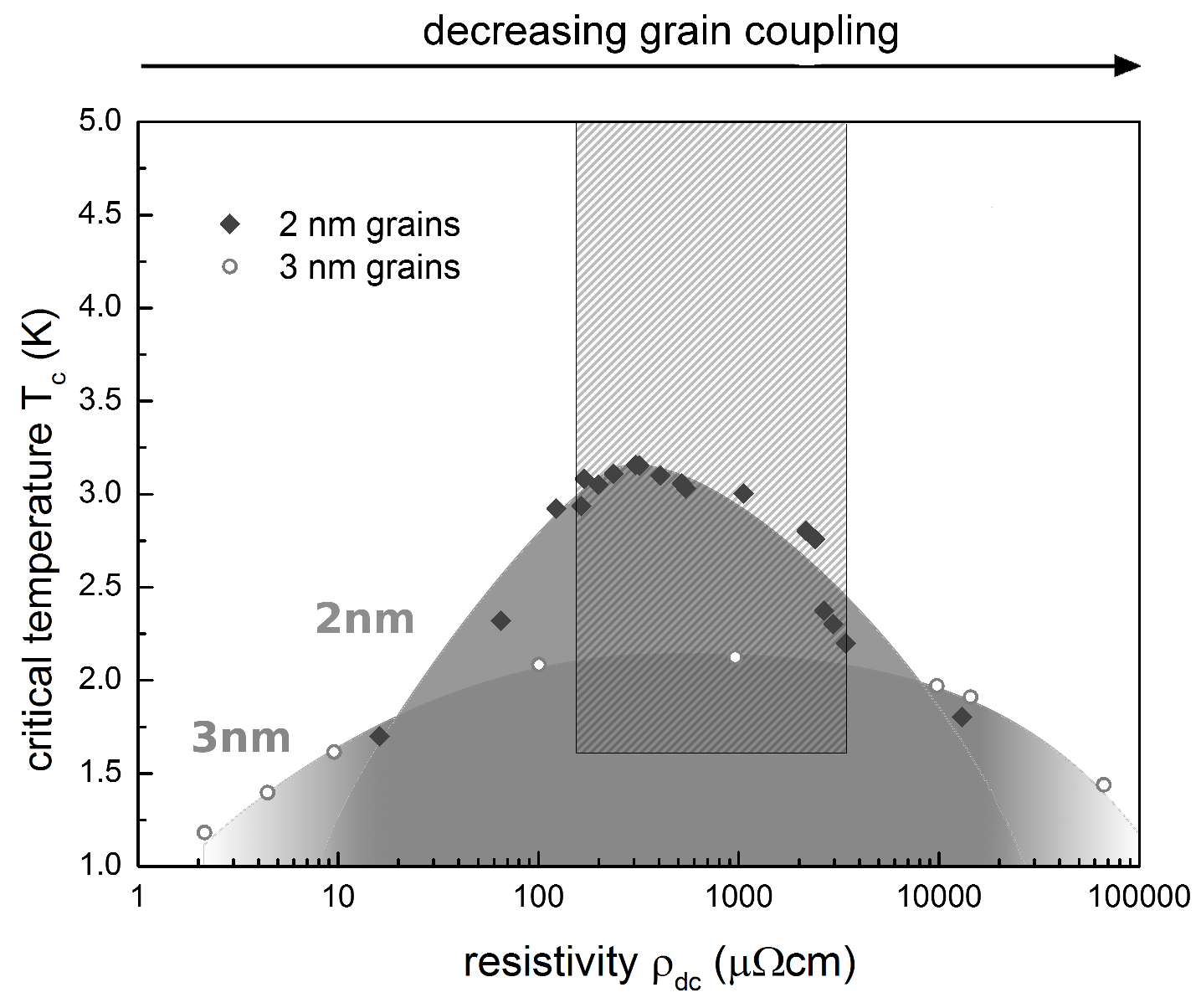}
\caption{\label{sketch_domeAl}\textbf{Superconducting domes in granular Al} thin films composed of nanoscaled grains experience first an enhancement of $T_c$ as the inter-grain coupling (measured by $\rho_\mathrm{dc}$) decreases and pass a maximum for optimal coupling, before superconductivity eventually ceases at high resistivities. For grains with average diameter of 3\,nm the maximum  $T_c$ is around twice the bulk value of 1.19\,K, whereas for smaller grains (2\,nm), $T_c$ can be pushed to 3.2\,K - a tremendous enhancement by around 300\%. The shaded area marks the regime of temperature and $\rho_\mathrm{dc}$ studied in this work. (Parts of the shown data are taken from Ref. \cite{Bachar2014PhD, Deu73}).} 
\end{centering}
\end{figure}

Bulk samples of Al become superconducting below a critical temperature of $T_c = 1.19$\,K and are considered prototypical superconductors, whose properties are well-understood within the conventional BCS theory. It has been realized in the late 1960's that thin superconducting Al films can readily be produced with a wide range of normal-state electrical transport resistivities $\rho_\mathrm{dc}$, ranging from metallic films with low resistivity (LR) and superconducting low-temperature ground state to high-resistivity (HR) films with activated resistivity behavior and insulating low-temperature ground state\sidenote{\footnotesize{An excellent review including the arguably most complete list of references on superconductivity in granular Al is found in Ref.\,\cite{Bachar2014PhD}}}. For disordered metallic thin films, a suppression of superconductivity due to pair breaking, e.g. resulting from spin-flip or strong potential scattering due to disorder, was already discussed soon after the introduction of BCS theory \cite{Abrikosov58,Abrikosov59,Abrikosov60}. However, thermally evaporated superconducting thin Al films were found to act quite differently: comparing films with increasing $\rho_\mathrm{dc}$ shows that $T_c$ first starts to rise above the bulk-value, passes a maximum with $T_c = 2$\,K at around $\rho_\mathrm{dc}$ = 1000\,$\mu\Omega$cm and subsequently decreases and eventually vanishes, enclosing a superconducting dome in the phase diagram, see Fig.\,\ref{sketch_domeAl}. The enhanced superconductivity is even more pronounced when the substrate, onto which the Al film is evaporated, is held at liquid-nitrogen temperatures during growth. Here, at $\rho_\mathrm{dc}$ of a few 100\,$\mu\Omega$cm, a maximum $T_c$ of 3.2\,K is found - almost three times the bulk value!     
With the advent of methods visualizing the morphological structure at the nanoscale it was possible to identify the granular nature of the thin Al films, i.e. macroscopic arrays of microscopic grains electrically coupled through thin oxide barriers covering the grains \cite{Shapira68,Deutscher73}. The thickness of the barrier naturally dictates the macroscopic resistivity, which, in turn, can be understood as the degree of inter-grain coupling. Apparently, the strength of coupling is the microscopic control parameter that tunes granular aluminum through the superconducting dome. Furthermore, it was now possible to attribute the differently shaped domes for room- and liquid-nitrogen temperature growth to the average grain size: the lower the substrate temperature, the smaller the grains, i.e. 3\,nm grains for growth at 300\,K and 2nm at 77\,K, in particular.\\

Comprehensive transport measurements on 2\,nm-grain samples \cite{Bachar2014PhD} show the rapid suppression of $T_c$ on the HR side of the superconducting dome. Films with a room-temperature resistivity\sidenote{\footnotesize{Unfortunately, the characterization of films by means of $\rho_\mathrm{dc}$ is handled inconsistently throughout literature. Some works refer to $\rho_\mathrm{dc}(300\,\mathrm{K})$, while others (e.g. this work) use $\rho_\mathrm{dc}$ right above $T_c$. As for poorly coupled films $\rho_\mathrm{dc}$ rises strongly between 300\,K and $T_c$, this ambiguity should be considered carefully. }} $\rho_\mathrm{rt}$ exceeding $\sim 1.3\times 10^4\,\mu\Omega$cm display the onset of a resistive transition at around 1.8\,K, yet also a pronounced resistive tail so that a true zero-resistance state is not observed for temperature higher than at least 500\,mK. For films with $\rho_\mathrm{rt}\geq 2.1\times 10^4\,\mu\Omega$cm this resistive tail turns into a exponential rise signaling insulating behavior. By means of muon spin-rotation ($\mu$SR) and magneto-resistance (MR) experiments, Bachar \emph{et al.} gathered evidence \cite{Bac15} that granular Al undergoes a metal-to-(Mott)-insulator transition (MIT), when the Coulomb blockade $E_C$ exceeds the (effective) Fermi energy $E_F$. This  transition is driven by the grain coupling and so it will take place even at zero temperature by virtue of the quantum nature of the transition.  With these insights on the nature of granular Al at very high resistivity, one can propose a phase diagram, that, in a broader context, suggests to put granular Al alongside a few of the most fascinating correlated-electron systems in the field of condensed matter.

\subsection{The broader context:\\superconducting domes and energy scales}
\begin{marginfigure}
\begin{centering}
\includegraphics[width=\marginparwidth]{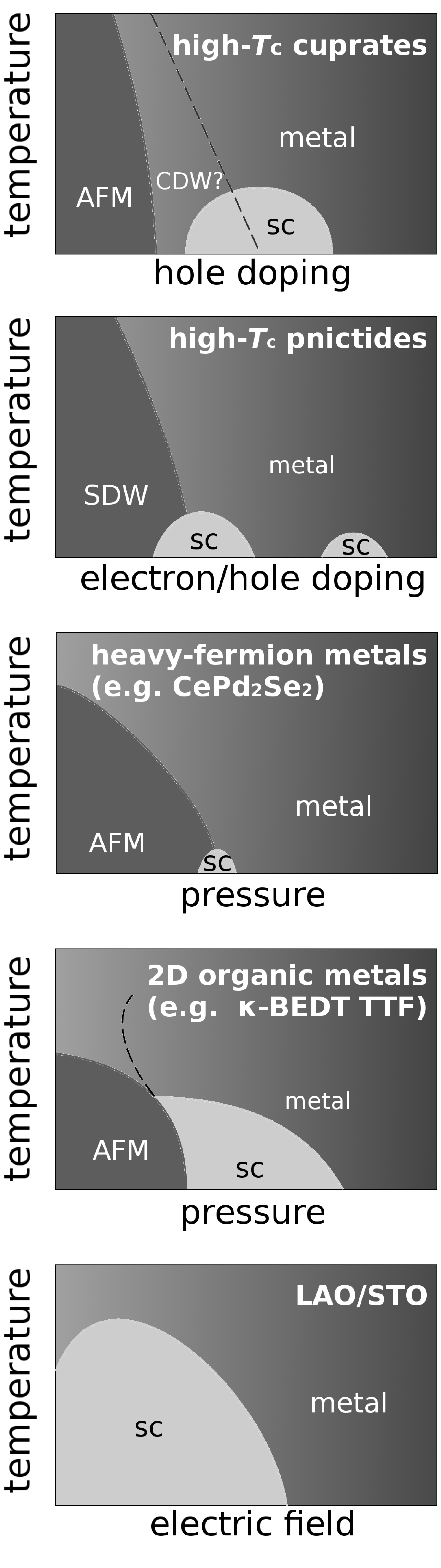}
\caption{\label{phase_overview}Schematic (and simplified) phase diagrams of various correlated-electron superconductors. Reproduced from Refs.\,\cite{Basov2011,Mathur98,Caviglia08,Dumm2009}}
\end{centering}
\end{marginfigure}
A superconducting dome is commonly found in the phase diagram of superconducting materials if their critical temperature $T_c$ depends on a control parameter such as electron- or hole doping, external magnetic fields, pressure, or chemical composition. The list of materials with a dome-like appearance of superconductivity reads like the \emph{who is who} of contemporary correlated electron systems: high-$T_c$ cuprates, pnictides and chalcogenides tuned by pressure or doping \cite{Basov2011}, heavy-fermion compounds such as CePd$_2$Se$_2$ tuned by pressure \cite{Mathur98}, certain quasi-2D organic metals such as the $\kappa$-BEDT-TTF salts tuned by interlacing ionic molecules \cite{Dumm2009}, or the two-dimensional LAO/STO interface tuned by electric gating \cite{Caviglia08}. Understanding the mechanisms that govern such phase diagrams is one of the major challenges in present solid state physics. The question of the origin of the dome is eventually equivalent to the fundamental question: What is the hidden mechanism that, upon tuning of a control parameter, initially favors the strengthening of superconductivity up to some optimum and afterwards leads to its suppression?

Despite the obvious differences in composition and structure, the phase diagrams of above systems share certain commonalities. Fig.\,\ref{phase_overview} schematically reproduces the generic phase diagrams. Neglecting the details, all superconducting domes appear in vicinity of a magnetically ordered phase (expect for the LAO/STO interface) which disappears with increasing control parameter. In many cases, the magnetic phase is believed to be terminated by a $T=0$ quantum critical point (QCP) around which superconductivity evolves as favorable ground state.     
\begin{marginfigure}
\begin{centering}
\includegraphics[width=\marginparwidth]{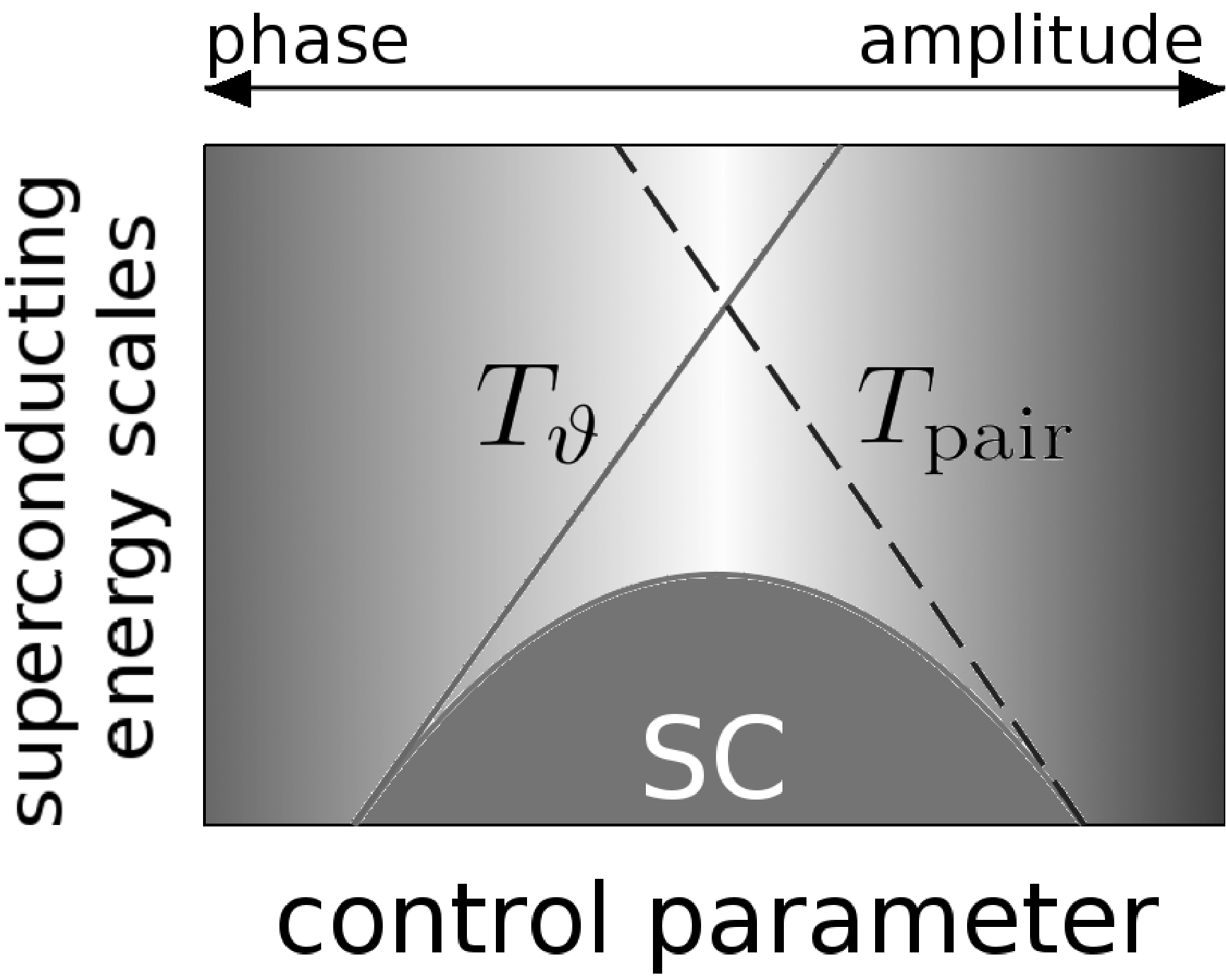}
\caption{\label{generic_dome}Generic scheme for a superconducting dome emergent from competing energy scales, i.e. pairing amplitude $\Delta$ and superfluid stiffness $J$, oppositely affected by an increasing control parameter. Reproduced from Ref.\,\cite{emery94}.}
\end{centering}
\end{marginfigure}
From the experimental point of view, any attempt to explain the particular shape of a superconducting dome eventually needs to identify the mechanism that destroys superconductivity at $T_c$. Being a phase-coherent condensate of pairs, this happens by virtue of either the loss of pairing amplitude or the loss of phase coherence - two mechanisms each of which can be quantified by energy scales, namely the superconducting energy gap $\Delta$ and the superfluid stiffness $J$, respectively. While $\Delta$ simply measures the energy gain associated with pairing up two quasi particles, the \emph{superfluid stiffness} $J$ measures the robustness of the global superconducting phase field $\vartheta(\mathbf{r})$ against fluctuations. The latter quantity is closely related to the density of coherently condensed quasiparticles, the \emph{superfluid density} $n_s$. Clearly, the superconducting state requires $\Delta$ and $J$ to be finite. It is, however, not a priori clear, which of both vanishes and defines $T_c$ for a given system. For BCS superconductors, $J\sim E_F\gg \Delta$ and $T_c$ equals the temperature $T_\mathrm{pair}$, where pairing  no longer becomes energetically favorable, while the phase lock would, in principle, withstand up to much higher temperatures $T_\vartheta$. Emery and Kivelson \cite{emery94} have demonstrated that in a large class of unconventional superconductors the situation is different in a sense that $J$ is comparable or even smaller than $\Delta$ pointing towards a strong susceptibility for phase fluctuations. In these materials, superconductivity ceases at $T_\vartheta$ as consequence of lost phase coherence, while pairing remains energetically favorable up to temperatures $T_\mathrm{pair}>T_c$.

\begin{marginfigure}
\begin{centering}
\includegraphics[width=\marginparwidth]{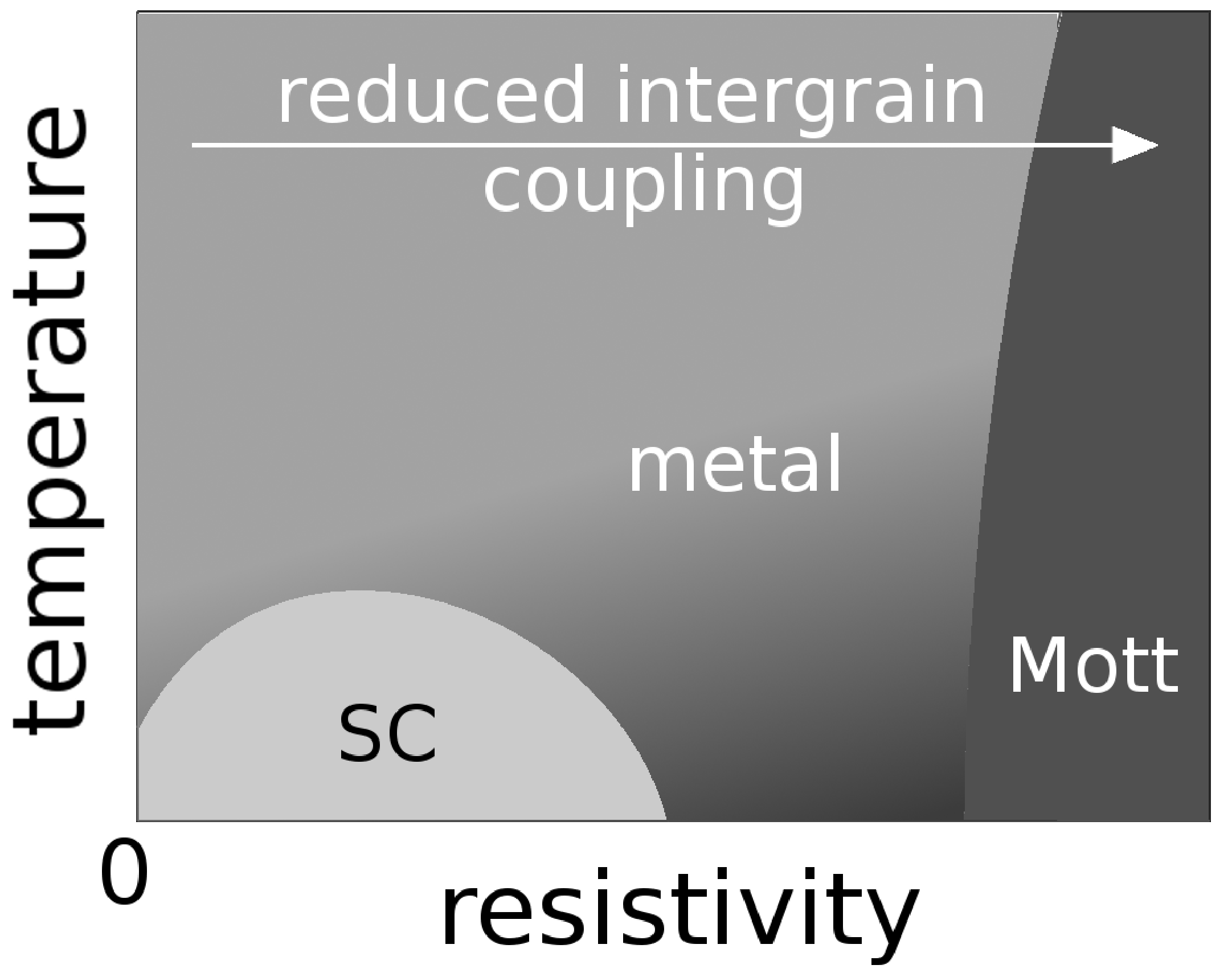}
\caption{\label{phaseAl_generic}Schematic phase diagram of granular Al. Note the close resemblance with the generic diagrams of established unconventional superconductors.}
\end{centering}
\end{marginfigure}

The vicinity to a magnetically ordered state is believed to play a key role in the non-phononic pairing mechanism and $\Psi(\mathbf{k})$-symmetry other than plain $s$-wave suggested for these unconventional superconductors. Roughly speaking, one can think of the melting magnetic order to enable long-range spin fluctuations which live on all length scales at the QCP and may act as bosonic pairing glue. This puts superconductivity in granular Al in another light: Although bulk Al is not magnetic and the overwhelming amount of experimental studies give no reason to assume anything else than phonon-mediated $s-$wave BCS superconductivity, the presence of localized spins and spin-flip scattering channels is confirmed by $\mu$SR and MR experiments \cite{Bac13,Bac15, Bachar2014PhD}. With increasing decoupling, the spin-flip relaxation rate grows and transport behavior reminiscent of Kondo systems as well as an increase of the effective electron mass $m^\ast$ is observed. In total, the dome-like appearance of superconductivity in presence of spin-flip scattering towards a Mott-like state \cite{Bac15}, see Fig.\,\ref{phaseAl_generic}, opens an intriguing new perspective on granular Al which serves as outline for experimental studies presented and discussed in the remainder of this chapter.        

\section{Measurements of the transport and  dynamical conductivity}\label{Measurements of the transport- and  dynamical conductivity}
\subsection{Resistive transition and paraconductivity}\label{paraconductivity}
\begin{figure}[t!]
\begin{centering}
\includegraphics[width=\textwidth]{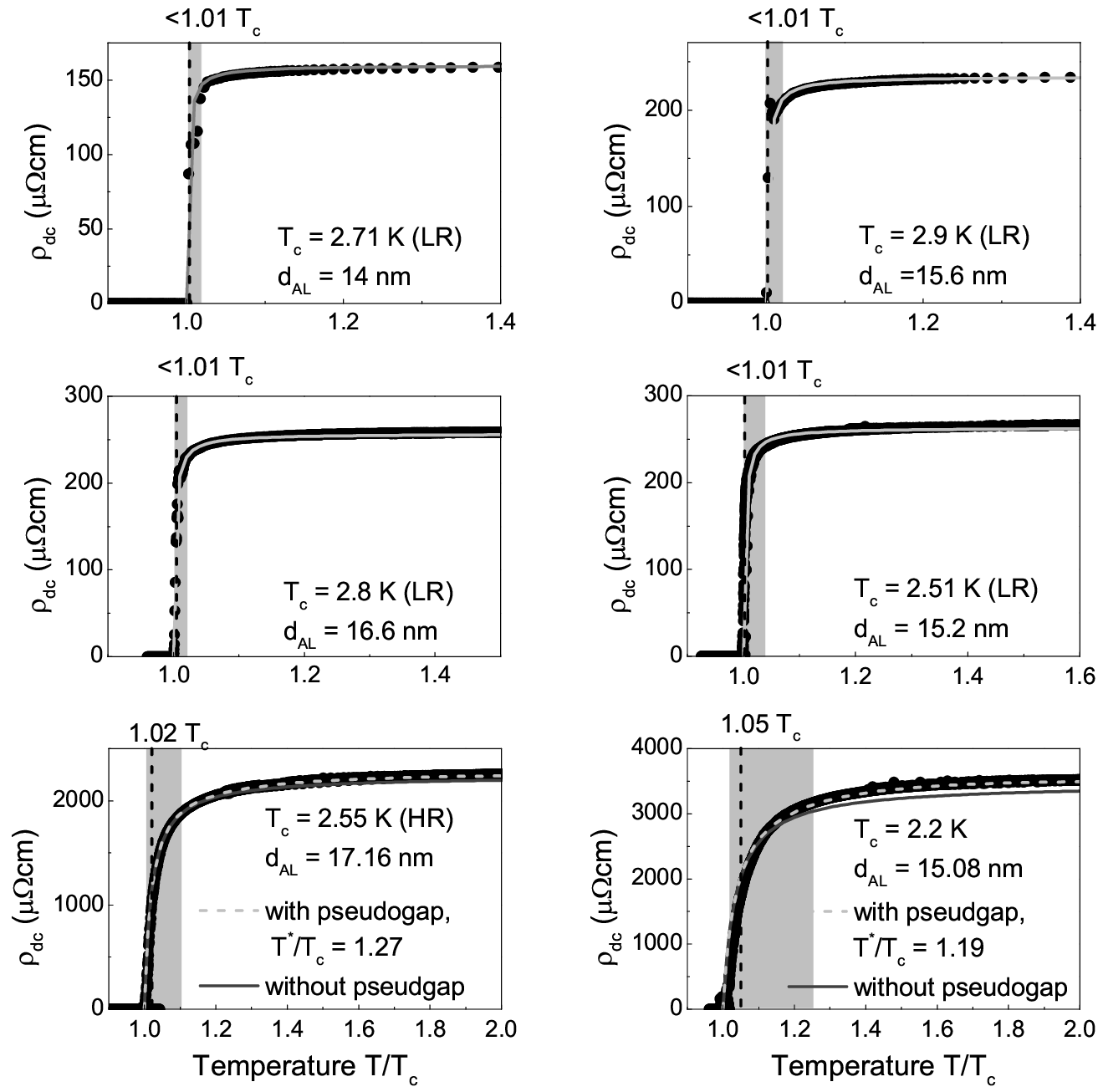}
\caption{\label{GrAl_res}\textbf{Resistive transitions for several representative granular Al samples} studied in this work together with fits of the resistivity with the AL formula (\ref{al2d}) for LR samples and with the modified AL formula (\ref{al2dpg}) for the HR samples. The parameters of the fit are shown in each panel. We define $T_c$ as the temperature, where $\rho_\mathrm{dc}$ becomes immeasurably small. The vertical dashed line denotes the midpoint of the transition as an alternative definition of $T_c$, being only marginally higher than our definition. The gray area shows the range in which $\rho_\mathrm{dc}$ is reduced by 10\% to 90\% from the value at 5\,K.} 
\end{centering}
\end{figure}
Granular Al thin films of 40\,nm thickness and $\sim 2$\,nm grain diameter  ~\cite{Bachar2014PhD,Deu73,Deutscher73} were deposited via thermal evaporation on cold substrates\sidenote{\footnotesize{See Sec.\,\ref{sampleAl} for more details on the sample preparation}}. The growth parameters were adjusted to yield granular Al with a dc-transport resistivity $\rho_\mathrm{dc}$ to cover the high- and low 
resistivity sides of the superconducting dome\sidenote{\footnotesize{Further information on sample growth and -characteristics are found in Sec. \ref{sampleAl}}}.\\ 
We start our discussion with the results of the dc-transport measurements of the superconducting transition. The measurements of the surface resistance $R(T)$ in the vicinity of the crucial transition regime were carried out with the sample and temperature sensor being immersed in liquid helium to guarantee a perfect thermal coupling\sidenote{\footnotesize{See the Appendix for low-temperature characteristics}}. The calculation of the resistivity from the surface resistance was done at $\sim 6$\,K well above $T_c$ within the van-der-Pauw analysis via
\begin{equation}
\rho_\mathrm{dc}=\frac{\pi d }{\mathrm{ln}2}\frac{(R_1+R_2)}{2}f
\end{equation}  
with $d=20$\,nm the film thickness and $R_{1,2}$ the surface resistances for different current-injection orientations, and $f$ a correction factor that varied between $1-0.9$ for the studied samples. This value of $\rho_\mathrm{dc}$ was afterwards used to scale the $R(T)$ curve to obtain $\rho_\mathrm{dc}(T)$. Figure\,\ref{GrAl_res} displays $\rho_\mathrm{dc}(T)$ for a set of granular Al films.
Here, $T_c$ is defined as temperature where the measured film resistance becomes vanishingly small. The transitions are comparably sharp as shown by the gray areas in  Fig.~\ref{GrAl_res} delimiting the regime, in which $\rho_\mathrm{dc}$ is reduced to 90\% and 10\% of the value well above the fluctuation regime at 5\,K. The alternative definition of $T_c$ as midpoint of the transition leads to only marginally larger values than the above definition: In case of the (LR) samples, the deviation $\delta T_c/T_c$ amounts to less than 1\% while it does not exceed 5\% for HR samples in agreement with previous works on similar films \cite{Bac13,Bac14b}.
To analyze the effect of the superconducting fluctuations above $T_c$ (the so-called \emph{paraconductivity}) we focus on the Aslamazov-Larkin (AL) contribution $\Delta\sigma$ \cite{VarlamovBook}, that is the most relevant near $T_c$. In agreement with previous work \cite{Bac14b}, the paraconductivity has the temperature dependence expected in 2D, so that \cite{VarlamovBook}:
\begin{eqnarray}
\label{al2d}
\frac{\Delta\sigma_{AL}}{\sigma_N} =\frac{e^2}{16\hbar d_{AL}\sigma_N}\frac{1}{\epsilon}=\frac{R_\square}{16 R_c}\frac{1}{\epsilon}, 
\end{eqnarray}
where $R_c=\hbar/e^2$, $R_\square=\rho_{dc}/d_{AL}$ and $d_{AL}$ is a transverse length scale, of the order of the film thickness, that determines the effective 2D unit for superconducting fluctuations. The parameter $\epsilon$ contains the temperature dependence, and in the BCS limit it is given by
\begin{equation}
\label{epsbcs}
\epsilon=\ln \frac{T}{T_c}.
\end{equation}
All data is fitted with the 2D AL formula (\ref{al2d}), using $d_{AL}$ and $T_c$ as a free parameters. As one can see in Fig.\,\ref{GrAl_res} the agreement with the data in the LR regime is remarkably good, with an effective thickness $d_{AL}\simeq 14-17$ nm as an adjustable fit parameter, that is within a factor of 2-3 from the real film thickness. Given the granular nature of the film this is a reasonable approximation, considering that the fit reproduces the data up to temperatures as large as twice $T_c$ without any other adjustable parameter. 
On the other hand, when one analyzes the HR films, two remarkable differences arise: (i) the resistivity is not completely saturated up to temperatures as large as twice $T_c$; (ii) the fit with the AL formula fails around $T\simeq 1.2 T_c$, since the experimentally measured paraconductivity decays faster than predicted by   Eq.\ (\ref{al2d}). Interestingly, the very same behavior has been observed also in underdoped cuprates for  samples in the pseudogap regime. In this case, the phenomenological function \cite{caprara2005}:
\begin{equation}
\label{al2dpg}
\frac{\Delta\sigma_{AL}}{\sigma_N} =\frac{R_\square}{16 R_c}\frac{1}{\epsilon_0\sinh(\epsilon/\epsilon_0)}, \quad \epsilon_0=\ln\frac{T^*}{T_c},
\end{equation}
turned out to be a very good model that reduces to the usual one (\ref{al2d}) when $\epsilon\ll \epsilon_0$, so that $\sinh(\epsilon/\epsilon_0)\simeq \epsilon/\epsilon_0$, but decays faster for $\epsilon\gg \epsilon_0$. As it has been discussed in the context of cuprates \cite{caprara2005,varlamov2011}, such a suppression of paraconductivity with respect to the standard formula (\ref{al2d}) can be indeed explained assuming that a pseudogap survives in the electronic Green's function up to a temperature $T^*$ larger than $T_c$. In the case of granular Al the formula (\ref{al2dpg}) works remarkably well for the two most disordered films, see Fig.\,\ref{GrAl_res}, where, as discussed below, pseudogap signatures are evident also from the analysis of the dynamical conductivity. Finally, Fig.\,\ref{dome+sig12_grAl}(a) unifies the results for all samples under study and sketches the superconducting dome, whose origin we will unravel in the next section by a systematic study of the superconducting energy determining $T_c$ by virtue of the dynamical conductivity.
\subsection{The dynamical conductivity and superconducting energy scales}\label{sec:Al_dyn}
\begin{figure}
\begin{centering}
\includegraphics[scale=0.75]{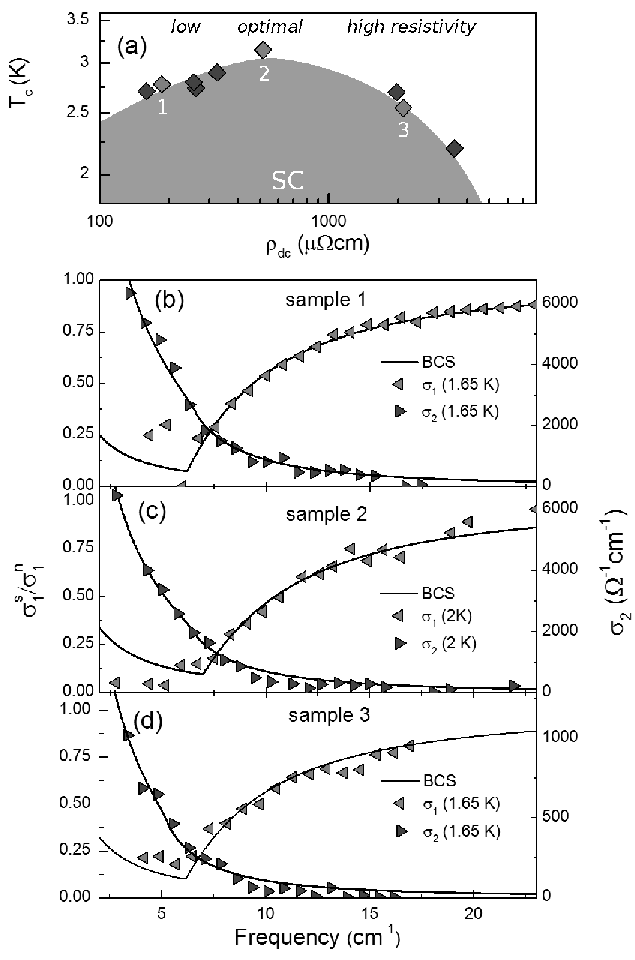}
\caption{\label{dome+sig12_grAl}\textbf{Superconducting dome and dynamical conductivity.}(a) Critical temperature $T_c$ as a function of the normal-state resistivity (measured at 5\,K) of granular Al films studied in this work. Yellow symbols refer to the samples displayed in panels below.  $T_c$ encloses a dome-like superconducting phase with low-, optimal- and high-resistivity regimes. (b-d) (Normalized) spectra of $\sigma_1(\nu)$ and $\sigma_2(\nu)$ of samples located on the left (sample 1), the right (sample 3), and at the maximum (sample 2) of the SC dome. The solid lines are fits to the Mattis-Bardeen theory. Note that the fit on $\sigma_1$ disregards the low-frequency range due to excessive conductivity beyond Mattis-Bardeen theory. }
\end{centering}
\end{figure}

We now turn to the experimental results on the real and imaginary parts of the dynamical conductivity $\hat{\sigma}(\nu)=\sigma_1(\nu)+i\sigma_2(\nu)$ obtained by virtue of phase-sensitive THz spectroscopy\sidenote{\footnotesize{See the Appendix for details on the experimental technique and raw-data treatment.}}. 
Figure \ref{dome+sig12_grAl}(b-d) displays conductivity spectra for a set of three representative samples spanning from the LR to the HR side of the dome at temperatures $T\approx 0.6 T_c$. Note that here and later on we discuss the conductivity normalized to the normal-state conductivity. The reason is that with a substrate thickness of 2\,nm the Fabry-Perot peaks of the superconducting transmission are extremely sharp and, thus, easily affected by standing waves possibly mimicking a too high or too low conductivity. This effect can be removed by normalization as the standing wave pattern does not substantially change with temperature leaving a smooth curve we can compare within the BCS predictions. In all three cases, $\sigma_1(\nu)$ is strongly suppressed for $T<T_c$. At the same time, $\sigma_2(\nu)$ displays a strong decay with increasing frequency reflecting the response of the superfluid.  

To make further progress on the data, we fit $\sigma_1(\nu)$ and $\sigma_2(\nu)$ to the Mattis-Bardeen (MB) functional for dirty-limit\sidenote{\footnotesize{i.e. that the $\Delta/\hbar$ is much less than the quasi particle relaxation rate $\Gamma$. Contrarily, in the clean-limit there is no dissipative conductivity between the $\delta$-response of the condensate at $\nu=0$ (at $T=0$)and, if at all, high energy interband transitions.}} superconductors \cite{mb58,Pracht2016}
\begin{eqnarray}
\label{s1}
\frac{\sigma_1^\mathrm{MB}(\nu)}{\sigma_n}&=&\frac{\pi e^2n_s}{m^*\sigma_n}\delta(\nu)\nonumber\\
&&+\frac{2}{h\nu} \int\limits_\Delta^\infty \mathrm{d}\epsilon\, g(\epsilon)\left[ f(\epsilon)- f(\epsilon-h\nu)\right]\nonumber\\
&&+\frac{\Theta}{h\nu}\int\limits_{\Delta-h\nu}^{-\Delta}  \mathrm{d}\epsilon\, g(\epsilon)) \left[1-2 f(\epsilon+h\nu)\right]\nonumber\\&&\\
\label{s2}
\frac{\sigma_2^\mathrm{MB}(\nu)}{\sigma_n}&=&\frac{1}{h\nu} \int\limits_{-\Delta,\Delta-h\nu}^\Delta  \mathrm{d}\epsilon\,\Big(g(\epsilon)\left[ 1-2f(\epsilon+h\nu)\right]\nonumber\\
&&\times\frac{\epsilon(\epsilon+h\nu)+\Delta^2}{\sqrt{\Delta^2-\epsilon^2}\sqrt{(\epsilon+h\nu)^2-\Delta^2}}\Big),
\end{eqnarray}
where $\sigma_n$ is the normal-state conductivity, $f(\epsilon)$ the Fermi-Dirac distribution, $\Theta=\Theta(h\nu-2\Delta)$, and the function $g(\epsilon)$ is defined as
\begin{equation}
\label{ge}
g(\epsilon)=\frac{\epsilon(\epsilon+h\nu)+\Delta^2}{\sqrt{\epsilon^2-\Delta^2}\sqrt{(\epsilon+h\nu)^2-\Delta^2}}\\
\end{equation}   
which is the explicit form of the general response function (\ref{FominovCond}) discussed in Sec. \ref{Sec:Green} for the case of BCS Green's functions.   
The accuracy of the MB fit of $\sigma_1(\nu)$ and $\sigma_2(\nu)$ via $\Delta$ as the only free parameter is shown in Fig.\,\ref{dome+sig12_grAl}(b-d). Even though the fit captures well the increase of conductivity at $\nu>2\Delta/(hc)$ (where $h$ is the Planck constant and $c$ is the speed of light), it underestimates $\sigma_1(\nu)$ at low frequencies. Such an excess conductivity resembles the one observed, e.g., in disordered NbN\sidenote{\footnotesize{see the previous chapter for an comprehensive discussion}} and InO films \cite{crane2007,sherman15} and in cuprate films \cite{corson2000}, and is attributed to SC collective modes \cite{corson2000,stroud2000,cea14,Swa14,sherman15,Cea2015}, not included in the MB theory. In the case of granular Al, where the Josephson coupling between grains is expected to be spatially inhomogeneous, this excess conductivity may be attributed to SC phase fluctuations, made optically active by disorder \cite{stroud2000,cea14,Swa14}. A comprehensive discussion of this anomalous absorption is found in Sec.\,\ref{Goldstone modes}, where it is identified as the Goldstone mode. The first energy scale of interest, the energy gap measuring the pairing amplitude $\Delta$ is obtained from $\sigma_1(\nu)$. To access the second scale, the superfluid stiffness $J$, we first fit $\sigma_2(\nu)$ to MB, see Fig.\,\ref{dome+sig12_grAl}(b-d), and construct the expression 
\begin{equation}
n_s=\frac{2\pi m^\ast}{e^2} \lim_{\nu\to 0}\nu\sigma^\mathrm{MB}_2(\nu),
\end{equation}  
where $m^\ast$ and $e$ are the electron effective mass\sidenote{\footnotesize{Studies \cite{Bac15} of the magneto resistance in granular Al films showed, that $m^\ast/m_0=(\rho_{dc}/\bar{\rho}_{dc})^{0.44}$ where $m_0$ is the free-electron mass and $\bar{\rho}_{dc}\approx 50\,\mu\Omega$cm is the resistivity, above which a mass enhancement sets in. We used this formula to estimate the mass enhancement for the samples of this study}} and charge. In a second step, we define \cite{Ben09}
\begin{equation}
\label{js}
J=\frac{\hbar^2 n_s a}{4m^*}=0.62 \times \frac{a}{\lambda^2} [K]
\end{equation}
Here, $a$ is a transverse length scale, expressed in \AA, $\lambda$ is the penetration depth in $\mu$m  and $n_s/m^\ast=1/\lambda^2 \mu_0 e^2$.
In an isotropic 3D system, the length scale $a$ in equation\ (\ref{js}) is the SC coherence length  $\xi_0$, which is the natural cut-off for phase fluctuations, while it is the film thickness in the 2D limit. Measurements of the upper critical field in similar samples \cite{Bac14b,Deu77} gave an estimate of $\xi_0\simeq 10$\,nm,  while the analysis \cite{Deu77} of the paraconductivity above $T_c$ indicates a 2D character with an effective 2D thickness for superconducting fluctuations of the order of $\simeq 15$\,nm throughout the phase diagram. For the sake of simplicity, we assume a constant value $a=10$\,nm in  (\ref{js}) for all samples under study. 
\begin{figure}
\begin{centering}
\includegraphics[width=\textwidth]{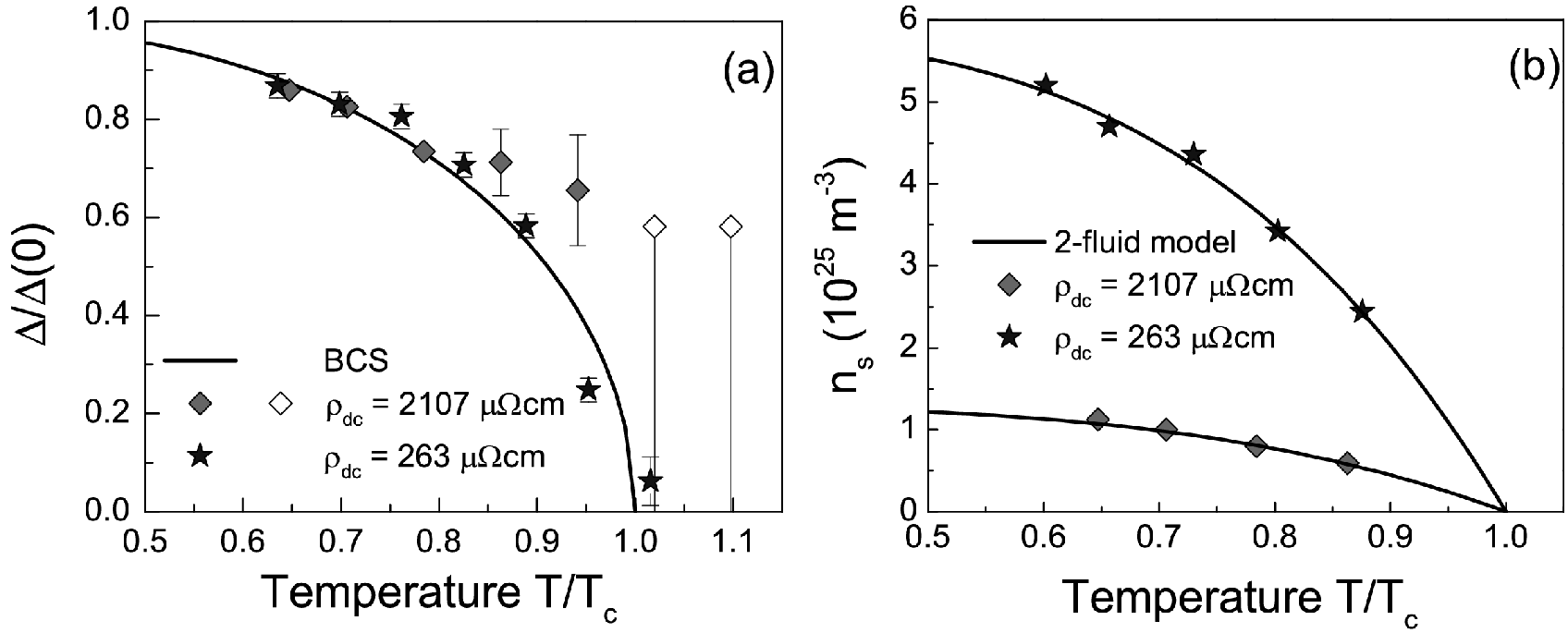}
\caption{\label{grAl_scales_vs_T}\textbf{Temperature dependence of the energy gap and superfluid density} for two samples on the LR and HR sides of the dome. (a) $\Delta(T)$ of both samples follows the BCS prediction closely, while, in case of the HR sample, $\Delta(T)$ decays weaker towards $T_c$ and tends to survive into the normal state (empty symbols). The persistence of a finite pairing above $T_c$ is in striking resemblance with the pseudogap phase of, e.g., the cuprates. (b) $n_s$ of both samples is perfectly described by the 2-fluid approximation and agrees with BCS theory.} 
\end{centering}
\end{figure}
While most measurements were performed at the base temperature $T=1.65$\,K of the cryogenic system, the relative temperature $T/T_c$ is different across the dome. For a reasonable comparison, we account for the different relative temperatures by calculating the zero-temperature expectations $\Delta(0), n_s(0)$ (and $J(0)$) according to the BCS self-consistency equation\sidenote{\footnotesize{where the ratio $\Delta(0)/k_BT_c$ (with $k_B$ the Boltzmann constant) is not constrained to the weak-coupling value 1.78}} \cite{Likharev1979}
\begin{equation}
\label{usadelSelfCons}
\ln\frac{T_c}{T}=2\pi k_B T \sum_{\omega_n}\left[\frac{1}{\hbar \omega_n}-\frac{1}{\sqrt{(\hbar \omega_n)^2+\Delta^2}}\right]
\end{equation}
where $\omega_n=\pi k_B T(2n+1)$ with $n\in \mathds{N}$ are the Matsubara frequencies and the two-fluid model
\begin{equation}
\frac{n_s}{n_{s}(0)}\approx 1-\left(\frac{T}{T_c}\right)^4\label{eq:2fluid},
\end{equation} 
respectively. We checked the applicability of the models (\ref{usadelSelfCons}) and (\ref{eq:2fluid}) for two representative samples on the LR and HR side, see Fig.\,\ref{grAl_scales_vs_T}, for which a temperature series was measured.\
The separate analysis of $\sigma_2(\nu)$ to obtain $n_{s}(0)$ is in principle redundant, since within the MB theory we could use directly the value of $\Delta(0)$ extracted from the $\sigma_1$ fits to determine the zero-temperature inductive response. Indeed, from Eq.\ (\ref{s2}) one immediately sees that  $\sigma_2(\nu\to 0,T=0)=\pi \Delta(0)\sigma_n/(h\nu)$. Since in the dirty limit $\sigma_n$ coincides with $1/\rho_\mathrm{dc}$ in the THz frequency range, we can estimate $n_{s}(0)$ as
\begin{equation}
n_{s}^\Delta(0)=\frac{2\pi m^*}{e^2}\frac{\pi \Delta(0)}{\hbar \rho_{dc}}\label{nSDelta}
\end{equation}
so that the corresponding estimate $J_\Delta$ of the stiffness follows as
\begin{equation}
\label{mb}
J_{\Delta}(0)=\frac{R_\mathrm{c}}{R_\square}\frac{\pi\Delta(0)}{4},
\end{equation}
where $R_\mathrm{c}=\hbar/e^2$ and $R_\square=\rho_\mathrm{dc}/a$ with same scale $a$ as used in equation (\ref{js}). 
However, since the deviations of $\sigma_1(\nu)$ from the MB behavior occurs exactly below $2\Delta$, we analyzed $\sigma_2(\nu)$ independently of $\sigma_1(\nu)$, and we cross-check afterwards the consistency between the two approaches, see Fig.\ \ref{fig:lambda}. We note that the given estimate of $J(0)$ should be taken as an upper bound, since it neglects the additional reduction due to inhomogeneous phase fluctuations \cite{cea14,Swa14,mayoh14}. However, the comparison with previous SQUID measurements \cite{Abr78,Ger82} of the penetration depth suggests that this effect is still quantitatively small for the samples under consideration. Here the extracted values of $n_{s}(0)$ and $n_{s}^\Delta(0)$ are converted to the penetration depth $\lambda=1/\sqrt{\mu_0e^2n_s/m^*}$ (with $\mu_0$ the vacuum permeability), in order to compare them with direct measurements of $\lambda$  done in previous works \cite{Abr78,Ger82}, see Fig.\,\ref{fig:lambda}. Both estimates of $\lambda$ are consistent with each other and they are in very good agreement with previous findings. This shows also that the quantitative suppression of $n_{s,0}$ due to the collective-mode contribution below $2\Delta$ in $\sigma_2(\nu)$ is relatively small, and it justifies the use of the MB formula to extrapolate $\sigma_2$ to zero frequency. 
\begin{marginfigure}
\begin{centering}
\includegraphics[width=\marginparwidth]{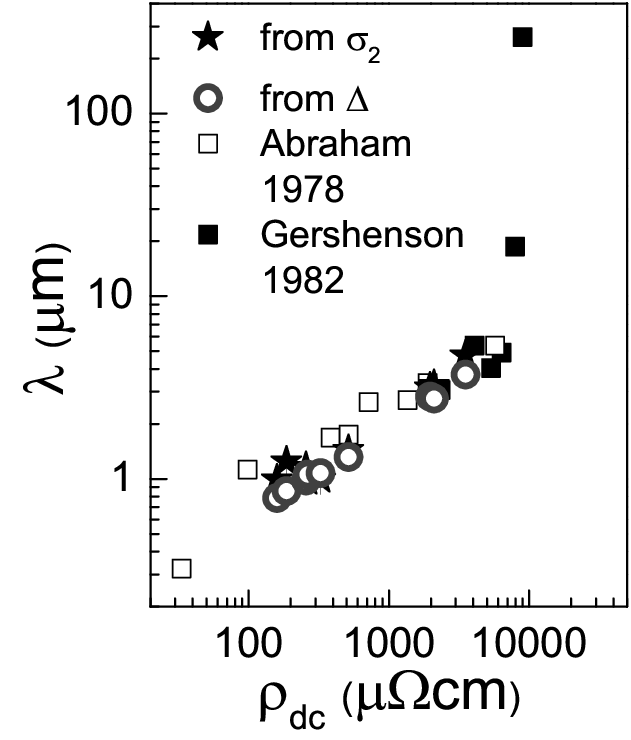}
\caption{\label{fig:lambda}
Penetration depth $\lambda$ versus normal-state resistance. The present work (colored dots) very well reproduces results obtained in previous works \cite{Abr78, Ger82} shown as empty and filled squares, where the inverse penetration depth has been directly measured. In addition, we note that the absolute numbers of $\lambda$ obtained from the inductive response (stars) are nearly identical with the calculation of $\lambda$ from $\Delta$ (circles) within MB theory.}
\end{centering}
\end{marginfigure}
With knowledge of the crucial energy scales $\Delta(0)$ and $J(0)$ we can now attempt to explain the particular shape of the superconducting dome of granular Al.
\begin{figure}[t!]
\begin{centering}
\includegraphics[width=\textwidth]{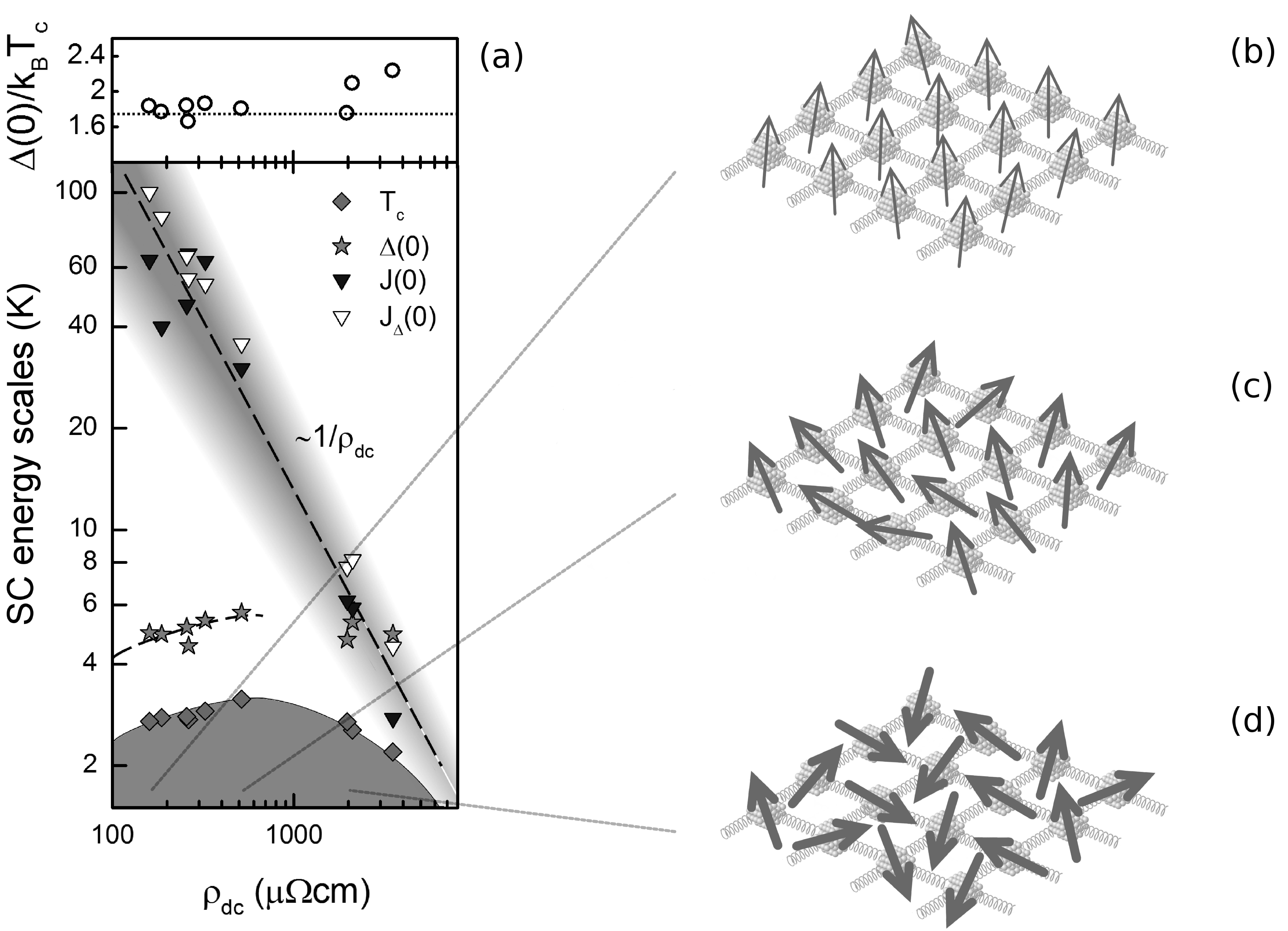}
\caption{\label{dome_exp}\textbf{Superconducting energy scales and local order parameters} (a) $T_c,\,\Delta(0)$, $J(0)$, $J_\Delta(0)$ (expressed in units of temperature) and $\Delta(0)/k_BT_c$ as a function of normal-state resistivity (measured at 5\,K) of granular Al films. $\Delta(0)$ (olive stars) follows the increase of $T_c$ on the left side of the dome for LR samples while it saturates in the HR regime. This is reflected in the ratio $\Delta(0)/k_BT_c$  which increases from the weak-coupling value 1.78 (dotted line) to 2.25 when crossing from the left to the right side of the dome. The calculation of superfluid stiffness from $\sigma_2(\nu)$ and from $\Delta(0)$, i.e. $J(0)$ and $J_\Delta(0)$, is subject to an uncertainty reflected by the shaded area. Dashed lines are guides to the eye. See the text for more details. (b-d) Illustration of the grain array for (b) strong,  (c) moderate, and (d) weak inter-grain coupling. The arrows indicate the phase (direction) and amplitude (thickness) of the local superconducting oder parameter. } 
\end{centering}
\end{figure}
\clearpage

\section{Enhanced pairing versus \\suppressed coherence}\label{Enhanced Cooper pairing versus suppressed coherence}
\subsection{Shaping the superconducting dome}
 We express $\Delta(0)$ and $J(0)$ in units of Kelvin and plot them together with $T_c$ as a function of $\rho_\mathrm{dc}$ in one frame, see Fig.\,\ref{dome_exp}(a). Starting on the LR side, we find $\Delta(0)<J(0)$ as common for most superconductors and $T_c=T_\mathrm{pair}$, i.e. superconductivity is destroyed by the cease of Cooper pairs. In terms of absolute numbers, however, J(0) is surprisingly small: For BCS superconductors, $J(0)\sim E_F$ which, for Al, is of the order of a few $10^4$\,K which is orders of magnitude more than realized in (even strongly coupled) granular Al. On a (simplifying) square lattice of grains the situation can be illustrated using local order parameters $\psi_i$ for each lattice site $i$. We schematically visualize the energy gap $\Delta_i$ (pairing amplitude) and the superconducting phase $\phi_i$ in the thickness and direction or the arrows in the right panels of Fig.\,\ref{dome_exp}: The situation for strong coupling on the LR side is sketched in Fig.\,\ref{dome_exp}(b) where all local $\psi_i$'s point roughly in the same direction. Consequently, the phase gradient between neighboring sites is small and $J(0)$ is large. Upon decoupling the grains and tuning towards higher resistivity, the rise of $T_c$ is accompanied by an identically rising $\Delta(0)$, which can be inferred from the constant ratio $\Delta(0)/k_BT_c\approx 1.78$ being in perfect agreement with BCS theory for weak-coupling superconductors. Microscopically, decoupling promotes the individual character of each grain such that quantum confinement becomes stronger and the shell effects more efficient boosting the pairing amplitude (illustrated by the thicker arrows in Fig.\,\ref{dome_exp}(c)). Up to intermediate resistivity, we also find $\Delta(0)<J(0)$ implying that superconductivity remains limited by the pairing and $T_c=T_\mathrm{pair}$. At the same time, however, we observe a remarkable decline of $J(0)\propto 1/\rho_\mathrm{dc}$, that can be viewed as consequence of growing phase fluctuations across the grain array, see Fig.\,\ref{dome_exp}(c). For even higher resistivity on the HR side, inter-grain coupling becomes so weak that phase fluctuations push $J(0)$ to values comparably to or, ultimately, even smaller than $\Delta(0)$. Even though the shell effect remains at play keeping $\Delta(0)$ strongly enhanced compared to the bulk\sidenote{\footnotesize{where,  according to BCS,  one can estimate\\ 
 $\Delta(0)=1.78T_c^\mathrm{bulk}=2.1\,\mathrm{K}$}}, see Fig.\,\ref{dome_exp}, $T_c$ is reduced again. In other words, superconductivity on the HR side is limited not by the loss of Cooper pairs, but the loss of the coherent superconducting phase, i.e. $T_c=T_\vartheta$. Consistently, the ratio $\Delta(0)/k_BT_c$ increases up to 2.25 in case of the least-coupled sample under study. In summary, the superconducting dome of granular Al and the underlying crossover from amplitude- to phase driven superconductivity can completely be ascribed to the competing interplay of pairing enhancement and suppressed phase coherence originating from quantum confinement and enhanced phase fluctuations, respectively.    \\

\subsection{The pseudogap for phase-driven \\superconductivity}\label{The pseudogap for phase-driven superconductivity}

The above suggested scenario has an intriguing consequence for the normal state entered via a phase-driven transition, which, at the same time, serves as important consistency check. When $T_c$ is suppressed by an overly small $J$, the non-zero $\Delta$ right above $T_c$ should imply a finite pairing amplitude up to some temperature $T_\mathrm{pair}>T_c$ even without a coherent condensate. Or, from another perspective, when temperature is reduced between $T_\mathrm{pair}>T>T_c$, a quasiparticle excitation gap should form, which equals the energy required to break \emph{preformed} pairs. As soon as $T=T_c$, these pairs condense and superconductivity appears. The notion of an excitation gap above $T_c$ due to preformed pairs has become very popular with unconventional cuprate superconductors, commonly referred to as \emph{pseudogap}\sidenote{\footnotesize{To which extent this enigmatic phase of the cuprates is related to superconductivity has been subject of numerous works, yet no consensus is achieved. For instance, the pseudogap phase may live 'behind' the superconducting dome in a sense that there are, loosely speaking, two kinds of Cooper pairs of which only one actually condenses below $T_c$ \cite{Geshkenbein1997,Perali2000,Ranninger1995,Ranninger1996}}}. 
\begin{figure}
\begin{centering}
\includegraphics[scale=0.6]{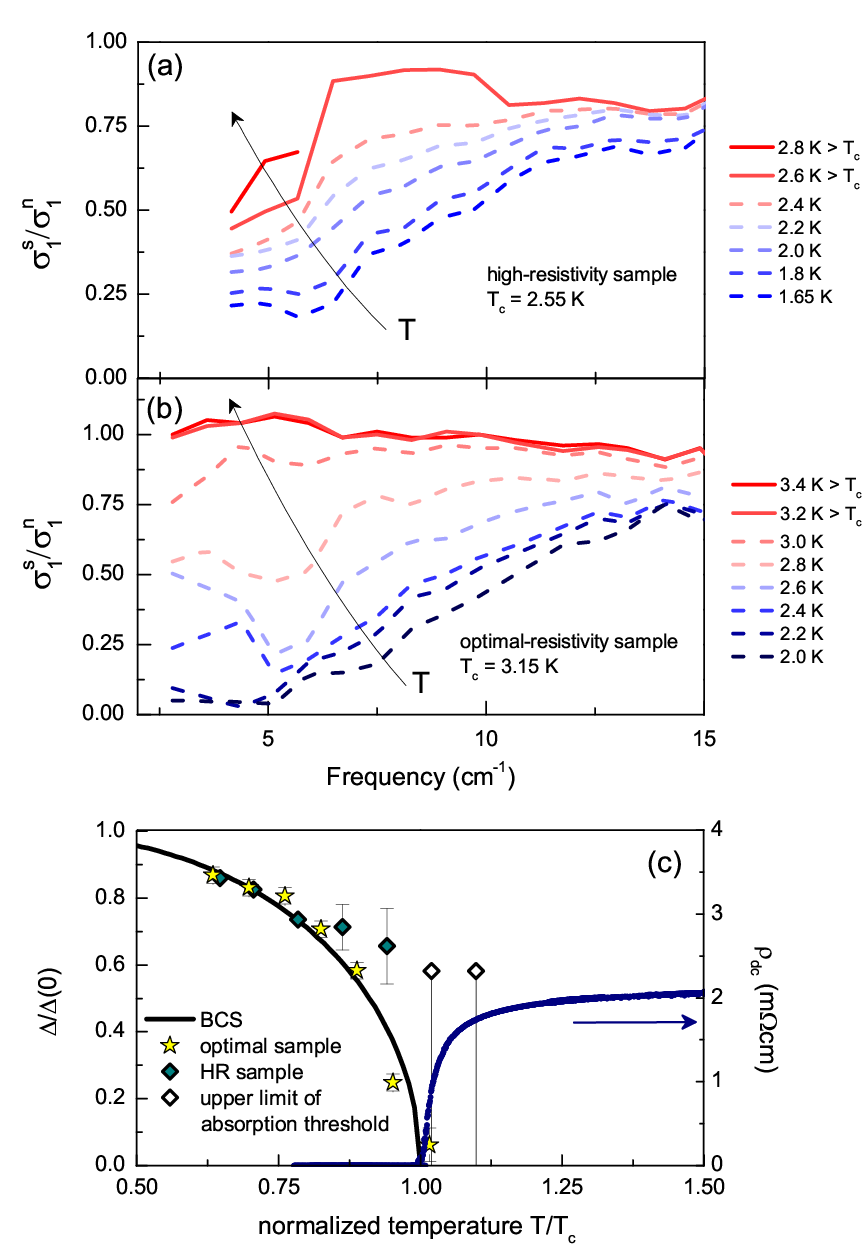}
\caption{\label{PG}\textbf{Temperature evolution of spectral gap.} (a-b) Temperature dependence of normalized  $\sigma_1(\nu)$  of a granular Al sample in the high- and optimal resistivity regimes. In case of the HR sample, the suppression of $\sigma_1(\nu)$  below $T_c=2.55$\,K (dashed lines) persists up to $T=2.8$\,K (solid lines), whereas the spectral gap closes right at $T_c$ in the LR regime. (c) Temperature dependence of the spectral gap for samples from the optimal (stars, sample 2) and high resistivity regimes (diamonds, sample 3). The blue data traces $\rho_{dc}(T)$ of the HR sample. For the HR sample, deviations from the BCS prediction for $\Delta(T)/\Delta(0)$ (black solid line) appear already at $T/ T_c< 1$, where $\Delta$ is anomalously large. The persistence of a gap across $T_c$ (empty diamonds) is in striking resemblance with strongly disordered or correlated superconductors.} 
\end{centering}
\end{figure}
While the situation in the cuprates is immensely complicated by correlation and competing phases, the nature of an pseudogap in phase-driven granular Al would be comparably clear. As discussed in Sec.\,\ref{paraconductivity}, a indirect signature of a pseudogap can be inferred from dc-transport measurements: While on the LR side the paraconductivity agrees with the 2D Azlamazov-Larkin (AL) model for superconducting fluctuations, the paraconductivity on the HR side requires a AL term which accounts for a pseudogap (see Ref.\,\cite{caprara2005} and references therein). However, this phenomenological model was discussed in context of unconventional cuprates, and its applicability to other systems is not a priori given. A more direct verification of the pseudogap would be given by an spectral gap in the dissipative conductivity $\sigma_1(\nu)$ similar as the superconducting gap $2\Delta$. In this sense, the depletion of states around $E_F$ should suppress the electromagnetic absorption up to energies corresponding to excitations across the pseudogap.  Indeed, such a feature was found for a HR sample. In Fig.\,\ref{PG}(a,b) we compare $\sigma_1(\nu)$ spectra of amplitude- and phase-driven samples recorded at various temperatures below and above $T_c$. For an ordinary dirty-limit BCS superconductor, any superconducting correlation vanishes at\sidenote{\footnotesize{or, more precisely, vanish above the fluctuation regime, which for the samples under consideration is limited to 1.02$T_c$ and can be neglected here, see Fig.\,\ref{GrAl_res}.}} $T_c$ and the normal state $\sigma_1(\nu)$ is frequency independent. In fact, at temperatures $T>T_c$ the spectra of the amplitude-driven sample, panel (b), are basically indistinguishable and flat within the experimental resolution. This is contrasted with the phase-driven sample, panel (a), where a strong suppression of $\sigma_1(\nu)$ is observed up to temperatures substantially higher than $T_c$ and the fluctuation regime. The spectral gap is of the same order as the superconducting gap and we can try to estimate it from the absorption threshold. Well below $T_c$, the fit of $\sigma_1(\nu)$ within the MB theory gives $\Delta(T)$ closely following the universal BCS curve, see  Fig.\,\ref{PG}(a). In approach of $T_c$, the obtained $\Delta$ exceeds the BCS prediction. Taking the minimum of $\sigma_1(\nu)$ at $T>T_c$ as upper estimate for the absorption threshold, i.e. the pseudogap, a spectral gap seems to persist into the normal state, in contrast with the amplitude-driven sample, where $\Delta(T)$ closely follows the BCS prediction all the way to zero at $T_c$ . As depicted in Fig.\,\ref{PG}(c) both gaps below and above $T_c$ seem to be smoothly connected. To analyze the gapped structure of $\sigma_1(\nu)$ in the normal state, a generalization of the MB functional is clearly required as for $\Delta=0$ Eq.\,\ref{s1} becomes
\begin{eqnarray}
\frac{\sigma_1(\nu)}{\sigma_0}&=&\frac{2}{h\nu}\int\limits_0^\infty\mathrm{d}\epsilon[f(\epsilon)-f(\epsilon+h\nu)]\nonumber\\
&&+\frac{1}{h\nu}\int\limits_{-h\nu}^0\mathrm{d}\epsilon[1-2f(\epsilon+h\nu)]
\end{eqnarray}
(where $\sigma_0$ denotes the normal state conductivity above any superconducting correlation), which after analytic integration equals to 1 without any frequency dependence. This can be understood as consequence of an energy independent density of states $\mathcal{D}$ considered in the standard MB theory. A more general treatment should include variations of $\mathcal{D}(\epsilon)$ on a energy scale comparable to $\Delta$ also in the normal state. Seibold \emph{et al.} re-derived the MB functional explicitly allowing for an energy-dependent, yet symmetric density of states \cite{Seibold2017}

\begin{eqnarray}
\label{MBmodified}
\frac{\sigma_1(\nu)}{\sigma_n}&=&\frac{2}{h\nu} \int\limits_\Delta^\infty \mathrm{d}\epsilon\,\frac{\mathcal{D}\left[\Lambda_1(\epsilon)\right]\mathcal{D}\left[\Lambda_2(\epsilon)\right]}{\Lambda_1(\epsilon)\Lambda_2(\epsilon)}\nonumber\\
&&\times\left[\epsilon(\epsilon+h\nu)+\Delta^2\right]\left[ f(\epsilon)- f(\epsilon+h\nu)\right]\nonumber\\
&&+\frac{\Theta(h\nu-2\Delta)}{h\nu}\int\limits_{\Delta-h\nu}^{-\Delta}  \mathrm{d}\epsilon\, \frac{\mathcal{D}\left[\Lambda_1(\epsilon)\right]\mathcal{D}\left[\Lambda_2(\epsilon)\right]}{\Lambda_1(\epsilon)\Lambda_2(\epsilon)}\nonumber\\
&&\times\left[\epsilon(\epsilon+h\nu)+\Delta^2\right] \left[1-2 f(\epsilon+h\nu)\right]\nonumber\\
\end{eqnarray}
where 
\begin{eqnarray}
\Lambda_1&=&\sqrt{\epsilon^2-\Delta^2}\nonumber\\ \Lambda_2&=&\sqrt{(\epsilon+h\nu)^2-\Delta^2}.
\end{eqnarray}
In the normal state this expression simplifies to
\begin{eqnarray}
\label{MBmodifiedNormal}
\frac{\sigma_1(\nu)}{\sigma_n}&=&\frac{2}{h\nu} \int\limits_0^\infty \mathrm{d}\epsilon\,\mathcal{D}(\epsilon)\mathcal{D}(\epsilon+h\nu)\left[ f(\epsilon)- f(\epsilon+h\nu)\right]\nonumber\\
&&+\frac{1}{h\nu}\int\limits_{-h\nu}^0 \mathrm{d}\epsilon\,\mathcal{D}(\epsilon)\mathcal{D}(\epsilon+h\nu) \left[1-2 f(\epsilon+h\nu)\right]\nonumber\\
&&
\end{eqnarray}
Using the functional (\ref{MBmodifiedNormal}) we can now attempt to fit the spectra above $T_c$. As a model, we employ the Altshuler-Aronov type\sidenote{\footnotesize{which is known to model the tunneling spectra of many disordered systems}} pseudogap \cite{Altshuler85} 
\begin{equation}\label{AA}
\mathcal{D}(\epsilon)=\alpha+(1-\alpha)\tanh^2\left(\frac{\epsilon}{\Omega}\right)
\end{equation} 
with the depletion parameter $\alpha$ and $\Omega\sim\Delta$ the amplitude of the pseudogap. Figure \ref{sig1_Pgapped}(a) displays $\sigma_1(\nu)$ at temperatures $T>T_c$ together with a curve calculated from Eq.~(\ref{MBmodifiedNormal}) including Eq.~(\ref{AA}) with $\Omega/hc=1.66$ and 1.45\,cm$^{-1}$ at 2.6 and 2.8\,K, respectively, and depletion $\alpha=0.1$. Furthermore, the theory curves were scaled by a factor 1.1. Although one cannot expect the simple model DOS (\ref{AA}) to reproduce the experimental $\sigma_1(\nu)$ entirely, it serves as a fairly good general description capturing the suppression at low frequencies and the rise with increasing frequency. At the same time, the pronounced feature at intermediate frequencies cannot be modeled appropriately and hence was excluded from the fit.  We can now study two distinct models assuming (i) a competing and (ii) common nature of the spectral gaps.\\
In the first scenario we assume, that the normal state gap opens at $T\gg T_c$ and acts as temperature independent background upon which the superconducting gap opens. This scenario resembles a model suggested\sidenote{\footnotesize{See the comprehensive review \cite{Hashimoto2014} by Hashimoto \emph{et al.} on the competition between superconducting order and pseudogap in unconventional cuprate superconductors.}} for cuprate superconductors, where the pseudogap characterizes an electronic phase distinct from the superconducting order. Here, for the sake of simplicity, we assume a temperature independent pseudogap with $\Omega/hc=1.66$\,cm$^{-1}$ (and the above depletion $\alpha=0$ and scaling factor 1.1), leading to the curve displayed in Fig.~\ref{sig1_Pgapped}(a). Below $T_c$, we let the superconducting gap open in the weak-coupling mean-field fashion according to Eq.~(\ref{usadelSelfCons}). Without any further free parameter, the generalized MB model (\ref{MBmodified}) results in the dot-dashed curves of the panels (b-f). Clearly, the separate treatment of two distinct gaps, $\Delta$ and $\Omega$, leads to a too strong depletion of states and, in turn, a too small absorption for all temperatures $T<T_c$.     \\
 \begin{figure}[b!]
\begin{centering}
\includegraphics[width=\textwidth]{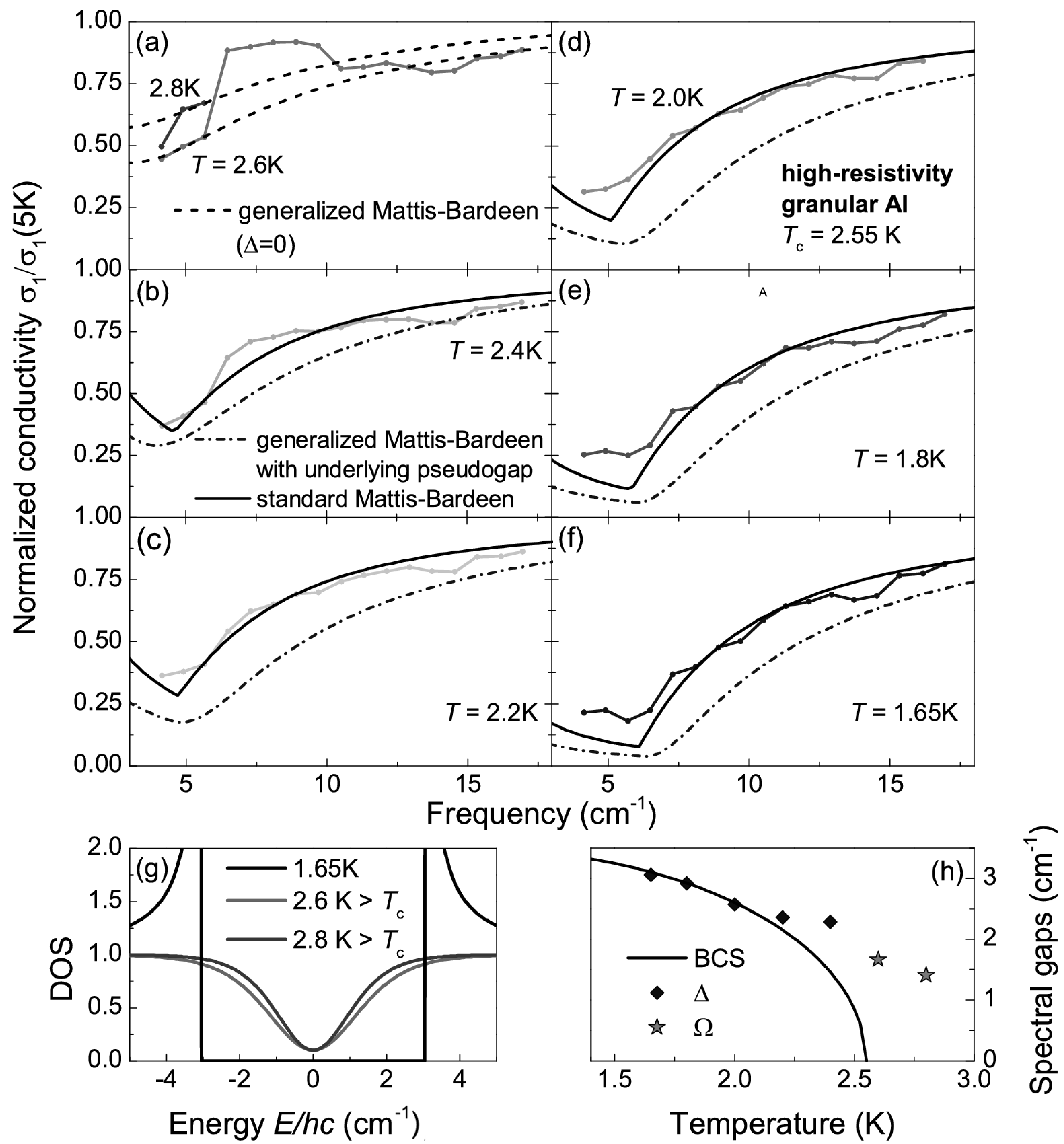}
\caption{\label{sig1_Pgapped}\textbf{Evolution of spectral gaps} below and above $T_c$ as apparent from (a-f) the dissipative conductivity $\sigma_1(\nu)$ of high-resistivity granular Al. (g) Fits to the standard and generalized Mattis-Bardeen functionals (see the main text for more information) include hard-gapped BCS-DOS and Aranov-Altshuler type pseudogap, respectively. (h) temperature dependence of the spectral gaps $\Delta$ and $\Omega$ for $T<T_c$ and $T>T_c$, respectively, together with the mean-field BCS prediction for the superconducting gap.} 
\end{centering}
\end{figure}
For the second scenario, we assume that the spectral gaps above and below $T_c$ are smoothly connected, i.e. there is no competition between superconducting order and another electronic phase. This scenario is somewhat more plausible than the above multi-phase model, as the latter arises by virtue of strong electronic correlations which play a minor role in granular Al. This model, however, leaves open the question, how $\Omega$ evolves into $\Delta$ across $T_c$. For the sake of simplicity, we model the experimental $\sigma_1(\nu)$ above $T_c$ with Eq.~(\ref{MBmodifiedNormal}), while below $T_c$ we skip the energy dependence of $\mathcal{D}$ and use the standard MB functional (\ref{s1}), where we treat $\Delta$ as fit parameter. The resulting curves are displayed in Fig.~\ref{sig1_Pgapped}(a-f). The corresponding values of the spectral gaps are shown in panel (h) together with the BCS $T$-dependence of $\Delta$ for arbitrary coupling strength. Apparently, this scenario of smoothly connected spectral gaps yields a much better description of the experimental data. Although the above treatment of spectral gaps right at $T_c$ is an oversimplication (which is reflected in the rather poor fit at 2.4\,K), the agreement is remarkable. In this sense, the all Cooper pairs preformed in the pseudogap phase condense into the superfluid condensate below $T_c$ - in contrary to the competing-order model, where only Cooper pairs of a certain Fermi surface fraction actually condense. Far below $T_c$, the system does not feature signatures of the pseudogap anymore and $\Delta(T)$ follows BCS theory.

\subsection{Multi-fractal wave functions - an alternative to enhancing $T_c$?}
To conclude this section we will briefly present an alternative mechanism to enhance $T_c$ based on the multi-fractal nature of the electron wavefunctions near localization. By a simple renormalization-group treatment of the conductivity discussed in Sec.~\ref{Sec:Anderson} we have seen that in 3D a metal-insulator transition occurs at a critical conductance $g_c$ or, equivalently, a critical disorder $p_c$ defining the \emph{mobility edge}. Following the reasoning of Aoki \cite{Aoki83}, for a system with $p_c+\epsilon$ all electron wavefunctions are localized so that the state fills a vanishingly small fraction of the total volume of the system. Above the mobility edge, $p_c-\epsilon$, the electrons are completely delocalized and the wavefunctions spread across the entire system. Consequently, for a system right at the mobility edge, both requirements have to be met which is accomplished by a multi-fractal nature of the wave function. Indeed, models based on multi-fractality \cite{Feigelman2007,Feigelman2010} have quite successfully been employed to explain the inhomogeneous superconducting correlations, a pseudogap at $T>T_c$, and a hard-gapped insulating state observed in systems close to the superconductor-insulator transition underlining the importance of multi-fractality on superconductivity. In a recent work \cite{Mayoh2015}, Mayoh and Garcia-Garcia analytically study the energy dependence and spatial distribution of the order parameter for different degrees of multi-fractality. Using percolative techniques, this enables the computation of the global $T_c$ defined as temperature, where a supercurrent starts to flow. The central result is a strong suppression of $T_c$ with disorder compared to the case of a homogeneous order parameter (i.e. in the clean limit), except for the case of very weak-coupling superconductors, such as Al, leading to the highly counterintuitive prediction of modest disorder actually \emph{enhancing} $T_c$. This scenario allows an alternative view on superconductivity in granular Al. At low resistivity, the electrons may easily hop from grain to grain so that the staying time within one grain is, pictorially speaking \cite{GarciaPers}, too short to feel the confinement and the shell effect. The grains simply act as scatterers, as long as the staying time remains sufficiently small, whereas the confinement-induced $T_c$ enhancement takes over only at higher resistivities. A clear separation between both effects and a determination of the relevant time scales is an important problem left open for further theoretical and experimental studies.            
\clearpage{}

\section{Goldstone modes}\label{Goldstone modes}
In this section, we closely examine a peculiar deviation from Mattis Bardeen theory apparent in Fig.~\ref{dome+sig12_grAl} we encountered in our discussion  of the low-frequency real conductivity $\sigma_1(\nu)$ of granular Al in Sec.~\ref{sec:Al_dyn}. In greater detail, we will show that the excessive conductivity can be interpreted in terms of collective excitations of the superconducting phase field or \emph{Goldstone} modes - a fascinating phenomenon if only because it bears on the nature of spontaneous $U$(1) symmetry breaking.\\
As we have seen in Sec.\,\ref{Sec:FieldTheoforSC}, the topology of the superconducting order parameter can be used to link the zero-resistance transport of a charged superfluid to the  phase rigidity, which itself is a direct consequence of the $U$(1) symmetry breaking. The Goldstone theorem\sidenote{\footnotesize{originally formulated in the context of high energy physics}} states that in any system with a spontaneously broken continuous symmetry a certain number (given by the rank of the broken-symmetry group) of low-energy collective excitations will appear. For a charged superfluid, we consider breaking of $U$(1) symmetry and a single Goldstone mode\sidenote{\footnotesize{This is similar to the case of ferromagnetism, where the spontaneous magnetization below the Curie temperature breaks rotational symmetry giving rise to Goldstone modes termed \emph{Magnons}. Another example is quantum chromodynamics, where the vacuum breaks chiral $SU$(2)$_L\times SU$(2)$_R$ symmetry causing the emergence of the light pions. Note that in both cases the symmetry of the Lagrangian is only approximate and thus the Goldstone excitations are light, but not massless as for exact symmetries. }}. Although the relationship between particle physics and superconductivity was recognized before the BCS milestone, the superconducting Goldstone modes remained elusive and only a theoretical postulate. This is remarkable, as the gapped superconductor should enable an observation of low-energy modes without contaminations of quasiparticle excitations. The early work of Anderson \cite{Anderson58} attributes the disappearance to Coulomb interactions, which shift the Goldstone modes to the plasma frequency $\omega_p\gg \Delta$, where the detection is a hopeless task. At the same time, this restriction defines a model system where the observation of a Goldstone mode should become possible. First, Anderson's argument only holds for long-range Coulomb interactions, but fails for short-range interactions common to disordered metals. Second, due to the nature of Coulomb interaction in 2D, the $\omega_p$ is shifted to zero such that Goldstone modes should acquire spectral weight inside the superconducting gap. Finally, Anderson's argument assumes a spatially uniform superconducting order - a requirement clearly violated in disordered or granular superconductors as evident from STM measurements. These considerations envision granular Al as prototypical system, where a Goldstone mode might be detectable by means of optical spectroscopy. \\
An instructive way to understand, how excitations of the superconducting phase can contribute to the conductivity, is the London equation
\begin{equation}
\mathbf{j}=\frac{qn}{m}\nabla\vartheta
\end{equation}
(with charge $q$ and mass $m$) we derived from the effective low-energy superconducting Lagrangian \ref{eq:LowBosonL} after integrating-out the energetic fields from \ref{eq:HiggsL1}. A gradient $\nabla\vartheta$ of the phase field causes a current that - without going into details - we can relate \cite{Swa14} to the conductivity\sidenote{\footnotesize{here as the response function for co-linear currents and fields in $x$-direction }} 
\begin{equation}
\sigma(\omega)\sim \frac{-\chi_{xx}(\mathbf{q}=0,\omega+i0^+)}{i(\omega+i0^+)}
\end{equation} 
obtained after analytic continuation of the the current-current correlation function on the lattice (with Matsubara frequencies $\omega_n=2n\pi k_BT$)
\begin{equation}
\chi_{xx}(\mathbf{q},i\omega_n)=\sum_\mathbf{r}\int_0^\beta d\tau\langle j_x(\mathbf{r},\tau)j_x(\mathbf{0},0)\rangle e^{i\mathbf{qr}}e^{i\omega_n\tau}
\end{equation}
in the limit $\mathbf{q}\to 0$. 
\begin{figure}[b!]
\begin{centering}
\includegraphics[width=\textwidth]{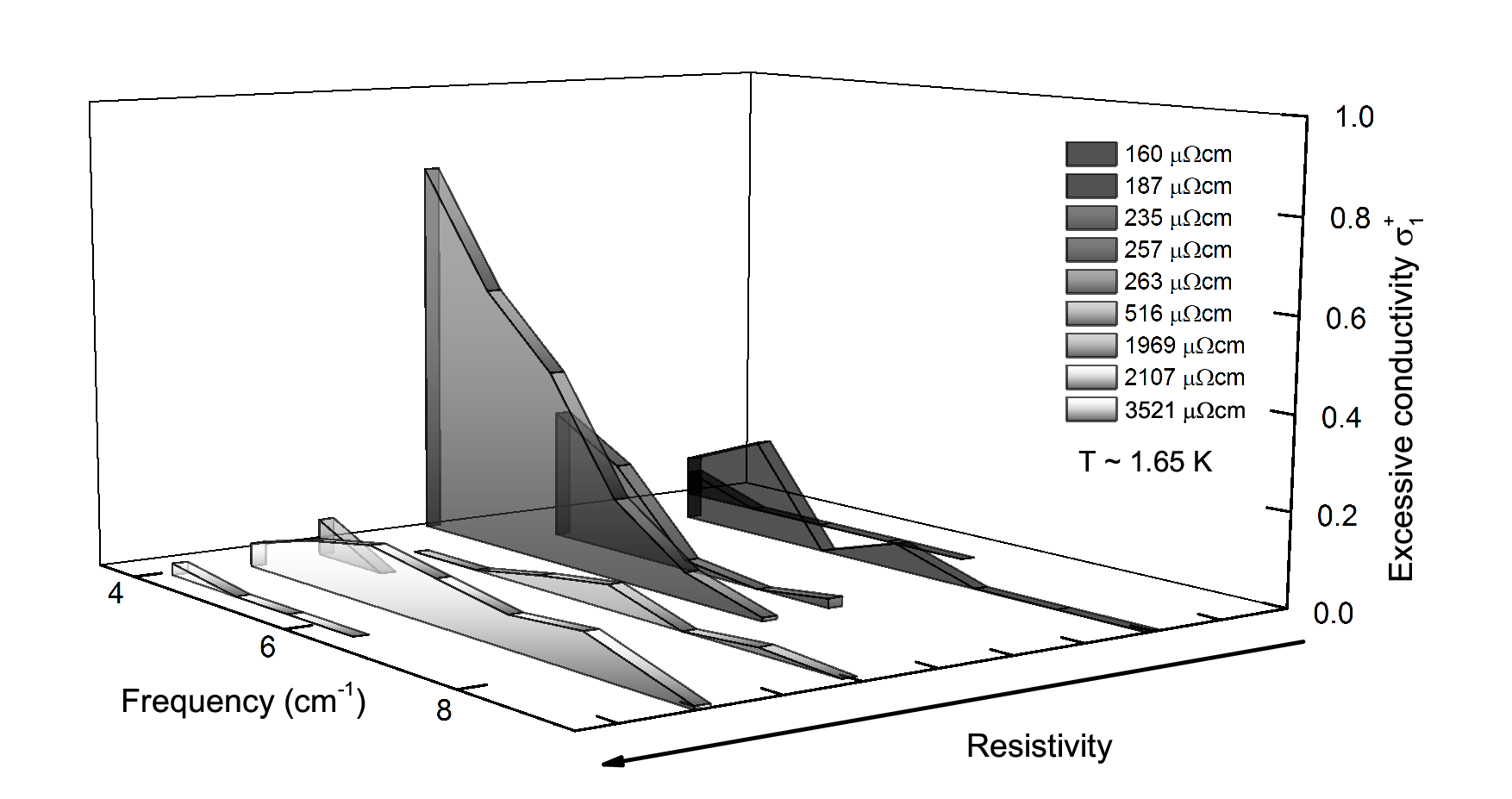}
\caption{\label{Al_sub-gap_TempDep}\textbf{Excessive sub-gap conductivity $\sigma_1^+(\nu)$ in granular Al} for various temperatures below $T_c$ for samples on the LR and HR sides of the superconducting dome.} 
\end{centering}
\end{figure}
\noindent To obtain a deeper quantitative understanding of how the Goldstone modes contribute to the conductivity, it is important to start from a microscopic Hamiltonian for superconductivity. Cea \emph{et al.} \cite{cea14} considered the fermionic attractive Hubbard model on the square lattice with on-site energy $U<0$, local disorder $V_i$,  and hopping $t$ where they define
\begin{equation}
j_\alpha(\mathbf{q},\omega)=-q^2 K_{\alpha\beta}(\mathbf{q},\omega)A_\beta(\mathbf{q},\omega)
\end{equation} 
linking the total current in $\alpha$-direction to a gauge field in $\beta$-direction. Clearly, the physics are contained in the response kernel $K_{\alpha\beta}$ which can further be split into diamagnetic and paramagnetic responses
\begin{eqnarray}
K_{\alpha\beta}(\mathbf{q},\omega)&=&K^\mathrm{dia}_{\alpha\beta}(\mathbf{q},\omega)+K^\mathrm{para}_{\alpha\beta}(\mathbf{q},\omega)\nonumber\\
&=&D\delta_{\alpha\beta}-\chi_{\alpha\beta}(\mathbf{q},\omega)
\end{eqnarray}
with the diamagnetic term $D$ being given by the spectral weight. While the diamagnetic term accounts for the response of the superfluid, the paramagnetic term gives the quasiparticle conductivity (i.e. particle-hole excitations) but allows to treat collective modes by inclusion of higher-order vertex corrections of MB theory. The authors of Ref.~\cite{cea14} supplemented the bare quasiparticle excitation ($\chi^{(0)}$) with a higher order contribution
where an incoming photon couples to a particle-hole excitations (the so-called BCS bare-bubble), then excites an intermediate collective mode, another particle-hole excitation, and eventually leaves. The possibility of such a process, however, does not automatically imply its observability as the collective modes in clean superconductors only contribute to the finite-$\mathbf{q}$ longitudinal response and the Goldstone mode cannot be excited by a photon \cite{cea14,Schrieffer1964}. In disordered and hence inhomogeneous thin films, however, the coupling becomes non-zero even at $\mathbf{q}=0$ and should lead to a finite absorption. Calculation of the real conductivity via
\begin{equation}
\sigma_1(\omega)=-q^2\mathrm{Re}\frac{K_{xx}(\mathbf{0},\omega)}{i(\omega+i0^+)}
\end{equation}
\begin{figure}[t!]
\begin{centering}
\includegraphics[width=\textwidth]{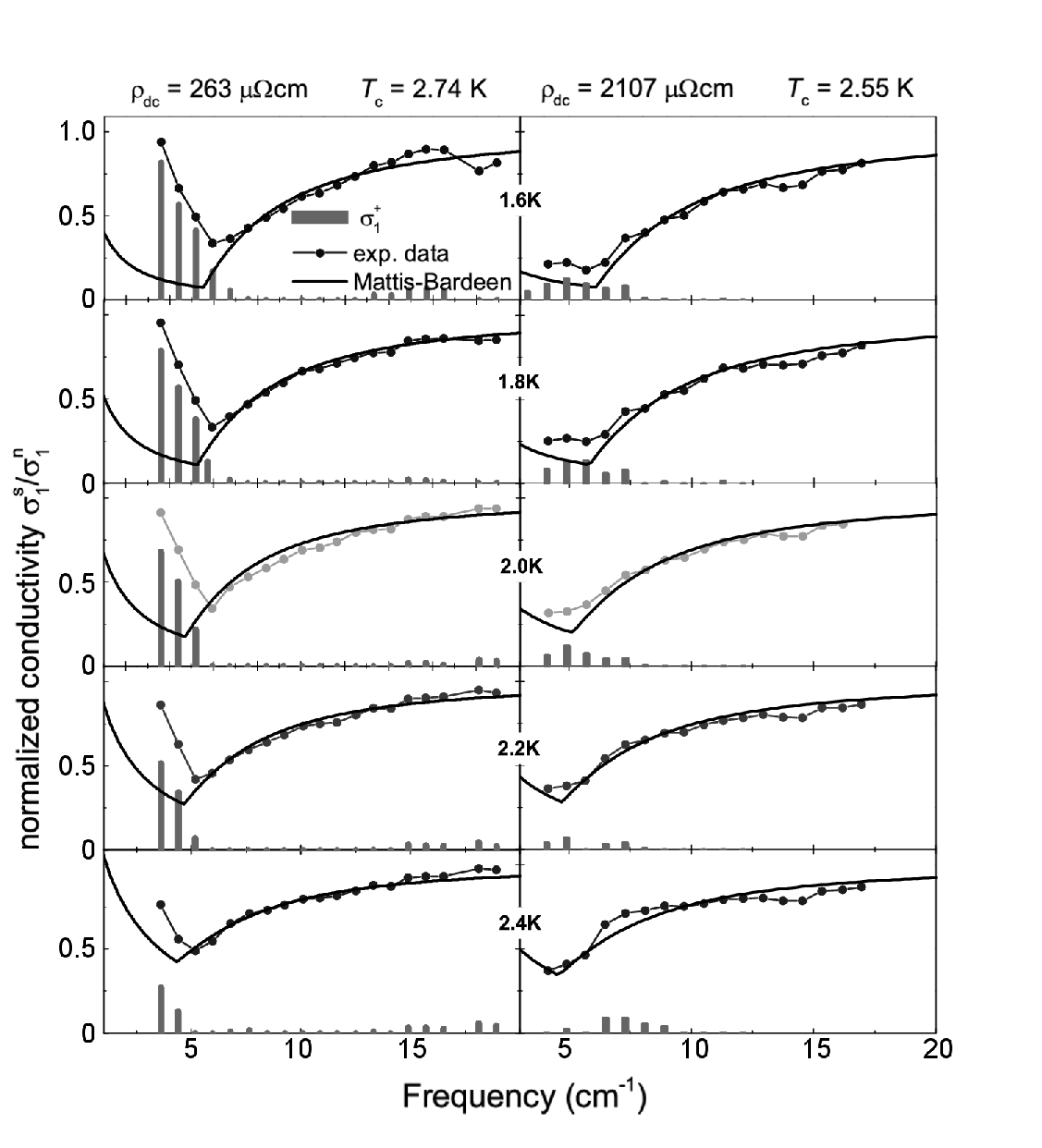}
\caption{\label{Al_sub-gap_all}\textbf{Decomposition of $\sigma_1(\nu)$} in terms of a quasiparticle background following MB theory (solid lines) and additional sub-gap contribution $\sigma_1^+(\nu)$ (bars) for two samples measured during the second cool-down.} 
\end{centering}
\end{figure}
\noindent reproduces the MB result on the bare-bubble level, while the Goldstone mode indeed appears in $\sigma_1$ by virtue of vertex corrections. Depending on the strength of localization/delocalization tendencies set by the relative energy scales $U/t$ and $V/t$, a notable amount of spectral weight amounts below the quasiparticle threshold $2\Delta$.\\
In the light of these considerations, we can revisit the experimental results on the low-frequency $\sigma_1$ obtained in the course of this work. For a number of samples, a significant amount of excessive conductivity $\sigma^+_1(\nu)$ (normalized to the $\sigma_1(\nu)$ at 5\,K) is isolated by subtracting the MB curve from the experimental data\sidenote{\footnotesize{see the previous sections for plots and how the fits were performed}}, as displayed in Fig.~\ref{Al_sub-gap_all}. At the time of writing it remains unclear, why some samples do and some do not display a notable $\sigma^+_1(\nu)$. It is worthwhile noting that the strongest sub-gap absorption was found for samples, which had been cooled down before. Interestingly, when cooled to $^4$He temperatures the first time, only a minor $\sigma^+_1(\nu)$ contribution could be isolated (see Fig.~\ref{dome+sig12_grAl}b), while $\sigma^+_1(\nu)$ increased drastically as evident from a subsequent measurement a few months later, see Fig.~\ref{Al_sub-gap_TempDep}. At the same time, the normal-state resistivity as measure of disorder increased from 189 to 263\,$\mu\Omega$cm and a slight reduction of $T_c$ from 2.78 to 2.74\,K. As the spectral weight associated with the Goldstone mode sensitively depends on the amount of disorder (rendering the superconducting state spatially inhomogeneous), the strong enhancement of $\sigma^+_1(\nu)$ after the first cool down could possibility be attributed to a higher level of disorder caused by the exposure to humidity when removed from the cryostat in the first place. A similar enhancement of $\sigma^+_1$ is observed for another sample, where $\rho_\mathrm{dc}$ increased from 1969 to 2110\,$\mu\Omega$cm with a concomitant drop of $T_c$ from 2.78 to 2.74\,K and 2.7 to 2.55\,K. A systematic study is required to explain the selective appearance of $\sigma^+_1(\nu)$ and left for future work\sidenote{\footnotesize{Further insights might be drawn from the observation of electron-glass behavior and a memory effect in strongly disordered insulting granular Al \cite{Grenet07,Delahaye08,Delahaye09,Delahaye11}}}.\\
We will, in what follows, focus on one sample, where the frequency and temperature dependencies of $\sigma^+_1(\nu)$ allows a particularly comprehensive discussion within the framework of optically active Goldstone modes calculated within a effective bosonic model.\\
A Hamiltonian, that is somewhat more applicable to real systems such as granular Al is the disordered quantum $XY$ spin-$\nicefrac{1}{2}$ model in a transverse random field. This model is inspired from a Heisenberg 2D-lattice of classical spins $s_j=|s_j|e^{i\theta_j}$ on each lattice site $j$ serving as simple model for the phase degrees of freedom in a superconductor with local order parameters $\psi_j=\Delta_j e^{i\theta_j}$. The analogy to a magnetic system is two-fold: First, the Heisenberg system has a non-zero expectation value of the net magnetic moments  (i.e. a finite magnetization $\mathbf{M}$) below the Curie-temperature, in correspondence with a non-zero modulus $|\Delta|$ (i.e. the superconducting energy gap) of the global order parameter $\Psi=\Delta e^{i\theta}$ below the critical temperature $T_c$. In both cases, a rotational $U$(1) symmetry is spontaneously broken as soon as $\mathbf{M}$ become finite and $\Psi$ acquires a certain phase. Second, the strength exchange interaction between spins $s_is_j\sim \cos(\theta_i-\theta_j)$ dictates the robustness of the magnetization against phase fluctuations similar as the superfluid stiffness $J$ is governed by the gradient of the global phase field. In practice, the fermionic spin degrees-of-freedom of the classical Heisenberg Hamiltonian are mapped to operators $S^x, S^y$, and $S^z$ leading to the quantum $XY$ model on a $N\times N$ lattice \cite{cea14,Ma1985}
\begin{equation}\label{XYmodel}
\mathcal{H}=-2\sum_i\xi_iS_i^z-2\sum_{i,j}J_{ij}\left(S^+_iS^-_j+\left(S_j^-\right)^\dagger\left(S_i^+\right)^\dagger\right)
\end{equation}  
with the \emph{pseudospin} operators defined as  
\begin{equation}
S_i^+=S_i^x+iS_i^y= c_{i\downarrow}^\dagger c_{i\uparrow}^\dagger, \qquad S^-_i=\left(S_i^+\right)^\dagger
\end{equation}
and
\begin{equation}
S_i^z=\frac{1}{2}\sum_\sigma c_{i,\sigma}^\dagger c_{i\sigma}-\frac{1}{2}.
\end{equation}
We see that $S_i^z=\pm \nicefrac{1}{2}$ represents the site $i$ being occupied (+) or unoccupied (-) by a Cooper pair and a finite in-plane magnetization $\langle S_i^x\rangle\neq 0$ represents superconducting order. The transverse random field $\xi_i$ takes the form of a on-site localization energy and represents disorder, while $J_{ij}$ is the energy required for pair-hopping. This model has proven its applicability to disordered superconductors as it accurately reproduces the emergent inhomogeneous superconducting order towards the SIT \cite{Lemarie2013}.  More importantly, this modes allows to include phase fluctuations above the mean-field level. In what follows, we will retrace the succession of steps as done by Cea \emph{et al.}, see Ref.\,\cite{cea14}. By means of a Holstein-Primakov (HP) transformation, the model (\ref{XYmodel}) is cast into a quadratic Hamiltonian $\mathcal{H}_\mathrm{PS}$ we can diagonalize into a bi-linear Hamiltonian $\mathcal{H}^\prime_\mathrm{PS}$ in the usual Bogoliubov scheme\sidenote{\footnotesize{that means an interacting system (four-operator term) is mapped to a non-interacting (two-operator term) by replacing the particles with energies $E$ by non-interacting quasiparticles with renormalized energies $E_\alpha$ that account for the interaction.}}.
\begin{equation}
\mathcal{H}^\prime_\mathrm{PS}=\frac{1}{2}\sum_i\sum_{\mu=x,y}J^\mu_i\left[\Delta_\mu\Phi_i\right]^2+\frac{1}{2}\sum_{i,j}\chi_{i.j}^{-1}L_iL_j\label{HP}
\end{equation}

\begin{figure}[b!]
\begin{centering}
\includegraphics[width=\textwidth]{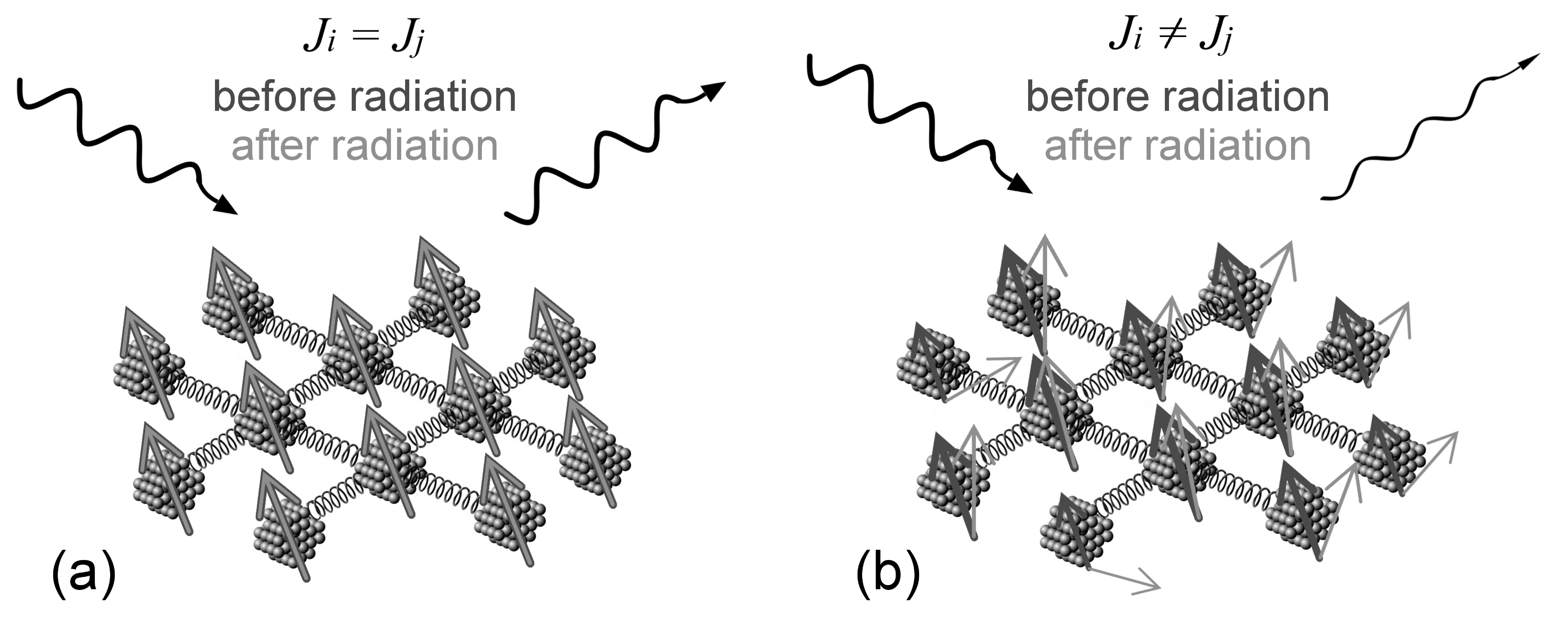}
\caption{\label{generic_phasemode}Sketch of the optical response of an array of Josephson junction, modeled as in Eq.\,(\ref{XYmodel}). Here the arrows represent the local spins on the array sites, connected by springs representing the local stiffnesses $J_{ij}$. To visualise its inhomogeneity, we set the arrow length proportional to the strength of the local stiffness and compare the spin orientation before (dark) and after (light) the photon irridation. In the clean case, panel (a), the phase modes are decoupled from the transverse electromagnetic field, so the spins preserve their orientation (i.e.  the phase of the order parameter is unchanged) and the radiation is not absorbed. On the other hand in the disordered case, panel (b), the spins respond to the incoming radiation with a local
change of their relative direction that is larger when the system has lower phase rigidity (i.e. lower local $J_{ij}$). This leads to an inelastic response which absorbs part of the incoming
radiation.}
\end{centering}
\end{figure}

Here, $\Delta_\mu$ is a discrete derivative in $\mu$-direction and the Bogoliubov quasiparticle operators $\gamma_\alpha,\gamma_\alpha^\dagger$ are made explicit in the phase and number operators $\Phi_i$ and $L_i$
\begin{eqnarray}
\Phi_i&=&\sum_\alpha\frac{\phi_{\alpha_i}}{\sqrt{2}}\left(\gamma_\alpha^\dagger-\gamma_\alpha\right)\\
L_i&=&\sum_\alpha\frac{\ell_{\alpha_i}}{\sqrt{2}}\left(\gamma_\alpha^\dagger+\gamma_\alpha\right)
\end{eqnarray} 
where $\phi_{i\alpha}$, $\ell_{i\alpha}$, and $\chi^{-1}_{i,j}$ are determined from the Bogoliubov transformation and $J^\mu_i=J\sin\theta_i\sin\theta_{i+\mu}$ is the local superfluid stiffness. The quasiparticle excitations of $\mathcal{H}^\prime_\mathrm{PS}$ introduced by the Bogoliubov transformation are equivalent to phase excitations providing a route to calculate the dispersion of the optically active Goldstone modes. The coupling to the gauge field $A_\mu$ is done in the minimal-coupling scheme replacing $\Delta_\mu\to\Delta_\mu-2eA_\mu$ in Eq.\,(\ref{HP}) and the conductivity eventually follows as 
\begin{equation}
\sigma=e^2\pi\delta(\omega)D_s+\sigma_\mathrm{reg}(\omega)\label{modecond}
\end{equation}
consisting of the superfluid response and a regular part including the effects of Goldstone modes. In more detail, $D_s=D-N^{-1}\sum_\alpha Z_\alpha$ is the non-disordered superfluid density $D$ reduced by the spectral weight of the phase excitations each of which has an energy $E_\alpha$ and comes with an effective dipole 
\begin{equation}
Z_\alpha=E^{-1}_\alpha\left[\sum_i 2J^\mu_i\Delta_\mu\phi_{i\alpha}\right]^2\label{dipole}
\end{equation}
The regular contribution in Eq.\,(\ref{modecond}) is given by the sum of all individual phase excitations
\begin{equation}
\sigma_\mathrm{reg}(\omega)=\frac{e^2\pi}{2N}\sum_\alpha Z_\alpha\left(\delta(\omega+E_\alpha)+\delta(\omega-E_\alpha)\right)
\end{equation}   
amounting to the Goldstone mode.    
As shown in Ref.\,\cite{cea14}, the crucial ingredient that renders the Goldstone mode optically active is the disorder-induced inhomogeneity of $J_i^\mu$, see Fig.\ref{generic_phasemode}. This can be understood by examination of Eq.\,(\ref{dipole}): for a homogeneous superfluid stiffness, $J_i^\mu=J_j^\mu=J$, the summation extends over all gradients giving the total phase gradient, that equals to zero for periodic boundary conditions, and hence $Z_\alpha=0$.

To make a connection between the model and $\sigma_1^+$ of the experiment is essentially reduced to the problem of finding the particular form of the inhomogeneity. A simple approach is the so-called diluted $XY$-model with a binomial distribution $P(J_{ij})=(1-p)\delta(J_{ij}-J)+p\delta(J_{ij})$ with $J>0$, i.e. a for a given $p\in[0,1]$ we find $(1-p)$ of the connections to have a coupling constant $J_{ij}=J$, while $p$ links are not superconducting as $J_{ij}=0$. For a particular choice of disorder parts of the lattice will segregate into superconducting islands disconnected from the percolative superconducting path. If now a phase gradient occurs over such an island, a charge inbalance will form\sidenote{\footnotesize{Which becomes clear by remembering that the charge-carrier number is the conjugated variable of the phase, see Sec.\,\ref{number-phase}}} between two ends of the islands effectively turning it into an optically active nano antenna.  \\
In the simple binomial approach, the problem of meeting experiment and theory is reduced to set the energy scale of the model $J$ and the value of disorder $p$. We will now see, that we can infer a solid approximation of $J$ from the experimental data. The Hamiltonian under consideration, Eq.\,(\ref{HP}), describes a phase-only system, which can be described equivalently by a low-energy action\sidenote{\footnotesize{This low-energy action can be derived from a interacting Lagrangian by integrating-out the high-energy fields in analogy to Sec.\,\ref{Sec:FieldTheoforSC}, however, working with Grassmannian rather than simple Heisenberg fields \cite{Rama1989}.}} for constant $J_{ij}=J$ \cite{Rama1989}
\begin{figure}[h]
\begin{centering}
\includegraphics[width=\textwidth]{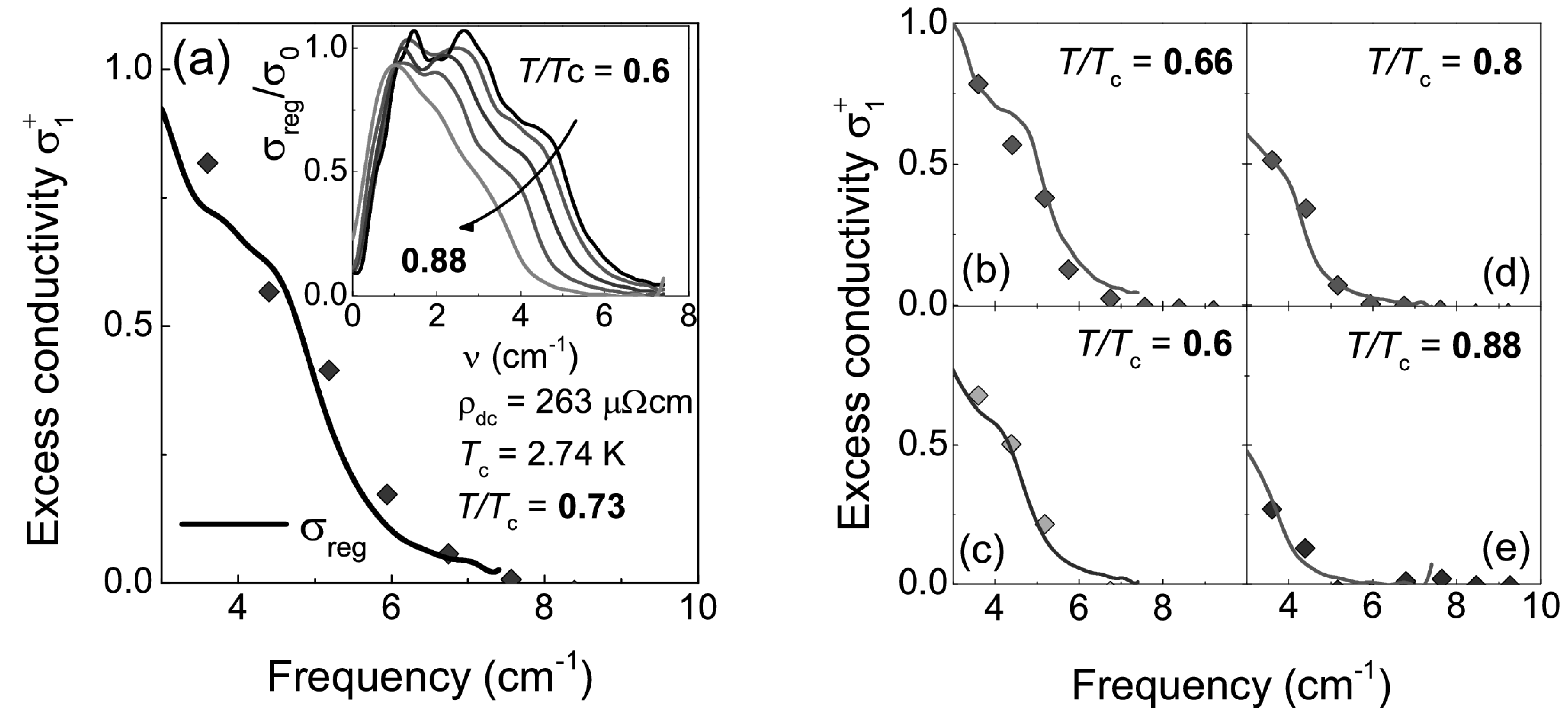}
\caption{\label{sig_reg_Al}\textbf{Excessive sub-gap conductivities} $\sigma_1^+(\nu)$ (symbols) extracted from the optical data and $\sigma_\mathrm{reg}(\nu)$ (lines) calculated within the diluted $XY$ model for $J=0.9$\,cm$^{-1}$ and a dilution $p=0.17$. The inset of (a) compares temperature evolution of $\sigma_\mathrm{reg}(\nu)$ shown in the other panels in detail. } 
\end{centering}
\end{figure}
\begin{equation}
S_\mathrm{HP}=\frac{1}{2}\int d\omega d\mathbf{k}\left[-J\mathbf{k}^2+\frac{1}{16J\xi_0^2}\omega^2\right]|\theta(\mathbf{k},\omega)|^2
\end{equation}
where $\xi_0$ is the typical length scale over which the discrete lattice-$\mathcal{H}_\mathrm{HP}$ is valid. The dispersion of the phase field $\theta(\mathbf{k},\omega)$ is obtained via the Euler-Lagrange formalism as 
\begin{equation}
\omega_\mathbf{k}=4J\xi_0|\mathbf{k}|\label{dispersion}
\end{equation}  
which is a ordinary sound-wave like dispersion and sets the typical energy scale of the Goldstone modes. In a disordered system without translational invariance, the modes will have a characteristic finite momentum $\xi_0\bar{|\mathbf{k}|}\sim 1$ so that Eq.\,(\ref{dispersion}) simplifies to
\begin{equation}
\bar{\omega}\sim 4J
\end{equation}
The strongest absorption is found at the lowest temperature at around 3.6\,cm$^{-1}$ so that we set $J=0.9$\,cm$^{-1}$ as starting point for the numerical calculations of $\sigma_\mathrm{reg}(\nu)$. Although this approximation may appear somewhat crude, it reproduces the experimental situation surprisingly well as we will discuss below. 
Far from criticality $p<p_c$ the temperature evolution of $J$ can be computed within the model Eq.\,\ref{HP}. To compare $\sigma_1^+(\nu)$ and $\sigma_\mathrm{reg}(\nu)$ for a given temperature,  $\sigma_\mathrm{reg}$ given in units of $e^2/\hbar$, is multiplied with a dimensionless factor $\alpha$  that defines an effective thickness  $d=e^2/(\hbar\alpha \sigma_\mathrm{dc})$ required to convert the 2D conductivity into a 3D one measured in the experiment. For the sample under study, we have $\sigma_\mathrm{dc}=3802\,(\Omega\mathrm{cm})^{-1}$ and $d=0.64/\alpha$\,nm which defines a length scale of the order of the lattice spacing as expected for a 2D$\to$3D conversion. The result is shown in Fig.\,\ref{sig_reg_Al}. The best agreement is obtained for a dilution $p=0.17$ and $\alpha=5.5$. The agreement between theory and experiment is remarkable: both the temperature and frequency dependence of $\sigma_1^+$ are accurately captured by the model. The agreement between theory and experiment and hence the interpretation in terms of the Goldstone mode can be justified even further by an indirect use of the inductive response $\sigma_2(\nu)$ via the penetration depth $\lambda$. As done earlier in this work, see Eq.\,(\ref{js}), we can relate $J$ and $\lambda$ in appropriate units via 
\begin{equation}
J[K]=0.62 \times \frac{d[\mathrm{\AA}]}{\lambda^2[(\mu\mathrm{m})^2]}
\end{equation}
For $J=0.9\,\mathrm{cm}^{-1}\hat{=}1.29$\,K and $\alpha=5.5$ we have an effective thickness of $d=1.6$\,$\mathrm{\AA}$ leading to $\lambda=0.88$\,$\mu$m. From analysis of the inductive response, we find $\lambda=1.05$\,$\mu$m for this sample, see Fig.\,\ref{fig:lambda}, in very good agreement with the value predicted for the particular choice of $\alpha$. Consequently, $p$ is the only remaining free parameter of the entire model strongly speaking in favor of our interpretation of $\sigma_1^+$ in terms of Goldstone modes $\sigma_\mathrm{reg}$.\\
\begin{figure}
\begin{centering}
\includegraphics[width=\textwidth]{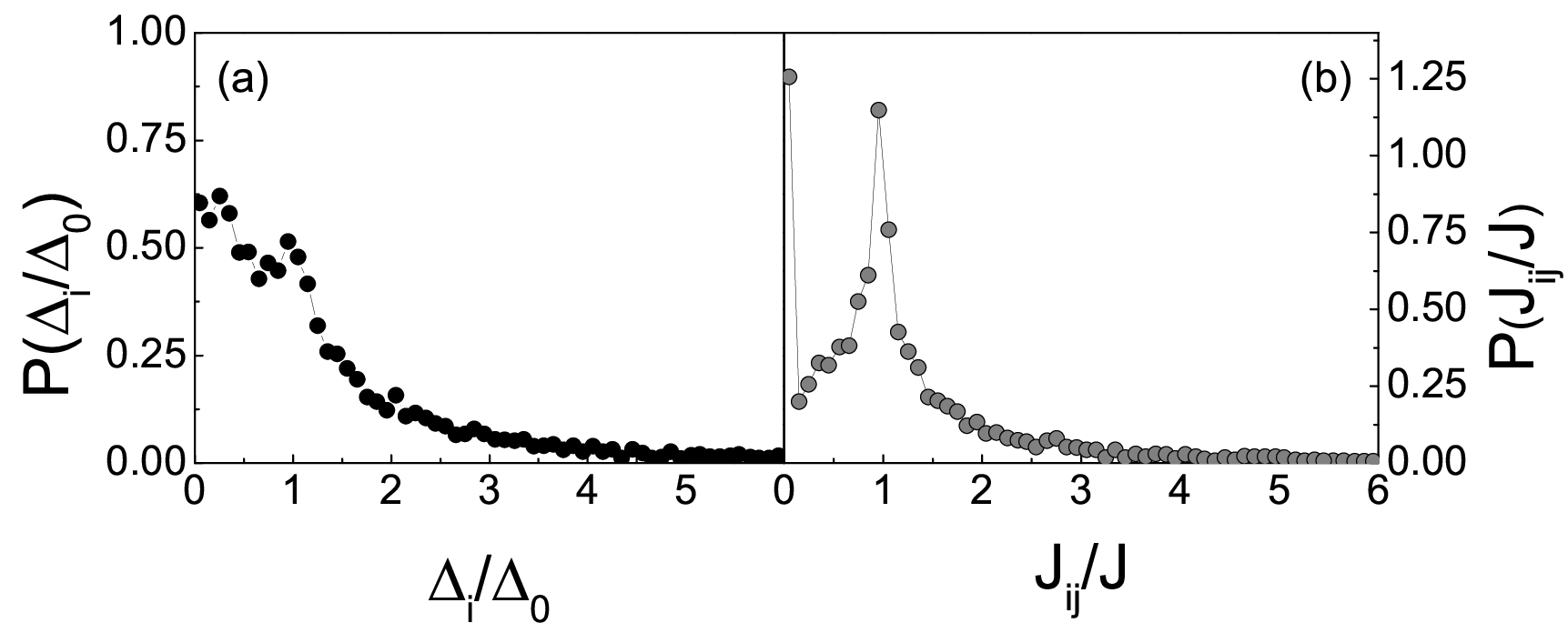}
\caption{\label{distro}\textbf{Probability distributions (not normalized) of $\Delta$ and $J_{ij}$} calculated for a granular superconductor with a Gaussian distribution of grain diameters and a resistance ratio $R_N/R_q=0.01$. Note that the distribution of couplings $J_{ij}$ resembles the simple binomial distribution used to calculate the solid lines in Fig.\,\ref{sig_reg_Al}} 
\end{centering}
\end{figure}
What may strike surprising is the series of rather crude approximations and assumptions, yet leading to a remarkable agreement between theory and experiment. For the sake of persuasiveness, the equivocally simple binomial distribution $P(J_{ij})$ should be examined in more detail aiming for a justification within the particular context of granular superconductors. The system that comes closest to granular Al is a disordered 3D network of coupled superconducting spheres with a distribution of radii where global superconductivity emerges in a percolative manner studied in Ref.\,\cite{mayoh14}. This model predicts a broad distribution $P(\Delta/\Delta_0)$ with $\Delta_0$ the bulk value and a shape depending on the number of broken links measured by the normal-state resistance $R_N$ normalized to the quantum resistance $R_q=h/4e^2$. For the sample under study, we find $R_N/R_q=0.01$ and the corresponding distribution $P(\Delta/\Delta_0)$ takes a form displayed in Fig.\,\ref{distro}(a)  \cite{LaraPers}. To translate $P(\Delta/\Delta_0)$ into the desired $P(J_{ij}/J_0)$ (with $J_0$ being the stiffness of the bulk) we assume $J_{ij}=(\Delta_i+\Delta_j)/2$, see  Fig.\,\ref{distro}(b). Surprisingly, the distribution is sharply peaked at $J_{ij}/J_0=1$ in close resemblance with the simple binomial distribution considered in our analysis. Understanding the tails as subordinate corrections, what remains, is a profound microscopic justification of our above simple model based on the shell-effect in granular Al.

\section{The effect of magnetic sub-gap bands?}

When we introduced superconductivity of granular Al we highlighted the existence of localized spins in granular Al as evident from MR and $\mu$SR studies and, in striking similarity to unconventional superconductors, pointed towards the possibility of a nontrivial interplay of superconducting pairing and spin-flip scattering. Although the appearance of the superconducting dome in granular Al can be explained by competing mechanisms, namely the shell effect and suppressed phase coherence, without referring to unconventional pairing or nearby magnetic ordering, for the sake of completeness, a possible mechanism for the sub-gap absorption related to spin-flip scattering should be carefully excluded.\\

Indeed, soon after BCS Abrikosov and Gor'kov (AG) have shown in seminal works \cite{Abrikosov58,Abrikosov59,Abrikosov60}, that superconductivity is strongly affected by magnetic impurities. Although the Anderson theorem guarantees a strong robustness of superconductivity against potential scattering, due to the time-reversal symmetry breaking nature of spin-flip scattering the Anderson theorem is escaped. AG realized that for a certain amount of magnetic impurities the threshold of absorption is pushed below $2\Delta$ and can vanish although superconductivity persists\sidenote{\footnotesize{At this point, we again stress that the superconducting order parameter $\psi$ is \emph{not} a priori the same entity as the energy gap $\Delta$}}.     \\
In Sec.\,\ref{Sec:Green} we introduced a powerful Green's function approach to include pair-breaking effects resulting, e.g., from spin-flip scattering off paramagnetic impurities. We can here use the versatile and straight formalism developed recently \cite{Fominov2011} to test various scenarios of sub-gap bands that may develop in presence of impurities. The basic idea is that transitions between sub-gap bands or transitions into the quasiparticle continuum become possible at low energies, which, in contrast with the BCS DOS, may enhance the low-energy conductivity. The bare existence of a sub gap band, however, does not a priori guarantee additional absorption channels restricted to sub-gap energies, which is obvious from the general form of the conductivity Eq.\,(\ref{FominovCond}) (see Sec.\,\ref{Sec:Green})   
\begin{eqnarray}
\hat{\sigma}(\omega)&=&\frac{\pi e^2 n_s}{m}\delta(\omega)\nonumber\\
&&-\frac{\sigma_0}{2\omega}\int\limits_{0}^\infty\mathrm{d}E\Big\{\tanh\frac{ E-\frac{\hbar\omega}{2}}{2k_\mathrm{B}T}\big[G^{+}G^{-}_1-iF^{+}F^{-}_2\big]\nonumber\\
&&-\tanh\frac{E+\frac{\hbar\omega}{2}}{2k_\mathrm{B}T}\big[(G^{-})^*G^{+}_1+i(F^{-})^*F^{+}_2\big]\Big\}\nonumber
\end{eqnarray}
where the \emph{integrated} DOS ($G^\pm_1$) is relevant for $\hat{\sigma}$. \\
To test the above idea, we infer the matrix Green's function $\check{g}$ by solving the Usadel equation Eq.\,(\ref{Usadel}) 
\begin{equation}
iE[\hat{\tau}_3\otimes\hat{\sigma}_0, \check{g}]-\Delta[\hat{\tau}_1\otimes\hat{\sigma}_0, \check{g}]+i[\widehat{\Sigma},\check{g}]=0\nonumber
\end{equation}  
(see Sec.\,\ref{Sec:Green}) with the self-energy term 
\begin{equation}
i[\widehat{\Sigma},\check{g}]=-\frac{i}{\tau_s}\hat{\tau}_2\otimes \hat{\sigma}_0\frac{\sin2\theta}{1+\xi^4+2\xi^2\cos2\theta}\nonumber
\end{equation}
for various pair breaking parameters $\tau_s$ and explicitly keeping $\xi\neq 0$ to unlock the center of impurity band from the continuum onset\sidenote{\footnotesize{where it resides in the AG limit broadening the coherence peaks as, e.g., discussed in Sec.\,\ref{Sec:ExpNbN} of this work.}}. 
\begin{figure}[t!]
\begin{centering}
\includegraphics[width=\textwidth]{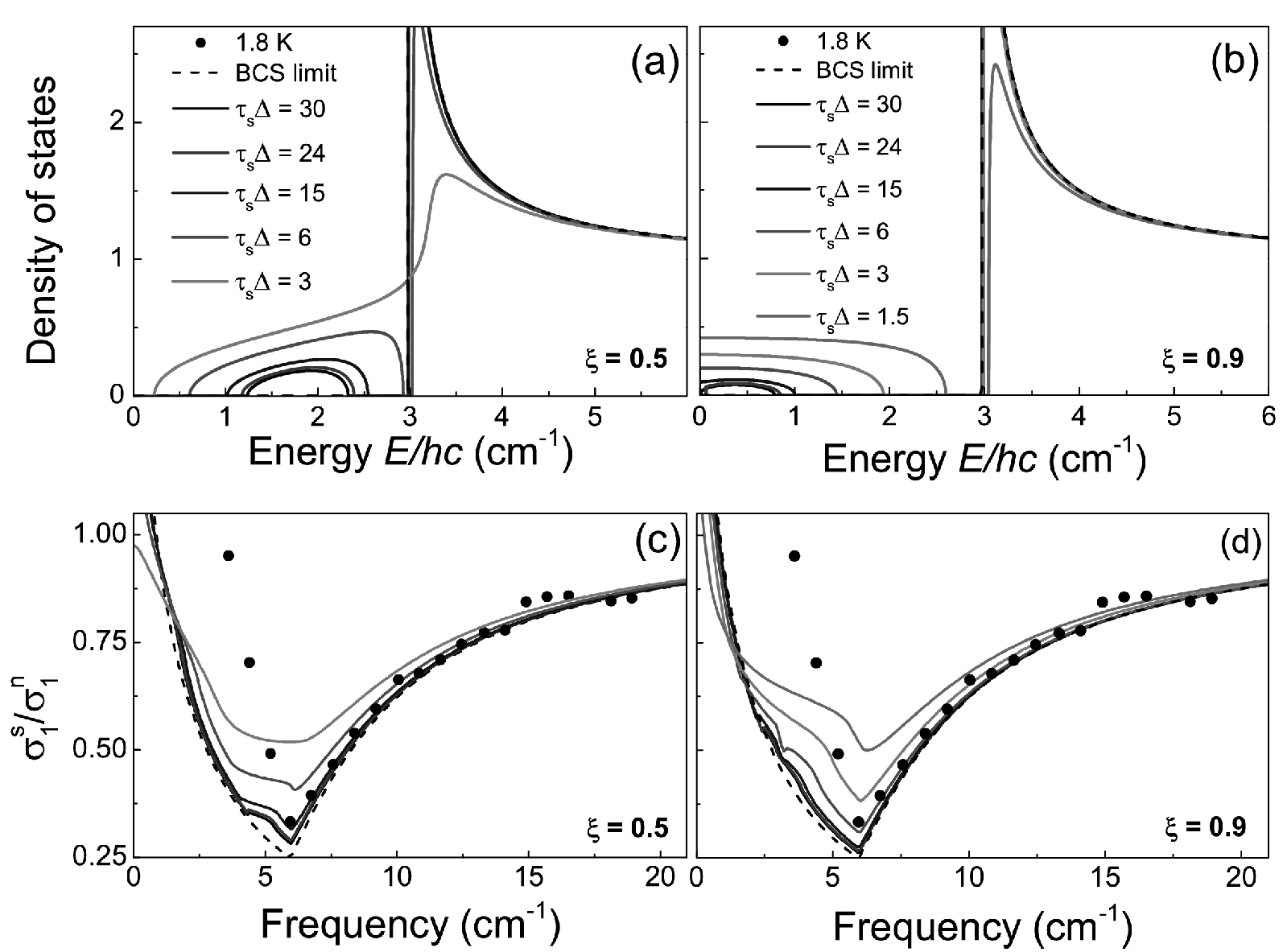}
\caption{\label{subbands}\textbf{Various realizations and strengths of sub-gap bands} (a,b) and to corresponding $\sigma_1(\nu)$ spectra (c,d) together with experimental data (sample $\rho_\mathrm{dc}=263$\,$\mu\Omega$cm, $T_c=2.78$\,K at 1.8\,K). No combination of parameters determining the position and size of the sub-gap band can equally fit both the excessive absorption at low energies and the BCS-like part of the experimental data.} 
\end{centering}
\end{figure}
Figure \ref{subbands}(a,b) displays the DOS for two realizations of a sub-gap band with $\xi=0.5$ and 0.9 and different values of pair breaking parameters $\tau_s$. For $\xi=0.5$ the band is located completely inside the gap, grows as $\tau_s\Delta$ is reduced and merges with the continuum below $\tau_s\Delta\approx 6$. In the other case, $\xi=0.9$, there is no hard gap and the DOS is finite even at 0 (or more precisely $E=E_F$). The corresponding spectra of (normalized) $\sigma_1(\nu)$ are displayed in the panels (c,d) together with the experimental data for a granular Al sample with a pronounced sub-gap absorption. While, indeed, the modified DOS gives rise to an enhancement of $\sigma_1$ at low energies, no combination of parameters can properly fit the strong rise and, at the same time, maintain the BCS like shape at higher energies. Also in the AG limit, these two requirements cannot be met simultaneously, as can be inferred from the plots of Sec.\,\ref{Sec:ExpNbN}. In conclusion, the spin-flip scattering found in granular Al can certainly be ruled out as origin of the sub-gap absorption discussed in this work.    
\clearpage{}
\section{Outlook}
The results of this work present a conclusive answer to the longest-standing question associated with superconductivity in granular Al: what shapes the superconducting dome? Yet they also add new questions which, together with other open problems, hopefully trigger a experimental continuation.\\
In this work we presented a conclusive and fully consistent scenario, attributing the enhancement of $T_c$ beyond the bulk value to quantum confinement and the shell effect. An independent confirmation of the shell effect could be done by means of scanning tunneling microscopy (STM) similar to what has been done on individual Sn and Pb nano islands \cite{bose10}. The differential conductance into single Al nanograin with various inter-grain coupling strengths would provide valuable insights in how the local pairing amplitude  $\Delta$ is affected, especially on the HR side, where $T_c$ is already suppressed by phase fluctuations. In addition, STM measurements could provide further insights to possible modifications of the DOS caused by localized spins, e.g. sub-gap bands or exotic so-called Yu-Shiba-Rusinov bound states \cite{Shiba1968,Rusinov1969,Balatsky2006} beyond the accessibility of optical probes. \\
\begin{marginfigure}
\begin{centering}
\includegraphics[width=\marginparwidth]{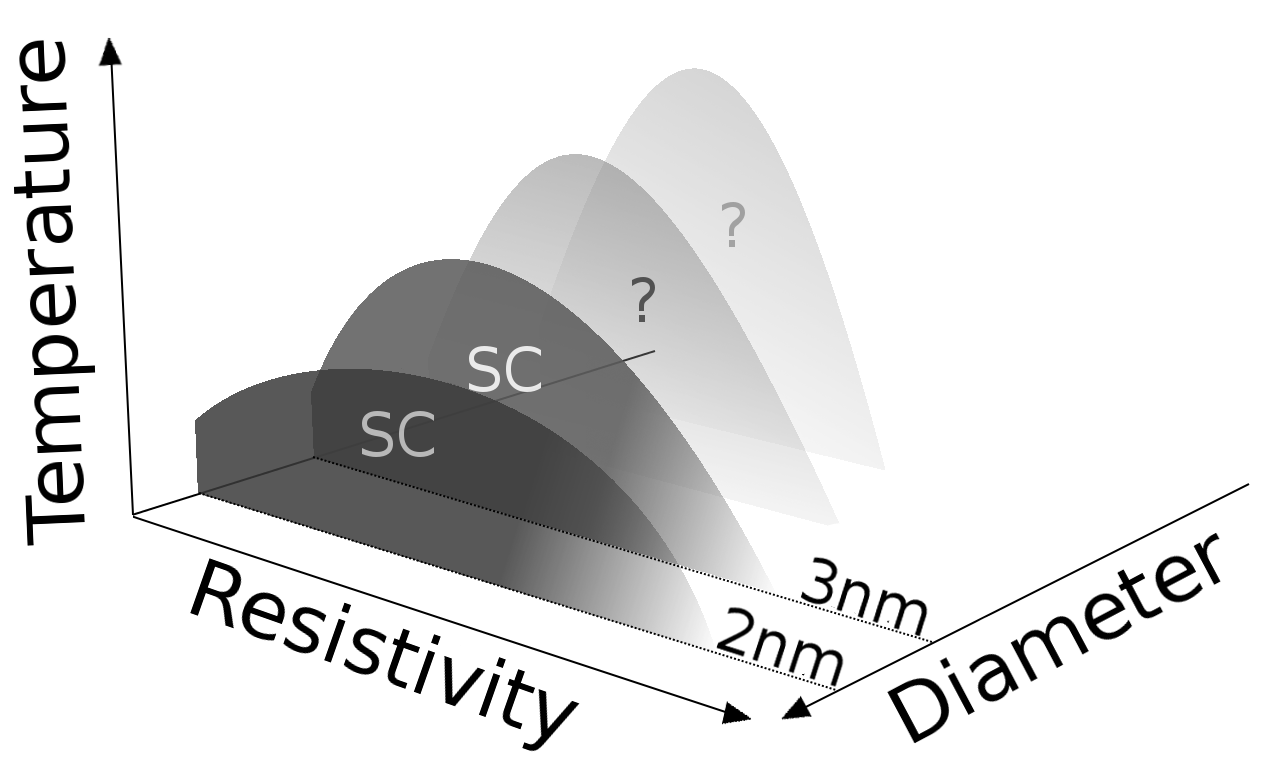}
\caption{\label{domes3D}Proposal for the superconducting domes of granular Al with different grain diameters. The two domes in the front (3\,nm and 2\,nm grains, labeled 'SC') with a maximum $T_c$ of 2.2\,K and 3.\,K, respectively, are experimentally established. The two domes in the back (labeled '?') with an even higher $T_c$ enhancement are envisioned theoretically for grain diameters less than 2\,nm.} 
\end{centering}
\end{marginfigure}
\noindent In addition to the fundamental questions raised above concerning the presently established phenomenology of 2\,nm Al grains, another important open problem widens this framework further: based on the nature of quantum confinement, one should expect the shell effect to be even more pronounced for grains which are smaller than the ones studied so far. Following the route employed to create 2\,nm grains, Al evaporated on substrates held at temperatures between 77 and 4.2\,K should consequently yield yet smaller grains. Indeed it was shown \cite{Strongin1968,Sixl1974} that for Al evaporated onto substrates held at 4.2\,K an increase up to at least $T_c=$4.5\,K can be achieved. Although Ref.\,\cite{Strongin1968, Sixl1974} neither give normal-state resistivity values nor grain sizes nor seek optimization thereof, it can be seen as strong evidence for an even stronger $T_c$ enhancement analogous to the one discussed above. Based on Ref.\,\cite{Strongin1968}, additional superconducting domes similar to the ones sketched in Fig.\,\ref{domes3D} can be envisioned. On the one hand, the smaller the grain, the stronger the electronic quantum confinement becomes. On the other hand, with decreasing diameter, the volume-to-surface ratio decreases and the coupling between grains could be reduced. Whether the resulting phase fluctuations impede the enhancement of $T_c$ by overcompensation of the shell effect is an open question of fundamental interest that experiments so far were not able to answer. \\
\begin{marginfigure}
\begin{centering}
\includegraphics[width=\marginparwidth]{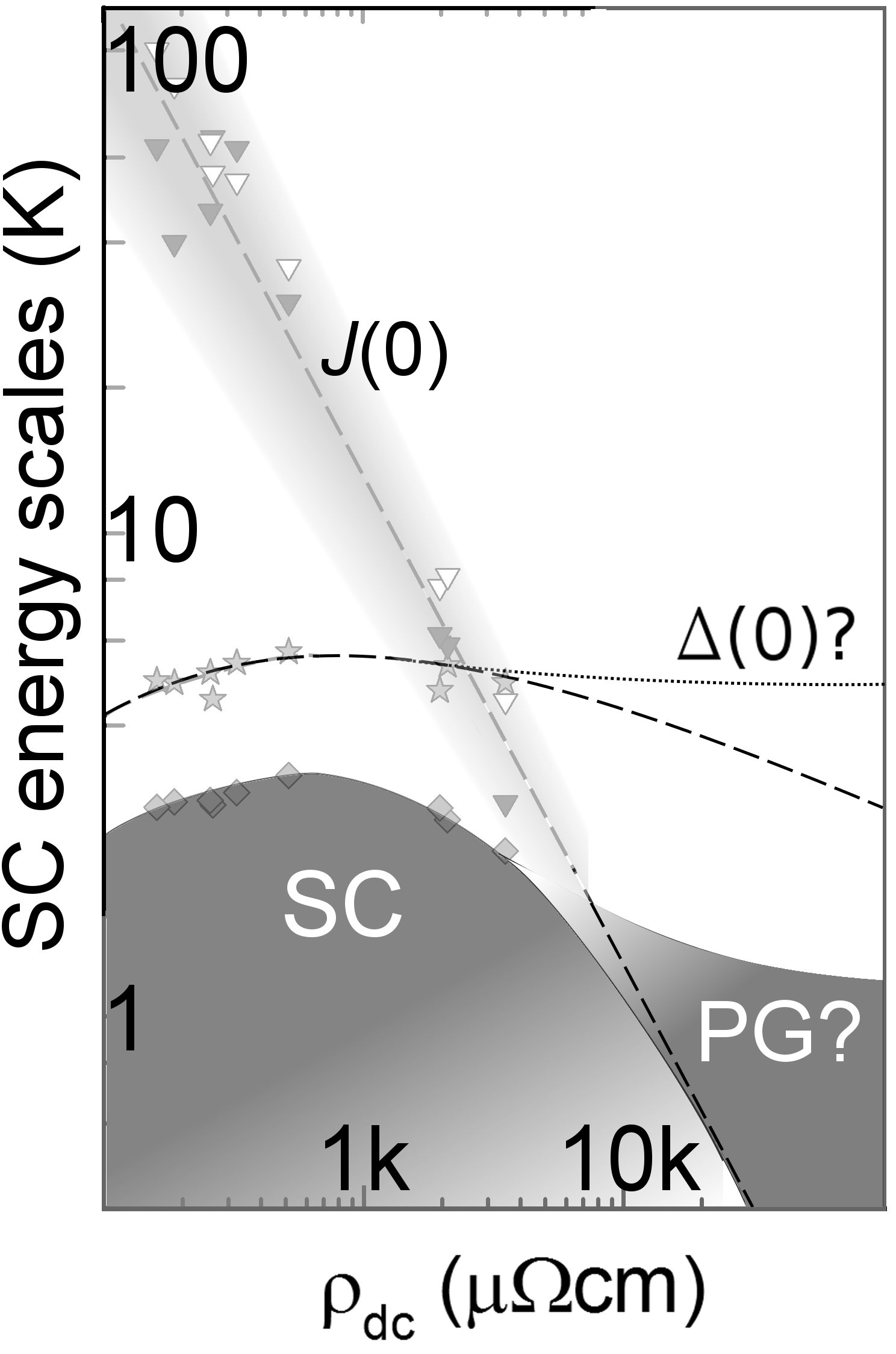}
\caption{\label{proposal}A possible extended phase diagram based on actual measurements (faint) discussed in this work, see Fig.\,\ref{dome_exp} and Sec.\,\ref{Enhanced Cooper pairing versus suppressed coherence}.} 
\end{centering}
\end{marginfigure}
\noindent Another set of open questions worth dedicated studies is related to the superconducting energy scales (and, thus, the phase diagram) at resistivities higher than in this work, see Fig.\,\ref{proposal}. The appearance of a pseudogap on the HR side is a natural consequence of the phase-driven nature of the transition, yet also deserves a closer examination. This would include systematic studies with samples far on the HR side with a resistivity of several 1000\,$\mu\Omega$cm in approach of the Mott insulating regime. In particular interest are the questions whether the pseudogap remains smoothly connected to the superconducting gap, how the temperature $T^*$, where it closes, evolves and if it exists also in samples, which do not show a full superconducting transition even in the $T=0$ limit. A related question is whether even weaker coupling pushes $\Delta$ to higher and higher values or if limiting mechanisms exist imposing a constraint on the shell-effect induced growth.  \\
Finally, another open question is whether similar effects can be found in nanograins of conventional superconductors other than Al. A cryogenic STM study on individual Sn nano particles showed drastic variations of the single-particle gap that were interpreted in terms of the shell effect \cite{bose10}, but since these nano particles were isolated, they did not involve a macroscopic superconducting condensate. However, previous work on Sn films \cite{Strongin1968}, though much less systematic than those on Al, suggests that similar phenomena can also enhance $T_c$ of Sn substantially beyond the bulk value of 3.7\,K.
\chapterend

\thispagestyle{empty}
\mbox{}

\chapter{Experimental studies on the Heavy Fermion metal {{CeCoIn$_5$}}}
\thispagestyle{empty}
\begin{flushright}
\footnotesize{
Bring den Vorschlaghammer mit\marginnote{\footnotesize{\textcolor{gray}{Bring a sledgehammer,\\ when you come over tonight, and\\we'll smash everything to pieces.}}}\\
Wenn du heute Abend kommst\\
Dann hauen wir alles kurz und klein \\[8pt]
\emph{Element of Crime}}
\end{flushright}
\emph{In this chapter, we  present measurements of the dynamical and transport conductivity of the heavy fermion (HF) metal CeCoIn$_5$ at THz energies. We start the discussion to the rich phenomenology of CeCoIn$_5$ in Sec.\,\ref{CCIintro} with an introduction to the the mid-$T$ Kondo- and low-$T$ Heavy-Fermion (HF) regimes, focusing on the experimental search for quantum-critical phenomena and the controversial discussion of their origin. Closing with look at previous IR studies, we then proceed to the discussion of the HF electrodynamics at THz energies. In Sec.\,\ref{Measurements of the dynamical and transport conductivity} we turn to the dc resistivity $\rho_\mathrm{dc}(T)$ and the dynamical response $\sigma_1+i\sigma_2$, we parametrize within the generalized Drude-model (GDM), Sec.\,\ref{Optics on correlated systems}. The GDM analysis reveals a strong $T$- and $\nu$-dependence of the complex resistivity $\rho=\sigma^{-1}$, and hence identifies the typical energy scale of the heavy quasiparticles (QP) at the studied $h\nu$ and $k_BT$ ranges. Section \ref{FLNFL} provides the ground for discussing the QP effective-mass enhancement $m^\ast/m_b$ and relaxation rate $\Gamma^\ast$ within Landau's Fermi-liquid (FL) theory, and makes a connection between the FL selfenergy $\Sigma(\nu,T)$ and the QP renormalization function $Z(\nu,T)$ on the one side and the memory function $M(\nu,T)$ of the GDM on the other. We introduce the hidden FL model which, as it turns out by close examination of $Z^{-1}$ and $\Gamma^\ast$ in Sec. \ref{hidden}, provides an interpretation of the optical response and previously reported thermodynamic non-FL properties. In Sec,\,\ref{scaling}, we study the scaling-behavior of $m^\ast(\nu,T)/m_b$ and finish with an outlook in Sec.\,\ref{CCIout}.    }

\clearpage{}

\section{A walk-through introduction to C\MakeLowercase{e}C\MakeLowercase{o}I\MakeLowercase{n}$_5$}\label{CCIintro}
A list set out in attempt to characterize the physics of CeCoIn$_5$ reads like a who-is-who of condensed-matter concepts whose importance to the field of modern solid state physics can hardly be overestimated. What is more, a notable amount of these concepts is essentially found on the simple fact that each Ce$^{3+}$ ion has a single 4$f$ electron with unpaired spin. Depending on thermal energy, the conduction electrons of CeCoIn$_5$ behave as a simple Drude metal, display the Kondo effect, hybridize with local moments and forming a coherent Heavy-Fermion (HF) state, and  eventually condense into an unconventional $d$-wave superfluid which, presumably, hides a quantum critical point at $T=0$ and contains a Fulde-Ferell-Larkin-Ovchinnikov state close to the upper critical fields \cite{Bianchi2003}. Retracing the $\rho_\mathrm{dc}(T)$ curve sketched in Fig.\,\ref{CCI_phase}, in what follows we will focus on the normal-conducting states above $T_c=2.3$\,K and concentrate on the HF regime where energies $k_BT$ and $h\nu$ are of the same order\sidenote{\footnotesize{A convenient (approximate) conversion between energies reads \begin{eqnarray}1\,\mathrm{K}&\hat{=}&100\,\mu\mathrm{eV}\nonumber\\&\hat{=}&1.5\,\mathrm{cm}^{-1}\nonumber\\&\hat{=}&20\,\mathrm{GHz}\nonumber\end{eqnarray}}} setting out the field for the experimental search for quantum critical behavior within this work.

\subsection{Metallic and Kondo regimes}\label{Metallic and Kondo scattering regimes}
\begin{figure}[h!]
\noindent \begin{centering}
\includegraphics[width=\textwidth]{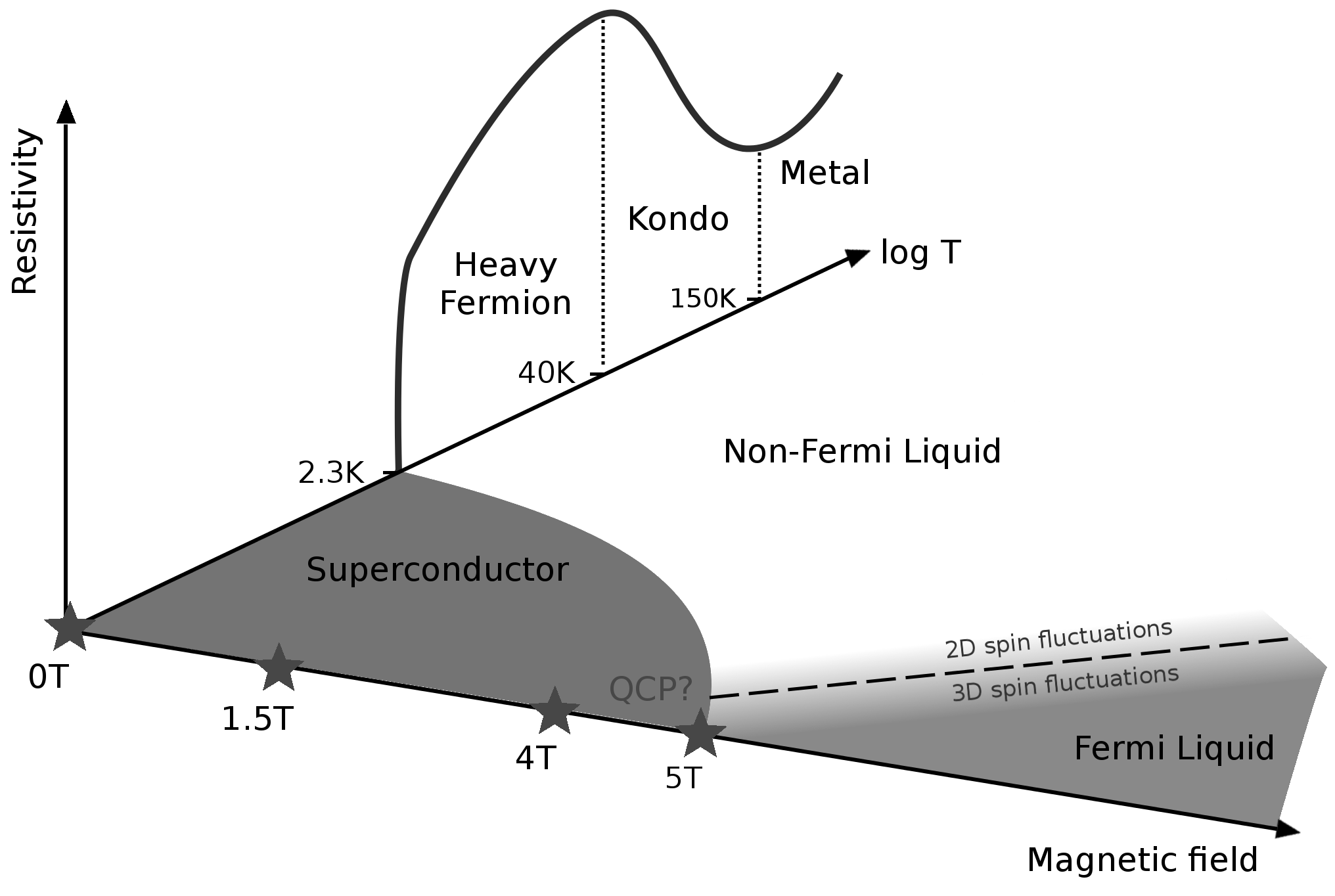}
\par\end{centering}
\caption{\label{CCI_phase}\textbf{Schematic phase diagram and dc resistivity of CeCoIn$_5$} as functions of temperature and magnetic field. Starting at room temperature, the system is a uncorrelated Drude-metal, crosses at $\sim$150\,K to the regime of Kondo scattering, before it enters the coherent heavy fermion state at $T_\mathrm{coh.}\approx40$\,K and eventually becomes superconducting below $T_c=2.3$\,K. As function of magnetic field, superconductivity is suppressed monotonically to zero at $H_{c2}=5$\,T and CeCoIn$_5$ turns into a FL at low temperatures. Above the SC dome and the dashed line demarcating a dimensional crossover 3D $\to$ 2D the system is a non-FL. The location of the QCP (contradicting proposals based on different experimental techniques are depicted as red stars) as well as the nature of the non-FL state are under heavy debate at the time of writing.  }
\end{figure}

Starting at room temperature, the dc transport resistivity of CeCoIn$_5$ (see the sketch in Fig.\,\ref{CCI_phase}) is reduced as temperature decreases exhibiting ordinary metallic properties. Here, $\rho_\mathrm{dc}$ is mainly governed by phonon scattering. This changes drastically at intermediate temperatures $\sim 150$\,K, where $\rho_\mathrm{dc}$ starts to rise again, i.e. a second scattering with an opposite temperature dependence as phonon appears. The anomalous behavior is attributed to an interaction between uncompensated spins of electrons localized to the Ce$^{3+}$ $4f$ shells\sidenote{\footnotesize{the notion of a atomic shell inside of a metal may irritate, yet it is a valid approximation due to the small volume of the atomic Ce 4$f$ shell leaving them basically intact even in a solid.}} and the conduction electrons giving rise to a certain scattering process contributing to $\rho_\mathrm{dc}$. This mechanism was suggested by Kondo \cite{Kondo1964} in attempt to describe the peculiar resistivity minimum of simple metals doped with magnetic ions. Using a diagrammatic approach, Kondo has shown that, if one goes beyond 1. order perturbation theory, the exchange scattering via an intermediate virtual state has leading to a log-$T$ growth of $\rho_\mathrm{dc}$ as temperature goes down. While this picture captures the resistance minimum, it also predicts a troublesome divergence in the $T\to 0$ limit. However, there is an elegant escape to this caveat: As temperature is reduced, the exchange interaction increases and tends to the formation of Kondo-singlets, i.e. a bound state of an conduction- and $f$ electron with antiparallel spins effectively removing the local moments from the system. Somewhat surprisingly, Kondo's ideas can also be applied to the dense lattice of 'impurities' in inter-metallic rare-earth compounds with unpaired 4$f$ or 5$f$ electrons interacting via a RKKY-type\sidenote{\footnotesize{originally formulated by Ruderman, Kittel, Kasuya, and Yosida for nuclear-spin exchange via the conduction sea.}} interaction \cite{Mott1974,Doniach1977}. In such systems, as for instance CeCoIn$_5$, the flattening crosses over to a massive reduction of resistivity at even lower temperatures, as schematically displayed in Fig.\,\ref{CCI_phase}. The reason lies in the antiferromagnetic interaction between the spins that tends to magnetic ordering at low temperatures. The emergent translational symmetry of the spin lattice ensures momentum conservation and elastic scattering (while the Kondo-scattering off disordered spin lattice or magnetic impurities does not conserve momentum and hence enhances resistivity). The massive reduction of resistivity due to coherent scattering is, however, challenged by a strongly enhanced effective mass of the conduction electrons coining the name heavy-fermion. The hybridization between the $f$ states and the conduction band leads to a band- or hybridization gap which, depending on the Fermi energy, leaves behind either a Kondo insulator or heavy-fermion metal as low-$T$ ground states. In case of the latter, we can imagine the emergence of 'heaviness' lying in the flatness of the band\sidenote{\footnotesize{Remember that the inverse curvature of the band defines the effective mass.}} or, more from the quantum mechanical point of view, in the indiscernibility of electrons: Before Kondo-singlet formation sets in, we deal with $f$-electron states which are strongly localized in space (and, from the transport perspective, can be though of having a infinitely high mass), and delocalized conduction states with a (light) band mass. Once having formed a Kondo singlet, the constituents loose their former identities to some extent, and, the $f$-electron has a finite probability to find itself in a conduction state and vice versa. Consequently, the $f$-electrons merge into the Fermi sea, however, at the expense of a heavy (yet not infinite) masses. Or, looking at the problem from the other side, the light conduction electrons acquire a finite probability to occupy a $f$-state and become localized letting them appear to slow down as due to an enhanced effective mass.

\subsection{Heavy-Fermion state and quantum criticality}\label{QCP}
The HF state in CeCoIn$_5$ has a number of properties, that cannot be explained within the framework of canonical FL theory: The electrical resistivity is linear in temperature \cite{Sidorov2002}, the electronic component of specific heat varies logarithmically as function of temperture \cite{Nakatsuji2002}, fractional $T$-powerlaws of the spin susceptibility \cite{Kim2001} and nuclear spin-lattice relaxation time \cite{Kawasaki2008}. CeCoIn$_5$ does not order magnetically down to lowest temperatures in contrast with the closely related isovalent CeRhIn$_5$ \cite{Sidorov2002}. This compound has an antiferromagnetic ground state below 3.8\,K  which can be suppressed by pressure terminated at a QCP separating the AFM state from an pressure-induced superconducting one. Sidorov \emph{et al.} noted that applying pressure to the Rh-compound is similar to replacing Rh by Co and, from the other perspective, CeCoIn$_5$ is located near an AFM instability reached by slightly negative pressure \cite{Sidorov2002}. While applying negative pressure to suppress superconductivity is more of a theoretical tuning knob, a more viable way are magnetic fields. At around $H_\mathrm{c2}=5$\,T, superconductivity ceases, however, not giving way to a magnetically ordered phase, but a FL at low and non-FL at elevated temperatures as schematically shown in Fig.\,\ref{CCI_phase}. From scaling behavior of the specific heat at $H_\mathrm{c2}$ and above as well as resistivity measurements, Bianchi \emph{et al.} concluded a field-induced QCP near $H_\mathrm{c2}$ \cite{Bianchi2003}. They further argued, that the anomalous non-FL properties are consistent with theoretical predictions for AFM spin fluctuations, yet a AFM ordering is less favorable than a superconducting ground state. The picture of an AFM QCP and the intimate relation between AFM fluctuations and non-FL behavior was later on substantiated by Ronning \emph{et al.} observing a logarithmically diverging specific heat and $T^2$-coefficient of the resistivity near $H_\mathrm{c2}$ \cite{Ronning2005}. Donath \emph{et al.} employed thermal-expansion measurements to discriminate between the conventional spin-density wave (SDW) and unconventional Kondo-breakdown (\emph{local} QCP) mechanisms characterized by 3D- or 2D-type fluctuations, respectively \cite{Donath2008}. The study revealed a dimensional crossover 3D $\to$ 2D with increasing temperature (dashed line in Fig.\,\ref{CCI_phase}) signaling a peculiar change in criticality, yet also supporting the view of an SDW mechanism and AFM QCP. This scenario for quantum criticality, however, is not free of contradictions: Hall effect \cite{Singh2007} and thermal-expansion studies \cite{Zaum2011} located the QCP at fields inside the superconducting dome at around 4\,K, resistivity measurements with the current applied along the $c$-axis at 1.5-3\,T \cite{Malinowski2005} and, oddly enough,  measurements of the Gr\"{u}neisen ratio as most precise probe for quantum criticality strongly suggest a zero-field QCP \cite{Tokiwa2013}. In the same way the location of the QCP is unclear, the reasoning concerning conventional or unconventional criticality becomes questionable and, as of yet, the nature of the quantum criticality in CeCoIn$_5$ remains an unsolved puzzle which to solve is apparently an experimental and intellectual challenge.

\subsection{A brief review of optical studies on C\MakeLowercase{e}C\MakeLowercase{o}I\MakeLowercase{n}$_5$}
\begin{figure}[h!]
\begin{centering}
\includegraphics[width=\textwidth]{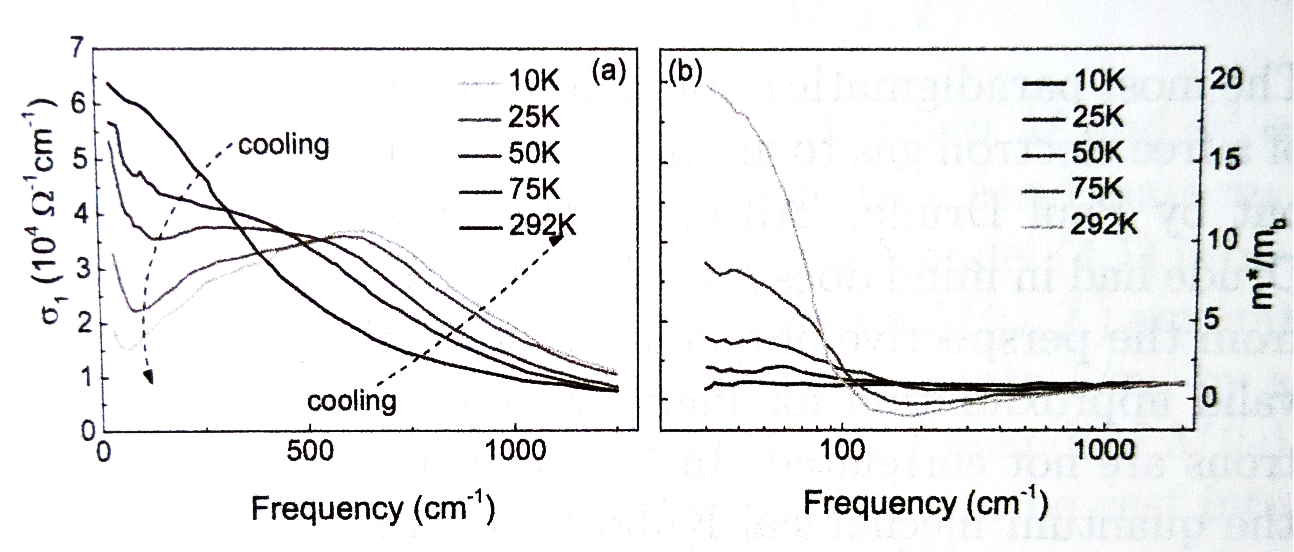}
\caption{\label{BasovIR}\textbf{IR properties of CeCoIn$_5$} obtained from single-crystal reflectance measurements. (a) $\sigma_1(\nu)$ for various temperatures. Below 50\,K a hybridization gap opens reflecting the hybridization of conduction and $f$ electrons. (b) Mass enhancement $m^\ast/m_b$ as function of frequency and temperature. Towards low energies, the IR light probes the heavy electrons as evident from a enhancement by an factor $\sim 20$ at lowest temperature. Data taken from Ref.\,\cite{singley2002}. }
\end{centering}
\end{figure}

At the time of writing, comparably little is known about the charge carrier dynamics of CeCoIn$_5$ in the normal state\sidenote{\footnotesize{Previous GHz and THz studies on CeCoIn$_5$ have explicitly addressed the superconducting state \cite{Ormeno2002a,Nevirkovets2008a,SudhakarRao2009,Truncik2013a}}} and how, if at all, the dynamical conductivity at finite temperatures can elucidate the zero-temperature QCP. Soon after the discovery of this compound, Singley \emph{et al.} measured the IR-reflectance between 30 and few 1000\,cm$^{-1}$ by means of IR-spectroscopy on single crystals \cite{singley2002}. At room temperature, $\sigma_1(\nu)$ shows the characteristic behavior of an uncorrelated Drude metal with a relaxation rate $\Gamma/2\pi$ at a few 100\,cm$^{-1}$, see Fig.\,\ref{BasovIR}(a). Below 100\,K, $\sigma_1(\nu)$ starts to develop a notch situated in the Drude roll-off that develops into a strong suppression as $T$ is reduced further. The appearence of the so-called hybridization gap can be seen as direct consequence of the hybridization between conduction and the 4$f$ electrons \cite{mena2005}. The essentially same result is found by Mena \emph{et al.} revisiting the IR-conductivity of CeCoIn$_5$ in the broader context of the Ce\emph{M}In$_5$ family, where $M$ is either Co, Rh, or Ir \cite{mena2005}: A hybridization gap $2\Delta(10\,K)\approx 600$\,cm$^{-1}$ and signatures of heavy electrons as narrow tail at low frequencies stemming from intraband transitions. Problematic with both approaches is the low-frequency limit of  30\,cm$^{-1}$. While Singley observes the tail of the heavy-electrons response only marginally and reconstruct its low-frequency extrapolation by spectral-weight arguments, direct\sidenote{\footnotesize{The presumably clearest demonstration of the narrow Drude roll-off due to \emph{slow} heavy fermions was measured on $U$-based HF metals \cite{Scheffler2013a, Scheffler2005,Ostertag2011,Dressel2002} combining microwave and THz spectroscopy}} optical evidence of the heavy electrons is completely missing from the spectra discussed in Ref.\,\cite{mena2005}. By means of extended-Drude analysis and spectral weight considerations, both works obtain the frequency dependence of the effective-mass ratio $m^\ast/m_B$, where $m_B$ is the (unknown) band mass, see Fig.\,\ref{BasovIR}(b). While $m^\ast$ does not show any frequency dependence above $\sim 40$\,K, it strongly increases below 100\,cm$^{-1}$ towards lower frequencies and temperatures inside the HF regime reaching values of 15-20 times the band mass. In a similar fashion, the electronic relaxation rate $\Gamma$ acquires a strong frequency dependence. \\
\noindent Regarding the open questions of quantum criticality and non-Fermi-liquid behavior in CeCoIn$_5$, the above works and IR-spectroscopy in general can only contribute little. With the HF state constrained between $T_c\approx 3$\,K and  $T_\mathrm{coh}\approx 40$\,K, the corresponding energy range to judiciously search for signs of quantum criticality is 0.25 - 3.5\,meV, or in units of frequency, 2 - 30\,cm$^{-1}$ just below the range of above IR studies. The lack of studies covering the quantum-critical regime  may strike surprising, as these ranges of a few ten Kelvin and the corresponding thermal frequencies $\hbar\omega\sim k_BT$ of a few ten wavenumbers can conveniently be covered with THz transmission spectroscopy, which however essentially depends on the availability of millimeter-sized high-quality films of not more than a few ten nanometer thickness. Only with the recent advances in the growth of large high-quality thin-film samples with sufficient chemical stability comprehensive studies of CeCoIn$_5$ in the quantum critical energy regimes have become feasible substantially advancing previous attempts \cite{Scheffler2013}.
\clearpage

\section{Interludium I: Optics on correlated systems}\label{Optics on correlated systems}
The most paradigmatic approach to describe the response of a free electron gas to an incident light field was worked out by Paul Drude. Although the microscopic picture Drude had in mind does not withstand thorough scrutiny from the perspective of quantum mechanics, it serves as valid approximation for many systems, where the electrons are not correlated. In fact, it can be derived from the fully quantum mechanical Kubo formula as free-electron limit \cite{Dre02}. The original Drude model predicts the dynamical conductivity to follow 
\begin{equation}
\sigma_D(\omega)=\frac{i\omega_p^2}{4\pi}\frac{1}{\omega-i/\tau_D}\label{DrudeEq}.
\end{equation}
Using this model, we assume a unique time scale $\tau_D$, on which carriers scatter, and a constant mass $m$ irrespective of the frequency they are probed with. Physically, this implies that the response of the system ($\sigma$) to an external perturbation at time $t$ does not depend on the systems history at times $t^\prime<t$. While this assumption certainly is a valid model for the free-electron gas in simple metals, omitting these constraints will lead to a generalized Drude model (GDM), whose applicability extends beyond those simple metals to the realm of correlated ones. To construct a GDM, G\"otze and W\"olfle \cite{Gotze1977} suggested to introduce a relaxation- or memory function defined as
\begin{equation}
M(\omega)=M_1(\omega)+iM_2(\omega)=\frac{\omega\chi(\omega)}{\chi_0-\chi(\omega)}\label{memory}
\end{equation}  
where $\chi_0=N/m$ and $\chi(\omega)$ is a correlation function, here explicitly for the current $j(t)$  
\begin{equation}
\chi(\omega)=-i\int\limits_0^\infty dt e^{i\omega t}\langle[j(t),j(0)]\rangle
\end{equation}
which is clearly non-local in time. Using the statistical-mechanics framework of Kubo \cite{Kub057}, the correlation function is related to the conductivity via \cite{Gotze1977}
\begin{eqnarray}
\sigma(\omega,T)&=&-i\frac{e^2}{\omega}\chi(\omega,T)+ i\frac{\omega_p^2}{4\pi\omega}\nonumber\\
&=&\frac{i\omega_p^2}{4\pi}\frac{1}{\omega+M(\omega,T)}\label{CondMem}
\end{eqnarray} 
where we have used Eq.\,(\ref{memory}) to link $\sigma$ and $M$ and $\omega_p^2=4\pi Ne^2/m_b$ contains the ordinary electron band mass $m_b$. The resemblance to the original Drude model (\ref{DrudeEq}) is obvious: introducing the memory function $M(\omega,T)$ amounts to replacing the single-valued relaxation rate $1/\tau_D$ by a complex and frequency dependent one. Expanding $M(\omega)$ into real and imaginary parts, Eq.(\ref{CondMem}) can be cast into the standard form of the GDM \cite{Dre02,BasovRev2011}
\begin{equation}
\sigma(\omega,T)=\frac{\omega_p^2}{4\pi}\frac{1}{-i\omega Z^{-1}(\omega,T)+\Gamma(\omega,T)} \label{GDM}
\end{equation}
where we defined the renormalization $Z^{-1}(\omega,T)\equiv 1+M_1(\omega,T)/\omega=m^\ast(\omega,T)/m_b$ \cite{Coleman2001} introducing the \emph{effective} mass $m^\ast$ and the (GDM) relaxation rate $\Gamma(\omega,T)=M_2(\omega,T)$. There are two remarkable things to note here. First, it is the frequency dependence of these two quantities that constitutes the 'generalized' description applicable to correlated electron systems beyond a simple Drude behavior. Second, the factor $\omega_p^2/4\pi$ has not changed during the above reasoning, i.e. it remains constant for a temperature independent carrier density $N$ and is inversely proportional to the \emph{band} rather than the \emph{effective} mass.
To access the functions $Z(\omega,T)$ and $\Gamma(\omega,T)$, we invert Eq.\,(\ref{GDM}) and define the complex dynamical resistivity $\rho(\omega)=\rho_1(\omega)+i\rho_2(\omega)=\sigma^{-1}(\omega)$ leading to \cite{Dre02,BasovRev2011}
\begin{eqnarray}
\rho_1(\omega,T)&=&\frac{\sigma_1(\omega,T)}{|\sigma(\omega,T)|^2}=\frac{4\pi}{\omega_p^2}\Gamma(\omega,T)\label{rho1}\\
-\rho_2(\omega,T)&=&\frac{\sigma_2(\omega,T)}{|\sigma(\omega,T)|^2}=\frac{4\pi\omega}{\omega_p^2}Z^{-1}(\omega,T)\label{rho2}
\end{eqnarray} 
Loosely speaking, the above identifications suggest to interpret $\rho_1$ as relaxation rate and $\rho_2/\omega$ as effective mass. The problem here is two-fold. First, without knowledge of $\omega_p$ these assignments are proportionalities instead of equalities. Second, it remains rather unclear, how these quantities should be interpreted in a physical context. In particular, it is not clear how to bring together the normal electrons in the Drude parameter $\omega_p$ and the apparently non-Drude frequency dependence. To unwind this conceptual problem, we rearrange  Eq.\,(\ref{GDM}) such that we obtain the same functional form of the original Drude model \cite{Dre02,BasovRev2011,deng2014} 
\begin{equation}
\sigma(\omega,T)=\frac{(\omega_p^\ast)^2}{4\pi}\frac{1}{-i\omega+\Gamma^\ast(\omega,T)} \label{QP}
\end{equation}
and define the starred \emph{quasiparticle} plasma frequency and relaxation rate 
\begin{eqnarray}
(\omega_p^\ast)^2&=&Z(\omega,T)\omega_p^2 \\
\Gamma^\ast(\omega,T)&=&Z(\omega,T)\Gamma(\omega,T)
\end{eqnarray}
The ingenious concept of quasiparticles (QP) as constituents of a conductor goes back to Landau's theory of Fermi-Liquids we will examine in more detail in Sec.\,\ref{FLNFL}.
To experimentally access the QP relaxation rate $\Gamma^\ast$ we invert Eq.\,(\ref{QP}) as before and obtain
\begin{eqnarray}
\rho_1(\omega,T)=\frac{\sigma_1(\omega,T)}{|\sigma(\omega,T)|^2}&=&\frac{4 \pi}{[\omega_p^\ast(\omega,T)]^2}\Gamma^\ast(\omega,T)\nonumber\\
&=&\frac{4 \pi}{\omega_p^2}Z^{-1}(\omega,T)\Gamma^\ast(\omega,T)\\
&& \label{help}
\end{eqnarray} 
where we have explicitly isolated the nontrivial frequency- and temperature dependence from $\omega_p^\ast$. To remove $\omega_p$ from the right side of Eq.\,(\ref{help}), we make usage of Eq.\,(\ref{rho2}) and obtain eventually
\begin{equation}
\Gamma^\ast(\omega,T)=\frac{\omega \sigma_1(\omega,T)}{\sigma_2(\omega,T)}\label{QPscat}.
\end{equation}
The importance of the above reasoning, especially the role of the renormalization $Z$, is hard to grasp at the first sight. In words, the QP relaxation rate $\Gamma^\ast$ equals to the one of the GDM-$\Gamma$ scaled with the renormalization factor. What rather appears as a semantic problem arising from different physical pictures, plays a substantial role for the question, whether or not we classify an electronic system as FL - a big deal for more elaborate interpretations in the context of strongly-correlated electron theories, as we will examine later in this work.

\section{Measurements of the transport and dynamical conductivity}\label{Measurements of the dynamical and transport conductivity}
\begin{figure}[b!]
\begin{centering}
\includegraphics[width= \textwidth]{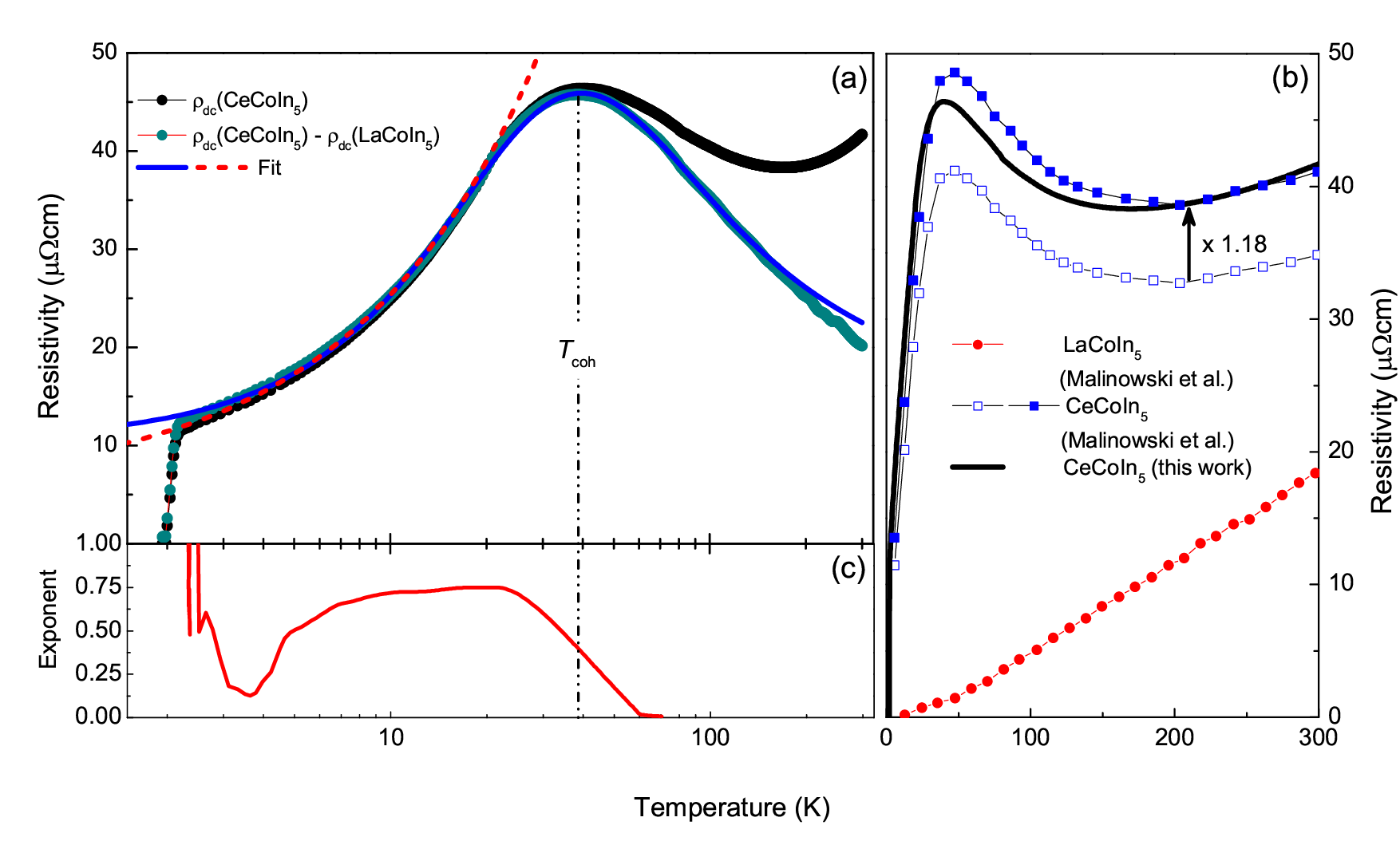}
\caption{\label{CCImagnRes}\textbf{Decomposition of transport resistivity in CeCoIn$_5$ } (a) total resistivity and magnetic contribution obtained after subtraction of the phononic contribution inferred from the non-magnetic metal LaCoIn$_5$ (data taken from Ref.\,\cite{Malinowski2005}) shown in panel (b). In order to meet measurements on thin films and single crystals, the resistivity of the latter were scaled with a factor 1.18. In the coherent HF regime, phonon scattering barely contributes to $\rho_\mathrm{dc}$ and thus can safely be neglected. The solid line in (a) is a fit that interpolates between the critical $T-$linear HF and the log$-T$ single-ion Kondo regimes. The dashed line is a simple power law $\rho_\mathrm{dc}\propto T^\alpha$. (c) Exponent $\alpha$ as function of the temperature, up to which the simple power-law was fitted.}
\end{centering}
\end{figure}

The samples under study are two 70\,nm thick films of CeCoIn$_5$ deposited via molecular beam epitaxy (MBE) \cite{Shimozawa2016} on a dielectric $10\times5\times0.5\,$\,mm$^3$ MgF$_2$ substrate both grown and measured in 2013 and 2015, respectively\sidenote{\footnotesize{Further information on sample growth and -characteristics are found in Sec. \ref{sampleCCI}}}. As the experimental outcome for both samples concerning the temperature and frequency dependencies is very similar, in what follows, we will discuss just one on behalf of both.
We start our discussion of the experimental results with the electrical transport resistivity $\rho_\mathrm{dc}(T)$ displayed in Fig.\,\ref{CCImagnRes}(a). The overall shape resembles the previously reported ones, see Panel (b) for a comparison to a measurement of Malinowski \cite{Malinowski2005}, clearly displaying the regimes of metallic behavior and incoherent Kondo-scattering, the coherent HF state, and, eventually, superconductivity. Before we can concentrate on the $\rho_\mathrm{dc}(T)$ of the HF state and, subsequently, examine the relaxation rate, we should quantify the phononic contribution to the resistivity.   

Generally, the low-$T$ behavior of $\rho_\mathrm{dc}(T)$ is expected to be governed by electronic scattering, yet it might be flawed by other processes such as phonon-scattering. Following Matthiessen´s rule\sidenote{\footnotesize{stating that impurity, phonon, electro-electron etc. relaxation rates linearly add up \begin{equation}\nonumber\tau^{-1}=\tau^{-1}_\mathrm{imp}+\tau^{-1}_\mathrm{Ph}+\tau^{-1}_\mathrm{ee}+...\end{equation}}}, it is sometimes possible to single out the electron-electron scattering by subtracting the phononic background inferred from a related compound where the considered scattering does not take place. In case of CeCoIn$_5$, for instance, the electronic scattering can be isolated by subtracting the phononic background of the non-magnetic LaCoIn$_5$ analogue. Here, we follow the procedure of Malinkowski \emph{et al.} and make use of their transport measurements \cite{Malinowski2005}: Figure\,\ref{CCImagnRes}(b) shows $\rho_\mathrm{dc}(T)$ of CeCoIn$_5$ studied in this work as well as of CeCoIn$_5$ and metallic LaCoIn$_5$ single crystals. Although the shape of CeCoIn$_5$ thin-film and single-crystal $\rho_\mathrm{dc}$ curves is slightly different and $T_c$ of the former is reduced by a few 100\,mK, we can match the high-temperature part, where phonon scattering is most dominant, by scaling of the single-crystal curve with a factor 1.18. Figure\,\ref{CCImagnRes}(a) displays, again, the CeCoIn$_5$ resistivity as measured and with the phononic contribution, i.e. the scaled LaCoIn$_5$ resistivity, subtracted. Above the coherent HF regime, the contribution from electron-electron scattering rapidly decays, while below barely no change upon phonon removal is found. Although comparing thin-film and single-crystal measurements should always be done with care\sidenote{\footnotesize{If in doubt, whether physics of an electronic system may substantially change when confined to thin films, the interested reader is invited to cast a glance at the previous chapters.}}, here we can safely regard the phonon-scattering contributions as subordinate in the $T$-range of interest and neglect them from now on.

In a previous work, Sidorov and Nakatsuji studied how the exponent $\alpha$ in $\rho_\mathrm{dc}(T)\propto \rho_0+AT^\alpha$ behaves as function of pressure and Ce$\to$ La substitution \cite{Sidorov2002,Nakatsuji2002}. For pure CeCoIn$_5$ without applied pressure, $\rho_\mathrm{dc}$ follows a $T$-linear behavior up $\sim$20\,K which is viewed as manifestation of NFL behavior and a first hint of a zero-temperature QCP in this compound. Instead of the previously reported $T$-linear behavior the sample shown in Fig.\,\ref{CCImagnRes} displays sub-linear behavior with an exponent $\sim0.75$ up to around 20\,K. One problem of this analysis is associated with the the mid-$T$ single-ion Kondo regime and the dominant log-$T$ behavior of $\rho_\mathrm{dc}$. Depending on which upper limit $T_\mathrm{max}$ is imposed to the fit, the obtained exponent $\alpha$ varies rather drastically. Figure \ref{CCImagnRes}(c) traces the powerlaw exponent $\alpha$ for $T_c<T_\mathrm{max}<300$\,K. Around $T_\mathrm{coh}$ we find $\alpha$ to be suppressed by the nearby Kondo regime resulting in a low value of $\sim$0.5. Below $\sim$20\,K the exponent tends to saturate to values between 0.7 and 0.75 before it drops again due to superconducting fluctuations. Irrespective of $T_\mathrm{max}$ we, on the one side, never reach the previously reported $T$-linear behavior. On the other side however, both the reduction of $T_c$ compared to the value found for single crystals, $T_c=2.3$\,K, and the anomalous sub-linear $T$ dependence was observed previously upon gradually replacing magnetic Ce$^{3+}$ with isovalent rare-earth (R) ions in Ce$_x$R$_{1-x}$CoIn$_5$ \cite{Paglione2007}. This effect was found to be strongest for non-magnetic Y$^{3+}$ substitution. Likewise, we attribute the sub-linear behavior and reduced $T_c$ to result from lattice defects which are common by-products of epitactical thin-film growth. 

A more elaborate way to cleanse the $T$-dependence of the non-FL from the one of the Kondo regime is an interpolation model treating both regimes on equal footings. Such an interpolation that crosses to marginal Fermi liquid \cite{Ruckenstein1991,Varma1993} towards low temperatues  has been suggested by W\"{o}lfle \cite{WolflePers} reading    
\begin{equation}\label{marginalcrossover}
\rho_\mathrm{dc}(T)=\rho_0+c_0\frac{T^{\alpha^\prime}}{\left(T^2-T^2_\mathrm{coh}\right)^{\alpha^{\prime}/2}}\left[\mathrm{ln}\left(\frac{T^2+T^2_\mathrm{coh}}{T^2_\mathrm{K}}\right)\right]^2
\end{equation}
where $T_\mathrm{K}$ is the Kondo temperature. A fit to this functional form is shown in Fig.\,\ref{CCImagnRes}(a) (blue line) giving $\alpha^\prime=1.08 -1.2$ depending on $T_\mathrm{max}$ nicely captures the resistivity peak and comes close to the previous linear-$T$ behavior \cite{Sidorov2002,Nakatsuji2002}. 
Having demonstrated the subordinate role of phonon-scattering in the relevant temperature range and the consistency with previous works, we now re-examine the commonly accepted NFL interpretation of CeCoIn$_5$ in the light of our optical measurements.\\

\begin{figure}[b!]
\begin{centering}
\includegraphics[width= \textwidth]{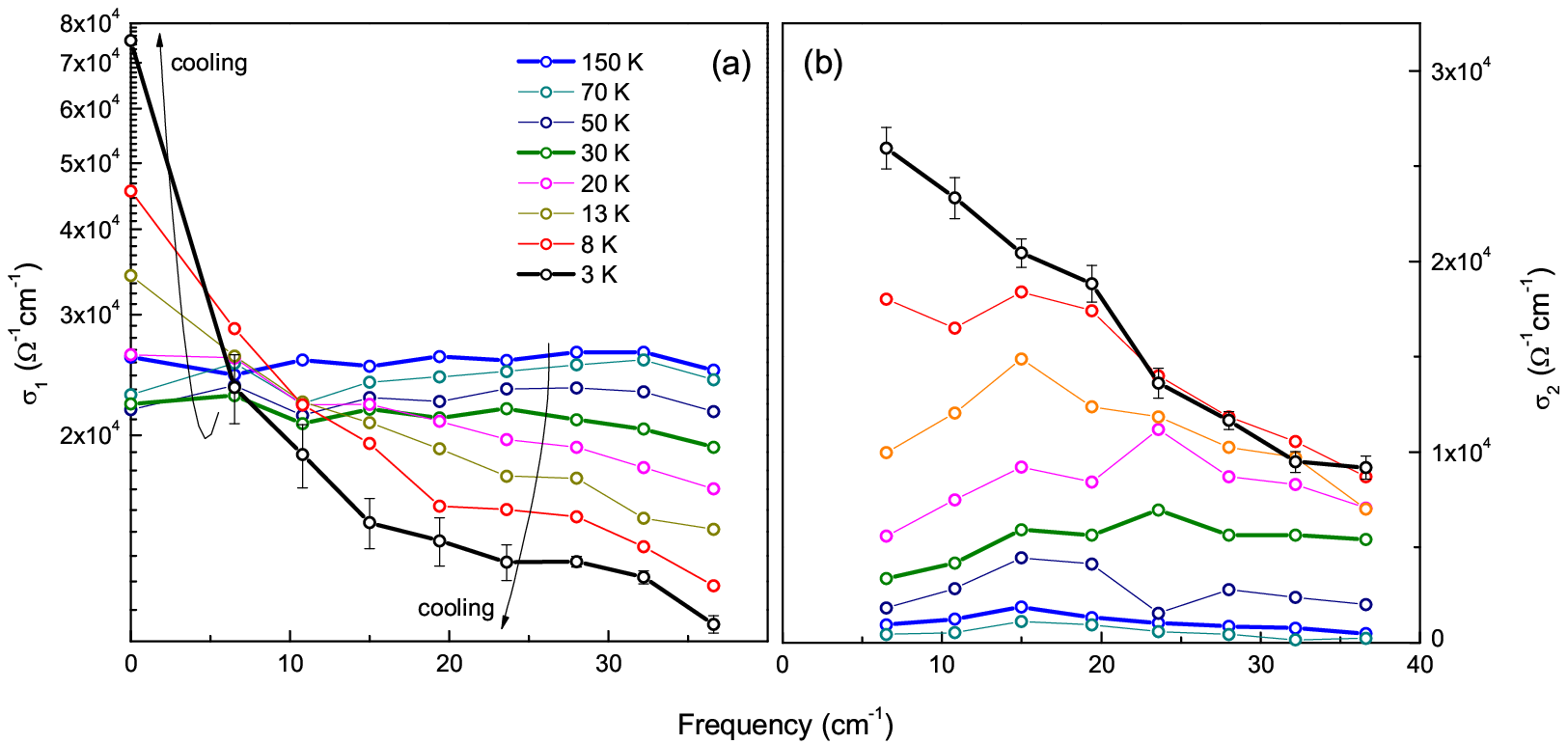}
\caption{\label{Fig2} \textbf{Complex conductivity $\sigma_1(\nu)+\mathrm{i}\sigma_2(\nu)$ of CeCoIn$_5$} at various temperatures. (a) Between 150 and 30\,K, $\sigma_1(\nu)$ is reduced slightly in agreement with the behavior of $\rho_\mathrm{dc}$ and shows a flat metallic behavior, which becomes strongly frequency-dependent below 30\,K. (b) $\sigma_2(\nu)$ acquires similar as $\sigma_1(\nu)$ a discernible frequency dependence only below 30\,K where the electrons become heavy. }
\end{centering}
\end{figure}
\begin{marginfigure}
\begin{centering}
\includegraphics[width=\marginparwidth]{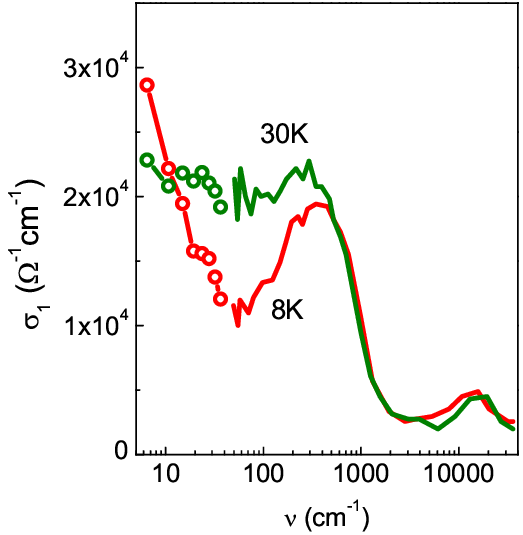}
\caption{\label{CCI_margin_sig1}Spectra of $\sigma_1$ measured at THz (this work) and IR frequencies taken from Ref.\,\cite{VdM2005} resolving the frequency dependencies over several orders of magnitude. The hybridization gap is clearly resolved in $\sigma_1(\nu)$ at 8\,K.}
\end{centering}
\end{marginfigure}

\begin{marginfigure}
\begin{centering}
\includegraphics[scale=0.3]{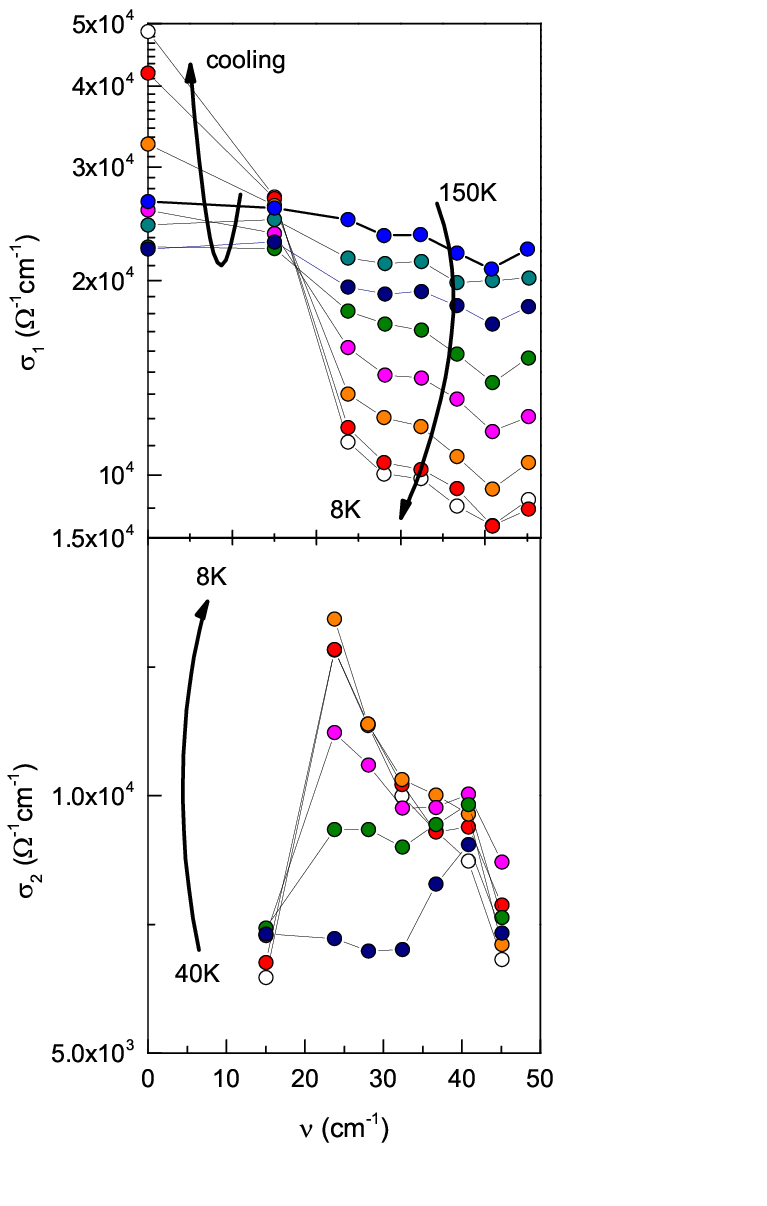}
\caption{\label{CCI_2nd}Spectra of $\sigma_{1,2}(\nu)$ of a second sample. The overall behavior agrees with the sample discussed in the main text. }
\end{centering}
\end{marginfigure}

\begin{figure}
\begin{centering}
\includegraphics[scale=0.35]{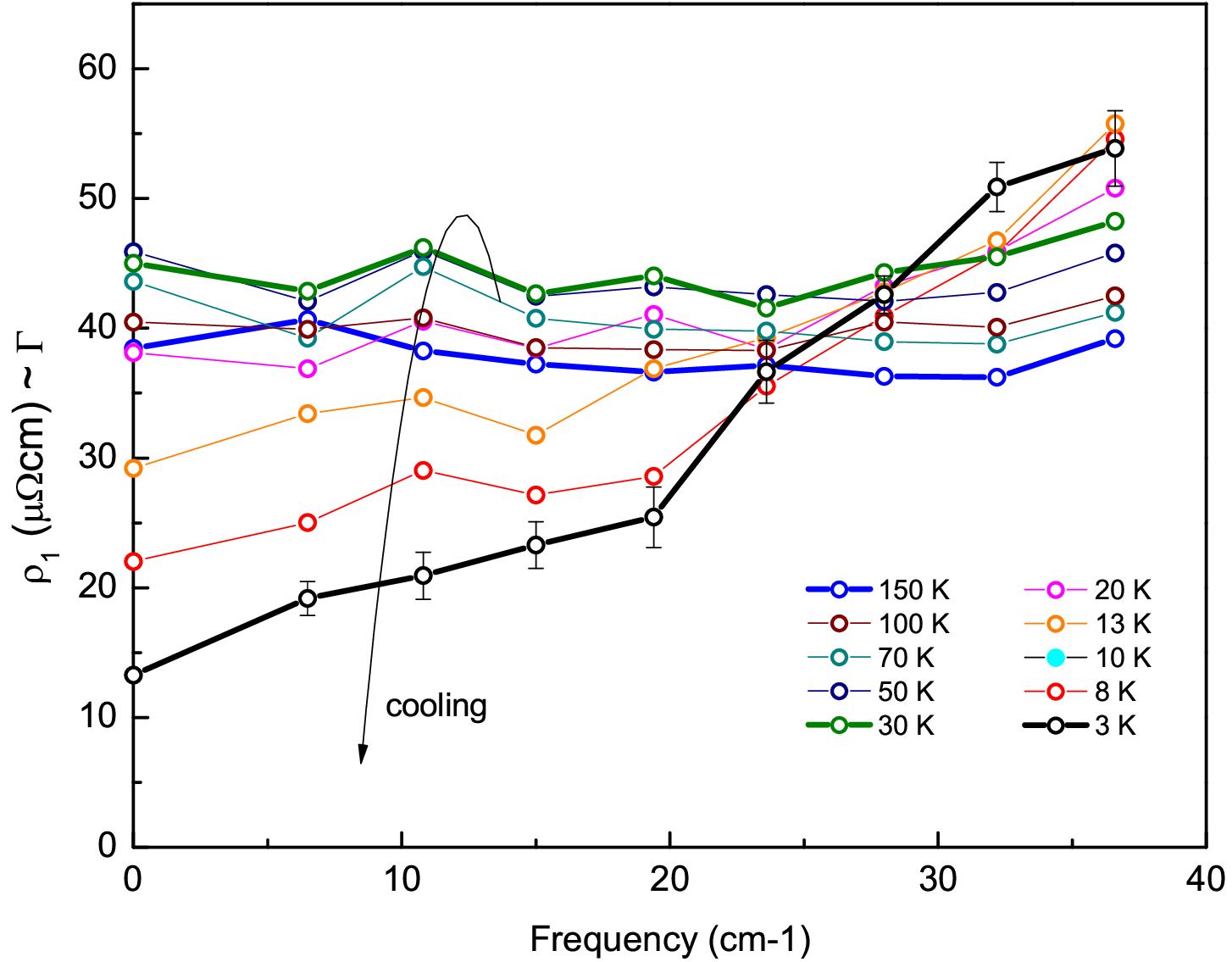}
\caption{\label{CCIFig3}\textbf{Real part $\rho_1(\nu)$ of the complex resistivity} at various temperatures as measure for the optical relaxation rate $\Gamma(\nu)$. Between 150 and 30\,K, $\Gamma$ is flat and acquires a frequency dependence only at lower temperatures in the HF state, where $\sigma_1(\nu)$ evolves a strong Drude peak.}. 
\end{centering}
\end{figure}

\begin{marginfigure}
\begin{centering}
\includegraphics[width=\marginparwidth]{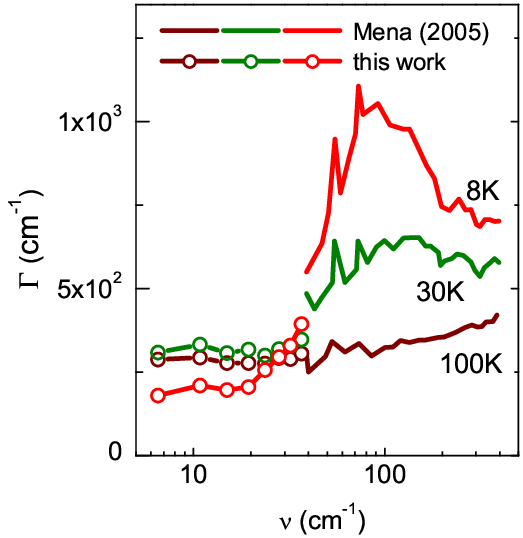}
\caption{\label{CCI_margin_rho1}Spectra of $\rho_s\propto \Gamma$ measured at THz (rescaled, this work) and IR frequencies taken from Ref.\,\cite{VdM2005} resolving the frequency dependencies over several orders of magnitude.}
\end{centering}
\end{marginfigure}

We now turn to the optical properties and display real and imaginary parts of the dynamical conductivity in Fig.~\ref{Fig2} at various temperatures (spectra for the second sample are displayed in Fig.\,\ref{CCI_2nd}). At the highest temperature 150\,K, the dissipative conductivity $\sigma_1(\nu)$ does not show any discernible frequency dependence as expected for a normal metal with the relaxation rate $\Gamma$ around 350\,cm$^{-1}$ at this temperature \cite{VdM2005}. Consequently, we also barely find a significant contribution to the out-of-phase conductivity $\sigma_2(\nu)$. Upon cooling to the maximum of   $\rho_\mathrm{dc}$ at around 35\,K, the flat metallic behavior of $\sigma_1(\nu)$ persists, the absolute values, however, shift to smaller values. 
This reduction amounts to $\sim15\%$ which is in excellent agreement with the increase in $\rho_\mathrm{dc}$. We attribute this reduction in $\sigma_1$ to the increased scattering between the conduction electrons and localized $f$-electrons. The values of $\sigma_2(\nu)$ increase slightly, but remain close to zero upon approaching $T_p$. Upon further reduction of temperature, the system crosses over to the coherent HF state which is accompanied by the opening of a hybridization gap in the IR regime \cite{singley2002,VdM2005} as displayed in Fig.~\ref{CCI_margin_sig1}. At THz frequencies, this gap causes a suppression of $\sigma_1(\nu)$ in the high-frequency limit, while $\sigma_1(\nu)$ tends to increase in the low-frequency limit following the suppression of $\rho_\mathrm{dc}$, phonon-, and electron-electron scattering. As the hybridization gap deepens, the decrease of $\sigma_1(\nu)$ with increasing frequency becomes stronger and covers nearly one order of magnitude at the lowest temperature (3\,K) in the spectral range studied. At the same time, $\sigma_2(\nu)$ also rises and acquires a frequency dependence shaping a broad peak that shifts towards lower frequencies and becomes more pronounced as temperature is reduced to 3\,K. \\
Figure \ref{CCIFig3} shows the optical relaxation rate $\Gamma$ as functions of temperature and frequency obtained from the GDM model via Eq.\,(\ref{rho1}). 
At 150\,K, $\Gamma$ does not vary with frequency within the studied spectral range, as expected for a simple metal without electron-electron interactions. As temperature is reduced to around 30\,K, $\Gamma$ increases, but remains flat. This rise is attributed to enhanced scattering between conduction and localized $f$-electrons in the Kondo regime. At lower temperatures, the HF state forms and we observe an emergent frequency-dependence of $\Gamma$. This can directly be seen as result of a increasing interaction between the carriers. We observe an increase of the frequency-dependence down to our lowest temperature. The overall trend is consistent with the flat dynamical conductivity with the Drude roll-off above the accessible THz range at high temperatures, and the opening of the hybridization gap and concomitant emergence of a narrow Drude peak of the heavy charge carriers. Towards higher frequencies, the THz data is smoothly connected to IR data \cite{VdM2005}, as depicted in Fig.\,\ref{CCI_margin_rho1}. 
\section{Interludium II\\ 
FL, non-FL, and hidden FL}\label{FLNFL} 
In his seminal work \cite{Landau56}, Landau showed theoretically that slowly turning on the interaction in an initially non-interacting Fermi gas of electrons will transform the old ground state adiabatically into the a new one of the Fermi \emph{Liquid} (FL) in the sense that the eigenstates of the former system remain intact, yet now acquire different energy. The occupied states will keep their quantum numbers (charge, spin, etc.), while the dynamic properties (mass, relaxation rate, magnetic moment, etc.) are renormalized. The great value of this one-to-one correspondence is that one can maintain certain descriptions valid in the non-interacting Fermi gas (such as the Drude formula) also for interacting systems. This elegant treatment, however, requires to abandon the notion of electrons (and holes) as entities carrying, e.g., the electrical current in favor of what is called Landau \emph{quasiparticles}\sidenote{\footnotesize{One should always keep in mind, that nature provides only one type of electron. In this sense, speaking of an quasiparticle-electron with, e.g., a mass $m\neq9.108...\times10^{-34}$\,kg can be misleading: it's a simple electron that \emph{appears to have} a mass other than $m_0$ to make the equations simpler.}} (QP). Due to the possibility of precise theoretical predictions for transport- and optical properties, FL theory has become one of the experimentally most comprehensively scrutinized models for the low temperature physics of correlated fermionic systems ranging from electrons in metals to $^3$He atoms above the superfluid phases. Optical spectroscopy and studies of the dynamical conductivity $\sigma(\omega)$ have turned out a particularly fertile testing ground for FL theory, as two of the central quantities of Landau QP, the renormalization $Z_\mathrm{qp}$  and the relaxation rate (inverse lifetime) $\Gamma^\ast_\mathrm{qp}$, can be accessed via the effective mass enhancement parametrization of $\sigma(\omega)$ within the GDM as we will see below.

Within FL theory, the energy contribution arising from the electron-electron interaction can be introduced in form of a self energy $\Sigma^\mathrm{FL}$, which at low energies and for $\mathrm{k}=0$ can be written\sidenote{\footnotesize{in units $\hbar=k_B=1$}} as as \cite{Berthod2013, Mravlje2016,Mravlje2011}
\begin{equation}
\label{FLselfenergy}
\Sigma^\mathrm{FL}(\omega,T)=\left(1-\frac{1}{Z_\mathrm{qp}}\right) \omega - \frac{i}{Z_\mathrm{qp}\pi T^\ast}\left[\omega^2+(\pi  T)^2\right]
\end{equation}
with $Z_\mathrm{qp}=(1-\partial\Sigma^\mathrm{FL}/\partial\omega)^{-1}$ the \emph{renormalization} factor and the coherence temperature $T^\ast\sim \epsilon_\mathrm{F}^\ast$, which for low energies uniquely sets energy scale for the physics of the Landau QP and which relates to the bare-electron Fermi energy $\epsilon_\mathrm{F}=Z_\mathrm{qp}^{-1}\epsilon_\mathrm{F}^\ast$. On general grounds, the QP relaxation rate is defined as 
\begin{equation}
\Gamma_\mathrm{qp}^\ast=-Z_\mathrm{qp}\mathrm{Im}\Sigma^\mathrm{FL}.
\end{equation}
Using the explicit form of the self energy, we obtain Landau's famous result\sidenote{\footnotesize{Note that the effect of impurity scattering and other scattering channels needs to be accounted for separately.}}
\begin{equation}
\Gamma_\mathrm{qp}^\ast(\omega,T)=\frac{1}{\pi T^\ast}\left[\omega^2+(\pi  T)^2\right]\label{FLscat}
\end{equation}
This can be compared to the optical relaxation rate $\Gamma^\ast$ obtained from the dynamical conductivity (see Sec.\,\ref{Optics on correlated systems})
\begin{eqnarray}
\Gamma^\ast(\omega,T)&=&\frac{M_2(\omega,T)}{1+M_1(\omega,T)/\omega}\\
&=&\frac{2}{3}\frac{1}{\pi T^\ast}\left[\omega^2+(2\pi  T)^2\right]\label{FLscatMemory}
\end{eqnarray}
based on the explicit form of the memory function for a FL \cite{Berthod2013}
\begin{equation}
M(\omega,T)=\left(\frac{1}{Z}-1\right)\omega+\frac{2}{3}\frac{i}{Z\pi T^\ast}\left[\omega^2+(2\pi  T)^2\right].
\end{equation}
displaying the same energy dependence as the QP relaxation rate $\Gamma^\ast_\mathrm{qp}$.
The connection between the GDM quantities $\Gamma^\ast$ and $Z^{-1}=1+M_1(\omega,T)/\omega$ and the Landau QP quantities $\Gamma _{qp}^{\ast }$
and $Z^{-1}_{qp}$ is only approximately known. As discussed by P. Allen \cite{Allen2015}, one has 

\begin{eqnarray}
\sigma (\omega ) &=&\frac{i\omega _{p}^{2}}{4\pi \omega }\int_{-\infty
}^{\infty }d\omega ^{\prime }\frac{f(\omega ^{\prime })-f(\omega ^{\prime
}+\omega )}{\omega -\Sigma _\mathrm{tr}(\omega ^{\prime }+\omega +i0)+\Sigma
_\mathrm{tr}(\omega ^{\prime }-i0)} \nonumber\\
&\approx &\frac{\omega _{p}^{2}}{4\pi \omega }\frac{1}{-i\omega /Z(\omega )+2%
\mathrm{Im}\Sigma _\mathrm{tr}(\omega +i0)}
\end{eqnarray}
where $Z^{-1}(\omega )=1-\mathrm{Re}[\Sigma _\mathrm{tr}(\omega )-\Sigma_\mathrm{tr}(0)]/\omega $. Here $\Sigma _\mathrm{tr}$ differs from the single particle self-energy $\Sigma^\mathrm{FL} (\mathbf{k},\omega )$ in two ways \cite{WolflePers}: (1) it is an average over the Fermi surface; (2) the internal momentum summations are weighted according to the effectiveness of a collision process in relaxing momentum. This weight function is $\propto q^{2}\propto 1-\cos \theta $, where $\mathbf{q}\ $is the momentum transferred in the process and $\theta $ is the scattering angle. It follows that $\Gamma ^{\ast }=2r_{1}\Gamma _\mathrm{qp}^{\ast }$ , and $Z^{-1}=r_{2}Z_\mathrm{qp}^{-1}$, where $r_{1,2}\approx \langle q^{2}\rangle /k_{F}^{2}$ are constants reflecting the fact that current dissipation involves finite transfer of momentum $q$ in QP scattering processes. On general grounds, it can take values from $r_{1,2}\ll1$ (predominantly forward scattering) to $r_{1,2}\approx 2$ (predominantly backward scattering). We will not dwell on a model description of the angle dependence of QP collisions, which would be required to estimate the parameter $r_{1,2}$, but will assume that $r_{1,2}\approx 1$ as one can expect for a system with AFM fluctuations \cite{WolflePers}.
In what follows, we will ignore the factor 2  when discussing the energy dependence of $\Gamma^\ast(\omega,T)$ (i.e. we set $\Gamma^\ast_\mathrm{qp}=\Gamma^\ast$) for the sake of simplicity.  Also, we will henceforth make no distinction between $Z^{-1}$, $Z^{-1}_\mathrm{qp}$, and $m^\ast/m_b$.

If scattering channels other than the one above can be excluded, the quadratic $T$- and $\omega$-dependence of $\Gamma^\ast$ is arguably the most fundamental signature of a FL. The conceptual simplicity of this criterion is contrasted with the challenge to unambiguously verify Eq.\,(\ref{FLscat}) it in an experiment. The main reason is that well-defined Landau QP exist only at temperatures extremely small compared to other electronic scales such as the bandwidth. Consequently, canonical FL behavior Eq.\,(\ref{FLscat}) appears - if at all - typically just at the scale of a few Kelvin and below. The demonstration of scattering according to Eq.\,(\ref{FLscat}) is therefore not only limited to low energies but also to clean metals (no impurities). Meeting these requirements in an optical experiment, however, is very challenging due to the high reflectivity of bulk samples or low transmittivity of thin films complicated by the notorious problems of optical spectroscopy at cryogenic conditions. Not surprising, it took almost 60 years to unambiguously verify Eq.\,(\ref{FLscat}) for the case of Sr$_2$RuO$_4$ by means of IR spectroscopy \cite{Stricker2014}. 
Without doubt, measuring the $\omega$ dependence of $\Gamma^\ast$ is significantly more complicated than the $T^2$-dependence, which can be inferred from a simple resistivity measurement as $\rho_\mathrm{dc}(T)$ is given by the imaginary part of the self energy \cite{Jacko2009}
\begin{equation}
\rho_\mathrm{dc}(T)\propto -\mathrm{Im}\Sigma^\mathrm{FL}(0,T)=\frac{(\pi  T)^2}{Z\pi T^\ast}
\end{equation}
and with Eq.\,(\ref{FLscat}) fully reflecting the $T$- dependence the QP relaxation rate
\begin{equation}
\Gamma^\ast(T)\propto Z\rho_\mathrm{dc}(T)\label{jelp}
\end{equation}
as $Z$ does not depend on $T$ (and $\omega$) for a FL. Instead of testing Eq.\,(\ref{FLscat}), a more accessible (yet less precise) criterion reads 
\begin{equation}\label{FLcrit}
\rho_\mathrm{dc}\propto   
\begin{cases}
    T^\alpha \text{ with }\alpha=2 \leftrightarrow \text{ FL} \\
    \text{else} \leftrightarrow \text{ non-FL } 
\end{cases}
\end{equation}  
This criterion is actually so well-established, that \emph{linear-T} has become a set phrase to characterize, e.g., the strange-metal phase of high-$T_c$ cuprates \cite{Phillips2011}. HF systems often show deviations of the $\rho_\mathrm{dc}\propto T^2$ behavior, although a consistent Fermi-liquid theory for HF systems can be derived from the Kondo-lattice model \cite{Auerbach86}. Similar as for cuprates, the non-FL is typically regarded a finite temperature effect stemming from magnetic instabilities at a zero-temperature QCP \cite{Prasanta2008}.

On the one side, it is clear that the quadratic growth of Eq.\,(\ref{FLscat}) levels off once the mean free path falls below the inter-atomic distance\sidenote{\footnotesize{This criterion is commonly expressed in terms of the Mott-Ioffe-Regel parameter $k_F\ell=1$, whereas conductors require a value above unity.}}. The crossover from a strange to a bad metal at a temperature $T_\mathrm{MIR}$ ultimately sets the seal on well-defined Landau QP \cite{Emery1995}, and thus, signals the ultimate breakdown of FL behavior. On the other side, however, the temperatures at which the bad metal regime is entered are typically very high \cite{Gunnarsson2003,Hussey2004}: In case of Sr$_2$RuO$_4$, the aforementioned prototypical FL system, $T_\mathrm{MIR}\approx 800$\,K while FL behavior according to (\ref{FLcrit}) is observed up to $T_\mathrm{FL}\approx 20$\,K \cite{Deng2013}. Motivated by the huge discrepancy between $T_\mathrm{MIR}$ and $T_\mathrm{FL}$ in Sr$_2$RuO$_4$ and similar compounds, Deng \emph{et al.} performed DMFT calculations on generic hole-doped Mott insulators resolving clear QP excitations surviving up to temperatures $T\approx T_\mathrm{MIR}$ \cite{Deng2013}. Following this idea, Xu \emph{et al.} demonstrated that the relaxation rate of these \emph{resilient} QP obeys the $T$-square behavior up to temperatures much higher than $T_\mathrm{FL}$, whereas $\rho_\mathrm{dc}(T)$ deviates from a parabola above $T_\mathrm{FL}$ \cite{Deng2013}. This seemingly paradox situation unwinds as $\rho_\mathrm{dc}(T>T_\mathrm{FL})$ no longer reflects $\Gamma^\ast(T)$ according to Eq.\,(\ref{jelp}), because $Z$ becomes a function of $T$. Showing a FL-like relaxation rate yet anomalous transport coefficients as due to the $T$-dependence of $Z$, these resilient QP form a quantum liquid tellingly referred to as \emph{hidden} FL \cite{Xu2013,deng2014,Anderson2008,Casey2011}. Indeed, in case of the prototypical correlated-electron system V$_2$O$_3$, where the DMFT picture of the Mott insulator is known to provide an accurate picture, such a hidden FL was unraveled from optical spectroscopy using the $\omega\to 0$ limits in the quasiparticle-interpretation of the GDM \cite{deng2014}. Furthermore, also for the Hund's metal CaRuO$_3$ the QP relaxation rate follows a $T^2$ behavior up to several 10\,K well above $T_\mathrm{FL}=1.5$\,K (see the Supplementary Material of Ref.\,\cite{deng2014}) in support of the hidden FL scenario to hold beyond the DMFT picture of the doped Mott insulator. This offers a fascinating new perspective on interacting electron systems and arises a far-reaching question: Can we understand some the numerous non-FL systems as hidden FL\sidenote{\footnotesize{To avoid confusion we stress that a hidden FL composed from resilient QP is actually a special kind of non-FL and should clearly be distinguished from the canonical low-$T$ FL composed from Landau QP.}}? In the remainder of this chapter, we will discuss the case of the HF metal CeCoIn$_5$ and, based on new measurements of the dynamical response and reconsideration of earlier experimental studies, answer this question affirmatively.   
\section{C\MakeLowercase{e}C\MakeLowercase{o}I\MakeLowercase{n}$_5$  - a hidden Fermi Liquid}\label{hidden}
The experimental status quo leaves little doubt that the HF state of CeCoIn$_5$ is not a canonical FL in zero and moderate magnetic fields $<5$\,T \cite{Nakatsuji2002,Kim2001,Kawasaki2008,Paglione2003,Bianchi2003,Donath2008,Ronning2005,Zaum2011,Sidorov2002}. While the breakdown of canonical FL behavior is often intimately related to the vicinity to a quantum critical point, the experimental situation could, as of yet, not provide an answer to the fundamental question: if it is apparently not a FL, what kind of quantum liquid is it then? Despite several attempts to explain the anomalous non-FL properties by quantum-critical fluctuations, no consistent picture has been reached to date as discussed in Sec.\,\ref{QCP}. Significantly, the urgent questions concerning the nature of the QCP as well as its location in the phase diagram are just two pieces of the entire CeCoIn$_5$ puzzle. Following the reasoning we outlined in Sec.\,\ref{FLNFL}, in what follows, we will examine the energy dependence of renormalization in CeCoIn$_5$ and demonstrate that the experimental anomalies find a ready explanation in the context of resilient QPs and a hidden FL. 
\begin{figure}[b!]
\begin{centering}
\includegraphics[scale=0.6]{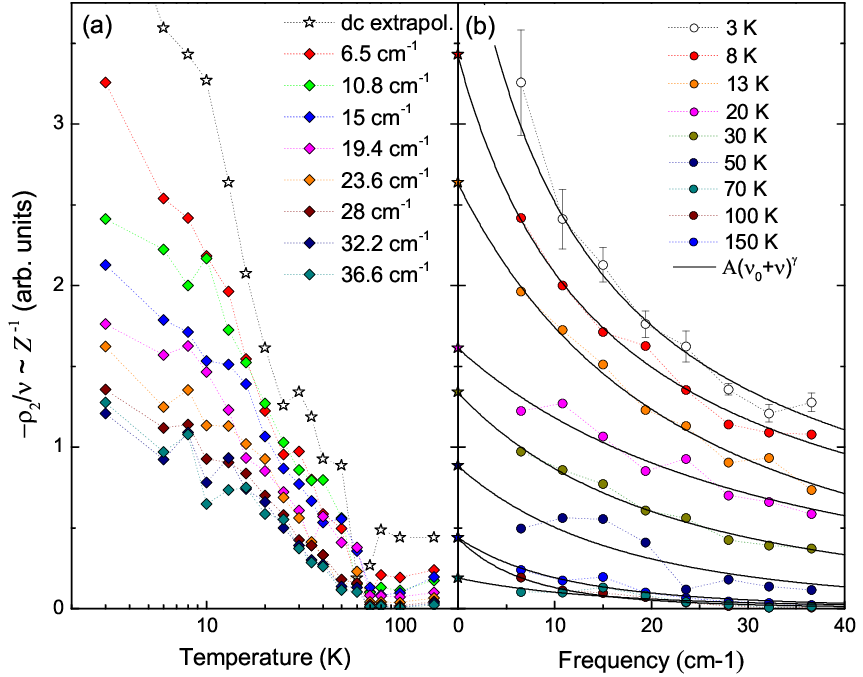} 
\caption{\label{3D_mass_mathematica}\textbf{Effective mass enhancement $m^\ast/m_b=Z^{-1}$} as measured by $-\rho_2/\nu$ versus (a) temperature and (b) frequency. Zero-frequency extrapolations $Z^{-1}(0)$ (stars) are estimated from fits to $A\left[\nu_0-\nu\right]^{-\gamma}$ via $A,\nu_0$, and $\gamma$ (solid lines). A strong energy dependence sets in at the crossover between Kondo and HF regimes, while absent at higher temperatures.}
\end{centering}
\end{figure}

Figure \ref{3D_mass_mathematica} displays $-\rho_2(\nu,T)/\nu$ which we interpret as measure of the renormalization $Z^{-1}=m^\ast/m_b$. The absence of frequency dependence of $\sigma_1(\nu)$ at high temperatures implies that the real part of $M(\nu)$ is essentially zero for $\nu\ll 350$\,cm$^{-1}$ (compare also to Fig.\,\ref{CCI_margin_sig1}) and therefore $Z=1$. We may then extract an approximate value for the plasma frequency 
\begin{equation}
\omega_p/2\pi=|2\pi c \nu/\epsilon_0 \rho_2(150\,\mathrm{K},\nu)|^{1/2}\approx 900\pm400\,\mathrm{THz}\nonumber
\end{equation}
which is about $3.7\pm1.6$\,eV. The approximate nature of this value in mind, however, we restrict our discussion to the energy dependence of the mass enhancement rather than $m^\ast(\nu,T)$ itself. With decreasing temperature and frequency, $Z^{-1}$ rises strongly: at the lowest temperature by a factor of $\sim$3 within the studied spectral range, and by a factor of 12 at the lowest frequency between 3\,K and the Kondo regime. Using the phenomenological ansatz $A\left[\nu_0-\nu\right]^{-\gamma}$ with $A,\nu_0$, and $\gamma$ being free parameters to model $-\rho_2(\nu)/\nu$ we can extrapolate the experimental to the transport limit. Figure \ref{3D_mass_mathematica} displays representative fits (red solid line) and the corresponding intersections with the dc-plane (red dots). The first important conclusion we can draw is the affirmation of CeCoIn$_5$ not being a canonical FL where $Z$ is a constant with respect to $T$ and $\nu$. The second one is that $Z(T,\nu)$ will have a notable impact on the renormalized relaxation rate $\Gamma^\ast=Z\Gamma$, which may help to elucidate the anomalous behavior of $\rho_\mathrm{dc}(T)$.\\ \\
\begin{figure}
\begin{centering}
\includegraphics[width=\textwidth]{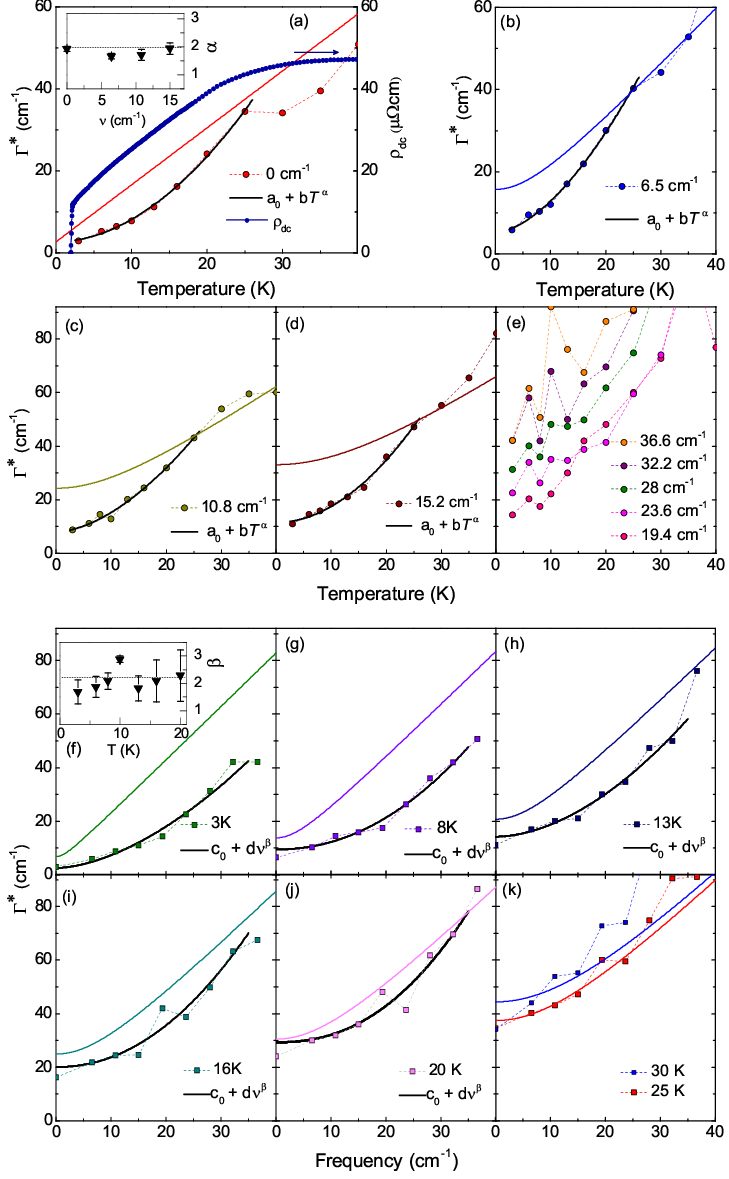}
\caption{\label{Fig3}QP relaxation rate $\Gamma^\ast$ versus temperature and frequency. Thick lines are powerlaw fits, thin lines demarcate the critical relaxation rate $\Gamma_\mathrm{crit.}$, only below which well defined QP exist. (a) While $\rho_\mathrm{dc}$ displays a non-FL sub-linear $T$-dependence, the relaxation rate of well-defined resilient QP $\Gamma^\ast<\Gamma_\mathrm{crit.}$ approximately follows the $T^2$- powerlaw of a hidden FL up to 25\,K. (b-d) $\Gamma^\ast(T)$ displays powerlaw behavior with exponents $\alpha\approx 2$, see inset in (a) up to 25\,K, where $\Gamma^\ast\geq\Gamma_\mathrm{crit.}$. (e) For $\nu>15.2$\,cm$^{-1}$, $\Gamma^\ast(T)$ no longer displays powerlaw behavior. (f-j) $\Gamma^\ast(\nu)$ at selected temperatures. Powerlaw fits yield exponents $\beta\approx 2$, see inset of (f), up $T=20$\,K, above which $\Gamma^\ast\geq \Gamma_\mathrm{crit}$ is found, see panel (k), and QP are no longer well-defined.}
\end{centering}
\end{figure}
Figure \ref{Fig3} displays the QP relaxation rate\sidenote{\footnotesize{We obtain absolute numbers for $\Gamma^\ast$ using Eq.\,(\ref{QPscat}) as the (unknown) $\omega_p$ cancels out. This also holds for the transport limit
\begin{eqnarray}
\rho_\mathrm{dc}&=&\frac{4\pi}{\omega_p^2}\Gamma_\mathrm{dc}\nonumber\\
Z_\mathrm{dc}\rho_\mathrm{dc}&=&\frac{4\pi}{\omega_p^2}Z_\mathrm{dc}\Gamma_\mathrm{dc}\nonumber\\
\frac{\rho_\mathrm{dc}}{\frac{4\pi}{\omega_p^2}\frac{m^\ast}{m_b}}&=&\Gamma_\mathrm{dc}^\ast\nonumber
\end{eqnarray}

where we apparently normalize $\rho_\mathrm{dc}$ to $(4\pi/\omega_p^2)\frac{m^\ast}{m_b}$ (rather than just $m^\ast/m_b$) - the quantity we obtain from the measurement via $\sigma_2/(\omega|\sigma|^2)$ }} $\Gamma^\ast(\nu,T)$ (including the energy independent impurity-scattering contribution) calculated using Eq.\,(\ref{QPscat}) at finite frequencies and via $\rho_\mathrm{dc}(T)Z(\nu\to 0, T)$ employing the dc-extrapolations. Starting with the dc limit (panel (a)), we observe a strong rise that due to the temperature dependence of $Z^{-1}$ does clearly not reflect the sub-linear $T$-dependence of $\rho_\mathrm{dc}(T)$. To the contrary, fitting a powerlaw $\Gamma^\ast(T<25\,\mathrm{K})\propto T^\alpha$ yields an exponent $\alpha=1.92\pm0.09$ close to the quadratic behavior of a genuine FL. The powerlaw behavior of $\Gamma^\ast(T)$ is also observed at finite frequencies $\nu\leq15$\,cm$^{-1}$ ($\hat{=}\,22\,K$), see panels (b-d), with exponents slightly smaller than $\alpha=2$ as shown in the inset of panel (a). At higher frequencies, panel (e), the Kondo scattering adds a non-trivial contribution to $\Gamma^\ast$ spoiling a simple powerlaw behavior. Looking at the $\nu$-dependence, see panels (f-j), we also observe powerlaw behavior $\Gamma^\ast(\nu)\propto \nu^\beta$ with exponents $\beta$ slightly smaller than 2, see the inset of panel (f), up to around 20\,K. At higher temperatures, see panel (k), fits do no longer identify reasonable powerlaw behavior. This upper limit is in good agreement with the temperature $T=25$\,K up to which $\Gamma^\ast(T)$ shows approximate $T^2$-behavior, see panel (c). \\

While the experimentally obtained exponents are sufficiently close to 2 for a reasonable interpretation within the hidden FL scenario, W\"{o}lfle has demonstrated \cite{Pracht2017} that values slightly \emph{smaller} than 2 can actually be attributed to the nature of resilient QP on a microscopic level: On general grounds, the rate at which two QP scatter from initial states with energies $\omega_{1,2}$ and momenta $\mathbf{k}_{1,2}$ into states with $\omega_{3,4}$ and $\mathbf{k}_{3,4}$ reads (at zero temperature)   
\begin{eqnarray}
\Gamma_\mathrm{qp}^{\ast }(\omega ,\mathbf{k;}T) &=&\sum_{\mathbf{k}_{2},\mathbf{k}_{3},
\mathbf{k}_{4}}\iiint \frac{\mathrm{d}\omega _{2}\mathrm{d}\omega _{3}\mathrm{d}\omega _{4}}{(2\pi)^3}
\Big[|U_{1,2;3,4}|^{2}\nonumber\\
&&\times A_{\mathbf{k}_{2}}(\omega _{2})A_{\mathbf{k}_{3}}(\omega _{3})A_{\mathbf{k}_{4}}(\omega _{4}) \nonumber\\
&&\times \pi \delta (\omega _{1}+\omega _{2}-\omega _{3}-\omega _{4})\nonumber\\
&& \times \delta (\mathbf{k}+\mathbf{k}_{2}-\mathbf{k}_{3}-\mathbf{k}_{4})\nonumber\\
&&\times \lbrack f_{2}(1-f_{3})(1-f_{4})+(1-f_{2})f_{3}f_{4}]\Big]\nonumber\\
&&
\end{eqnarray}
where $A_{\mathbf{k}}(\omega )=2\pi \delta (\omega -\epsilon _{\mathbf{k}}^{\ast })$ is the QP spectral function, $\epsilon _{\mathbf{k}}^{\ast }=Z_\mathrm{qp}\epsilon _{\mathbf{k}}$ is the renormalized Fermi energy,  $f_{2}=f(\omega _{2})$, etc. is the Fermi function, and $U_{1,2;3,4}$ is the interaction matrix element for scattering QP from $1,2$ into states $3,4$. After a sequence of simplifications the integrations can be carried out and cast into \cite{Pracht2017} 
\begin{equation}
\Gamma_\mathrm{qp}^{\ast }(\omega)\approx c(\omega)\frac{\omega^2}{\epsilon _{F}}
\end{equation}
where $N_{0}$ is the bare density of states at the Fermi level and $\epsilon_{F}$ is the bare Fermi energy and
\begin{equation}
c(\omega)\approx \frac{1}{\omega^{2}}\int_{0}^{\omega}d\epsilon
_{2}^{\ast }\int_{0}^{\epsilon _{2}^{\ast }}d\epsilon _{3}^{\ast }\frac{%
\epsilon _{F}}{\epsilon _{234}^{\ast }}|N_{0}U|^{2}
\end{equation}%
where $\epsilon _{234}^{\ast }$ is a linear function of $\epsilon _{2}^{\ast },\epsilon _{3}^{\ast },\epsilon _{4}^{\ast }$. At finite temperatures $T>0$, the integrals over the Fermi functions yield a slightly different result and we need to replace $\omega ^{2}\rightarrow \omega ^{2}+(\pi T)^{2}$ in above equations. In the FL regime, when $Z^{-1}=m^{\ast }/m$ is independent of energy, we get $c(\omega )=c_{FL}\propto \epsilon _{F}/\epsilon _{F}^{\ast }\propto m^{\ast}/m$, where $\epsilon _{F}^{\ast }$ is the renormalized Fermi energy (see Refs.\,\cite{Baym2004,Vollhardt2013}) and consequently restore the exact quadratic energy dependence of the relaxation rate. In the regime of resilient QP, the interaction $U$\ remains a weak function of energy, whereas integration over $\epsilon _{F}/\epsilon_{234}^{\ast }$  is approximately constant. In this way, W\"{o}lfle concluded $c(\omega)$ to be a slowly decreasing function of $\omega,T$ for resilient QP which serves as correction suppressing the powerlaw exponents slightly below 2 in agreement with the experimental observation.\\
\begin{figure}[b!]
\begin{centering}
\includegraphics[scale=0.45]{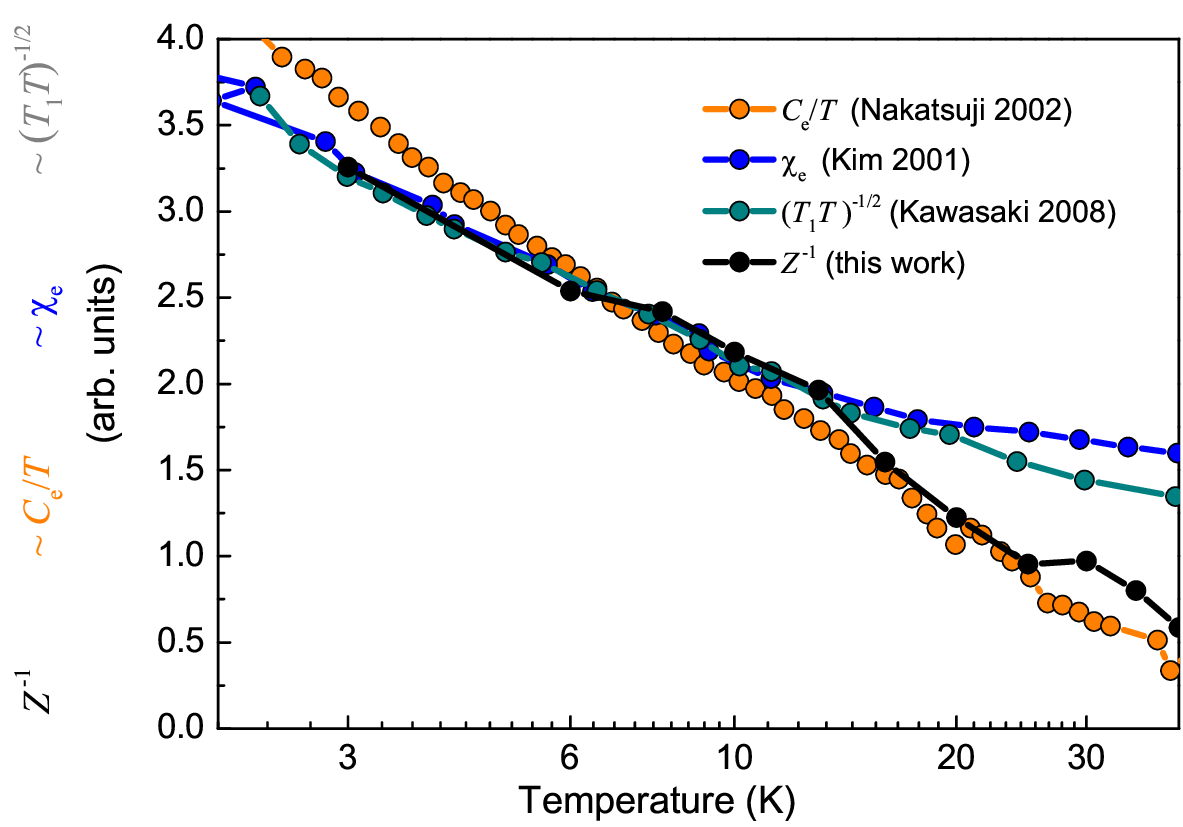}
\caption{\label{CCILogT} Renormalization factor $Z^{-1}(T)$ measured by optical spectroscopy (at 6.5\,cm$^{-1}$) and calculated from electronic specific heat, - susceptibility (at 0.1\,T), and nuclear spin-lattice relaxation (taken from Refs.\,\cite{Nakatsuji2002}, \cite{Kim2001}, and \cite{Kawasaki2008} respectively). The NFL $T$-dependence of $C_\mathrm{e}/T$, $\chi_\mathrm{e}$ and $T_1$ agree remarkable well with the one of $Z^{-1}$ and the scenario of resilient QPs constituting a hidden FL in CeCoIn$_5$.}
\end{centering}
\end{figure}
The limited range where $\Gamma$ varies approximately quadratically in energy can be attributed to the disappearance of resilient QP. One can  argue \cite{Pracht2017, WolflePers} that the existence of well-defined resilient QP is guaranteed as long as $\Gamma^\ast$ satisfies
\begin{equation}\label{FLcrit}
\hbar\Gamma^\ast \leq \hbar\Gamma_\mathrm{crit.}=2\sqrt{(\hbar\omega)^2+(k_BT)^2}+\hbar\Gamma_\mathrm{imp.}
\end{equation}
($\hbar,k_B$ are Planck's and Boltzmann's constants) where the factor 2 reflects our definition of the QP relaxation rate $\Gamma^\ast$ as twice $\Gamma^\ast_\mathrm{qp}=-Z\mathrm{Im}\Sigma$ where $\Sigma$ is the FL self-energy and $\Gamma_\mathrm{imp.}=2.7$\,cm$^{-1}$ is the residual impurity scattering estimated from Fig.\,\ref{Fig3}(a). Imposing the criterion Eq.\,(\ref{FLcrit}) to the experimental data (thin solid lines in  Fig.\,\ref{Fig3}(a-d) and (f-k)) leads to two conclusions: (i) The approximate $\nu^2$-behavior of $\Gamma^\ast(\nu)$ disappears as $\Gamma^\ast$ exceeds $\Gamma_\mathrm{crit.}$ above 20\,K. (ii) The temperature, up to which $\Gamma^{\ast}(0<\nu\leq15.2\mathrm{\,cm}^{-1},T)$ varies approximately as $T^2$ agrees roughly with the temperature for which $\Gamma^{\ast}(T)<\Gamma_\mathrm{crit}$ holds, see panels (b-d). Although $\Gamma^{\ast}(T)<\Gamma_\mathrm{crit}$ also holds at higher frequencies, a $T^2$-behavior is absent. This may be attributed to the vicinity to the Kondo regime at higher energies as (see the Supplementary Material for details) $\Gamma^\ast\propto T^2$  requires the QP interaction to be approximately independent of energy. This condition is no longer satisfied in the crossover regime to the high temperature independent Kondo ion lattice, when the  resilient QP no longer exist. Hence, we conclude that the optical response of CeCoIn$_5$ can be understood in terms of a hidden FL composed from heavy resilient QP, that expands over a major part of the HF state up to around $25-30$\,K.

In the above discussion, we silently assumed a constant charge carrier density $N$ as function of temperature. For HF systems, this assumption is, generally, not valid: When temperature falls below $T_\mathrm{coh}$, the previously localized $f$ electrons hybridize with the conduction electrons and, though heavy, become part of the Fermi surface and increase $N$. From our optical measurements, we cannot quantify the rise of $N$ below $T_\mathrm{coh}$, however, it will certainly affect the conductivity much less than the mass enhancement. With a carrier number of 1.5 per formula unit \cite{VdM2005}, the complete hybridization of one $f$ electron per unit would increase $N$ by a factor of $1.67$, which is about one order of magnitude less than the mass enhancement. In addition, the $T,\nu$-square regime is realized for temperatures lower than 25\,K, where a major portion of the $f$ electrons most likely has already merged into the Fermi sea. In further support, ARPES measurements \cite{Chen2016} have shown only a partial localized-itinerant transition of the $f$ electrons, down to temperatures $\sim$17\,K suggesting a even smaller change in $N$.   \\

Having elucidated the anomalous $\rho_\mathrm{dc}(T)$ of CeCoIn$_5$ from the perspective of resilient QPs and the $T$- and $\nu$- dependent $Z$ it is worthwhile to apply a similar reasoning to other observables to sharpen the non-FL classification towards a hidden FL. Aside from the resistivity, previous studies on the HF regime reported the electronic specific heat $C_\mathrm{e}\propto -T\mathrm{ln}T$ \cite{Nakatsuji2002,Kim2001}, the magnetic susceptibility $\chi_\mathrm{e}\propto T^{-0.42}$  \cite{Kim2001}, and the nuclear spin-lattice relaxation time $1/T_1\propto T^{0.25}$ \cite{Kawasaki2008}, whereas for a canonical FL one expects $ZC_\mathrm{e}^\mathrm{FL}\propto T$, $Z\chi_\mathrm{e}^\mathrm{FL}=$const.  \cite{Coleman2015}, and the Korringa relation $Z^2/T_1^\mathrm{FL}\propto T$. As for a FL $Z(T)$ is a constant, the above FL relations determine the entire $T$-dependencies. On the one side, the framework of hidden FL theory has yet not been worked out for the electronic specific heat, magnetic susceptibility, and spin-lattice relaxation time. On the other side, given the clear case of resistivity and QP relaxation rate, we can hypothesize a hidden FL-like $T$-dependence of these quantities masked by $Z(T)$. To resolve the influence of $Z(T)$, we rearrange the above FL predictions to read $C_\mathrm{e}/T\propto Z^{-1}$, $\chi_\mathrm{e}\propto Z^{-1}$, and $(T_1T)^{-1/2}\propto Z^{-1}$. Figure \ref{CCILogT} compares $Z^{-1}(T)$ at 6.5\,cm$^{-1}$ of this work to the corresponding forms calculated from the $C_\mathrm{e}$, $\chi_\mathrm{e}$ and $1/T_1$ data \cite{Kim2001,Nakatsuji2002,Kawasaki2008}. By rescaling the $y$-axes, all sets of $Z^{-1}(T)$ can remarkably well be collapsed on a mutual straight line over the relevant temperature ranges of the alleged non-FL behavior. In analogy with the QP relaxation rate above we come to the conclusion, that inclusion of $Z^{-1}(T)$ unveils FL-like behavior not only of optical but also thermodynamic quantities caused by resilient QPs.

\section{Scaling behavior of the effective mass}\label{scaling}
As much as there is hardly any doubt that CeCoIn$_5$ is a quantum-critical material, the nature and even the very localization of the QCP in the phase diagram poses a conundrum as outlined in Sec.\,\ref{QCP}. Hence, how (if at all) quantum criticality might serve as explanation for the observed non-FL behavior in a wide range of the phase diagram is an open question. The available experimental studies based on the temperature and magnetic-field dependence of resistivity, specific heat, thermal expansion, Gr\"{u}neisen parameter, Hall effect, nuclear spin-lattice relaxation, and magnetic susceptibility (to name the most important ones) document diverse and sophisticated approaches, yet at the same time do not constitute a coherent picture. 
A clear experimental access to quantum criticality is often complicated or even completely spoiled by non-critical contributions. This holds, in particular, for scaling behavior which is a typical signature of quantum critical regimes. Disorder and the temperature dependent phonon scattering, for instance, may completely overshadow electronic correlations effects in, e.g., the real conductivity making a scaling analysis thereof precarious. A quantity such as the effective mass, whose temperature- and frequency dependence solely results from electronic interactions, could be an alternative. The search for scaling based on $m^\ast$ is therefore less ambiguous, yet also barely employed approach due to the experimental difficulties to measure $m^\ast(\nu,T)$ over a sufficiently large parameter range. Typically, de-Hass van-Alphen (dHvA) or angle resolved photo emission spectroscopy (ARPES) measurements are performed both known for notorious difficulties at elevated temperatures and oxidized surfaces, respectively. Especially for CeCoIn$_5$, where, on the one side, degradation is a well-known problem due to the highly reactive Ce$^{3+}$-ions, and on the other side, the temperature range of interest extends to several 10\,K. Consequently, experimental studies are scarce: At the time of writing, the available dHvA studies concentrate on the superconducting state \cite{Hall2001}, the effect of Cd substitution in the Fermi surface topology, \cite{Capan2010}, high magnetic fields \cite{Kim2001} or applied pressures \cite{Shishido2004}. Similarly, ARPES studies focus on basic Fermi-surface topology \cite{Wen2011}, $f$-orbital occupancy for Yb doping \cite{Booth2011}, hybridization effects in the normal \cite{Koitzsch2007} and superconducting states \cite{Koitzsch2013} and do not explicitly address $m^\ast(T)$ over a wide range of temperature. \\

On general grounds, one has to discriminate between two main scenarios at the QCP. In the (conventional) spin-density-wave (SDW) model realized, e.g., in CeNi$_2$Ge$_2$ \cite{Kuchler2003} or CeIn$_{3-x}$Sn$_x$ \cite{Kuchler2006}, the Kondo temperature remains finite at the the QCP and the local moments are completely hybridized. SDW- or Hertz-Millis \cite{Hertz1976,Millis1993} quantum criticality is captured by a non-interacting Ginzburg-Landau-Wilson type field theory where Gaussian AFM fluctuations are the only relevant critical modes at play. In the (unconventional) case of local quantum criticality, the destruction of  AFM at the QCP goes hand in hand with the breakdown of Kondo screening such that in the vicinity of the QCP unscreened local moments persist down to lowest temperatures \cite{Si2001,Coleman2001,Zhu2003,Grempel2003,Senthil2004}. Indeed, these local moments providing a second type of critical modes appear to play a key role in the quantum critical properties of certain antiferromagnetic HF metals. In the paradigmatic case of CeCu$_{6-x}$Au$_x$ measurements of the spin susceptibility \cite{Schroeder2000} played a crucial role in revealing the nature of the QCP by demonstrating $\omega/T$ scaling in agreement with local quantum criticality rather than the SDW picture, where scaling takes a different form. Consequently, measurements covering the quantum critical regime in both temperature and frequency allowing to search of scaling behavior can be of great importance to discriminate between the two scenarios of quantum criticality. 

At the time of writing, none of both models have been worked out concerning scaling behavior of $m^\ast(T)$. Hence, in what follows, we use a heuristic approach inspired from previous work of Schroeder \emph{et al.} \cite{Schroeder2000} and conductivity studies on the layered Co-oxide \cite{Limelette2013} and certain high-$T_c$ cuprates \cite{VdM2003}, reading
\begin{equation}
m^\ast(\omega, T)=m_bT^{\alpha} g\left(\frac{\omega}{T^\beta}\right)\label{scalinglaw}
\end{equation} 
where $\alpha<0$, $\beta>0$ and the universal scaling function $g$ is defined as
\begin{equation}
g\left(\frac{\omega}{T^\beta}\right)=\left(y+\frac{2\pi h c\omega}{(k_B T)^\beta}\right)^{-1}\label{g}
\end{equation}
with $y$ a dimensionless parameter. Usually, the optimal collapse is found by minimizing the distance between data and a universal function\sidenote{\footnotesize{See, e.g., Ref.\,\cite{Stricker2014} for an instructive example of FL scaling behavior in SrRuO$_3$}}. Without knowledge of such a universal function, identifying the optimal collapse becomes a more delicate problem. Here, we employ an algorithm that optimizes a sufficiently good initial guess such as Eq.\,(\ref{g}). Consider the right side of Eq.\,(\ref{scalinglaw}) given as an analytic scaling function 
\begin{equation}
\chi=\chi(\omega,T,\{\alpha_k\})
\end{equation}
with $\{\alpha_k\}$ an arbitrary set of real scaling parameters. Further, consider indices $i=1,~...~n$ labeling $n$ discrete frequencies and $j=1,~...~N$ labeling $N$ temperatures. 
Suppose the experimental data to be arranged in sets 
\begin{equation}
s^j=\left\{(x_1^j,m^\ast(T^j,\omega_1));~(x_2^j,m^\ast(T^j,\omega_2));~...~(x_n^j,m^\ast(T^j,\omega_n))\right\}
\end{equation}
where $x_i^j=\chi(\omega_i,T_j,\{\alpha_k\})$ is the scaling variable and $x_1^j<x_2^j<...<x_n^j$. For each set $s^j$ we construct a continuous (linear) interpolation function $f^j(x)$. To judge the quality of the data collapse for a particular choice of parameters $\{\alpha_k\}$ we define the quantity
\begin{equation}
\delta^j(\{\alpha_k\})=\frac{1}{C^j-c^j}\int\limits_{c^j}^{C^j}|f^j(x)-f^{j+1}(x)|\mathrm{d}x
\end{equation}
measuring the area between two neighboring interpolations $f^j$ and $f^{j+1}$ in their $overlap$ interval defined by $c^j\equiv \mathrm{max}[x_1^j,x_1^{j+1}]$ and $C^j\equiv \mathrm{min}[x_n^j,x_n^{j+1}]$, weighted by the length of the interval. The collapse between two adjacent sets $s^j$ and $s^{j+1}$ is best for minimal $\delta^j$. Therefore, the collapse between all sets is optimal for  a particular choice of $\{\alpha_k\}$ minimizing the expression  
\begin{eqnarray}
\Delta(\{\alpha_k\})&=&\sum\limits_{j=1}^{N-1}\frac{1}{C^j-c^j}\int\limits_{c^j}^{C^j}|f^j(x)-f^{j+1}(x)|\mathrm{d}x\nonumber\\
&&\label{Algo}
\end{eqnarray}
which can easily be implemented numerically. Importantly, this procedure does not require knowledge of the universal function \emph{a priori}. The applicability of the suggested algorithm is restricted to sufficiently \emph{close} data sets with a finite overlap $C^j-c^j>0$ and care must be taken to avoid negative-valued $\delta^j$ leading to an ill-defined minimal $\Delta$. Here, we include an inverse step-function $1/\Theta^j(C^j-c^j)$ to the sum in Eq.~(\ref{Algo}) returning $\Delta\to \infty$ in case of  negative-valued $\delta^j$.   
\begin{figure}[t!]
\begin{centering}
\includegraphics[scale=0.36]{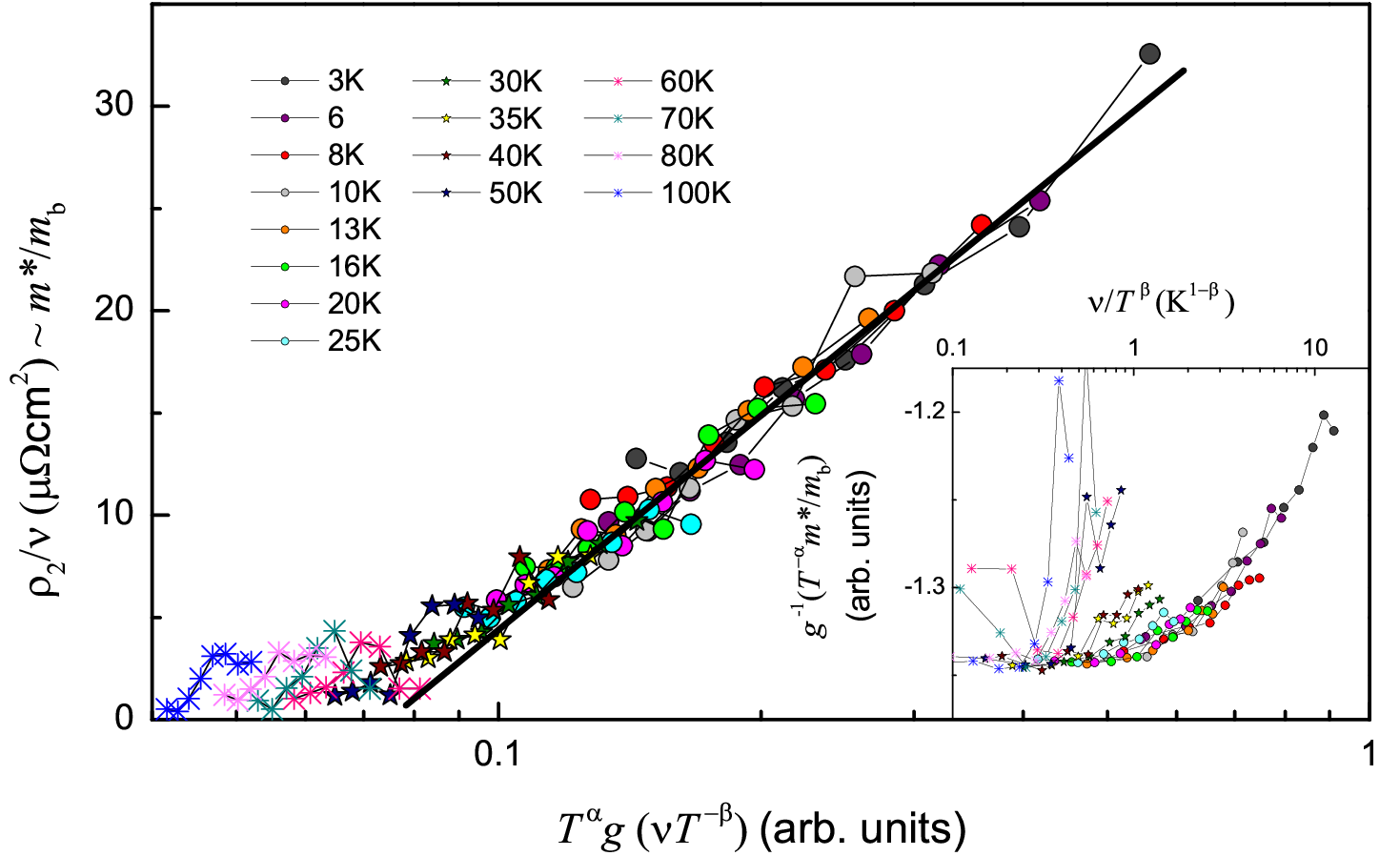}
\caption{\label{CCI_scaling}\textbf{Scaling behavior of the effective mass} measured by $\rho_2/\nu$ according to Eq.~(\ref{scalinglaw}) and (\ref{g}). The optimal collapse is achieved for $\alpha=-0.94$, $\beta=0.96$, and $y=0.24$ revealing an approximate $\nu/T$ scaling of the effective mass. The Inset shows the same data collapse plotted versus $\nu/T^\beta$.}
\end{centering}
\end{figure}
The best collapse of data is displayed in Fig.~\ref{CCI_scaling} and achieved for $\alpha=-0.94$, $\beta=0.96$ and $y=0.24$ forming three distinct temperature regimes of scaling. First, at temperatures above 50\,K (high-temperature side of the $\rho_\mathrm{dc}(T)$ peak), $m^\ast$ does not depend on energy and $m^\ast/m_b\sim 1$ scales trivially (asterisks). Second, between 40 and 20\,K, $m^\ast$ becomes energy dependent and the universal curve bends upwards (stars). Third, down to 3\,K (full circles) we find $m^\ast$ to follow a straight line in the log-linear plot and the data collapse to become sharper with decreasing temperature. The third range coincides with the regime where the hidden FL state is revealed from the approximate $T^2$ behavior of the QP relaxation rate. The scaling of $m^\ast$, however, can be traced up to 40\,K, exceeding the hidden-FL regime by nearly a factor of two. The exponents $\alpha=-0.94$ and $\beta=0.96$  can be compared to the two paradigmatic predictions for the susceptibility, namely $\alpha=-1.5$ and $\beta=1.5$ for a  SDW-QCP within Hertz-Millis theory and local quantum criticality characterized by $\beta=1$. Although neither of both predictions is exactly met in our experimental data, it is tempting to interpret the approximate $\nu/T$ scaling as signature of local quantum criticality in agreement with the aforementioned recent proposal of zero-field QCP and 2D critical fluctuations.\\
The above scaling only expands over roughly one order of magnitude in energy. The search for scaling on a larger parameter range, however, is deemed to fail. Towards higher temperatures (and corresponding IR frequencies) above the HF state, the energy dependence of $m^\ast$ vanishes \cite{VdM2005,singley2002} and scaling becomes obsolete. At lower temperatures (and microwave frequencies), superconducting fluctuations and superconductivity itself strongly affect the charge dynamics \cite{Ormeno2002a, Nevirkovets2008a, SudhakarRao2009, Truncik2013a} and make studies of $m^\ast$ difficult. Also the suppression of superconductivity with magnetic fields is problematic, as the resulting normal state is canonical FL \cite{Paglione2003} with a different dimensionality \cite{Donath2008}. Understanding the relation (if there is) between the observed scaling behavior, quantum criticality, and the hidden-FL state in CeCoIn$_5$ and beyond is a challenge for future theoretical work.   \\
\newpage
\section{Outlook}\label{CCIout}
\begin{marginfigure}
\begin{centering}
\includegraphics[scale=0.15]{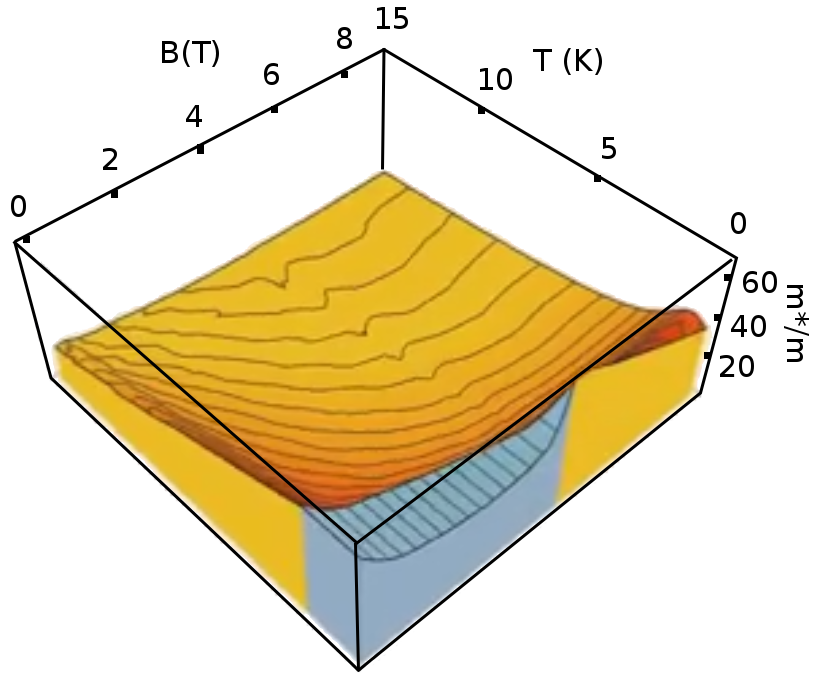}
\caption{\label{CCI_broun} $m^\ast(T,B)/m_b$ of CeCoIn$_5$ at a fixed GHz frequency measured with a microwave resonator \cite{Broun2016}. Note flattening in the high-field and low-$T$ FL regime.}
\end{centering}
\end{marginfigure}

In this work we have demonstrated how the dynamical response at matching energy scales $k_BT\sim h\nu$ yields valuable information about the nature of the HF state in CeCoIn$_5$ advancing our current understanding not inly incrementally, but providing a complete new perspective on its enigmatic non-FL behavior. The identification of a hidden FL over a substantial part of the HF state is an important finding on its own, yet also unravels just a small part of the entire phase diagram. Thinking in terms of resilient QP of a hidden FL might be a fruitful starting point to reconsider the quantum critical behavior, especially the anomalous transport coefficients in approach of the conjectured field-tuned QCPs, as discussed in Sec.\,\ref{QCP}, although the hidden FL theory has (if possible at all) not been worked out for finite magnetic fields yet. Given the high-field and low-$T$ recovery of canonical FL and the zero-field hidden FL, we hypothesize that, first, the non-FL regime in between is also characterized by a strong $T,\nu$ dependent renormalization and, second, consideration thereof will reveal a parabolic QP relaxation rate $\Gamma^\ast(\nu,T)$. Indeed, Broun \emph{et al.} measured the $B$- and $T$-dependence of the effective mass at a fixed GHz frequency using a microwave resonator approach revealing a $T$-dependence of $m^\ast/m_b$, see Fig.\,\ref{CCI_broun}, that expands over the entire phase diagram but levels off in the FL state and, hence, hints the importance of renormalization not only in zero-field but the entire non-FL regime.
Measuring the dynamical response at finite fields will also help to clarify the relevance of the approximate $\nu/T$ scaling of the effective mass. More desirably, however, is a better understanding of the scaling behavior within the different models for quantum criticality.

\begin{marginfigure}
\begin{centering}
 \vspace{\stretch{10}}
\includegraphics[scale=0.2]{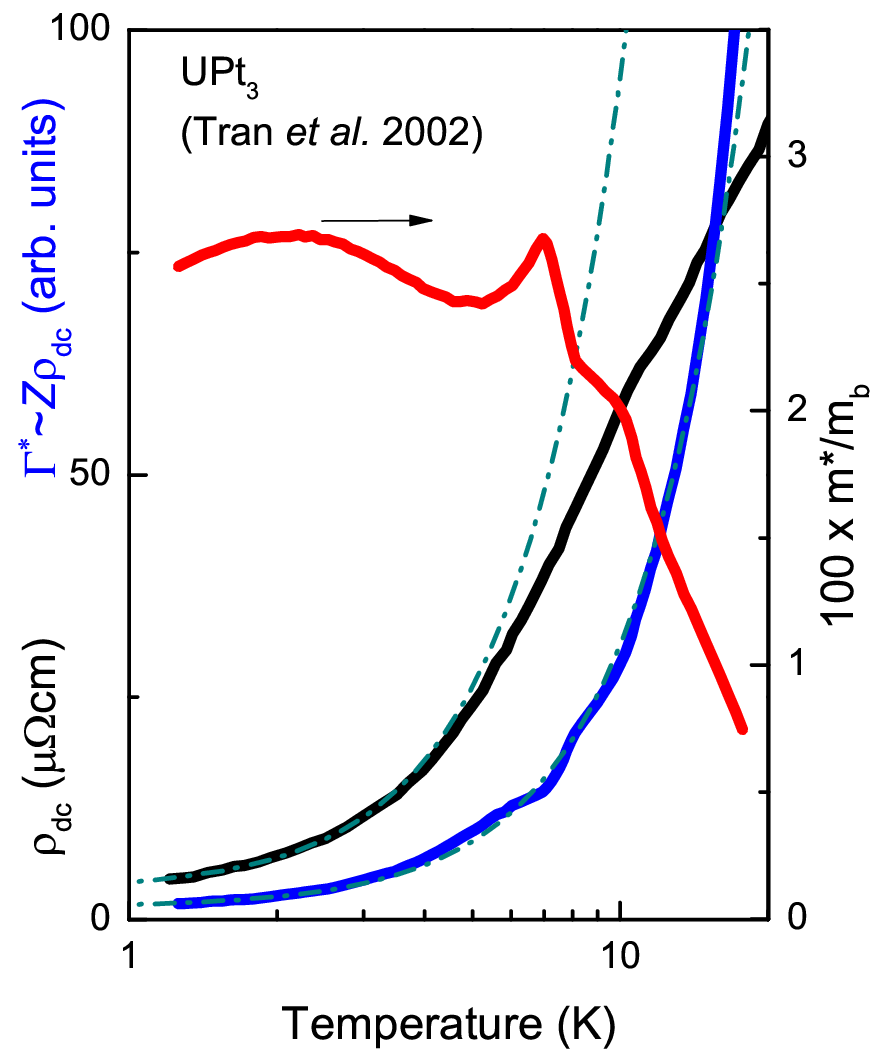}
\caption{\label{CCI_UPt3} $\rho_\mathrm{dc}$, $\Gamma^\ast$, and $m^\ast/m_b$ versus $T$ for the HF superconductor UPt$_3$. Dashed lines are $T^2$ fits. Data taken from \cite{Tran2002}.}
\end{centering}
\end{marginfigure}

After the Mott insulator V$_2$O$_3$ and the Hund's metal CaRuO$_3$, we add the HF metal CeCoIn$_5$ as third entry to the list of known hidden-FL materials. Although  formulated for the doped Mott insulators, the hidden FL concept seems robust and might serve as model for further non-FL systems that so far escaped solid understanding. In some cases we can test this hypothesis on basis of earlier measurements. Referring to work  \cite{Tran2002} of Tran \emph{et al.}, we suggest the U-based HF metal UPt$_3$ to be a hidden FL: UPt$_3$ displays a $T$-square rise of $\rho_\mathrm{dc}$ up to about $T_\mathrm{FL}=5$\,K, just where $m^\ast/m_b$ acquires a notable $T$-dependence, see Fig.\,\ref{CCI_UPt3}. Taking the renormalization into account, one finds a $T^2$ law for $\Gamma^\ast$ up to at least $17$\,K$\gg T_\mathrm{FL}$, consistent with a hidden FL. However, applying this analysis to measurements of the almost quantum-critical HF metal CeFe$_2$Ge$_2$  by Bosse \emph {et al.} \cite{Bosse2016} does not uncover a quadratic $\Gamma \ast$ behind $\rho_\mathrm{dc}\propto T^{1.5}$. For most correlated electron systems, however, $m^\ast(T)/m_b$ has not been measured yet in the relevant temperature range and the question of possible hidden FL behavior remains open.\\
\begin{marginfigure}
\begin{centering}
\includegraphics[scale=0.23]{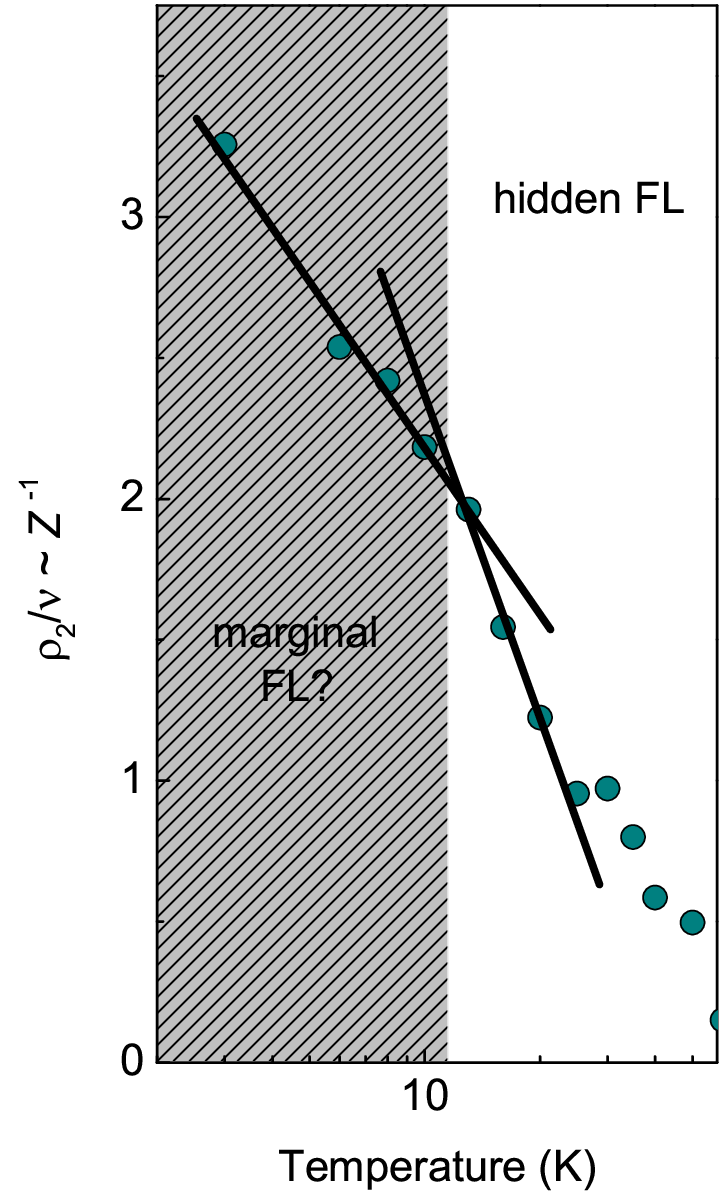}
\caption{\label{CCI_marginal}Renormalization $Z^{-1}$ versus $T$ at 6.5\,cm$^{-1}$. The solid lines are guides to the eyes demarcating a crossover at around 10\,K with increasing temperature - from marginal FL to hidden FL?}
\end{centering}
\end{marginfigure}
As silently hypothesized using Eq.\,(\ref{marginalcrossover}) on $\rho_\mathrm{dc}(T)$, we note \cite{WolflePers} that the crossover from linear to sub-linear $\rho_\mathrm{dc}(T)$ and weak to stronger (log-like) decline of $m^\ast(T)/m_b\sim Z^{-1}$ at 6.5\,cm$^{-1}$, see Fig.\,\ref{CCI_marginal}, at around 10\,K could be interpreted as a crossover from a hidden FL with resilient QP to a so-called marginal FL with critical QP \cite{Ruckenstein1991,Varma1993}. Then, however, also the quadratic behavior of $\Gamma^\ast(\nu,T)$ should give way to $\Gamma^\ast\propto-|x|/\mathrm{ln}(x)$ with $x=\nu,T$ \cite{WolflePers}. Unfortunately, such a small difference in functional form cannot be inferred with certainty based on the available data. The question, if a marginal FL serves as bridge between the (presumably) non-critical hidden FL regime and the QCP therefore remains an experimental and intellectual challenge for future research.  \\

To put the behavior of CeCoIn$_5$ in the broader context, THz studies studies of the related compounds CeRhIn$_5$ and CeIrIn$_5$ as well as the parent compound CeIn$_3$ are highly desirable. At the time of writing, thin-film growth of these materials is still a major challenge \cite{Shimozawa2016}. As for CeIn$_3$, Matsuda \emph{et al.} managed to grow sufficiently thin films of a few ten nanometer on MgF$_2$ substrates. First measurements of the THz response reveal a frequency dependent $\Gamma$ in the HF regime similar as for CeCoIn$_5$, but inconclusive energy dependence of $m^\ast/m_b$. In addition, the films tend to rapid degradation at room temperature even in He atmosphere. Although from in-situ monitoring of $\rho_\mathrm{dc}(T)$ we can rule out a degradation at low temperatures on a measurement scale of $\sim$36\,hours, the situation is apparently much more delicate as for CeCoIn$_5$. The growth of CeRhIn$_5$ films with a thickness less than $\sim$120\,nm and without a CeIn$_3$ buffer layer is still not possible. Due to their high conductivity, these films suppress the THz transmission amplitude to $10^{-4}-10^{-5}$ where a reliable data analysis is borderline. With the recent advances in MBE growth of these compounds, it can be hoped that the current obstacles will be removed and reliable THz studies of the Ce$Me$In$_5$ become feasible.

\chapterend

\clearpage{}

\appendix
\chapter{Experiment and \\Analysis}
\thispagestyle{empty}
\section{Sample preparation}\label{Samples}
\subsection{NbN thin films}\label{sampleNbN}
The majority of  superconducting NbN samples were supplied by P. Raychaudhuri (Mumbai, India) and grown via reactive magneto sputtering (Ar/N$_2$ atmosphere at 5\,mTorr) as thin films between 20 and 30nm thickness on dielectric sapphire or MgO substrates held at 600$^\circ$C. By careful control of the Ar/N$_2$ ratio (varying between 90/10 and 70/30) while sputtering, the number of atomic defects (i.e. Nb vacancies) was adjusted producing NbN films covering a wide range of disorder. Clean-limit samples with a low dc-transport resistivity $\rho_\mathrm{dc}$ have high critical temperatures up to $T_c\approx15$\,K whereas $T_c$ and $\rho_\mathrm{dc}$ and $T_c$ are rapidly increased and suppressed, respectively, in the approach of quantum criticality. The most-critical sample\sidenote{\footnotesize{Note that the plot Fig.\,\ref{Fig:Densities} comprises data for a even more critical sample with $T_c=3.1$\,K inferred from microwave spectroscopy done by M. Mondal \emph{et al.}, see Ref.~\cite{Mondal13}}} studied towards its THz and tunneling response has $T_c=4$\,K. A minor number of supplementary clean-limit samples grown in the identical fashion where supplied by M. Siegel and K. Illin (Karlsruhe, Germany). Here, the film thickness is of the order of 5\,nm but values of $\rho_\mathrm{dc}$ and $T_c$ are in line with the samples of the aforementioned batch so that no separate treatment is required. Accordingly, in our discussion we will not distinguish between the batches, and will use the value of $T_c$ as a straight-forward quantity measuring the distance from quantum criticality. More details of the growth process and transport characterization can be found in \cite{Chand2012PhD} and references therein.    

\subsection{Granular Al}\label{sampleAl}
Thin films of granular Al were deposited using thermal evaporation, see Ref.\,\cite{Bachar2014PhD} for details. In order to obtain high-quality transmission measurements, films of 40\,nm thickness were deposited on 2\,mm thick MgO substrates. The substrates were kept at liquid nitrogen temperature during growth. Clean Al pellets (purity of 99.999\%) were evaporated from alumina-coated Mb boat in O$_2$ partial pressure of about $10^{-5}$\,Torr. Samples of different resistivity $\rho_{\mathrm{dc}}$ ranging between a few 100 and 1000\,$\mu\Omega$cm at 300\,K were produced by carefully varying the evaporation rate and the O$_2$ partial pressure. While the grains in the low-resistivity (LR) limit can be considered well-coupled, the opposite case of completely decoupled grains is realized in the high-resistivity limit (HR). The shape, size, and size distribution of grains in granular Al has been studied by Deutscher \emph{et al.} \cite{Bachar2014PhD,Deu73,Deutscher73} by means of dark-field microscopy: Low-resistivity films deposited on substrates held at room temperature display spherical grains with a broad distribution of diameter peaked at around 5\,nm that sharpens sightly and shifts towards 3\,nm with increasing resistivity. Substrates kept at liquid nitrogen temperatures while deposition favor the growth of films with a narrow size distribution with a mean of 2\,nm irrespective of resistivity within the scope of this work \cite{Lerer2013PhD}. The critical influence of O$_2$ concentration and substate temperature on the structural properties of granular Al can be explained following the early ideas of Shapira and Deutscher \cite{Shapira68}. Starting from a nucleation seed, grains grow spherically by accumulation of Al atoms and dielectric Al$_2$O$_3$ molecules. As growth continues, Al$_2$O$_3$ is expelled to the grain's surface where it eventually terminates growth of the metallic bulk, when forming a complete insulating oxide shell. Clearly, a high concentration of O$_2$ favors the formation of Al$_2$O$_3$ and tends to thicken the insulating barriers between grains lowering the electronic inter-grain coupling and increasing the transport resistivity. The reduction of grain size with decreasing substrate temperature results from the suppression of coalescence, i.e. the formation of larger structures from the 'melting' of smaller ones.      

\subsection{CeCoIn$_5$}\label{sampleCCI}
The samples under study are two 70\,nm thick films of CeCoIn$_5$ deposited via molecular beam epitaxy (MBE) \cite{Shimozawa2016} on a dielectric $10\times5\times0.5\,$\,mm$^3$ MgF$_2$ substrate both grown and measured in 2013 and 2015, respectively. The low growth rate of 0.01 - 0.02\,nm/s allows a very precise control of the layer thickness. The growth process and high crystalline quality was monitored \emph{in situ} via the reflection high-energy electron diffraction (RHEED) technique. MBE growth of Ce-based heavy fermion thin films \cite{shishido2010,mizukami2011,shimozawa2012,goh2012,shimozawa2014} has already been employed for previous THz studies \cite{Scheffler2013}, however high-quality films were obtained only with additional metallic buffer layers. Despite the advanced MBE growth THz transmission measurements nevertheless remain challenging for several reasons. First, the film needs to be thin enough to allow for a detectable transmission signal, while the sample quality favors thick films. Here we have chosen a thickness of 70\,nm, which is a compromise between sample quality and suitability for our experimental technique. Second, thin films of CeCoIn$_5$ tend to rapid degradation in ambient air conditions so that exposure time must be cut to a minimum \cite{Scheffler2013} Third, the aperture, through which the focused THz radiation passes before it is transmitted through the sample, needs to have a diameter $d_a$ smaller than the sample. We have chosen $d_a=3$\,mm which restricts our accessible spectral range to wavelengths shorter than $\sim d_a/2=1.5$\,mm ($\sim 6$\,cm$^{-1}$) due to diffraction effects.

After deposition in Kyoto, each film was sealed in a glass tube under vacuum conditions before it was shipped to Stuttgart. Right after removal from the glass tube, the samples were mounted onto the THz sample holder, transferred to the cryostat, and rapidly cooled down in He-gas atmosphere so that the exposure time to ambient air was less than 5 minutes\sidenote{\footnotesize{After deposition, the samples were immediately sealed in the glass tubes, so that the overall exposure time is of $\sim$10 minutes.}}. The entire optical measurements were subsequently performed during a period of about 36 hours. During this time, the samples were always kept below 150\,K. Afterwards, they were removed from the cryostat and contacted in standard 4-point geometry in order to measure the dc-sheet resistivity $\rho_\mathrm{dc}$.\\
\begin{figure}[t!]
\begin{centering}
\includegraphics[scale=0.58]{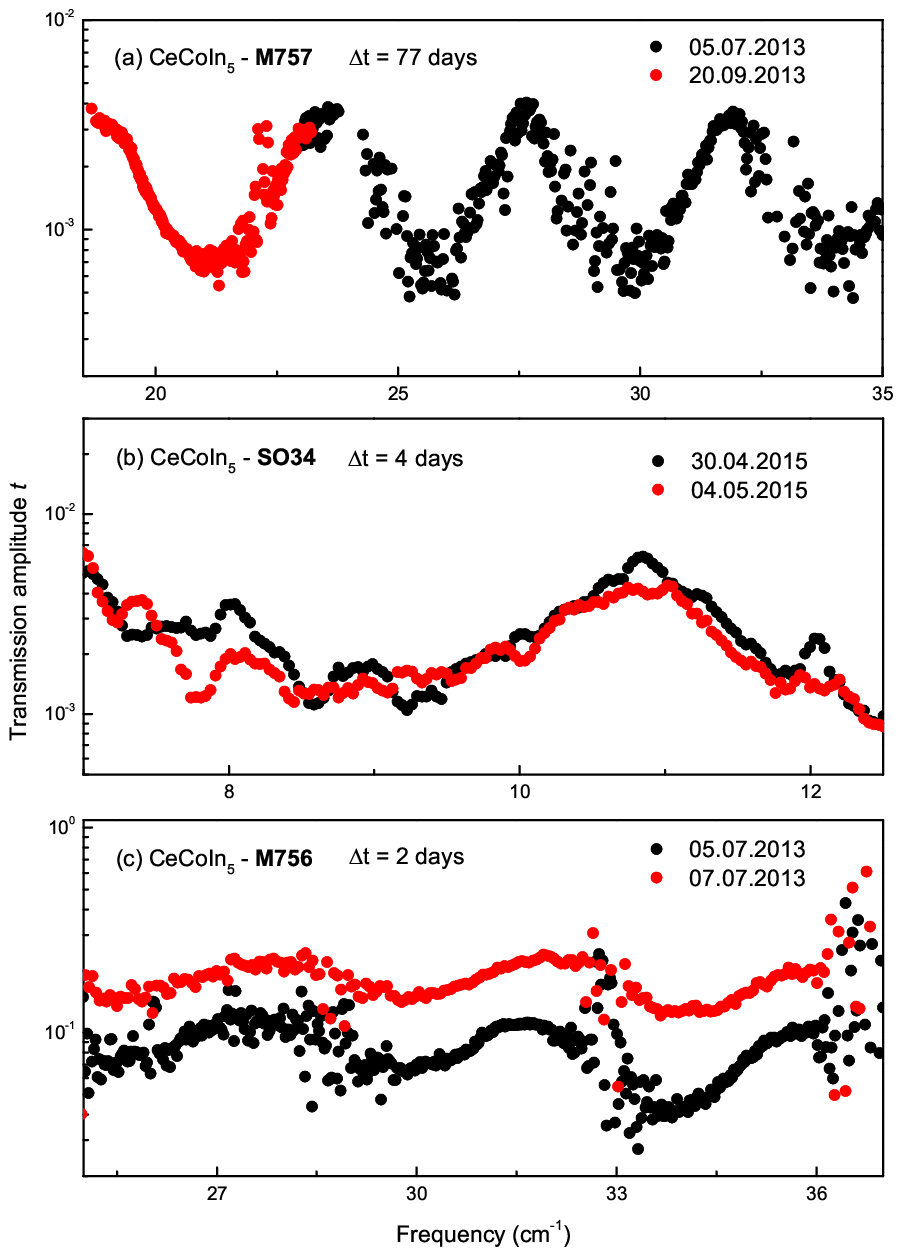}
\caption{\label{CCI_degradation}\textbf{Transmission-amplitude spectra for different CeCoIn$_5$ samples as function of time} (a,b) Spectra of samples M757 (examined in the remainder of this chapter) and SO34 do not show signs of degradation (i.e. a enhanced transmission amplitude) after 77 and 4 days, respectively. (c) For comparison, sample M756 displays a notable rise of $t(\nu)$ after only 2 days even though being kept under high-vacuum conditions, and thus, experimental data for this sample has not been considered for further analyses.}
\end{centering}
\end{figure}
Containing highly reactive Ce$^{3+}$ ions, thin films of CeCoIn$_5$ are notoriously susceptible for oxidation and require a careful handling. While the first generation of thin films suffered from poor quality and rapid degradation even if kept at low temperatures and under vacuum conditions \cite{Weig2011}, the advances in MBE growth have led to a considerably enhanced crystalline quality and chemical stability of the samples studied within this work. Nevertheless, ruling out any effects degradation is a crucial prerequisite for a reliable data interpretation. By comparing spectra of the transmission amplitude, in what follows, we use a sensitive probe to measure the degree of degradation over time. 
Starting with sample M757 (i,e, the sample on which major parts of the discussion in this Chapter is based) Fig.\,\ref{CCI_degradation}(a) displays $t(\nu)$ as measured right after mounting the sample inside the cryostat at 225\,K and at 300\,K after being kept at high-vacuum conditions for 77 days. Although the frequency ranges are different, the spectra connect smoothly and do not exhibit a jump towards higher values as it would be expected for a (partially) oxidized film. Consequently, we can rule out sample degradation during measurement time of $\sim 60$ hours.
The same conclusion can be drawn for the second film (SO34) discussed in this work, see panel (b). After a time span of 4 days, during which the optical and transport measurements were performed, no changes in $t(\nu)$ (as measured at 120\,K) appear which would point towards film degradation. Note that the spectra slightly differ, which, however, has to be attributed to a marginally different optical alignment.
The above exclusion of degradation can be contrasted with a sample of poor quality. Panel (c) compares $t(\nu)$ of sample M756 measured at 300\,K after removal from the glass tube and after being stored in high-vacuum for 2\,days. First, the low quality can be inferred from the transmission amplitude being about 50 times higher than for the samples shown in panels (a-b) despite the nominally same film thickness of 70\,nm. Second, even if kept in high-vacuum, $t(\nu)$ rose significantly over a time span of 2\,days due to oxidation of the film.
Even though the processes during growth, that eventually dictate the fate of the film, are unclear, the above considerations allow a discussion where any spurious contamination of degradation effects can certainly be ruled out.

\section{Frequency-domain THz spectroscopy}
\begin{figure}
\noindent \begin{centering}
\includegraphics[width=\textwidth]{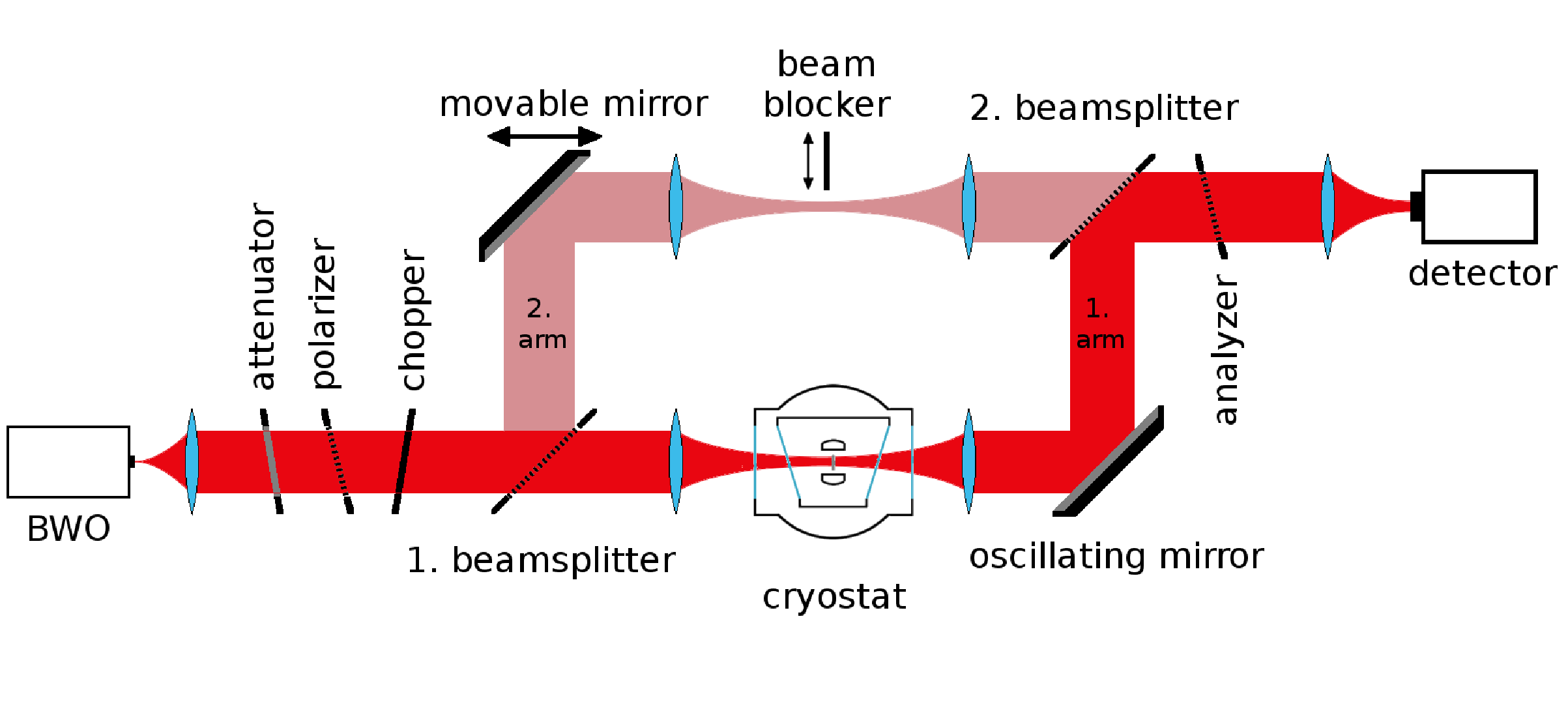} \\
\includegraphics[width=\textwidth]{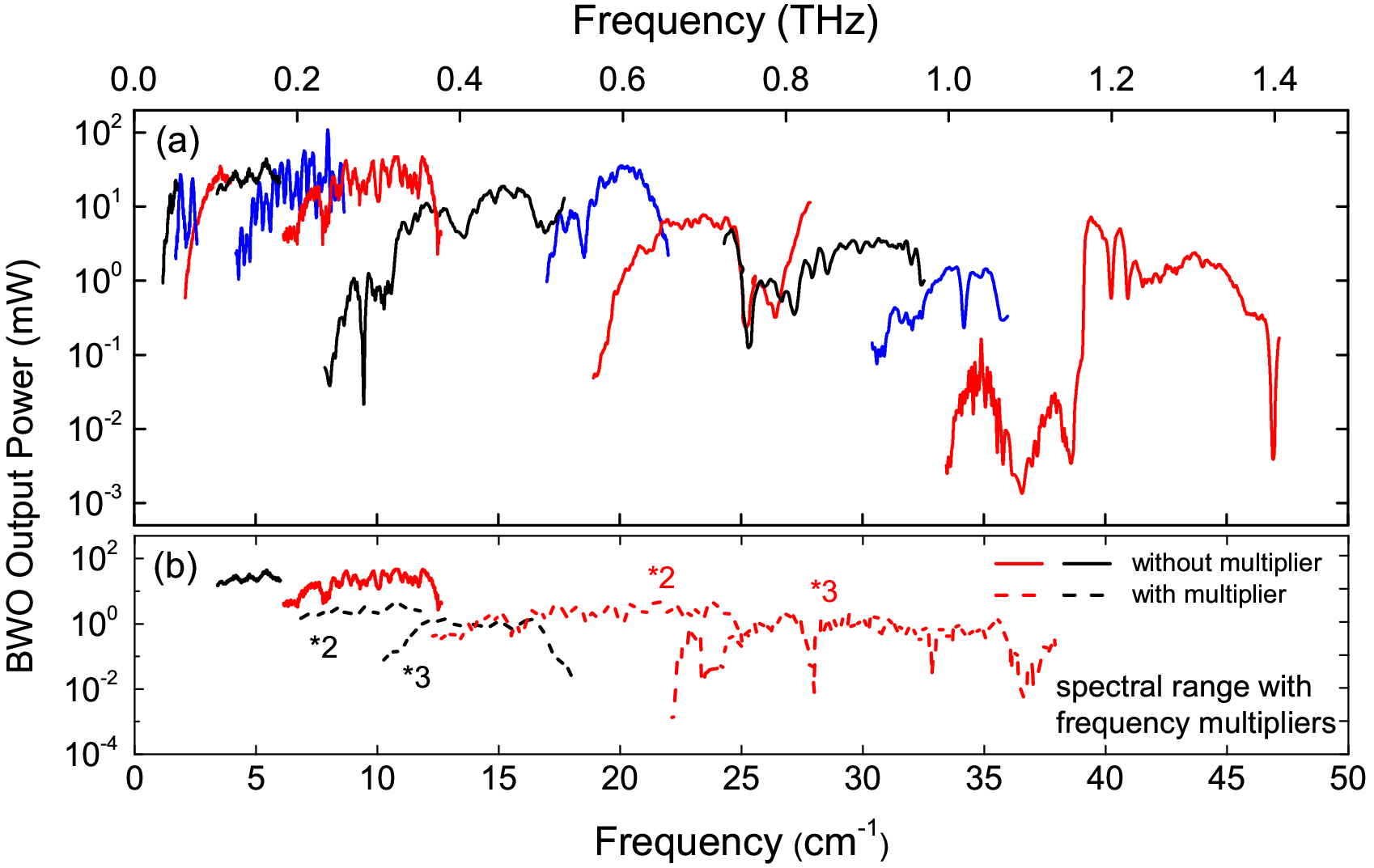}
\par\end{centering}
\caption{\label{fig:machzehnder}Top: Simplified sketch of the utilized
Mach-Zehnder interferometer. Attenuators, polarizer, chopper, and analyzer are tilted with respect to the beam propagation axis to suppress standing waves between these parts. Measuring the phaseshift requires both arms of the interferometer, whereas the second arm is blocked for transmission measurements. Bottom: (a) BWO output power of several sources versus frequency. The range from 1 to 47\,cm$^{-1}$ is covered entirely. (b) The available spectral range of a source can be increased using passive frequency doubler and tripler (dashed lines). The signal intensity in the doubled/tripled spectral range, however, is 3 to 4 orders of magnitude smaller and may require use of a $^4$He-cooled bolometer for detection.}
\end{figure}

Our THz studies are based on the frequency-domain spectroscopy technique. Coherent, monochromatic and continuous THz radiation is generated by frequency-tunable backward-wave-oscillators (BWO) that provide frequencies from 1 to 47\,cm$^{-1}$ \cite{Kozlov1998}. This spectral range is covered by a number of different BWO radiation sources (in the following just \textit{sources}), see bottom of Fig. \ref{fig:machzehnder}. The power of the radiation strongly depends on the
frequency and may be as large as 100\,mW for low-frequency and significantly smaller (0.1\,mW)
for high-frequency sources. As a consequence, the low-frequency radiation power might be too \textit{strong} and requires use of attenuators to avoid sample heating or detector overload. The great benefit of FDS compared to time-domain spectroscopy (TDS) is to process frequencies step by step with a stability and resolution exceeding $\Delta\omega/\omega=10^{-6}$ depending on the BWO power supply. 

While TDS excites all available energies simultaneously and has to struggle with minor signal strength, FDS can directly probe narrow-band excitations and resolve detailed line shapes. The high output power leads to high signal-to-noise ratio that can reach values up to $10^6$. Frequency doublers and triplers (Virginia Diodes) that allow extending the frequency range of a separate source can effectively be used. Loss of the radiation intensity is easily 
compensated by use of a $^4$He bolometer (see below). After the radiation is generated by the BWO, it is collimated by a Teflon, quartz, or polyethylene lens and guided by aluminum mirrors and wire grids, see top of Fig. \ref{fig:machzehnder}, which act as beam splitters depending on polarization. Other lenses are used to focus the radiation on the sample under study and the detector. Attenuators, polarizer, chopper, and analyzer are tilted with respect to the beam propagation axis to suppress standing waves between these parts. The quasi-optical part of the experiment is a Mach-Zehnder interferometer which allows us to measure both amplitude and phaseshift of THz radiation passing through a sample.

The radiation is detected by either a Golay cell or a $^{4}$He-cooled bolometer (Infrared Laboratories), depending on signal strength. For transmission measurements, the radiation is mechanically chopped to make it detectable using a lock-in technique.

Determining the spectra of optical parameters of the plane-parallel sample involves measurements of the transmission and phaseshift spectra from which the spectra of real and imaginary parts of complex permittivity, conductivity, etc., are calculated directly and without use of any additional analysis (such as Kramers-Kronig analysis).

Recording a transmission spectrum is performed with the second arm of the interferometer blocked. It consists of two separate steps. First, the signal intensity versus frequency is recorded with no sample in the beam path. This transmission spectrum is regarded 100\% transmission and used as measurement calibration. Second, the transmission spectrum is recorded with the sample in the beam path and the absolute transmission spectrum is obtained as a result of division of the two corresponding data arrays. In this way we can account for the frequency-dependent output power of the radiation passing through the measurement channel.
To measure phaseshift spectra, the second arm has to be unblocked. Again, the measurement consists of two separate steps: calibration without and measurement with the sample in the beam path. In order to obtain reliable data, one has to guarantee optimal interference of the two beams at the analyzer. This is provided by aligning the spectrometer elements and achieving maximal ratio between minimal (destructive interference) and maximal (constructive interference) signal strength. The interference signal is detected using a lock-in technique. However, the signal is modulated not using the chopper (amplitude modulation) but the oscillating mirror (phase modulation), as shown in Fig. \ref{fig:machzehnder}. During a measurement process, the movable mirror is automatically kept in a position corresponding to zero-order destructive interference (minimal signal) while the frequency range is swept. The mirror position versus frequency is recorded when the sample is out of (calibration scan) and in (measurement) the optical path, and the phaseshift spectrum is calculated from the difference between the two spectra\sidenote{\footnotesize{the measured quantity is \textit{phaseshift} $\phi$ (rad), however, in this paper we will always refer to \textit{relative phaseshift} $\phi/\nu$ (rad/cm$^{-1}$) with $\nu=\omega/(2\pi c $) and $c$ the speed of light in vacuum}} 

Optical anisotropy is an issue that has to be taken into account carefully. E.g., the Fabry-P\'erot resonances can become more complex \cite{Scheffler2009}. Many common substrates for thin-film deposition are birefringent such as Al$_2$O$_3$ or NdGaO$_3$. To align the polarization direction \textbf{E} of the radiation to sample axes of interest, one can adjust either the sample or the polarization angle of the 1. beam splitter (followed by corresponding turning of the 2. beam splitter, setting it to 90$^{\circ}$ relative to the 1. beam splitter). In case of optical anisotropy of both substrate and film and misaligned optical axes, the optical set-up and analysis has to be customized \cite{Ostertag2011}. 

\section{Analysis of the complex transmission data}\label{sec:singlepeak}
Typical spectra of transmission amplitude $t$ and phase shift normalized to frequency $\phi/\nu$ of a representative superconductor (granular Al) and metallic (CeCoIn$_5$) samples are shown in Fig.~\ref{fig:raw_fit} and \ref{fig:raw_CeCoIn5}. The pronounced oscillation pattern stems from multiple reflections inside the substrate, which acts as Fabry Perot (FP) resonator. 
\begin{figure}[t!]
\includegraphics[width=\textwidth]{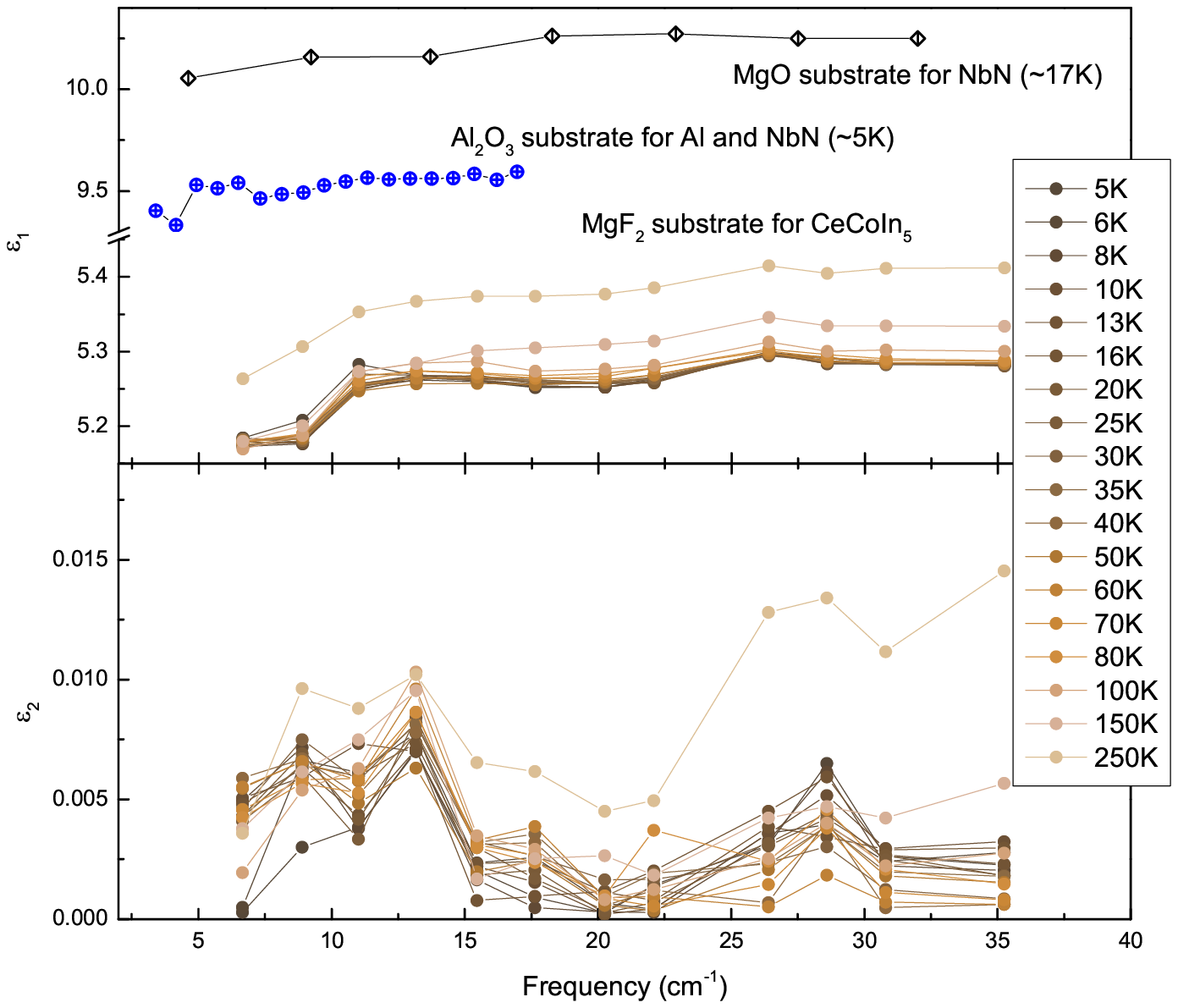}
\caption{\label{fig:substrate}\textbf{Optical properties of the substrates employed in this work} The upper panel displays the real part $\epsilon_1$ of the complex permittivity of Al$_2$O$_3$ (used as substrate for granular Al and NbN), MgO (substrate of some NbN films) and MgF$_2$ (substrate for CeCoIn$_5$). The lower panel displays the imaginary part $\epsilon_2$ of $MgF_2$ for various temperatures. }
\end{figure}
To model the particular behavior of $t$ and $\phi/\nu$ we use the Fresnel equations for multiple reflections \cite{Dre02}, where the thickness $d$ and dielectric function $\hat{\epsilon}(\nu)=\epsilon_1(\nu)+i\epsilon_2(\nu)$ of substrate (s) and thin film (f) directly enter
\begin{eqnarray}
t&=&t(d_\mathrm{s},\epsilon_1^\mathrm{s}(\nu),\epsilon_2^\mathrm{s}(\nu);\,d_\mathrm{f},\epsilon_1^\mathrm{f}(\nu),\epsilon_2^\mathrm{f}(\nu))\\
 \phi/\nu&=&\phi(d_\mathrm{s},\epsilon_1^\mathrm{s}(\nu),\epsilon_2^\mathrm{s}(\nu);\,d_\mathrm{f},\epsilon_1^\mathrm{f}(\nu),\epsilon_2^\mathrm{f}(\nu))/\nu
\end{eqnarray} 
Equivalently, this can be expressed in terms of the dynamical conductivity $\hat{\sigma}(\nu)=\sigma_1(\nu)+i\sigma_2(\nu)$, which is directly related to $\hat{\epsilon}$ via
\begin{equation}
\hat{\epsilon}(\nu)=1+\frac{i}{2\pi}\frac{\hat{\sigma}(\nu)}{\nu\epsilon_0}
\end{equation}
with $\epsilon_0$ the permittivity of the vacuum. With the thicknesses $d_\mathrm{f}$ and $d_\mathrm{s}$ regarded as constant (i.e. we neglect thermal contraction or expansion) this general formalism leaves 4 parameters left to be determined. Substrate materials employed in this work were Al$_2$O$_3$ (Sapphire) for granular Al and NbN films, MgO for some NbN films, and MgF$_2$ for CeCoIn$_5$, see Fig.~\ref{fig:substrate}. In case of the superconducting materials, we are only interested in the low-T properties of Al$_2$O$_3$ and MgO, covering the range up a few K. Here, the loss in the substrate can be neglected. In fact, the transmission amplitude of bare substrates is so close to unity that any fit converges to $\epsilon_2=0$ irrespective of frequency. The other part, $\epsilon_1$, turns out to be finite, yet temperature independent in the relevant regime. A slight increase with frequency is observed for both Al$_2$O$_3$ and MgO which can be attributed to absorption processes, e.g. due to optically active IR phonons. In case of the displayed Al$_2$O$_3$ curve a deviation from the smooth dispersion is observed in the low-frequency limit. This effect does not necessarily reflect an intrinsic property of the material, but is more likely to stem from disturbances in the optical transmission spectra, e.g. due to standing waves or parasitic-radiation effects. Still, from the viewpoint of a subsequent film analysis, these somewhat 'non-physical' $\epsilon_1$ values are important as the inclusion thereof in the film analysis can account for the above mentioned optical disturbances leaving alone the desired film properties. Consequently, the effective $\epsilon_1$ dispersion has to be determined for each measurement and fed into the corresponding film analysis. In case of MgF$_2$, which was used as substrate material of CeCoIn$_5$, more caution was required, as the measurements expanded over a vast temperature range 3 - 150\,K, where the temperature dependence of $\epsilon_{1,2}$ of the substrate may be non-trivial. In fact, between roughly 250 - 80\,K we find $\epsilon_1$ to feature a temperature dependence as shown in Fig.\,\ref{fig:substrate}. For lower temperatures (which are more relevant to the physical questions addressed in this work), this $T$-dependence levels of and becomes negligible. What remains is a slight increase with increasing frequency, which most likely can be attributed to IR-processes similar as in the cases of  Al$_2$O$_3$ and MgO. At elevated temperatures, also a significant dielectric loss is apparent, which is reduced continuously towards low temperatures. The peculiar dispersion of $\epsilon_2$ may strike surprising, yet it does not affect the film analysis significantly as it becomes obvious when considering the thickness of substrate ($d_\mathrm{s}$) and film ($d_\mathrm{f}$): with a ratio of $d_\mathrm{s}/d_\mathrm{f}\approx 5\times 10^{4}$ one should expect equal contributions to the overall optical response when $\epsilon_{2}$ of the film is $5\times 10^4$ higher than $\epsilon_{2}$ of MgO. At 150\,K, however, the ratio $\epsilon_\mathrm{2,f}/\epsilon_\mathrm{2,s}$ already exceeds $10^7$ and becomes even greater as temperature is reduced. Consequently, the dielectric loss can safely be neglected in the analysis of the film as it influences the absorption by less than 0.1\%. 
 
\begin{figure}[t!]
\includegraphics[width=\textwidth]{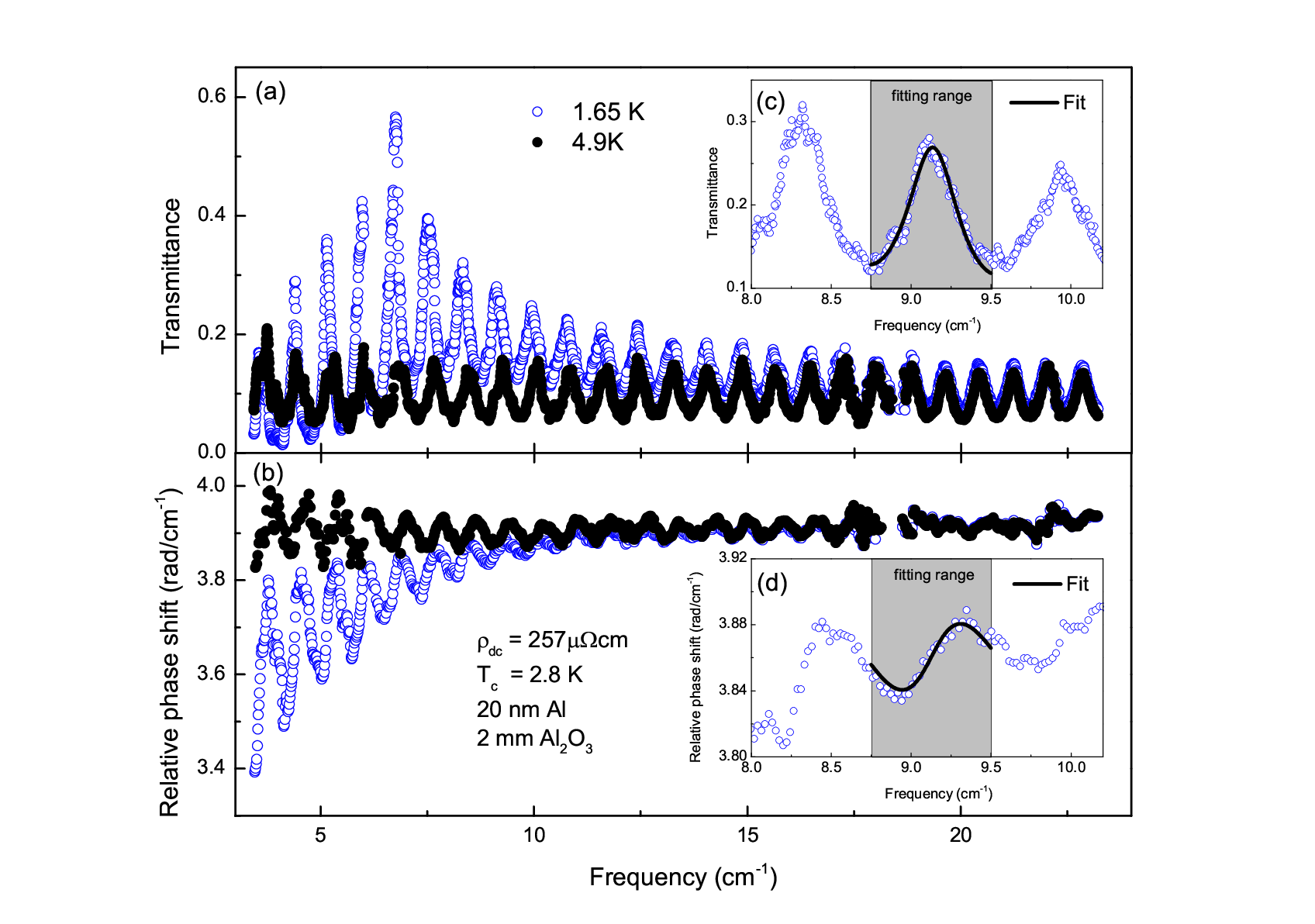}
\caption{\label{fig:raw_fit}Raw (a) transmittance $t$ and (b) relative phase shift $\phi/\nu$ spectra of a granular Al sample in the normal (black) and superconducting state (blue). The pronounced oscillation pattern arises from multiple reflections inside the substrate which acts as a Fabry-Perot resonator. Insets (c) and (d) show the same data in the superconducting state together with a fit to Fresnel equations simultaneously performed on $t$ and $\phi/\nu$) via $\sigma_{1,2}$ of the film. Note that each FP transmittance peak is fitted separately in a narrow range (gray box) to obtain the frequency dependence of $\sigma_{1,2}$ in the most unbiased way.}
\end{figure}
\begin{figure}[t!]
\includegraphics[width=\textwidth]{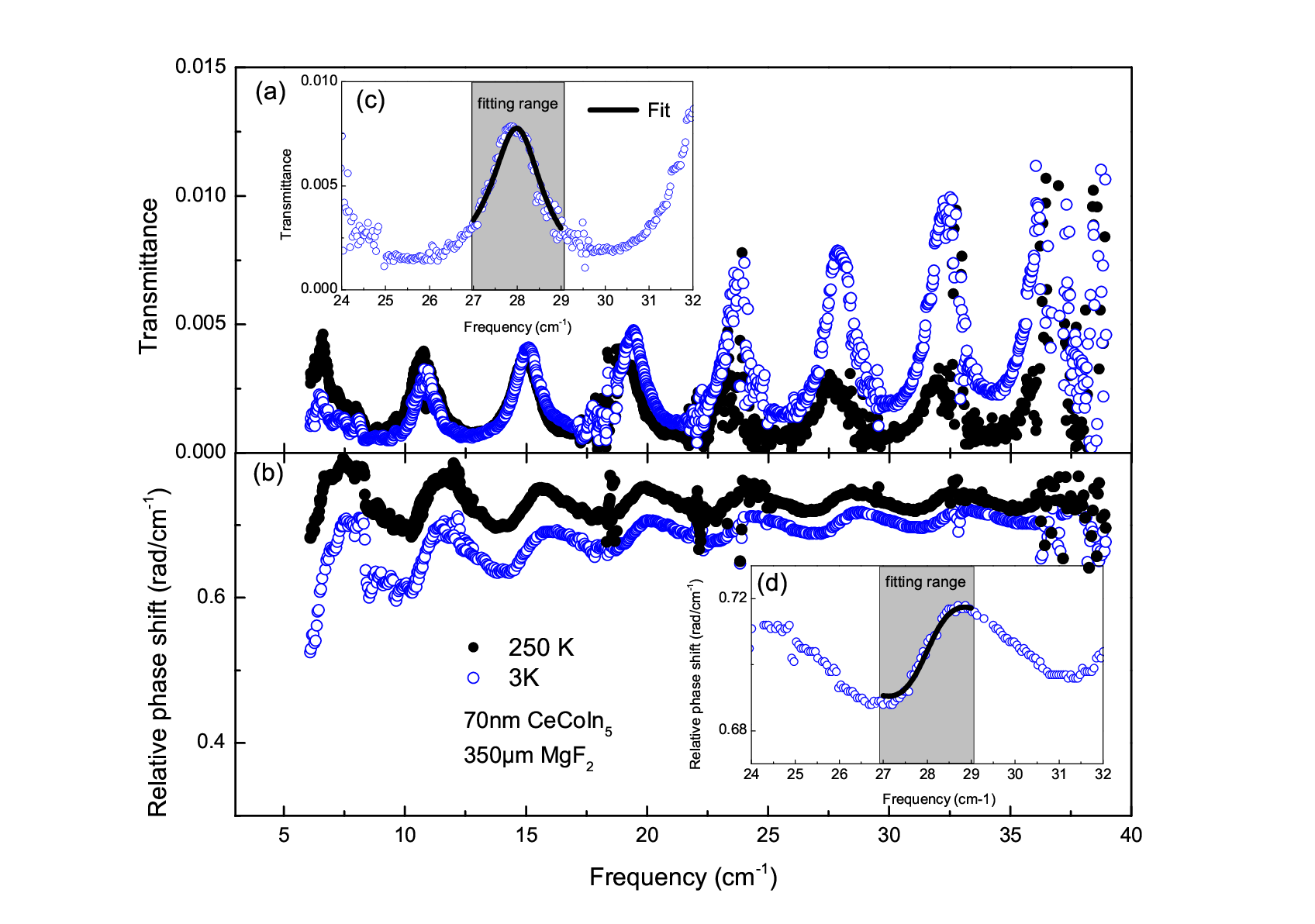}
\caption{\label{fig:raw_CeCoIn5}Raw (a) transmittance $t$ and (b) relative phase shift $\phi/\nu$ spectra of a CeCoIn$_5$ sample in the normal-metal (black) and Heavy-Fermion states (HF, blue).  Insets (c) and (d) show the same data in the HF state state together with a fit to Fresnel equations simultaneously performed on $t$ and $\phi/\nu$ via $\sigma_{1,2}$ of the film. The optical properties of the substrate for a given temperature have been inferred from an additional measurement of a bare MgF$_2$ sample and were subsequently fed into the film analysis.}
\end{figure}
\subsubsection*{Dirty-limit superconductors}
To calculate the conductivity of a superconducting film, see  Fig.~\ref{fig:raw_fit}, we process each FP peak separately: We start with the normal state, where we fit each FP peak (and the corresponding inflection range in $\phi/\nu$) in a narrow frequency window, see Fig.~\ref{fig:raw_fit}(c-d), via $\sigma_1$ of the film and $\epsilon_1$ of the substrate. This is possible because the relaxation rate is far above the THz frequency range so that $\sigma_2$ of the film is approximately zero, and the substrate is perfectly transparent at THz frequencies ($\epsilon_2=0$) as we checked beforehand with a bare reference substrate. Afterwards, we keep $\epsilon_1$ of the substrate constant, proceed with the corresponding FP peak at lower temperatures, and fit via $\sigma_{1,2}$ of the film. This procedure yields a pair of $\sigma_{1,2}$ for the center frequency of each FP peak for each temperature. Note, that this analysis does not incorporate any microscopic model and, thus, can be regarded completely unbiased.

\subsubsection*{Heavy-Fermion metals}
At elevated temperatures, the optical properties of the substrate have to be included more carefully to account for a possible frequency dependence. For a given temperature, the complex transmission of a bare reference substrate was fitted to $\epsilon_1$ and $\epsilon_2$ within each BWO spectral range and afterwards set as constant substrate parameters in the single-peak fits of the film-on-substrate analysis, see  Fig.~\ref{fig:raw_CeCoIn5}.  In case of MgF$_2$ used as substrate for the heavy-fermion thin-films discussed in this work, no measurable changes with temperatures were observed below $\sim 80$\,K. 
\clearpage{}
\section{Measurement protocols and low-$T$ characteristics}
Many measurements discussed in this work are performed at low temperatures, where the correct sensing of the actual sample temperature is a notoriously difficult task. In case of THz spectroscopy as performed in this work, the correct measurement of temperature is challenged by two main issues: First, the set up for a transmission measurement requires most of the sample to be uncovered such that a sensor usually cannot be mounted directly on the sample, and second, thermal coupling between the sample and the sensor (usually mounted on the solid brass sample holder) is established by He gas with a pressure of a few 100\,mbar. The comparably poor thermal coupling becomes even weaker at temperatures below $4.2$\,K, where $^4$He liquefies leaving even less exchange gas. As many of the measurements featured in this work have been performed below 4.2\,K, in particular the ones on granular Al, special care has been taken to characterize the low-temperature performance and quantify the error in $T$-sensing, as we will elucidate below.  \\
First, we describe the protocol for $R(T)$ and THz measurements at the base temperature $T=1.65K$  
\begin{enumerate}
\item{Place the temperature sensor either on the sample or the sample holder at the $z$-position of the aperture.}
\item{Align the sample under study horizontally with the temperature sensor}
\item{Condense liquid He into the sample chamber (inner vacuum chamber, IVC) up to $\sim 4$\,cm above the sample.}
\item{After the condensation is terminated, the sample quickly cools down to $\sim 2$\,K across the superconducting transition. During this cool down, the $R(T)$ measurement is performed.}
\item{Pumping on the IVC He bath reduces the temperature of the sample (and sensor) quickly to $\sim 1.65$\,K, where the temperature stabilization is maintained by careful control of the pumping speed. The stability reached is of the order of mK.}
\end{enumerate}
The protocol for $R(T)$ and THz measurement at $T=1.65K$ and elevated temperatures is
\begin{enumerate}
\item{Place the temperature sensor on the sample and follow steps 2-4 of the above protocol.}
\item{To raise the temperature between 2 and 4.2\,K using the PID feedback control.}
\item{Pump on the IVC He bath to lower the temperature between 2 and 1.58\,K. Switch off the the PID control and stabilize at the desired temperature by regulating the pumping speed. In this way of operation, the temperature stability is of the order of a few mK.} 
\end{enumerate}
Placing the sensor on the film is only possible for sufficiently large samples. Measurements between 2 and 4.2\,K with the sensor mounted on the sample holder are still possible, yet subject to a greater uncertainty of the temperature. The following quantification may serve as guide to estimate the temperature uncertainty for metallic thin films on insulating substrates (in what follows, 5\,nm TiN on a silicon substrate). 
\begin{figure}
\begin{centering}
\includegraphics[width=\textwidth]{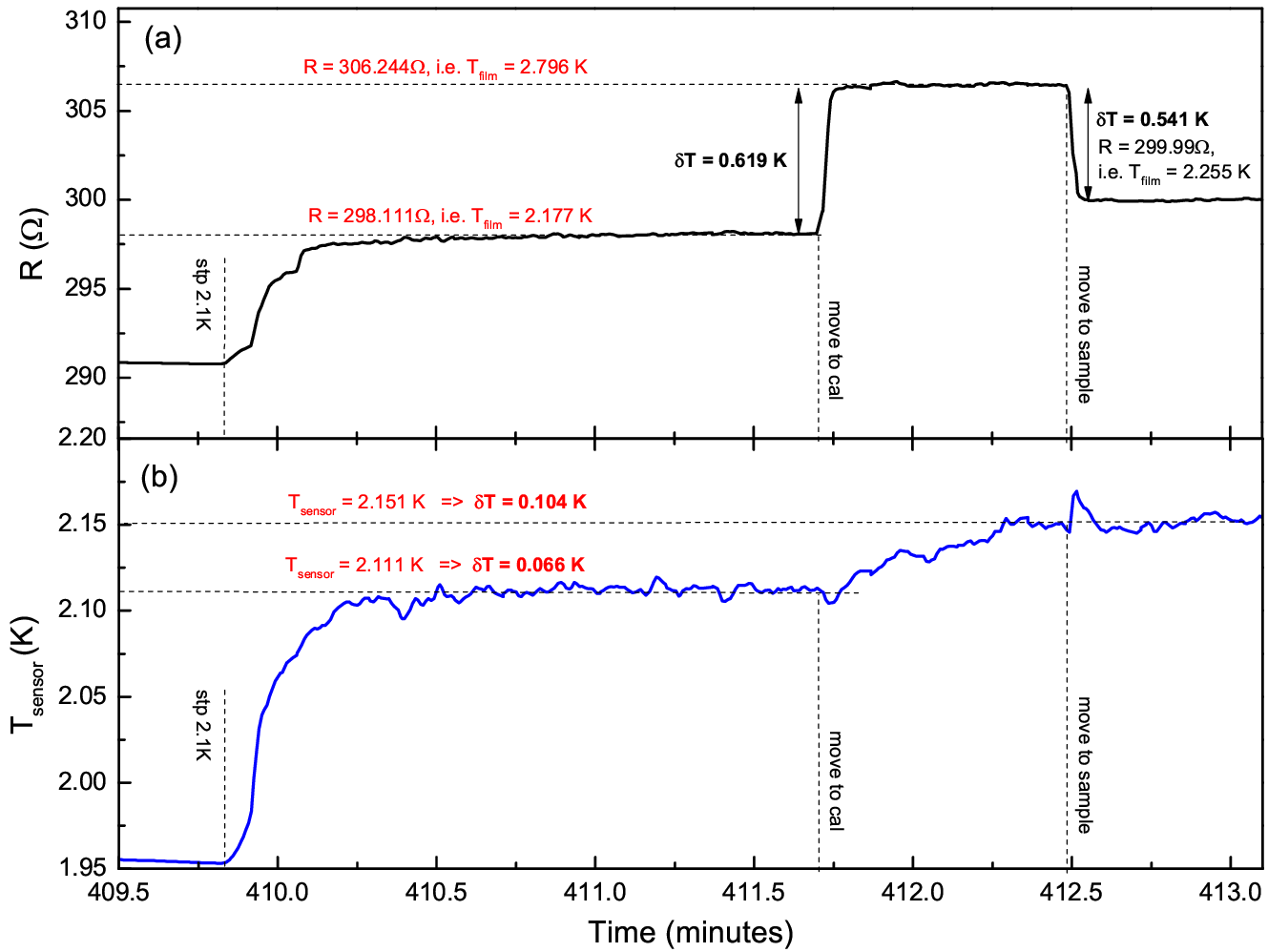}
\includegraphics[width=\textwidth]{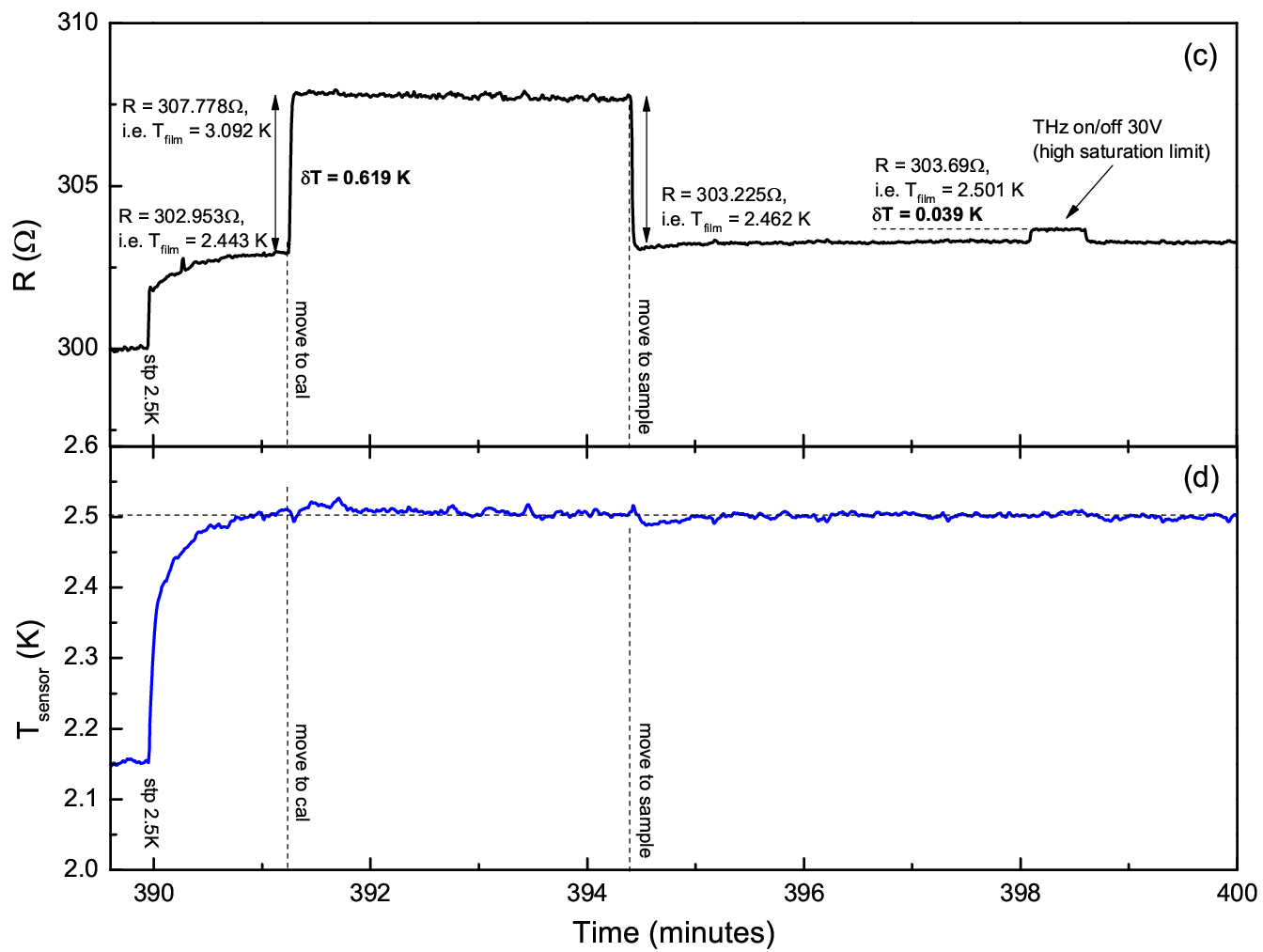}
\caption{\label{Tevolution} (a) time-evolution of resistance $R$ and inferred temperature $T_\mathrm{film}$ of a metallic thin-film test sample and (b) the temperature read-out $T_\mathrm{sensor}=2.11$\,K with the sensor mounted to the sample holder. The discrepancy between $T_\mathrm{film}$ and $T_\mathrm{sensor}$ amounts to 100\,mK. Vertical displacements lead to significant changes of the sample temperature, however, almost no time lag is observed. (c,d) For $T_\mathrm{sensor}=2.5$\,K the error in temperature amounts to roughly $\delta T = 60$\,mK. } 
\end{centering}
\end{figure}
\subsubsection{The influence of the $T$-sensor position for the $T$-measurement}
To estimate the difference between actual sample temperature and sensor readout, a $R(T)$ \emph{calibration} of the thin film was recorded between 2 and 4.2\,K with sample and sensor being thermally coupled by liquid He. After removal of the liquid He as contact medium, changes in $R$ can then directly be translated in changes in temperature. Figure \ref{Tevolution}(a,b) displays $R$ and the sensor read-out as function of time. After the temperature read-out has reached $T_\mathrm{sensor}=2.11$\,K, (set point 2.1\,K) the film stabilizes at a resistance that translates to $T_\mathrm{film}=2.18$\,K, i.e. too high by $\delta T=70$\,mK, implying that the sample is coupled more effectively to the heater than the sensor. A vertical translation of the sample by $\sim 2$\, cm upwards (i.e. moving the sample frame to calibration position) leads to a sudden rise of $\delta T = 620$\,mK.  The subsequent back shift, however, causes an almost immediate thermalization to $T_\mathrm{film} = 2.26$\,K  whilst $T_\mathrm{sensor}=2.15$\,K leaving a difference of $\delta T = 110$\,mK. Figure \ref{Tevolution}(c,d) displays the identical procedure at a set point temperature of $2.5$\,K gives a similar deviation (however with the opposite sign) $T_\mathrm{sensor}-T_\mathrm{film}=2.5\,\mathrm{K}-2.44\,\mathrm{K}=60$\,mK. Also here basically no thermal hysteresis is observed upon changing the sample position. Furthermore, Fig. \ref{Tevolution}(c) quantifies heating by THz irradiation to $\delta T =40$\,mK - however only for an intensity 10 times higher than used for spectroscopy. For realistic intensities for spectroscopic usage, the heating by THz radiation is of the order of a few mK and negligible in this temperature range. We note here again, that the crucial measurements on granular Al discussed in this work were performed with the sensor mounted directly on the sample so that the above considerations and errors in temperature do not apply.

\subsubsection{Cooling efficiency and $T$-sensing accuracy at the base temperature}
\begin{figure}[t!]
\begin{centering}
\includegraphics[width=\textwidth]{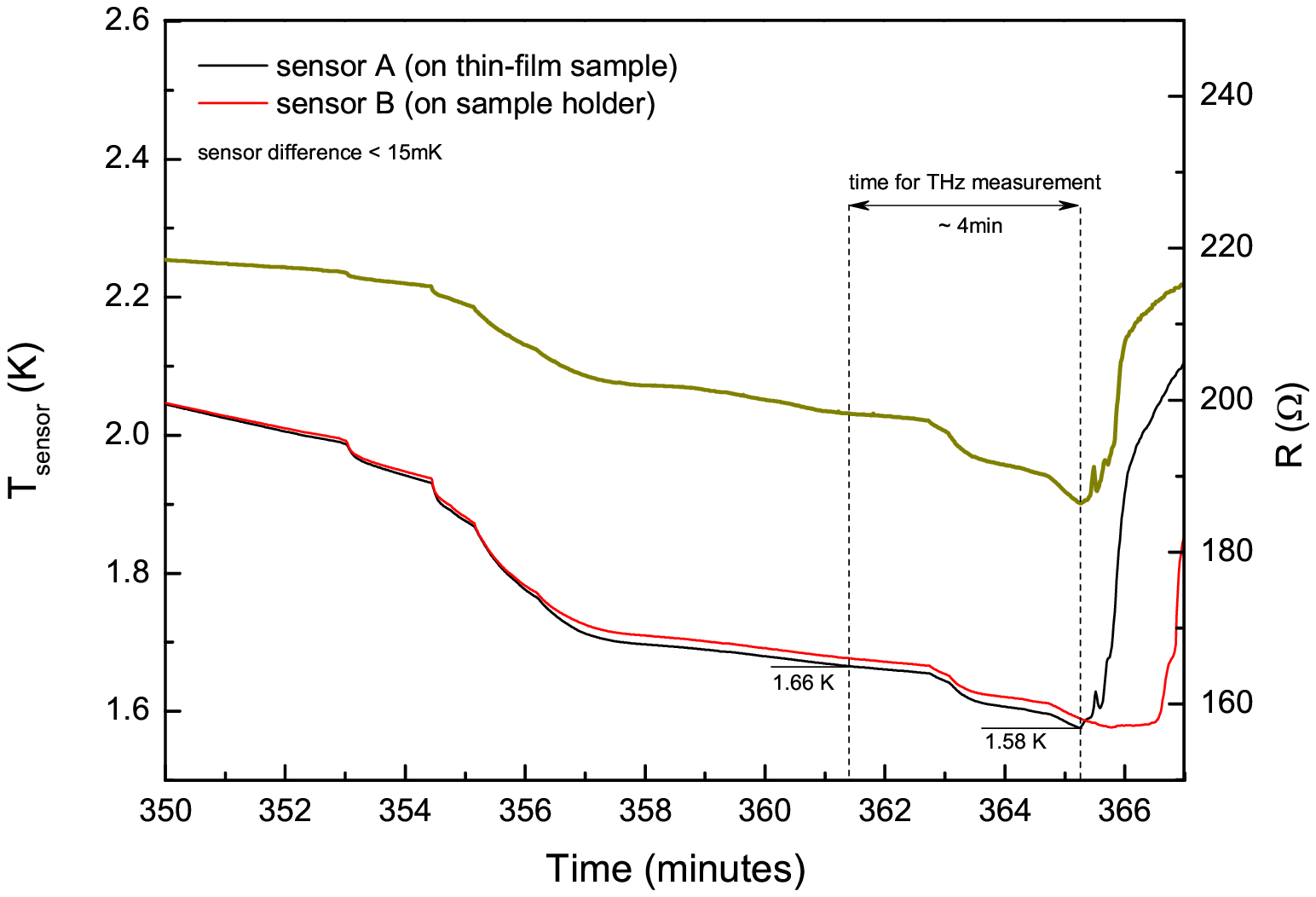}
\caption{\label{lowT_Perform} (a) time-evolution of resistance $R$ and inferred temperature $T_\mathrm{film}$ of a metallic thin-film test sample and temperature measured at the sample position (sensor A, black) and on the sample holder (B, red). The total error between actual sample temperature and $T$-measurement at the sample holder is determined to be less than 23\,mK at the lowest temperatures. } 
\end{centering}
\end{figure}

We now prove that the temperature of a thin film sample actually follows the sensor read-out and estimate the error in temperature at lowest temperatures. For validation, a sensor was mounted directly on a metallic thin-film sample and another sensor was mounted to the sample holder at the same vertical position. The time evolution of both sensor read-outs as well as the film resistance is displayed in Fig.\,\ref{lowT_Perform}. 
We can draw several important conclusions
\begin{itemize}
\item{Once the level of liquid He falls below the sample, a reduction in temperature is still observed, despite the poor thermal coupling. This might be attributed to a film of 
superfluid He thermally anchoring the sample and sensor to the He IVC-bath}
\item{The persistent cooling can also be approved for the film itself as evident from the concomitant reduction of surface resistance (Note that here we are in a regime, where $R(T)$ is reduced with decreasing temperature)}
\item{The difference in the temperature readout for the sensors being positioned either on the sample holder or on the thin-film sample amounts to less than 15\,mK when the level of liquid He has been reduced below the sample (i.e. to allow for the optical measurement. The same difference is observed to the $^4$He normal-superfluid transition at $T_\lambda$ (thin-film sensor 2.198\,K, sample holder sensor 2.183\,K) when both sensors are coupled perfectly well by superfluid He.}
\item{Taking the mean of both read-outs at the normal-superfluid transition gives a calibration uncertainty of $7.5$\,mK. Together with the above mentioned read-out difference of $15$\,mK at the base temperature, we estimate the maximum error to be less than 23\,mK between the sensor mounted on the sample holder and the thin film under study at the base temperature of 1.65\,K.}
\item{Down to the lowest temperatures, a change in $T$ leads to a simultaneously change in $R$ implying a good thermal coupling between sensor on the sample holder and the sample.}
\end{itemize}

\clearpage{}

\bibliographystyle{ieeetr}
\bibliography{thesis_literature.bib}

\begin{thebibliography}{100}

\bibitem{Lancaster2014}
T.~Lancaster and S.~T. Blundell, {\em Quantum Field Theory for the Gifted
  Amateur}, vol.~1.
\newblock OUP Oxford, 2014.

\bibitem{Cooper2011}
L.~N. Cooper and D.~Feldman, {\em BCS: 50 Years}, vol.~1.
\newblock World Scientific, 2011.

\bibitem{Gorkov58}
L.~P. Gor'kov {\em Sov. Phys. JETP}, vol.~7, p.~505, 1958.

\bibitem{Gorkov59}
L.~P. Gor'kov {\em Sov. Phys. JETP}, vol.~9, p.~1364, 1959.

\bibitem{eilenberger68}
G.~Eilenberger, ``Transformation of {G}or'kov equation for type {II}
  superconductors into transport-like equations,'' {\em Z. Physik}, vol.~214,
  p.~195, 1968.

\bibitem{Usadel70}
K.~D. Usadel, ``Generalized diffusion equation for superconducting alloys,''
  {\em Phys. Rev. Lett.}, vol.~25, pp.~507--509, Aug 1970.

\bibitem{Fominov2016}
Y.~V. Fominov private communication (2016).

\bibitem{Fominov2011}
Y.~V. Fominov, M.~Houzet, and L.~I. Glazman, ``Surface impedance of
  superconductors with weak magnetic impurities,'' {\em Phys. Rev. B}, vol.~84,
  p.~224517, Dec 2011.

\bibitem{Rammer86}
J.~Rammer and H.~Smith, ``Quantum field-theoretical methods in transport theory
  of metals,'' {\em Rev. Mod. Phys.}, vol.~58, pp.~323--359, Apr 1986.

\bibitem{Larkin86}
A.~I. Larkin and Y.~N. Ovchinnikov, {\em Nonequilibrium Superconductivity},
  vol.~1.
\newblock Elsevier, 1986.

\bibitem{Feigelman2012}
M.~V. Feigel'man and M.~A. Skvortsov, ``Universal broadening of the
  {B}ardeen-{C}ooper-{S}chrieffer coherence peak of disordered superconducting
  films,'' {\em Phys. Rev. Lett.}, vol.~109, p.~147002, Oct 2012.

\bibitem{Larkin1971}
A.~Larkin and Y.~N. Ovchinnikov, ``Density of state in inhomogeneous
  superconductors,'' {\em Zh. Eksp. Teor. Fiz.}, vol.~61, pp.~2147--2159, Nov.
  1971.

\bibitem{gant10}
V.~F. Gantmakher and V.~T. Dolgopolov, ``Superconductor--insulator quantum
  phase transition,'' {\em Phys.-Usp.}, vol.~53, no.~1, pp.~1--49, 2010.

\bibitem{goldman15}
Y.-H. {Lin}, J.~{Nelson}, and A.~M. {Goldman}, ``{Superconductivity of very
  thin films: The superconductor-insulator transition},'' {\em Physica C
  Superconductivity}, vol.~514, pp.~130--141, July 2015.

\bibitem{Dobrosavljevic2012}
V.~Dobrosavljevic, N.~Trivedi, and J.~M.~J. Valles, {\em Conductor Insulator
  Quantum Phase Transitions}.
\newblock OUP Oxford, 2012.

\bibitem{Abrahams1979}
E.~Abrahams, P.~W. Anderson, D.~C. Licciardello, and T.~V. Ramakrishnan,
  ``Scaling theory of localization: Absence of quantum diffusion in two
  dimensions,'' {\em Phys. Rev. Lett.}, vol.~42, pp.~673--676, Mar 1979.

\bibitem{anderson59}
P.~W. Anderson, ``Theory of dirty superconductors,'' {\em J. Phys. Chem.
  Solids}, vol.~11, no.~1, pp.~26--30, 1959.

\bibitem{Lee2009}
P.~Lee, ``Theory of solids 2.'' lecture notes available at http://ocw.mit.edu,
  Massachusetts Instituite of Technology, 2009.

\bibitem{Abrikosov58}
A.~A. Abrikosov and L.~P. Gor'kov {\em Zh. Eksp. Teor. Fiz.}, no.~35, p.~1558,
  1958.
\newblock [Sov. Phys. JETP \textbf{8}, 1090-1098 (1959)].

\bibitem{Abrikosov59}
A.~A. Abrikosov and L.~P. Gor'kov {\em Zh. Eksp. Teor. Fiz.}, vol.~36, p.~319,
  1959.
\newblock [Sov. Phys. JETP \textbf{9}, 220 (1959)].

\bibitem{Abrikosov60}
A.~A. Abrikosov and L.~P. Gor'kov {\em Zh. Eksp. Teor. Fiz.}, vol.~39, p.~1781,
  1960.
\newblock [Sov. Phys. JETP \textbf{12}, 1243 (1961)].

\bibitem{Fin94}
A.~M. Finkel'stein, ``Suppression of superconductivity in homogeneously
  disordered systems,'' {\em Physica B}, vol.~197, no.~1, pp.~636--648, 1994.

\bibitem{Graybeal84}
J.~M. Graybeal and M.~R. Beasley, ``Localization and interaction effects in
  ultrathin amorphous superconducting films,'' {\em Phys. Rev. B}, vol.~29,
  pp.~4167--4169, Apr 1984.

\bibitem{sacepe08}
B.~Sac{\'e}p{\'e}, C.~Chapelier, T.~I. Baturina, V.~M. Vinokur, M.~R. Baklanov,
  and M.~Sanquer, ``Disorder induced inhomogeneities of the superconducting
  state close to the superconductor-insulator transition,'' {\em Phys. Rev.
  Lett.}, vol.~101, no.~15, p.~157006, 2008.

\bibitem{Samuely2016}
P.~Szab\'o, T.~Samuely, V.~Ha\ifmmode~\check{s}\else \v{s}\fi{}kov\'a,
  J.~Ka\ifmmode \check{c}\else \v{c}\fi{}mar\ifmmode~\check{c}\else
  \v{c}\fi{}\'{\i}k, M.~\ifmmode \check{Z}\else
  \v{Z}\fi{}emli\ifmmode~\check{c}\else \v{c}\fi{}ka, M.~Grajcar, J.~G.
  Rodrigo, and P.~Samuely, ``Fermionic scenario for the destruction of
  superconductivity in ultrathin {M}o{C} films evidenced by {STM}
  measurements,'' {\em Phys. Rev. B}, vol.~93, p.~014505, Jan 2016.

\bibitem{Zemlicka2015}
M.~\ifmmode \check{Z}\else \v{Z}\fi{}emli\ifmmode~\check{c}\else \v{c}\fi{}ka,
  P.~Neilinger, M.~Trgala, M.~Reh\'ak, D.~Manca, M.~Grajcar, P.~Szab\'o,
  P.~Samuely, i.~c.~v. Ga\ifmmode~\check{z}\else \v{z}\fi{}i, U.~H\"ubner,
  V.~M. Vinokur, and E.~Il'ichev, ``Finite quasiparticle lifetime in disordered
  superconductors,'' {\em Phys. Rev. B}, vol.~92, p.~224506, Dec 2015.

\bibitem{Fisher90}
M.~P.~A. Fisher, G.~Grinstein, and S.~M. Girvin, ``Presence of quantum
  diffusion in two dimensions: Universal resistance at the
  superconductor-insulator transition,'' {\em Phys. Rev. Lett.}, vol.~64,
  pp.~587--590, Jan 1990.

\bibitem{Haviland89}
D.~B. Haviland, Y.~Liu, and A.~M. Goldman, ``Onset of superconductivity in the
  two-dimensional limit,'' {\em Phys. Rev. Lett.}, vol.~62, pp.~2180--2183, May
  1989.

\bibitem{Yazdani95}
A.~Yazdani and A.~Kapitulnik, ``Superconducting-insulating transition in
  two-dimensional $\mathit{a}$-{M}o{G}e thin films,'' {\em Phys. Rev. Lett.},
  vol.~74, pp.~3037--3040, Apr 1995.

\bibitem{kamlapure13}
A.~Kamlapure, T.~Das, S.~C. Ganguli, J.~B. Parmar, B.~S., and P.~Raychaudhuri,
  ``Emergence of nanoscale inhomogeneity in the superconducting state of a
  homogeneously disordered conventional superconductor,'' {\em Sci. Rep.},
  vol.~3, p.~2979, 2013.

\bibitem{Ghosal2001}
A.~Ghosal, M.~Randeria, and N.~Trivedi, ``Inhomogeneous pairing in highly
  disordered \textit{s} -wave superconductors,'' {\em Phys. Rev. B}, vol.~65,
  p.~014501, Nov 2001.

\bibitem{Ghosal1998}
A.~Ghosal, M.~Randeria, and N.~Trivedi, ``Role of spatial amplitude
  fluctuations in highly disordered $\mathit{s}$-wave superconductors,'' {\em
  Phys. Rev. Lett.}, vol.~81, pp.~3940--3943, Nov 1998.

\bibitem{sherman12}
D.~Sherman, G.~Kopnov, D.~Shahar, and A.~Frydman, ``Measurement of a
  superconducting energy gap in a homogeneously amorphous insulator,'' {\em
  Phys. Rev. Lett.}, vol.~108, p.~177006, Apr 2012.

\bibitem{sherman14}
D.~Sherman, B.~Gorshunov, S.~Poran, N.~Trivedi, E.~Farber, M.~Dressel, and
  A.~Frydman, ``Effect of {C}oulomb interactions on the disorder-driven
  superconductor-insulator transition,'' {\em Phys. Rev. B}, vol.~89,
  p.~035149, Jan 2014.

\bibitem{sacepe10}
B.~Sac\'ep\'e, B., C.~Chapelier, T.~I. Baturina, V.~M. Vinokur, M.~R. Baklanov,
  and M.~Sanquer, ``Pseudogap in an thin film of a conventional
  superconductor,'' {\em Nat. Commun.}, vol.~1, p.~140, 2010.

\bibitem{Benhabib2015}
S.~Benhabib, A.~Sacuto, M.~Civelli, I.~Paul, M.~Cazayous, Y.~Gallais, M.-A.
  M\'easson, R.~D. Zhong, J.~Schneeloch, G.~D. Gu, D.~Colson, and A.~Forget,
  ``Collapse of the normal-state pseudogap at a {L}ifshitz transition in the
  {B}i$_2${S}r$_2${C}a{C}u{O}$_{8+\delta}$ cuprate superconductor,'' {\em Phys.
  Rev. Lett.}, vol.~114, p.~147001, Apr 2015.

\bibitem{emery94}
V.~J. Emery and S.~A. Kivelson, ``Importance of phase fluctuations in
  superconductors with small superfluid density,'' {\em Nature}, vol.~374,
  pp.~434 -- 437, March 1994.

\bibitem{coumou13}
P.~C. J.~J. Coumou, E.~F.~C. Driessen, J.~Bueno, C.~Chapelier, and T.~M.
  Klapwijk, ``Electrodynamic response and local tunneling spectroscopy of
  strongly disordered superconducting {T}i{N} films,'' {\em Phys. Rev. B},
  vol.~88, p.~180505, Nov 2013.

\bibitem{driessen12}
E.~F.~C. Driessen, P.~C. J.~J. Coumou, R.~R. Tromp, P.~J. de~Visser, and T.~M.
  Klapwijk, ``Strongly disordered tin and {N}b{T}i{N} $s$-wave superconductors
  probed by microwave electrodynamics,'' {\em Phys. Rev. Lett.}, vol.~109,
  p.~107003, Sep 2012.

\bibitem{Chand2012PhD}
M.~Chand, {\em Transport, magneto-transport and electron tunneling studies on
  disordered superconductors}.
\newblock PhD thesis, Tata Insitute of Fundamental Research, Mumbai, 2012.

\bibitem{PratapPers}
P.~Raychaudhuri private communication, 2015.

\bibitem{Tinkham2004}
M.~Tinkham, {\em Introduction to Superconductivity}.
\newblock Dover Publications, New York, 2~ed., 2004.

\bibitem{Gazit2015}
S.~Gazit, {\em Dynamics Near Quantum Criticality in Two Space Dimensions}.
\newblock Springer Theses, 2015.

\bibitem{Chockalingam2008}
S.~P. Chockalingam, M.~Chand, J.~Jesudasan, V.~Tripathi, and P.~Raychaudhuri,
  ``Superconducting properties and {H}all effect of epitaxial {N}b{N} thin
  films,'' {\em Phys. Rev. B}, vol.~77, p.~214503, Jun 2008.

\bibitem{Sengupta2005}
K.~Sengupta and N.~Dupuis, ``Mott-insulator-to-superfluid transition in the
  {B}ose-{H}ubbard model: A strong-coupling approach,'' {\em Phys. Rev. A},
  vol.~71, p.~033629, Mar 2005.

\bibitem{Fisher1989}
M.~P.~A. Fisher, P.~B. Weichman, G.~Grinstein, and D.~S. Fisher, ``Boson
  localization and the superfluid-insulator transition,'' {\em Phys. Rev. B},
  vol.~40, pp.~546--570, Jul 1989.

\bibitem{Auerbach2015}
A.~Auerbach private communication (2015).

\bibitem{Sachdev1999}
S.~Sachdev, {\em Quantum Phase Transitions}, vol.~1.
\newblock Cambridge University Press, 1999.

\bibitem{Chakravarty1989}
S.~Chakravarty, B.~I. Halperin, and D.~R. Nelson, ``Two-dimensional quantum
  {H}eisenberg antiferromagnet at low temperatures,'' {\em Phys. Rev. B},
  vol.~39, pp.~2344--2371, Feb 1989.

\bibitem{Podolsky2011}
D.~Podolsky, A.~Auerbach, and D.~P. Arovas, ``Visibility of the amplitude
  ({H}iggs) mode in condensed matter,'' {\em Phys. Rev. B}, vol.~84, p.~174522,
  Nov 2011.

\bibitem{Gazit2013}
S.~Gazit, D.~Podolsky, and A.~Auerbach, ``Fate of the {H}iggs mode near quantum
  criticality,'' {\em Phys. Rev. Lett.}, vol.~110, p.~140401, Apr 2013.

\bibitem{Gazit2013B}
S.~Gazit, D.~Podolsky, A.~Auerbach, and D.~P. Arovas, ``Dynamics and
  conductivity near quantum criticality,'' {\em Phys. Rev. B}, vol.~88,
  p.~235108, Dec 2013.

\bibitem{Mondal13}
M.~Mondal, A.~Kamlapure, S.~C. Ganguli, J.~Jesudasan, V.~Bagwe, L.~Benfatto,
  and P.~Raychaudhuri, ``Enhancement of the finite-frequency superfluid
  response in the pseudogap regime of strongly disordered superconducting
  films,'' {\em Sci. Rep.}, vol.~3, p.~1357, 2013.

\bibitem{chock09}
S.~P. Chockalingam, M.~Chand, A.~Kamlapure, J.~Jesudasan, A.~Mishra,
  V.~Tripathi, and P.~Raychaudhuri, ``Tunneling studies in a homogeneously
  disordered $s$-wave superconductor: {N}b{N},'' {\em Phys. Rev. B}, vol.~79,
  p.~094509, Mar 2009.

\bibitem{chand2012}
M.~Chand, G.~Saraswat, A.~Kamlapure, M.~Mondal, S.~Kumar, J.~Jesudasan,
  V.~Bagwe, L.~Benfatto, V.~Tripathi, and P.~Raychaudhuri, ``Phase diagram of
  the strongly disordered $s$-wave superconductor {N}b{N} close to the
  metal-insulator transition,'' {\em Phys. Rev. B}, vol.~85, p.~014508, Jan
  2012.

\bibitem{Swa14}
M.~Swanson, Y.~L. Loh, M.~Randeria, and N.~Trivedi, ``Dynamical conductivity
  across the disorder-tuned superconductor-insulator transition,'' {\em Phys.
  Rev. X}, vol.~4, p.~021007, Apr 2014.

\bibitem{Dressel2002}
M.~Dressel, N.~Kasper, K.~Petukhov, B.~Gorshunov, G.~Gr\"uner, M.~Huth, and
  H.~Adrian, ``{{Nature of Heavy Quasiparticles in Magnetically Ordered Heavy
  Fermions ${\mathrm{UPd}}_{2}{\mathrm{Al}}_{3}$ and ${\mathrm{UPt}}_{3}$}},''
  {\em Phys. Rev. Lett.}, vol.~88, p.~186404, Apr 2002.

\bibitem{Trivedi2016}
N.~Trivedi private communication (2016).

\bibitem{sherman15}
D.~Sherman, U.~S. Pracht, B.~Gorshunov, S.~Poran, J.~Jesudasan, M.~Chand,
  P.~Raychaudhuri, M.~Swanson, N.~Trivedi, A.~Auerbach, M.~Scheffler, and
  M.~Dressel, ``The {H}iggs mode in disordered superconductors close to a
  quantum phase transition,'' {\em Nat. Phys.}, vol.~11, pp.~188--192, 2015.

\bibitem{Svistunov2014}
N.~Prokof'ev and B.~Svistunov, ``Superfluid-insulator transition in
  commensurate disordered bosonic systems: Large-scale worm algorithm
  simulations,'' {\em Phys. Rev. Lett.}, vol.~92, p.~015703, Jan 2004.

\bibitem{Cea2015}
T.~Cea, C.~Castellani, G.~Seibold, and L.~Benfatto, ``Nonrelativistic dynamics
  of the amplitude ({H}iggs) mode in superconductors,'' {\em Phys. Rev. Lett.},
  vol.~115, p.~157002, Oct 2015.

\bibitem{cea14}
T.~Cea, D.~Bucheli, G.~Seibold, L.~Benfatto, J.~Lorenzana, and C.~Castellani,
  ``Optical excitation of phase modes in strongly disordered superconductors,''
  {\em Phys. Rev. B}, vol.~89, p.~174506, May 2014.

\bibitem{Cheng2016}
B.~Cheng, L.~Wu, N.~J. Laurita, H.~Singh, M.~Chand, P.~Raychaudhuri, and N.~P.
  Armitage, ``Anomalous gap-edge dissipation in disordered superconductors on
  the brink of localization,'' {\em Phys. Rev. B}, vol.~93, p.~180511, May
  2016.

\bibitem{matsunaga13}
R.~Matsunaga, Y.~I. Hamada, K.~Makise, Y.~Uzawa, H.~Terai, Z.~Wang, and
  R.~Shimano, ``Higgs amplitude mode in the {BCS} superconductors
  {N}b$_{1-x}${T}i$_x${N} induced by terahertz pulse excitation,'' {\em Phys.
  Rev. Lett.}, vol.~111, p.~057002, Jul 2013.

\bibitem{Manske2015}
D.~Manske private communication (2016).

\bibitem{Carruthers1968}
P.~Carruthers and M.~M. Nieto, ``Phase and angle variables in quantum
  mechanics,'' {\em Rev. Mod. Phys.}, vol.~40, pp.~411--440, Apr 1968.

\bibitem{mayoh14}
J.~Mayoh and A.~M. Garc\'ia-Garc\'ia, ``Strong enhancement of bulk
  superconductivity by engineered nanogranularity,'' {\em Phys. Rev. B},
  vol.~90, p.~134513, Oct 2014.

\bibitem{Garcia2008}
A.~M. Garc\'{\i}a-Garc\'{\i}a, J.~D. Urbina, E.~A. Yuzbashyan, K.~Richter, and
  B.~L. Altshuler, ``Bardeen-{C}ooper-{S}chrieffer theory of finite-size
  superconducting metallic grains,'' {\em Phys. Rev. Lett.}, vol.~100,
  p.~187001, May 2008.

\bibitem{Olofsson2008}
H.~Olofsson, S.~\AA{}berg, and P.~Leboeuf, ``Semiclassical theory of
  {B}ardeen-{C}ooper-{S}chrieffer pairing-gap fluctuations,'' {\em Phys. Rev.
  Lett.}, vol.~100, p.~037005, Jan 2008.

\bibitem{Blatt1963}
J.~M. Blatt and C.~J. Thompson, ``Shape resonances in superconducting thin
  films,'' {\em Phys. Rev. Lett.}, vol.~10, pp.~332--334, Apr 1963.

\bibitem{Parmenter68b}
R.~H. Parmenter, ``Characteristic parameters of a granular superconductor,''
  {\em Phys. Rev.}, vol.~167, pp.~387--392, Mar 1968.

\bibitem{Parmenter68}
R.~H. Parmenter, ``Size effect in a granular superconductor,'' {\em Phys.
  Rev.}, vol.~166, pp.~392--396, Feb 1968.

\bibitem{bose10}
S.~Bose, A.~M. Garc\'{\i}a-Garc\'{\i}a, M.~M. Ugeda, J.~D. Urbina, C.~H.
  Michaelis, I.~Brihuega, and K.~Kern, ``Observation of shell effects in
  superconducting nanoparticles of {S}n,'' {\em Nat Mater}, vol.~9,
  pp.~550--554, 07 2010.

\bibitem{Bachar2014PhD}
N.~Bachar, {\em Spin-flip scattering in superconducting granular Aluminum
  films}.
\newblock PhD thesis, Tel Aviv University, 2014.

\bibitem{Deu73}
G.~Deutscher, M.~Gershenson, E.~Grunbaum, and I.~Y., ``Granular superconducting
  films,'' {\em Journal of Vacuum Science and Technology}, vol.~10, p.~697,
  Sept. 1973.

\bibitem{Shapira68}
Y.~Shapira and G.~Deutscher, ``Metal-insulator transition in composite thin
  films,'' {\em Thin Dolid Films}, vol.~87, pp.~444--450, Apr. 1968.

\bibitem{Deutscher73}
G.~Deutscher, H.~Fenichel, M.~Gershenson, E.~Grunbaum, and Z.~Ovadyahu,
  ``Transition to zero dimensionality in granular aluminum superconducting
  films,'' {\em Journal of Low Temperature Physics}, vol.~10, pp.~231--234,
  Jan. 1973.

\bibitem{Bac15}
N.~Bachar, S.~Lerer, A.~Levy, S.~Hacohen-Gourgy, B.~Almog, H.~Saadaoui,
  Z.~Salman, E.~Morenzoni, and G.~Deutscher, ``Mott transition in granular
  aluminum,'' {\em Phys. Rev. B}, vol.~91, p.~041123, Jan 2015.

\bibitem{Basov2011}
D.~N. Basov, R.~D. Averitt, D.~van~der Marel, M.~Dressel, and K.~Haule,
  ``Electrodynamics of correlated electron materials,'' {\em Rev. Mod. Phys.},
  vol.~83, pp.~471--541, Jun 2011.

\bibitem{Mathur98}
N.~D. Mathur, F.~M. Frosche, S.~R. Julian, I.~R. Walker, D.~M. Freye, R.~K.~W.
  Haselwimmer, and G.~G. Lonzarich, ``Magnetically mediated superconductivity
  in heavy fermion compounds,'' {\em Nature}, vol.~394, no.~6688, pp.~39--43,
  1998.

\bibitem{Caviglia08}
A.~D. Caviglia, S.~Gariglio, N.~Reyren, D.~Jaccard, T.~Schneider, M.~Gabay,
  S.~Thiel, G.~Hammerl, J.~Mannhart, and J.~M. Triscone, ``Electric field
  control of the {L}a{A}l{O}$_3$/{S}r{T}i{O}$_3$ interface ground state,'' {\em
  Nature}, pp.~624--627, July 2008.

\bibitem{Dumm2009}
M.~Dumm, D.~Faltermeier, N.~Drichko, M.~Dressel, C.~M\'ezi\`ere, and P.~Batail,
  ``Bandwidth-controlled {M}ott transition in
  $\kappa-$({BEDT-TTF})$_2${C}u[{N}({CN})$_2$]{B}r$_x${C}l$_{1-x}$: Optical
  studies of correlated carriers,'' {\em Phys. Rev. B}, vol.~79, p.~195106, May
  2009.

\bibitem{Bac13}
N.~Bachar, S.~Lerer, S.~Hacohen-Gourgy, B.~Almog, and G.~Deutscher,
  ``Kondo-like behavior near the metal-to-insulator transition of nanoscale
  granular aluminum,'' {\em Phys. Rev. B}, vol.~87, p.~214512, Jun 2013.

\bibitem{Bac14b}
S.~Lerer, N.~Bachar, G.~Deutscher, and Y.~Dagan, ``Nernst effect beyond the
  coherence critical field of a nanoscale granular superconductor,'' {\em Phys.
  Rev. B}, vol.~90, p.~214521, Dec 2014.

\bibitem{VarlamovBook}
A.~I. Larkin and A.~Varlamov, {\em Theory of Fluctuations in Superconductors}.
\newblock Clarendon Press, Oxford, 2005.

\bibitem{caprara2005}
S.~Caprara, M.~Grilli, B.~Leridon, and J.~Lesueur, ``Extended paraconductivity
  regime in underdoped cuprates,'' {\em Phys. Rev. B}, vol.~72, p.~104509, Sep
  2005.

\bibitem{varlamov2011}
A.~Levchenko, M.~R. Norman, and A.~A. Varlamov, ``Nernst effect from
  fluctuating pairs in the pseudogap phase of the cuprates,'' {\em Phys. Rev.
  B}, vol.~83, p.~020506, Jan 2011.

\bibitem{mb58}
D.~C. Mattis and J.~Bardeen, ``Theory of the anomalous skin effect in normal
  and superconducting metals,'' {\em Phys. Rev.}, vol.~111, no.~2,
  pp.~412--417, 1958.

\bibitem{Pracht2016}
U.~S. Pracht, J.~Simmendinger, M.~Dressel, R.~Endo, T.~Watashige, Y.~Hanaoka,
  M.~Shimozawa, T.~Terashima, T.~Shibauchi, Y.~Matsuda, and M.~Scheffler,
  ``Charge carrier dynamics of the heavy-fermion metal {C}e{C}o{I}n$_5$ probed
  by {TH}z spectroscopy,'' {\em Journal of Magnetism and Magnetic Materials},
  vol.~400, pp.~31 -- 35, 2016.

\bibitem{crane2007}
R.~Crane, N.~P. Armitage, A.~Johansson, G.~Sambandamurthy, D.~Shahar, and
  G.~Gr\"uner, ``Survival of superconducting correlations across the
  two-dimensional superconductor-insulator transition: A finite-frequency
  study,'' {\em Phys. Rev. B}, vol.~75, p.~184530, May 2007.

\bibitem{corson2000}
J.~Corson, J.~Orenstein, S.~Oh, J.~O'Donnell, and J.~N. Eckstein, ``Nodal
  quasiparticle lifetime in the superconducting state of
  ${\mathrm{bi}}_{2}{\mathrm{sr}}_{2}{\mathrm{cacu}}_{2}{O}_{8+\mathit{\delta}}$,''
  {\em Phys. Rev. Lett.}, vol.~85, pp.~2569--2572, Sep 2000.

\bibitem{stroud2000}
S.~Barabash, D.~Stroud, and I.-J. Hwang, ``Conductivity due to classical phase
  fluctuations in a model for high-${T}_{c}$ superconductors,'' {\em Phys. Rev.
  B}, vol.~61, pp.~R14924--R14927, Jun 2000.

\bibitem{Ben09}
L.~Benfatto, C.~Castellani, and T.~Giamarchi, ``Broadening of the
  {B}erezinskii-{K}osterlitz-{T}houless superconducting transition by
  inhomogeneity and finite-size effects,'' {\em Phys. Rev. B}, vol.~80,
  p.~214506, Dec 2009.

\bibitem{Deu77}
G.~Deutscher and S.~A. Dodds, ``Critical-field anisotropy and fluctuation
  conductivity in granular aluminum films,'' {\em Phys. Rev. B}, vol.~16,
  pp.~3936--3942, Nov 1977.

\bibitem{Likharev1979}
K.~K. Likharev, ``{Superconducting weak links},'' {\em Rev. Mod. Phys.},
  vol.~51, pp.~101--159, jan 1979.

\bibitem{Abr78}
D.~Abraham, G.~Deutscher, R.~Rosenbaum, and S.~Wolf, ``Magnetic penetration
  depth in granular {A}l-{A}l$_2${O}$_3$ films,'' {\em J. Phys. Coll. Paris
  (France)}, vol.~39, pp.~586--587, 1978.

\bibitem{Ger82}
M.~Gershenson and W.~McLean, ``Magnetic susceptibility of superconducting
  granular aluminum,'' {\em Low Temp. Phys.}, vol.~47, no.~1-2, pp.~123--136,
  1982.

\bibitem{Geshkenbein1997}
V.~B. Geshkenbein, L.~B. Ioffe, and A.~I. Larkin, ``Superconductivity in a
  system with preformed pairs,'' {\em Phys. Rev. B}, vol.~55, pp.~3173--3180,
  Feb 1997.

\bibitem{Perali2000}
A.~Perali, C.~Castellani, C.~Di~Castro, M.~Grilli, E.~Piegari, and A.~A.
  Varlamov, ``Two-gap model for underdoped cuprate superconductors,'' {\em
  Phys. Rev. B}, vol.~62, pp.~R9295--R9298, Oct 2000.

\bibitem{Ranninger1995}
J.~Ranninger, J.~M. Robin, and M.~Eschrig, ``Superfluid precursor effects in a
  model of hybridized bosons and fermions,'' {\em Phys. Rev. Lett.}, vol.~74,
  pp.~4027--4030, May 1995.

\bibitem{Ranninger1996}
J.~Ranninger and J.~M. Robin, ``Manifestations of the pseudogap in the
  boson-fermion model for {B}ose-{E}instein-condensation-driven
  superconductivity,'' {\em Phys. Rev. B}, vol.~53, pp.~R11961--R11963, May
  1996.

\bibitem{Seibold2017}
G.~Seibold, L.~Benfatto, and C.~Castellani, ``On the application of
  mattis-bardeen theory in strongly disordered superconductors.''
  arXiv:1702.01610, 2017.

\bibitem{Altshuler85}
B.~L. Altshuler, A.~G. Aronov, A.~L. Efros, and M.~Pollak, eds., {\em
  Electron-Electron interactions in disordered systems}.
\newblock North Holland: Elsevier, 1985.

\bibitem{Hashimoto2014}
M.~Hashimoto, I.~M. Vishik, R.-H. He, T.~P. Devereaux, and Z.-X. Shen, ``Energy
  gaps in high-transition-temperature cuprate superconductors,'' {\em Nature
  Physics}, vol.~10, pp.~483--495, 2014.

\bibitem{Aoki83}
H.~Aoki, ``Critical behaviour of extended states in disordered systems,'' {\em
  Journal of Physics C: Solid State Physics}, vol.~16, pp.~205--208, 1983.

\bibitem{Feigelman2007}
M.~V. Feigelman, L.~B. Ioffe, V.~E. Kravtsov, and E.~A. Yuzbashyan,
  ``Eigenfunction fractality and pseudogap state near the
  superconductor-insulator transition,'' {\em Phys. Rev. Lett.}, vol.~98,
  p.~027001, Jan 2007.

\bibitem{Feigelman2010}
M.~Feigelman, L.~Ioffe, V.~Kravtsov, and E.~Cuevas, ``Fractal superconductivity
  near localization threshold,'' {\em Annals of Physics}, vol.~325, no.~7,
  pp.~1390 -- 1478, 2010.
\newblock July 2010 Special Issue.

\bibitem{Mayoh2015}
J.~Mayoh and A.~M. Garc\'{\i}a-Garc\'{\i}a, ``Global critical temperature in
  disordered superconductors with weak multifractality,'' {\em Phys. Rev. B},
  vol.~92, p.~174526, Nov 2015.

\bibitem{GarciaPers}
A.~M. Garc\'ia-Garc\'ia private communication (2016).

\bibitem{Anderson58}
P.~W. Anderson, ``Absence of {D}iffusion in {C}ertain {R}andom {L}attices,''
  {\em Phys. Rev.}, vol.~109, pp.~1492--1505, Mar 1958.

\bibitem{Schrieffer1964}
R.~S. Schrieffer, {\em Theory of Superconductivity}.
\newblock Redwood City, CA: Addison-Wesley, 1964.

\bibitem{Grenet07}
T.~Grenet, J.~Delahaye, M.~Sabra, and F.~Gay, ``{Anomalous electric-field
  effect and glassy behaviour in granular aluminium: electron glass?},'' in
  {\em {12th International Conference on Transport in Interacting Disordered
  Systems (TIDS-12)}}, vol.~5, (Marburg, Germany), p.~680, {Wiley}, Aug. 2007.

\bibitem{Delahaye08}
J.~Delahaye and T.~Grenet, ``{Electrical glassy behavior in granular aluminium
  thin films},'' in {\em {5th International Conference on Electronic Crystals
  (ECRYS)}}, vol.~404, (Carg{\`e}se, France), p.~470, {Elsevier}, Aug. 2008.

\bibitem{Delahaye09}
J.~Delahaye and T.~Grenet, ``{Ageing in granular aluminium insulating thin
  films},'' in {\em {13th Conference on Transport and Interactions in
  Disordered Systems (TIDS 13)}}, vol.~18, (Rackeve, Hungary), p.~830, {Wiley},
  Aug. 2009.

\bibitem{Delahaye11}
J.~Delahaye, J.~Honor{\'e}, and T.~Grenet, ``{Slow Conductance Relaxation in
  Insulating Granular Al: Evidence for Screening Effects.},'' {\em {Physical
  Review Letters}}, vol.~106, p.~186602, May 2011.

\bibitem{Ma1985}
M.~Ma and P.~A. Lee, ``Localized superconductors,'' {\em Phys. Rev. B},
  vol.~32, pp.~5658--5667, Nov 1985.

\bibitem{Lemarie2013}
G.~Lemari\'e, A.~Kamlapure, D.~Bucheli, L.~Benfatto, J.~Lorenzana, G.~Seibold,
  S.~C. Ganguli, P.~Raychaudhuri, and C.~Castellani, ``Universal scaling of the
  order-parameter distribution in strongly disordered superconductors,'' {\em
  Phys. Rev. B}, vol.~87, p.~184509, May 2013.

\bibitem{Rama1989}
T.~V. Ramakrishnan, ``Superconductivity in disordered thin films,'' {\em
  Physica Scripta}, vol.~T27, pp.~24--30, 1989.

\bibitem{LaraPers}
L.~Benfatto private communication (2016).

\bibitem{Shiba1968}
H.~Shiba, ``Classical spins in superconductors,'' {\em Prog. Theor. Phys.},
  vol.~40, no.~3, pp.~435--451, 1968.

\bibitem{Rusinov1969}
A.~I. Rusinov, ``On the theory of gapless superconductivity in alloys
  containing paramagnetic impurites,'' {\em Sov. Phys. JETP}, vol.~29, p.~1101,
  1969.

\bibitem{Balatsky2006}
A.~V. Balatsky, I.~Vekhter, and J.-X. Zhu, ``Impurity-induced states in
  conventional and unconventional superconductors,'' {\em Rev. Mod. Phys.},
  vol.~78, pp.~373--433, May 2006.

\bibitem{Strongin1968}
M.~Strongin and O.~F. Kammerer, ``Superconductive phenomena in ultrathin
  films,'' {\em J. Appl. Phys.}, vol.~39, p.~2509, 1968.

\bibitem{Sixl1974}
H.~Sixl, J.~Gromer, and H.~C. Wolf, ``Tunnelspektroskopie an dünnsten
  supraleitenden {A}luminium-, {I}ndium- und {B}leischichten,'' {\em Z.
  Naturforsch.}, vol.~29a, pp.~319--331, 1974.

\bibitem{Bianchi2003}
A.~Bianchi, R.~Movshovich, C.~Capan, P.~G. Pagliuso, and J.~L. Sarrao,
  ``{{Possible Fulde-Ferrell-Larkin-Ovchinnikov Superconducting State in
  ${\mathrm{C}\mathrm{e}\mathrm{C}\mathrm{o}\mathrm{I}\mathrm{n}}_{5}$}},''
  {\em Phys. Rev. Lett.}, vol.~91, p.~187004, Oct 2003.

\bibitem{Kondo1964}
J.~Kondo, ``Resistance minimum in dilute magnetic alloys,'' {\em Progress of
  Theoretical Physics}, vol.~32, no.~1, pp.~37--49, 1964.

\bibitem{Mott1974}
N.~F. Mott, ``Rare-earth compounds with mixed valencies,'' {\em Philosophical
  Magazine}, vol.~30, no.~2, p.~403, 1974.

\bibitem{Doniach1977}
S.~Doniach, ``Kondo lattice and weak antiferromagnetism,'' {\em Physica},
  vol.~91B, p.~231, 1977.

\bibitem{Sidorov2002}
V.~A. Sidorov, M.~Nicklas, P.~G. Pagliuso, J.~L. Sarrao, Y.~Bang, A.~V.
  Balatsky, and J.~D. Thompson, ``Superconductivity and quantum criticality in
  $\mathrm{C}\mathrm{e}\mathrm{C}\mathrm{o}\mathrm{I}{\mathrm{n}}_{\mathrm{5}}$,''
  {\em Phys. Rev. Lett.}, vol.~89, p.~157004, Sep 2002.

\bibitem{Nakatsuji2002}
S.~Nakatsuji, S.~Yeo, L.~Balicas, Z.~Fisk, P.~Schlottmann, P.~G. Pagliuso,
  N.~O. Moreno, J.~L. Sarrao, and J.~D. Thompson, ``Intersite {C}oupling
  {E}ffects in a {K}ondo {L}attice,'' {\em Phys. Rev. Lett.}, vol.~89,
  p.~106402, Aug 2002.

\bibitem{Kim2001}
J.~S. Kim, J.~Alwood, G.~R. Stewart, J.~L. Sarrao, and J.~D. Thompson,
  ``Specific heat in high magnetic fields and non-{F}ermi-liquid behavior in
  {C}e\emph{{M}e}{I}n$_5$ (\emph{{M}e}={I}r, {C}o),'' {\em Phys. Rev. B},
  vol.~64, p.~134524, Sep 2001.

\bibitem{Kawasaki2008}
Y.~Kawasaki, S.~Kawasaki, M.~Yashima, T.~Mito, G.~qing Zheng, Y.~Kitaoka,
  H.~Shishido, R.~Settai, Y.~Haga, and Y.~ÅŒnuki, ``{A}nisotropic {S}pin
  {F}luctuations in {H}eavy-{F}ermion {S}uperconductor {C}e{C}o{I}n$_5$:
  {I}n-{NQR} and {C}o-{NMR} {S}tudies,'' {\em Journal of the Physical Society
  of Japan}, vol.~72, no.~9, pp.~2308--2311, 2003.

\bibitem{Ronning2005}
F.~Ronning, C.~Capan, A.~Bianchi, R.~Movshovich, A.~Lacerda, M.~F. Hundley,
  J.~D. Thompson, P.~G. Pagliuso, and J.~L. Sarrao, ``Field-tuned quantum
  critical point in {C}e{C}o{I}n$_5$ near the superconducting upper critical
  field,'' {\em Phys. Rev. B}, vol.~71, p.~104528, Mar 2005.

\bibitem{Donath2008}
J.~G. Donath, F.~Steglich, E.~D. Bauer, J.~L. Sarrao, and P.~Gegenwart,
  ``Dimensional crossover of quantum critical behavior in {C}e{C}o{I}n$_5$,''
  {\em Phys. Rev. Lett.}, vol.~100, p.~136401, Mar 2008.

\bibitem{Singh2007}
S.~Singh, C.~Capan, M.~Nicklas, M.~Rams, A.~Gladun, H.~Lee, J.~F. DiTusa,
  Z.~Fisk, F.~Steglich, and S.~Wirth, ``Probing the {Q}uantum {C}ritical
  {B}ehavior of {C}e{C}o{I}n$_5$ via {H}all {E}ffect {M}easurements,'' {\em
  Phys. Rev. Lett.}, vol.~98, p.~057001, Jan 2007.

\bibitem{Zaum2011}
S.~Zaum, K.~Grube, R.~Sch\"afer, E.~D. Bauer, J.~D. Thompson, and
  H.~v.~L\"ohneysen, ``Towards the {I}dentification of a {Q}uantum {C}ritical
  {L}ine in the ($p$, $b$) {P}hase {D}iagram of {C}e{C}o{I}n$_5$ with
  {T}hermal-{E}xpansion {M}easurements,'' {\em Phys. Rev. Lett.}, vol.~106,
  p.~087003, Feb 2011.

\bibitem{Malinowski2005}
A.~Malinowski, M.~F. Hundley, C.~Capan, F.~Ronning, R.~Movshovich, N.~O.
  Moreno, J.~L. Sarrao, and J.~D. Thompson, ``$c$-axis magnetotransport in
  $\mathrm{Ce}\mathrm{Co}{\mathrm{in}}_{5}$,'' {\em Phys. Rev. B}, vol.~72,
  p.~184506, Nov 2005.

\bibitem{Tokiwa2013}
Y.~Tokiwa, E.~D. Bauer, and P.~Gegenwart, ``Zero-field quantum critical point
  in {C}e{C}o{I}n$_5$,'' {\em Phys. Rev. Lett.}, vol.~111, p.~107003, Sep 2013.

\bibitem{Ormeno2002a}
R.~J. Ormeno, A.~Sibley, C.~E. Gough, S.~Sebastian, and I.~R. Fisher,
  ``Microwave conductivity and penetration depth in the heavy fermion
  superconductor ${\mathrm{cecoin}}_{5}$,'' {\em Phys. Rev. Lett.}, vol.~88,
  p.~047005, Jan 2002.

\bibitem{Nevirkovets2008a}
I.~Nevirkovets, O.~Chernyashevskyy, C.~Petrovic, J.~Ketterson, and B.~K. Sarma,
  ``Microwave absorption measurements of the heavy-fermion superconductor
  {C}e{C}o{I}n$_5$,'' {\em Physica C: Superconductivity}, vol.~468, no.~5,
  pp.~432 -- 434, 2008.

\bibitem{SudhakarRao2009}
G.~V. Sudhakar~Rao, S.~Ocadlik, M.~Reedyk, and C.~Petrovic, ``Low-frequency
  excitation in the optical properties of superconducting {C}e{C}o{I}n$_5$,''
  {\em Phys. Rev. B}, vol.~80, p.~064512, Aug 2009.

\bibitem{Truncik2013a}
C.~J.~S. Truncik, W.~A. Huttema, P.~J. Turner, S.~\"Ozcan, N.~C. Murphy, P.~R.
  Carri\`{e}re, E.~Thewalt, K.~J. Morse, A.~J. Koenig, J.~L. Sarrao, and D.~M.
  Broun, ``Nodal quasiparticle dynamics in the heavy fermion superconductor
  {C}e{C}o{I}n$_5$ revealed by precision microwave spectroscopy,'' {\em Nat.
  Commun.}, vol.~4, p.~2477, 2013.

\bibitem{singley2002}
E.~J. Singley, D.~N. Basov, E.~D. Bauer, and M.~B. Maple, ``Optical
  conductivity of the heavy fermion superconductor {C}e{C}o{I}n$_5$,'' {\em
  Phys. Rev. B}, vol.~65, p.~161101, Apr 2002.

\bibitem{mena2005}
F.~P. Mena, D.~van~der Marel, and J.~L. Sarrao, ``Optical conductivity of
  {C}e\emph{{M}}{I}n$_5$ (\emph{{M}}={C}o, {R}h, {I}r),'' {\em Phys. Rev. B},
  vol.~72, p.~045119, Jul 2005.

\bibitem{Scheffler2013a}
M.~Scheffler, K.~Schlegel, C.~Clauss, D.~Hafner, C.~Fella, M.~Dressel,
  M.~Jourdan, J.~Sichelschmidt, C.~Krellner, C.~Geibel, and F.~Steglich,
  ``{{Microwave spectroscopy on heavy-fermion systems: Probing the dynamics of
  charges and magnetic moments}},'' {\em Phys. Status Solidi B}, vol.~250,
  no.~3, pp.~439--449, 2013.

\bibitem{Scheffler2005}
M.~Scheffler, M.~Dressel, M.~Jourdan, and H.~Adrian, ``Extremely slow drude
  relaxation of correlated electrons,'' {\em Nature (London)}, vol.~438,
  no.~7071, pp.~1135--1137, 2005.

\bibitem{Ostertag2011}
J.~P. Ostertag, M.~Scheffler, M.~Dressel, and M.~Jourdan, ``Terahertz
  conductivity of the heavy-fermion compound {UN}i$_2${A}l${3}$,'' {\em Phys.
  Rev. B}, vol.~84, p.~035132, Jul 2011.

\bibitem{Scheffler2013}
M.~Scheffler, T.~Weig, M.~Dressel, H.~Shishido, Y.~Mizukami, T.~Terashima,
  T.~Shibauchi, and Y.~Matsuda, ``Terahertz {C}onductivity of the
  {H}eavy-{F}ermion {S}tate in {C}e{C}o{I}n$_5$,'' {\em Journal of the Physical
  Society of Japan}, vol.~82, no.~4, p.~043712, 2013.

\bibitem{Dre02}
M.~Dressel and G.~Gr\"{u}ner, {\em Electrodynamics of Solids - Optical
  Properties of Electrons in Matter}.
\newblock Cambridge University Press, Cambridge, 2002.

\bibitem{Gotze1977}
W.~G\"otze and P.~W\"olfle, ``Homogeneous dynamical conductivity of simple
  metals,'' {\em Phys. Rev. B}, vol.~6, pp.~1226--1238, Aug 1972.

\bibitem{Kub057}
R.~Kubo, ``Statistical-mechanical theory of irreversible processes. i. general
  theory and simple applications to magnetic and conduction problems,'' {\em J.
  Phys. Soc. Jpn.}, vol.~12, pp.~570--586, 1957.

\bibitem{BasovRev2011}
D.~N. Basov, R.~D. Averitt, D.~van~der Marel, M.~Dressel, and K.~Haule,
  ``Electrodynamics of correlated electron materials,'' {\em Rev. Mod. Phys.},
  vol.~83, pp.~471--541, Jun 2011.

\bibitem{Coleman2001}
P.~Coleman, C.~Pepin, Q.~Si, and R.~Ramazashvili, ``How do {F}ermi liquids get
  heavy and die?,'' {\em J. Phys.: Condens. Matter}, vol.~13, pp.~723--738,
  2001.

\bibitem{deng2014}
X.~Deng, A.~Sternbach, K.~Haule, D.~N. Basov, and G.~Kotliar, ``Shining {L}ight
  on {T}ransition-{M}etal {O}xides: {U}nveiling the {H}idden {F}ermi
  {L}iquid,'' {\em Phys. Rev. Lett.}, vol.~113, p.~246404, Dec 2014.

\bibitem{Shimozawa2016}
M.~Shimozawa, S.~K. Goh, T.~Shibauchi, and Y.~Matsuda, ``{{From Kondo Lattices
  to Kondo Superlattices}},'' {\em Rep. Prog. Phys.}, vol.~79, p.~074503, 2016.

\bibitem{Paglione2007}
J.~Paglione, T.~A. Sayles, P.-C. Ho, J.~R. Jeffries, and M.~B. Maple,
  ``Incoherent non-{F}ermi-liquid scattering in a {K}ondo lattice,'' {\em
  Nature Physics}, vol.~3, pp.~703--706, October 2007.

\bibitem{Ruckenstein1991}
A.~Ruckenstein and C.~Varma, ``A theory of marginal {F}ermi-liquids,'' {\em
  Physica C: Superconductivity}, vol.~185, pp.~134 -- 140, 1991.

\bibitem{Varma1993}
C.~Varma, ``Towards a theory of the marginal {F}ermi-liquid state,'' {\em
  Journal of Physics and Chemistry of Solids}, vol.~54, no.~10, pp.~1081 --
  1084, 1993.

\bibitem{WolflePers}
P.~W\"{o}lfle private communication (2016).

\bibitem{VdM2005}
F.~P. Mena, D.~van~der Marel, and J.~L. Sarrao, ``Optical conductivity of
  {C}e\emph{{M}}{I}n$_5$ (\emph{{M}}={C}o, {R}h, {I}r),'' {\em Phys. Rev. B},
  vol.~72, p.~045119, Jul 2005.

\bibitem{Landau56}
L.~Landau, ``Theory of a {F}ermi {L}iquid,'' {\em Zh. Eksp. Teor. Fiz.},
  vol.~30, p.~1058, 1956.

\bibitem{Berthod2013}
C.~Berthod, J.~Mravlje, X.~Deng, R.~\ifmmode~\check{Z}\else \v{Z}\fi{}itko,
  D.~van~der Marel, and A.~Georges, ``Non-{D}rude universal scaling laws for
  the optical response of local {F}ermi liquids,'' {\em Phys. Rev. B}, vol.~87,
  p.~115109, Mar 2013.

\bibitem{Mravlje2016}
J.~Mravlje private communication (2016).

\bibitem{Mravlje2011}
J.~Mravlje, M.~Aichhorn, T.~Miyake, K.~Haule, G.~Kotliar, and A.~Georges,
  ``Coherence-incoherence crossover and the mass-renormalization puzzles in
  {S}r$_2${R}u{O}$_3$,'' {\em Phys. Rev. Lett.}, vol.~106, p.~096401, Mar 2011.

\bibitem{Allen2015}
P.~B. Allen, ``Electron self-energy and generalized {D}rude formula for
  infrared conductivity of metals,'' {\em Phys. Rev. B}, vol.~92, p.~054305,
  Aug 2015.

\bibitem{Stricker2014}
D.~Stricker, J.~Mravlje, C.~Berthod, R.~Fittipaldi, A.~Vecchione, A.~Georges,
  and D.~van~der Marel, ``Optical {R}esponse of {S}r$_2${R}u{O}$_3$ {R}eveals
  {U}niversal {F}ermi-{L}iquid {S}caling and {Q}uasiparticles {B}eyond {L}andau
  {T}heory,'' {\em Phys. Rev. Lett.}, vol.~113, p.~087404, Aug 2014.

\bibitem{Jacko2009}
A.~C. Jacko, J.~O. Fj$\ae$restad, and B.~J. Powell, ``A unified explanation of
  the {Kadowaki}-{W}oods ratio on strongly correlated metals,'' {\em Nature
  Physics}, vol.~5, pp.~422--425, 2009.

\bibitem{Phillips2011}
P.~Phillips, ``Mottness collapse and \emph{{T}}-linear resistivity in cuprate
  superconductors,'' {\em Philosophical Transactions of the Royal Society of
  London A: Mathematical, Physical and Engineering Sciences}, vol.~369,
  no.~1941, pp.~1574--1598, 2011.

\bibitem{Auerbach86}
A.~Auerbach and K.~Levin, ``Kondo {B}osons and the {K}ondo {L}attice:
  {M}icroscopic {B}asis for the {H}eavy {F}ermi {L}iquid,'' {\em Phys. Rev.
  Lett.}, vol.~57, pp.~877--880, Aug 1986.

\bibitem{Prasanta2008}
M.~Prasanta, {\em Heavy-Fermion Systems}.
\newblock Elsevier, 2008.

\bibitem{Emery1995}
V.~J. Emery and S.~A. Kivelson, ``Superconductivity in bad metals,'' {\em Phys.
  Rev. Lett.}, vol.~74, pp.~3253--3256, Apr 1995.

\bibitem{Gunnarsson2003}
O.~Gunnarsson, M.~Calandra, and J.~E. Han, ``\textit{Colloquium} : Saturation
  of electrical resistivity,'' {\em Rev. Mod. Phys.}, vol.~75, pp.~1085--1099,
  Oct 2003.

\bibitem{Hussey2004}
N.~E. Hussey, K.~Takenaka, and H.~Takagi, ``Universality of the
  {M}ott-{I}offe-{R}egel limit in metals,'' {\em Philosophical Magazine},
  vol.~84, no.~27, pp.~2847--2864, 2004.

\bibitem{Deng2013}
X.~Deng, J.~Mravlje, R.~\ifmmode~\check{Z}\else \v{Z}\fi{}itko, M.~Ferrero,
  G.~Kotliar, and A.~Georges, ``How bad metals turn good: Spectroscopic
  signatures of resilient quasiparticles,'' {\em Phys. Rev. Lett.}, vol.~110,
  p.~086401, Feb 2013.

\bibitem{Xu2013}
W.~Xu, K.~Haule, and G.~Kotliar, ``Hidden {F}ermi {L}iquid, {S}cattering {R}ate
  {S}aturation, and {N}ernst {E}ffect: {A} {D}ynamical {M}ean-{F}ield {T}heory
  {P}erspective,'' {\em Phys. Rev. Lett.}, vol.~111, p.~036401, Jul 2013.

\bibitem{Anderson2008}
P.~W. Anderson, ``Hidden {F}ermi liquid: The secret of high-${T}_{c}$
  cuprates,'' {\em Phys. Rev. B}, vol.~78, p.~174505, Nov 2008.

\bibitem{Casey2011}
P.~A. Casey and P.~W. Anderson, ``Hidden fermi liquid: Self-consistent theory
  for the normal state of high-${T}_{c}$ superconductors,'' {\em Phys. Rev.
  Lett.}, vol.~106, p.~097002, Feb 2011.

\bibitem{Paglione2003}
J.~Paglione, M.~A. Tanatar, D.~G. Hawthorn, E.~Boaknin, R.~W. Hill, F.~Ronning,
  M.~Sutherland, L.~Taillefer, C.~Petrovic, and P.~C. Canfield,
  ``Field-{I}nduced {Q}uantum {C}ritical {P}oint in {C}e{C}o{I}n$_5$,'' {\em
  Phys. Rev. Lett.}, vol.~91, p.~246405, Dec 2003.

\bibitem{Pracht2017}
U.~S. Pracht, M.~Dressel, P.~MravlScheffler, M."{o}lfle, R.~Endo, T.~Watashige,
  Y.~Hanoka, M.~Shimozawa, T.~Terashima, T.~Shibauchi, Y.~Matsuda, and
  M.~Scheffler, ``{{Resilient heavy quasiparticles in CeCoIn$_5$: Indications
  from dynamical response for a hidden Fermi liquid}}.'' in preparation (2017).

\bibitem{Baym2004}
G.~Baym and C.~Pethick, {\em {{Landau Fermi-Liquid Theory: Concepts and
  Applications}}}.
\newblock WILEY-VCH Verlag, 2004.

\bibitem{Vollhardt2013}
D.~Vollhardt and P.~W\"{o}lfle, {\em {{The Superfluid Phases of Helium 3}}}.
\newblock Dover Publications, New York, 2013.

\bibitem{Chen2016}
Q.~Y. Chen, D.~F. Xu, X.~H. Niu, J.~Jiang, R.~Peng, H.~Xu, C.~H.~P. Wen, Z.~F.
  Ding, K.~Huang, L.~Shu, Y.~J. Zhang, H.~Lee, V.~N. Strocov, M.~Shi, F.~Bisti,
  T.~Schmitt, Y.~B. Huang, P.~Dudin, X.~C. Lai, S.~Kirchner, H.~Q. Yuan, and
  D.~L. Feng, ``Direct observation of how the heavy fermion state develops in
  {C}e{C}o{I}n$_5$.'' arXiv:1610.06724.

\bibitem{Coleman2015}
P.~Coleman, {\em Introduction to Many-Body Physics}.
\newblock Cambridge University Press, 1~ed., 2015.

\bibitem{Hall2001}
D.~Hall, E.~C. Palm, T.~P. Murphy, S.~W. Tozer, Z.~Fisk, U.~Alver, R.~G.
  Goodrich, J.~L. Sarrao, P.~G. Pagliuso, and T.~Ebihara, ``Fermi surface of
  the heavy-fermion superconductor {C}e{C}o{I}n$_5$: The de {H}aas-van {A}lphen
  effect in the normal state,'' {\em Phys. Rev. B}, vol.~64, p.~212508, Nov
  2001.

\bibitem{Capan2010}
C.~Capan, Y.-J. Jo, L.~Balicas, R.~G. Goodrich, J.~F. DiTusa, I.~Vekhter, T.~P.
  Murphy, A.~D. Bianchi, L.~D. Pham, J.~Y. Cho, J.~Y. Chan, D.~P. Young, and
  Z.~Fisk, ``Fermi surface evolution through a heavy-fermion
  superconductor-to-antiferromagnet transition: de {H}aas-van {A}lphen effect
  in {C}d-substituted {C}e{C}o{I}n$_5$,'' {\em Phys. Rev. B}, vol.~82,
  p.~035112, Jul 2010.

\bibitem{Shishido2004}
H.~Shishido, R.~Settai, S.~Hashimoto, Y.~Inada, and Y.~ÅŒnuki, ``De {H}aas
  van {A}lphen effect of {C}e{R}h{I}n$_5$ and {C}e{C}o{I}n$_5$ under
  pressure,'' {\em Journal of Magnetism and Magnetic Materials},
  vol.~272â€“276, Part 1, pp.~225 -- 226, 2004.
\newblock Proceedings of the International Conference on Magnetism (ICM 2003).

\bibitem{Wen2011}
J.~Xiao-Wen, L.~Yan, Y.~Li, H.~Jun-Feng, Z.~Lin, Z.~Wen-Tao, L.~Hai-Yun,
  L.~Guo-Dong, H.~Shao-Long, Z.~Jun, L.~Wei, W.~Yue, D.~Xiao-Li, S.~Li-Ling,
  W.~Gui-Ling, Z.~Yong, W.~Xiao-Yang, P.~Qin-Jun, W.~Zhi-Min, Z.~Shen-Jin,
  Y.~Feng, X.~Zu-Yan, C.~Chuang-Tian, and Z.~Xing-Jiang, ``{G}rowth,
  {C}haracterization and {F}ermi {S}urface of {H}eavy {F}ermion
  {C}e{C}o{I}n$_5$ {S}uperconductor,'' {\em Chinese Physics Letters}, vol.~28,
  no.~5, p.~057401, 2011.

\bibitem{Booth2011}
C.~H. Booth, T.~Durakiewicz, C.~Capan, D.~Hurt, A.~D. Bianchi, J.~J. Joyce, and
  Z.~Fisk, ``Electronic structure and $f$-orbital occupancy in {Y}b-substituted
  {C}e{C}o{I}n$_5$,'' {\em Phys. Rev. B}, vol.~83, p.~235117, Jun 2011.

\bibitem{Koitzsch2007}
A.~Koitzsch, S.~Borisenko, D.~Inosov, J.~Geck, V.~Zabolotnyy, H.~Shiozawa,
  M.~Knupfer, J.~Fink, B.~BÃ¼chner, E.~Bauer, J.~Sarrao, and R.~Follath,
  ``Observing the heavy fermions in {C}e{C}o{I}n$_5$ by angle-resolved
  photoemission,'' {\em Physica C: Superconductivity and its Applications},
  vol.~460â€“462, Part 1, pp.~666 -- 667, 2007.
\newblock Proceedings of the 8th International Conference on Materials and
  Mechanisms of Superconductivity and High Temperature SuperconductorsM2S-HTSC
  \{VIII\}.

\bibitem{Koitzsch2013}
A.~Koitzsch, T.~K. Kim, U.~Treske, M.~Knupfer, B.~B\"uchner, M.~Richter,
  I.~Opahle, R.~Follath, E.~D. Bauer, and J.~L. Sarrao, ``Band-dependent
  emergence of heavy quasiparticles in {C}e{C}o{I}n$_5$,'' {\em Phys. Rev. B},
  vol.~88, p.~035124, Jul 2013.

\bibitem{Kuchler2003}
R.~K\"uchler, N.~Oeschler, P.~Gegenwart, T.~Cichorek, K.~Neumaier, O.~Tegus,
  C.~Geibel, J.~A. Mydosh, F.~Steglich, L.~Zhu, and Q.~Si, ``Divergence of the
  {G}r\"uneisen {R}atio at {Q}uantum {C}ritical {P}oints in {H}eavy {F}ermion
  {M}etals,'' {\em Phys. Rev. Lett.}, vol.~91, p.~066405, Aug 2003.

\bibitem{Kuchler2006}
R.~K\"uchler, P.~Gegenwart, J.~Custers, O.~Stockert, N.~Caroca-Canales,
  C.~Geibel, J.~G. Sereni, and F.~Steglich, ``Quantum criticality in the cubic
  heavy-fermion system {C}e{I}n$_{3-x}${S}n$_x$,'' {\em Phys. Rev. Lett.},
  vol.~96, p.~256403, Jun 2006.

\bibitem{Hertz1976}
J.~A. Hertz, ``Quantum critical phenomena,'' {\em Phys. Rev. B}, vol.~14,
  pp.~1165--1184, Aug 1976.

\bibitem{Millis1993}
A.~J. Millis, ``Effect of a nonzero temperature on quantum critical points in
  itinerant fermion systems,'' {\em Phys. Rev. B}, vol.~48, pp.~7183--7196, Sep
  1993.

\bibitem{Si2001}
Q.~Si, S.~Rabello, K.~Ingersent, and J.~L. Smith, ``Locally critical quantum
  phase transitions in strongly correlated metals,'' {\em Nature}, vol.~413,
  pp.~804--808, 2001.

\bibitem{Zhu2003}
J.-X. Zhu, D.~R. Grempel, and Q.~Si, ``Continuous {Q}uantum {P}hase
  {T}ransition in a {K}ondo {L}attice {M}odel,'' {\em Phys. Rev. Lett.},
  vol.~91, p.~156404, Oct 2003.

\bibitem{Grempel2003}
D.~R. Grempel and Q.~Si, ``Locally {C}ritical {P}oint in an {A}nisotropic
  {K}ondo {L}attice,'' {\em Phys. Rev. Lett.}, vol.~91, p.~026401, Jul 2003.

\bibitem{Senthil2004}
T.~Senthil, M.~Vojta, and S.~Sachdev, ``Weak magnetism and non-{F}ermi liquids
  near heavy-fermion critical points,'' {\em Phys. Rev. B}, vol.~69, p.~035111,
  Jan 2004.

\bibitem{Schroeder2000}
A.~Schr\"{o}der, G.~Aeppli, R.~Coldea, M.~Adams, O.~Stockert, E.~L\"{o}hneysen,
  H. v.~Bucher, R.~Ramazashvili, and P.~Coleman, ``Onset of antiferromagnetism
  in heavy-fermion metals,'' {\em Nature}, vol.~407, pp.~351--355, September
  2000.

\bibitem{Limelette2013}
P.~Limelette, V.~Ta~Phuoc, F.~Gervais, and R.~Fr\'esard,
  ``$\ensuremath{\omega}/t$ scaling of the optical conductivity in strongly
  correlated layered cobalt oxide,'' {\em Phys. Rev. B}, vol.~87, p.~035102,
  Jan 2013.

\bibitem{VdM2003}
D.~v.~d. Marel, H.~J.~A. Molegraaf, J.~Zaanen, Z.~Nussinov, F.~Carbone,
  A.~Damascelli, H.~Eisaki, M.~Greven, P.~H. Kes, and M.~Li, ``Quantum critical
  behaviour in a high-\emph{{T}}$_c$ superconductor,'' {\em Nature}, vol.~425,
  pp.~0028--0836, September 2003.

\bibitem{Broun2016}
D.~M. Broun private communication (2016).

\bibitem{Tran2002}
P.~Tran, S.~Donovan, and G.~Gr\"uner, ``Charge excitation spectrum in
  {UP}t$_3$,'' {\em Phys. Rev. B}, vol.~65, p.~205102, Apr 2002.

\bibitem{Bosse2016}
G.~Boss\'e, L.~Pan, Y.~S. Li, L.~H. Greene, J.~Eckstein, and N.~P. Armitage,
  ``Anomalous frequency and temperature-dependent scattering and {H}und's
  coupling in the almost quantum critical heavy-fermion system
  {C}e{F}e$_2${G}e$_2$,'' {\em Phys. Rev. B}, vol.~93, p.~085104, Feb 2016.

\bibitem{Lerer2013PhD}
S.~Lerer, {\em Nernst effect in low dimensionality supercondctors}.
\newblock PhD thesis, Tel Aviv University, 2013.

\bibitem{shishido2010}
H.~Shishido, T.~Shibauchi, K.~Yasu, H.~Kontani, T.~Terashima, and Y.~Matsuda,
  ``Tuning the {D}imensionality of the {H}eavy {F}ermion {C}ompound
  {C}e{I}n$_3$,'' {\em Science}, vol.~327, p.~980, 2010.

\bibitem{mizukami2011}
Y.~Mizukami, H.~Shishido, T.~Shibauchi, M.~Shimozawa, S.~Yasumoto, D.~Watanabe,
  M.~Yamashita, H.~Ikeda, T.~Terashima, H.~Kontani, and Y.~Matsuda, ``Extremely
  strong-coupling superconductivity in artificial two-dimensional {K}ondo
  lattices,'' {\em Nat. Phys.}, vol.~7, p.~849, 2011.

\bibitem{shimozawa2012}
M.~Shimozawa, T.~Watashige, S.~Yasumoto, Y.~Mizukami, M.~Nakamura, H.~Shishido,
  S.~K. Goh, T.~Terashima, T.~Shibauchi, and Y.~Matsuda, ``Strong suppression
  of superconductivity by divalent ytterbium {K}ondo holes in
  {C}e{C}o{I}n$_5$,'' {\em Phys. Rev. B}, vol.~86, p.~144526, Oct 2012.

\bibitem{goh2012}
S.~K. Goh, Y.~Mizukami, H.~Shishido, D.~Watanabe, S.~Yasumoto, M.~Shimozawa,
  M.~Yamashita, T.~Terashima, Y.~Yanase, T.~Shibauchi, A.~I. Buzdin, and
  Y.~Matsuda, ``Anomalous {U}pper {C}ritical {F}ield in
  {C}e{C}o{I}n$_5$/{Y}b{C}o{I}n$_5$ {S}uperlattices with a {R}ashba-{T}ype
  {H}eavy {F}ermion {I}nterface,'' {\em Phys. Rev. Lett.}, vol.~109, p.~157006,
  Oct 2012.

\bibitem{shimozawa2014}
M.~Shimozawa, S.~K. Goh, R.~Endo, R.~Kobayashi, T.~Watashige, Y.~Mizukami,
  H.~Ikeda, H.~Shishido, Y.~Yanase, T.~Terashima, T.~Shibauchi, and Y.~Matsuda,
  ``Controllable {R}ashba {S}pin-{O}rbit {I}nteraction in {A}rtificially
  {E}ngineered {S}uperlattices {I}nvolving the {H}eavy-{F}ermion
  {S}uperconductor {C}e{C}o{I}n$_5$,'' {\em Phys. Rev. Lett.}, vol.~112,
  p.~156404, Apr 2014.

\bibitem{Weig2011}
T.~Weig, ``Electrodynamics of correlated cerium materials,'' Master's thesis,
  Universit\"{a}t Stuttgart, 2011.

\bibitem{Kozlov1998}
G.~V. Kozlov and A.~Volkov, {\em Millimeter and Submillimeter Wave Spectroscopy
  of Solids (Topics in Applied Physics)}, vol.~74.
\newblock Springer, Berlin, 1998.

\bibitem{Scheffler2009}
M.~Scheffler, J.~P. Ostertag, and M.~Dressel, ``{{Fabry-Perot resonances in
  birefringent YAlO$_3$ analyzed at terahertz frequencies}},'' {\em Opt.
  Lett.}, vol.~34, no.~22, pp.~3520--3522, 2009.

\end{thebibliography}
\thispagestyle{empty}
\chapterend

\newpage
\noindent\textbf{Publications and manuscripts related to this work}
\begin{enumerate}
\item{\sidenote{\footnotesize{See Chapter 4}}U. S. Pracht et al. \emph{Uncovering a Hidden Fermi Liquid in a Quantum-Critical Heavy-Fermion Metal}, in preparation (2017)}
\item{\sidenote{\footnotesize{See Chapter 3}}U. S. Pracht et al. \emph{Goldstone modes in granular aluminum}, in preparation (2017)}
\item{\sidenote{\footnotesize{See Chapter \ref{Sec:Al}}}U. S. Pracht et al. \textit{Shaping a superconducting dome: Enhanced Cooper pairing and suppressed phase coherence in coupled Al nano grains}, Physical Review B \textbf{93}, 100503 (2016)}
\item{\sidenote{\footnotesize{See Chapter 4}}U. S. Pracht et al. \textit{Charge carrier dynamics of the Heavy-Fermion Metal CeCoIn$_5$ probed by THz spectroscopy}, Journal of Magnetism and Magnetic Materials \textbf{400}, 31-35 (2016)}
\item{\sidenote{\footnotesize{See Chapter \ref{Sec:NbN}}}D. Sherman, U. S. Pracht, B. Gorshunov, S. Poran, J. Jesudasan, M. Chand, P. Raychaudhuri, M. Swanson, N. Trivedi, A. Auerbach, M. Scheffler, A. Frydman, and M. Dressel \textit{The Higgs mode in disordered superconductors close to a quantum phase transition}, Nature Physics \textbf{11}, 188-192 (2015)}
\item{\sidenote{\footnotesize{See Chapter \ref{Sec:Al}}}N. Bachar, U. S. Pracht, E. Farber, M. Dressel and M. Scheffler \textit{Signatures of unconventional Superconductivity in granular Aluminum}, Journal of Low Temperature Physics \textbf{179}, 83 (2014)}
\item{U. S. Pracht et al. \textit{Electrodynamics of the Superconducting State in Ultra-Thin Films at THz frequencies}, IEEE Transactions on THz Science and Technology \textbf{3}, 1-12 (2013)}
\end{enumerate}

\end{spacing}
\end{document}